\documentclass[superscriptaddress,groupedaddress,nofootnoteinbib,11pt,notoc]{article}
\pdfoutput=1
\usepackage{jcappub}

\usepackage{graphicx,color}
\usepackage[table]{xcolor}
\usepackage{appendix}
\usepackage{latexsym,amsmath,amssymb,graphicx,booktabs}
\usepackage{epsfig,latexsym}
\usepackage{hyperref}
 \usepackage{float}
\numberwithin{equation}{section}
\usepackage{url}
\usepackage{booktabs}

\definecolor{MyBlue}{rgb}{0.15,0.15,0.70}

\hypersetup{
colorlinks=true,
citecolor=MyBlue,
linkcolor=MyBlue,
urlcolor=MyBlue
}

\definecolor{lightgray}{gray}{0.9}

\newcommand{\dgw}{d_L^{\,\rm gw}}
\newcommand{\dem}{d_L^{\,\rm em}}
\newcommand{\dcom}{d_{\rm com}}
\newcommand{\dmax}{d_{\rm max}}
\newcommand{\hatO}{\hat{\Omega}}

\newcommand{\red}{} 

\newcommand{\nn}{\nonumber}

\renewcommand\({\left(}
\renewcommand\){\right)}
\renewcommand\[{\left[}
\renewcommand\]{\right]}

\newcommand{\ra}{\rightarrow}

\def\lsim{\raise 0.4ex\hbox{$<$}\kern -0.8em\lower 0.62
ex\hbox{$\sim$}}

\def\gsim{\raise 0.4ex\hbox{$>$}\kern -0.7em\lower 0.62
ex\hbox{$\sim$}}

\def\lbar{{\hbox{$\lambda$}\kern -0.7em\raise 0.6ex
\hbox{$-$}}}

\newcommand\eq[1]{eq.~(\ref{#1})}
\newcommand\eqs[2]{eqs.~(\ref{#1}) and (\ref{#2})}
\newcommand\Eq[1]{Equation~(\ref{#1})}

\newcommand\eqss[3]{eqs.~(\ref{#1}), (\ref{#2}) and (\ref{#3})}

\newcommand\eqst[2]{eqs.~(\ref{#1})--(\ref{#2})}

\newcommand\p{\partial}

\newcommand\ee{\end{equation}}
\newcommand\be{\begin{equation}}
\def\bea{\begin{array}}
\def\eea{\end{array}}\def\ea{\end{array}}
\newcommand\ees{\end{eqnarray}}
\newcommand\bees{\begin{eqnarray}}
\def\nn{\nonumber}

\def\eps{\epsilon}

\def\dslash{\hspace{-1mm}\not{\hbox{\kern-2pt $\partial$}}}
\def\Dslash{\not{\hbox{\kern-2pt $D$}}}
\def\pslash{\not{\hbox{\kern-2.1pt $p$}}}
\def\kslash{\not{\hbox{\kern-2.3pt $k$}}}
\def\qslash{\not{\hbox{\kern-2.3pt $q$}}}

\newcommand{\vk}{{\bf k}}
\newcommand{\vx}{{\bf x}}

\def\p1{{\bf p}_1}
\def\p2{{\bf p}_2}
\def\k1{{\bf k}_1}
\def\k2{{\bf k}_2}

\newcommand{\dddM}{\kern 0.2em \raise 1.9ex\hbox{$...$}\kern -1.0em \hbox{$M$}}
\newcommand{\dddQ}{\kern 0.2em \raise 1.9ex\hbox{$...$}\kern -1.0em \hbox{$Q$}}
\newcommand{\dddI}{\kern 0.2em \raise 1.9ex\hbox{$...$}\kern -1.0em\hbox{$I$}}
\newcommand{\dddJ}{\kern 0.2em \raise 1.9ex\hbox{$...$}\kern-1.0em
\hbox{$J$}}
\newcommand{\dddcalJ}{\kern 0.2em \raise 1.9ex\hbox{$...$}\kern-1.0em
\hbox{${\cal J}$}}

\newcommand{\dddO}{\kern 0.2em \raise 1.9ex\hbox{$...$}\kern -1.0em
\hbox{${\cal O}$}}
\def\dddz{\raise 1.5ex\hbox{$...$}\kern -0.8em \hbox{$z$}}
\def\dddd{\raise 1.8ex\hbox{$...$}\kern -0.8em \hbox{$d$}}
\def\dddbd{\raise 1.8ex\hbox{$...$}\kern -0.8em \hbox{${\bf d}$}}
\def\ddbd{\raise 1.8ex\hbox{$..$}\kern -0.8em \hbox{${\bf d}$}}
\def\dddx{\raise 1.6ex\hbox{$...$}\kern -0.8em \hbox{$x$}}

\newcommand{\msun}{M_{\odot}}

\newcommand{\ode}{\Omega_{\rm DE}}
\newcommand{\oma}{\Omega_{M}}
\newcommand{\ora}{\Omega_{R}}

\newcommand{\ola}{\Omega_{\Lambda}}

\newcommand{\rde}{\rho_{\rm DE}}
\newcommand{\wde}{w_{\rm DE}}

\graphicspath{ {./Graphs/} }

\title{Cosmology with LIGO/Virgo dark sirens:
Hubble parameter and modified gravitational wave propagation}
\author{Andreas Finke,}
\author{Stefano Foffa,}
\author{Francesco Iacovelli,}
\author{Michele Maggiore}
\author{and Michele Mancarella}

\affiliation{D\'epartement de Physique Th\'eorique and Center for Astroparticle Physics,\\
Universit\'e de Gen\`eve, 24 quai Ansermet, CH--1211 Gen\`eve 4, Switzerland}

\abstract{We present a detailed study of the methodology for correlating  
`dark sirens' (compact binaries coalescences without electromagnetic counterpart) with galaxy catalogs. We propose several  improvements on the current state of the art, and we apply them to the  GWTC-2 catalog of  LIGO/Virgo gravitational wave (GW)  detections, and the GLADE galaxy catalog, performing a detailed study of several sources of systematic errors that, with the expected increase in statistics,  will eventually become the dominant limitation. We provide a measurement of $H_0$ from dark sirens alone, finding as the best result $H_0=\red{67.3^{+27.6}_{-17.9}}\,\,{\rm km}\, {\rm s}^{-1}\, {\rm Mpc}^{-1}$ ($68\%$ c.l.) which is, currently, the most stringent constraint  obtained using only dark sirens.
Combining dark sirens with the counterpart for GW170817 we find  $H_0=
\red{72.2^{+13.9}_{-7.5}} \,{\rm km}\, {\rm s}^{-1}\, {\rm Mpc}^{-1}$.

We also study modified GW propagation,  which is a smoking gun of dark energy and modifications of gravity at cosmological scales, and we show that current observations of dark sirens already start to provide interesting limits.  From dark sirens alone, our best result for the parameter $\Xi_0$ that measures deviations from GR (with $\Xi_0=1$ in GR) is $\Xi_0=\red{2.1^{+3.2}_{-1.2}}$.
We finally discuss limits on modified GW  propagation under the tentative identification of the flare  ZTF19abanrhr as the electromagnetic counterpart of the binary black hole coalescence GW190521, in which case our most stringent result is $\Xi_0=\red{1.8^{+0.9}_{-0.6}}$. 

\vspace{1mm}
We release the publicly available code $\tt{DarkSirensStat}$, which is available under open source license at   \url{https://github.com/CosmoStatGW/DarkSirensStat}.}

\emailAdd{andreas.finke@unige.ch}
\emailAdd{stefano.foffa@unige.ch}
\emailAdd{francesco.iacovelli@unige.ch}
\emailAdd{michele.maggiore@unige.ch}
\emailAdd{michele.mancarella@unige.ch}

\begin{document}
\maketitle
\flushbottom

\section{Introduction}

In the last few years, gravitational wave (GW) astronomy and cosmology have become a reality.  The detection of the first  binary black hole (BBH) coalescence, GW150914~\cite{Abbott:2016blz}, was a historic moment; another milestone was the observation   of the first binary neutron star (BNS)  coalescence, GW170817, and of its electromagnetic counterpart~\cite{TheLIGOScientific:2017qsa,Goldstein:2017mmi,Savchenko:2017ffs,Monitor:2017mdv}. Since then,  many additional detections have taken place, to the extent that, during the recent O3 LIGO/Virgo run, BBH coalescences have been detected at a rate of about $1.5$ per week. The recent release of the results from the first part of the O3 run (O3a)  reports 39  candidate detections (of which, statistically, $\sim 3$ can be false alarms), including BBHs up to redshift $z\sim 0.8$~\cite{Abbott:2020niy}. 

For applications to cosmology, the crucial feature of compact binary coalescences is that, from their GW signal, one can reconstruct the luminosity distance to the source~\cite{Schutz:1986gp}, and for this reason they are referred to as `standard sirens'.  
Much work has been devoted to investigating the cosmological information that could be obtained from such  measurements, either when the redshift of the source is provided by the observation of  an  electromagnetic counterpart, or using statistical methods, see e.g.~\cite{Holz:2005df,Dalal:2006qt,MacLeod:2007jd,Nissanke:2009kt,Cutler:2009qv,DelPozzo:2011yh,Taylor:2012db,Chen:2017rfc,Feeney:2018mkj,Gray:2019ksv}.

In a flat $\Lambda$CDM model, the expression for the luminosity distance as a function of redshift is given by
\be\label{dLemLCDM}
d_L(z)=\frac{c}{H_0}\, (1+z) \, \int_0^z\, 
\frac{d\tilde{z}}{\sqrt{\oma (1+\tilde{z})^3+\ora (1+\tilde{z})^4+\ola }}\, ,
\ee
where $\oma$ and $\ora$  are the present matter  and radiation density fractions, respectively, and  $\ola$ is the energy density fraction associated to the cosmological constant, with $\oma+\ora+\ola=1$ in the flat case. For completeness we have  included the contribution from radiation; however, it is completely negligible at the redshifts relevant for standard sirens (we also neglect curvature, which can be trivially included). In the limit $z\ll  1$ we recover the Hubble law $d_{L}(z)\simeq (c/H_0)z$, so from a measurement at such redshifts we can extract $H_0$.

The first measurement of $H_0$ from a standard siren has been possible thanks to the first observed BNS coalescence, 
GW170817. Using the information on the redshift coming from its electromagnetic counterpart gives 
$H_0=74^{+16}_{-8}\,  {\rm km}\, {\rm s}^{-1}\, {\rm Mpc}^{-1}$ (median and symmetric $68.3\%$ credible interval), or 
 $H_0=70^{+12}_{-8}\,\, {\rm km}\, {\rm s}^{-1}\, {\rm Mpc}^{-1}$ (maximum a posteriori $68.3\%$ interval)~\cite{Abbott:2017xzu}. This was the first proof-of-principle that $H_0$ can be extracted from standard sirens. However, the error from this single detection is still too large to discriminate between the value of $H_0$ obtained from  late-Universe probes~\cite{Riess:2019cxk,Wong:2019kwg}, and that  inferred from early-Universe probes assuming $\Lambda$CDM~\cite{Aghanim:2018eyx,Abbott:2018xao}, which are currently  in disagreement at  $5.3\sigma$ level. One can estimate  that ${\cal O}(50-100)$ standard sirens with counterpart  are needed to reach the accuracy required  to arbitrate this discrepancy~\cite{Chen:2017rfc,Feeney:2018mkj}.

In the absence of a counterpart one can resort to  statistical methods, in particular correlating the GW signal with galaxy catalogs, as already proposed in the pioneering work~\cite{Schutz:1986gp}, and reformulated in a modern Bayesian framework in a number of more recent papers~\cite{DelPozzo:2011yh,Chen:2017rfc,Feeney:2018mkj,Gray:2019ksv} (see also \cite{Nair:2018ign,Mukherjee:2019wcg,Yu:2020vyy,Vijaykumar:2020pzn,Mukherjee:2020hyn,Bera:2020jhx,Mukherjee:2020mha} for recent related approaches exploiting galaxy clustering, and 
\cite{Borhanian:2020vyr} for possible improvements  from the detection of higher harmonics). In this context, compact  binary coalescences  without electromagnetic counterpart are often called `dark sirens'.
The first application of the statistical method to actual data has been performed  in  \cite{Fishbach:2018gjp}, where a measurement of  $H_0$ has been obtained from  GW170817 without making use of the known electromagnetic counterpart, leading to a value $H_0=77^{+37}_{-18}\,\, {\rm km}\, {\rm s}^{-1}\, {\rm Mpc}^{-1}$. While the error is obviously larger than that obtained by making use of the counterpart, still this provides a  proof-of-principle of the statistical method. A measurement of $H_0$ from a BBH dark siren was presented 
in \cite{Soares-Santos:2019irc}, correlating the event GW170814 with the galaxy catalog from the Dark Energy Survey (DES), obtaining $H_0=75^{+40}_{-32}\,\, {\rm km}\, {\rm s}^{-1}\, {\rm Mpc}^{-1}$ (for a  uniform prior in the range range $[20, 140] \,\, {\rm km}\, {\rm s}^{-1}\, {\rm Mpc}^{-1}$). In~\cite{Palmese:2020aof} a similar analysis has been performed for  the dark siren GW190814, which resulted from the coalescence of a $23\msun$ BH with a compact object with mass $2.6\msun$, leading to $H_0=66^{+55}_{-18}\,\, {\rm km}\, {\rm s}^{-1}\, {\rm Mpc}^{-1}$ when using  the dark siren GW190814 only.
In \cite{Abbott:2019yzh}
the LIGO/Virgo collaboration (LVC) has obtained the value   $H_0=\red{69^{+16}_{-8}}\,\, {\rm km}\, {\rm s}^{-1}\, {\rm Mpc}^{-1}$ by combining the  detections from the O1 and O2 runs, using GW170817 as a standard siren with counterpart, and the other events as dark sirens (with most of the improvement, with respect to using only GW170817, coming from GW170814, a well localized event that falls in the region of the sky covered by the DES catalog, which has a high level of completeness).
While a significantly larger number of dark sirens will be needed in order to eventually obtain stringent constraints on $H_0$, these works provide the first concrete applications of the statistical method to the determination of $H_0$.

In the present paper we further develop the formalism of the statistical treatment of `dark sirens', proposing improvements of various technical aspects. In particular, we include a direction-dependent notion of completeness of a galaxy catalog, we propose a novel method for dealing with incomplete catalogs, and we discuss in detail different approximations involved in the computation of the selection bias,  including in its computation both the correct prior given by the galaxy catalog and 
additional selection effects related to the choice of the relevant GW events. We show that these improvements outperform state of the art methods. We will also extend our analysis to the determination of the parameters that characterize modified GW propagation, that, as we will discuss below, is the potentially most promising signal of deviations from  General Relativity (GR) that can be accessed by standard sirens. We will finally apply the formalism to the most recent public data, including the O3a LIGO/Virgo run, both for the measurement of $H_0$ in the framework of $\Lambda$CDM, and for limits on modified GW propagation in the framework of modified gravity.

The paper is organized as follows. In sect.~\ref{sect:dLgw} we will review (following mostly \cite{Belgacem:2017ihm,Belgacem:2018lbp}) how, in modified gravity, the quantity measured from the coalescence of standard sirens is not the standard luminosity distance but a different quantity, that we called the `GW luminosity distance', and which is sensitive to modifications of the propagation equation of tensor modes over a cosmological background. Such modifications are always present in modified gravity theories and can give a potentially very large effect (reaching even $80\%$ at large redshifts, in some models that are phenomenologically viable). This makes modified GW propagation   a potentially very interesting observable for GW detectors, already in their current second-generation (2G) stage. In sect.~\ref{sect:Meth} we  present the formalism
that we will use for dark sirens, based on a hierarchical Bayesian framework, and we will discuss in great detail a number of technical aspects (in particular related to methods for dealing with catalog incompleteness and with crucial normalization factors), proposing various improvements. We will then discuss our results in sect.~\ref{sect:results}, applying this formalism to the most recent LIGO/Virgo data. Sect.~\ref{sect:Concl} contains our conclusions. Some further technical material is collected in various appendices.

\section{GW luminosity distance in modified gravity}\label{sect:dLgw}

\subsection{Modified GW propagation}

Even if, to date, the study of the LIGO/Virgo standard sirens has been mostly focused on obtaining from them a measurement of $H_0$, eventually  the most interesting result would be the observation of  effects that can be directly traced to dark energy (DE) and modifications of GR on cosmological scales. As usual, on cosmological scales it is convenient to perform a separation between the
homogeneous Friedmann-Robertson-Walker (FRW) background, and  scalar, vector and tensor perturbations over it. 
In a generic modified gravity theory, both the  background evolution and the perturbations differ from that in $\Lambda$CDM. At the background level, the effect of deviations from  $\Lambda$CDM induced by a dynamical dark energy is encoded into the DE equation of state $\wde(z)$, defined by $p_{\rm DE}(z)=\wde(z)\rde(z)$, where
$p_{\rm DE}(z)$ and $\rde(z)$ are the DE pressure and energy density, respectively. $\Lambda$CDM is recovered for $\wde(z)=-1$.
In a theory with a generic DE equation of state $\wde(z)$, the DE density is given as a function of redshift by 
\be\label{4rdewdeproofs}
\rde(z)  =\rho_0 \Omega_{\rm DE}\, e^{ 3\int_{0}^z\, \frac{d\tilde{z}}{1+\tilde{z}}\, [1+\wde(\tilde{z})]}\, ,
\ee
where $\Omega_{\rm DE}=\rde(0)/\rho_0$ is the DE density fraction and $\rho_0=3H_0^2/(8\pi G)$ is the critical density. The corresponding expression for the luminosity distance is
\be\label{dLemmod}
d_L(z)=\frac{c}{H_0}\, (1+z) \,\int_0^z\, 
\frac{d\tilde{z}}{\sqrt{\oma (1+\tilde{z})^3+\ora (1+\tilde{z})^4+ \rde(\tilde{z})/\rho_0}}\, .
\ee
Again, in the limit $z\ll  1$ this reduces to Hubble's law $d_{L}(z)\simeq (c/H_0)z$. However, 
at higher redshifts we get access to the possibility of measuring the DE equation of state. Any deviation from the $\Lambda$CDM value $\wde(z)=-1$ would  provide evidence for a dynamical dark energy.

On top of this effect, related to the background evolution, in modified gravity also the perturbations will be different. Scalar perturbations determine the evolution of large-scale structures or the propagation of photons in a perturbed Universe, and the study of
deviations from 
$\Lambda$CDM in the scalar sector is  among the targets of  the next generation of galaxy surveys. Vector perturbations only have decaying modes and are usually irrelevant, both in GR and in modified gravity theories. Tensor perturbations,  for modes well inside the horizon, are just GWs traveling over the FRW background and, in particular, the equation for tensor  perturbations  determines how GWs  propagate across cosmological distances. In GR, the free propagation of tensor perturbations over FRW is governed by the equation  
\be\label{4eqtensorsect}
\tilde{h}''_A+2{\cal H}\tilde{h}'_A+c^2k^2\tilde{h}_A=0\, ,
\ee
where $\tilde{h}_A(\eta, \vk)$ is the Fourier-tranformed GW amplitude.\footnote{We use standard notation:  $A=+,\times$ labels the two polarizations, the prime denotes the derivative with respect to cosmic time $\eta$, defined by $d\eta=dt/a(t)$, $a(\eta)$ is the FRW scale factor, and 
${\cal H}=a'/a$.} In  modified gravity both the `friction term' $2{\cal H}\tilde{h}'_A$ and the term $k^2\tilde{h}_A$ in the above equation can in principle be modified. A change in  the coefficient of the $k^2\tilde{h}_A$ term induces a speed of GWs, $c_{\rm gw}$, different from that of light. After the observation of GW170817, this is  now excluded  at a level  $|c_{\rm gw}-c|/c< {\cal O}(10^{-15})$ \cite{Monitor:2017mdv} (unless one invokes scale-dependent modifications, as in \cite{deRham:2018red}), and, in fact, this observation has ruled out a large class of modified gravity models~\cite{Creminelli:2017sry,Sakstein:2017xjx,Ezquiaga:2017ekz,Boran:2017rdn,Baker:2017hug}. However, the modified gravity models that pass this constraint still, in general, induce a change in the `friction term', so the propagation equation for tensor modes becomes~\cite{Saltas:2014dha,Lombriser:2015sxa,Nishizawa:2017nef,Arai:2017hxj,Belgacem:2017ihm,Amendola:2017ovw,Belgacem:2018lbp,Belgacem:2019pkk}, 
\be\label{prophmodgrav}
\tilde{h}''_A  +2 {\cal H}[1-\delta(\eta)] \tilde{h}'_A+c^2k^2\tilde{h}_A=0\, ,
\ee
for some function of time $\delta(\eta)$ that encodes the modification from GR.\footnote{More generally, one could have a scale-dependent function $\delta_{\vk}(\eta)$. Concrete models, however, usually predict a scale-independent function $\delta(\eta)$, basically because, in the absence of an explicit length scale in the cosmological model,  for dimensional reasons the dependence on $k$ enters only  through  the ratio $\lambda_k/H^{-1}_0$ (where $\lambda_k=2\pi/k$). It can then be  expanded in powers of $\lambda_k/H^{-1}_0$, and for GW modes well inside the horizon, as those of interest for ground based as well as space interferometers, we can stop  to the zeroth-order term. The same happens for the functions, usually denoted by  $\mu_{\vk}(z)$ and $\Sigma_{\vk}(z)$, that are commonly used to parametrize deviation in the scalar perturbation sector.}$^{,}$\footnote{Note that, in the context of Horndeski theories, instead of the function that we called $\delta(\eta)$, is used a function $\alpha_M(\eta)$, with $\alpha_M(\eta)=-2\delta(\eta)$.} In particular, the comprehensive study in \cite{Belgacem:2019pkk} shows that the phenomenon of modified GW propagation, encoded in the function $\delta(\eta)$, is completely generic  and, in fact, takes place in all  modified gravity models that have been investigated.

In GR, using \eq{4eqtensorsect}, one finds that the GW amplitude decreases over cosmological distances as the inverse of the FRW scale factor. As a consequence, one can show that the amplitude of the GWs emitted by a compact binary, after propagation from the source to the  observer, is proportional to the inverse of the luminosity distance to the source, with  coefficients that depend on the inclination angle of the orbit (see e.g.  Section 4.1.4 of \cite{Maggiore:1900zz} for derivation). This is at the basis of the fact that compact binaries are standard sirens, i.e. that the luminosity distance of the source can be extracted from their signal. However, when the GW propagation is rather governed by \eq{prophmodgrav}, this result is modified. It can then be shown that the quantity extracted from GW observations is no longer the standard luminosity distance $d_L(z)$ of the source [that, in this context, we will denote by $\dem(z)$, since this is the quantity that would be measured, for instance, using the  electromagnetic signal from a counterpart]. Rather, the quantity extracted from GW observation is a  `GW luminosity distance'  \cite{Belgacem:2017ihm} $\dgw(z)$, related to $\dem(z)$ by~\cite{Belgacem:2017ihm,Belgacem:2018lbp}
\be\label{dLgwdLem}
\dgw(z)=\dem(z)\exp\left\{-\int_0^z \,\frac{dz'}{1+z'}\,\delta(z')\right\}\, ,
\ee
where the function $\delta$ that appears in \eq{prophmodgrav} has now been written as a function of redshift. 

In summary, when interpreted in the framework of  modified gravity,  the observation of GWs from compact binaries  provides a measurement of 
$\dgw(z)$, whose expression in terms of the cosmological parameters and of the functions $\delta(z)$ and $\wde(z)$ is obtained from  
\eq{dLgwdLem}, with $\dem(z)$ given by \eqs{dLemmod}{4rdewdeproofs}.  Given a specific modified gravity  model and the values of its cosmological parameters, this measurement can then be translated into a prediction for the redshift $z$ of the source. This prediction could then be  compared with the value of $z$ obtained from an electromagnetic counterpart, if available, or can be used in a statistical approach, by comparing it with the redshift of the galaxies in a catalog (or, more precisely, using a galaxy catalog to define a prior on the redshift of the potential host, as we will discuss in details below).  In modified gravity, the prediction for the redshift of the source will differ from the 
$\Lambda$CDM prediction obtained from \eq{dLemLCDM}, for three distinct reasons:

\begin{enumerate} 
\item  First, the value of the cosmological parameters $H_0$, $\oma$ in a modified gravity theory, obtained by comparing the theory to standard cosmological datasets such as the cosmic microwave background (CMB), supernovae (SNe), baryon acoustic oscillations (BAO), structure formation, etc., will in general differ from the corresponding predictions obtained in $\Lambda$CDM. This is due to the fact that  both the background evolution and the scalar perturbations of the two theories are in general different.

\item  The modification of the background evolution, encoded in the DE equation of state $\wde(z)$ or, equivalently, in a non-constant DE density $\rde(z)$, affects  the (`electromagnetic') luminosity distance (\ref{dLemmod}). 

\item  On top of it, the modification of the  tensor perturbation sector leaves a further imprint in $\dgw(z)$, expressed by the function $\delta(z)$ in \eq{dLgwdLem}.

\end{enumerate}

As we will discuss in sect.~\ref{sect:persp},  in modified gravity theories the modification of the tensor perturbation sector, point (3) above, can give the dominant effect,  both because the size of the effect can be much larger, and because the effects in points (1) and (2) tend to compensate each other. Before entering into this, it is  useful to discuss a simple parametrization of  modified GW propagation. 

\subsection{The $(\Xi_0,n)$ parametrization}

In general, it is difficult to extract from the data a full function of redshift, such as $\wde(z)$ or $\delta(z)$, and parametrizations in terms of a small number of parameters are necessary. For the DE equation of state a  standard choice is the $(w_0,w_a)$ parametrization \cite{Chevallier:2000qy,Linder:2002et}, which in terms of the scale factor reads $w_{\rm DE}(a)=w_0+(1-a)w_a$, or, in terms of redshift,
\be\label{w0wa}
w_{\rm DE}(z)= w_0+\frac{z}{1+z} w_a\, .
\ee 
For modified GW propagation, a convenient parametrization, in terms of two parameters $(\Xi_0,n)$, has been proposed in \cite{Belgacem:2018lbp}. Rather than parametrizing $\delta(z)$, it is simpler to parametrize directly the ratio $\dgw(z)/\dem(z)$ (which is also the directly observed quantity), in the form
\be\label{eq:fit}
\frac{d_L^{\,\rm gw}(z)}{d_L^{\,\rm em}(z)}=\Xi_0 +\frac{1-\Xi_0}{(1+z)^n}\, .
\ee
This parametrization reproduces  the fact that, as $z\ra 0$, $d_L^{\,\rm gw}/d_L^{\,\rm em}$ must go to one, since,  as the distance to the source goes to zero, there can be no effect from modified propagation. 
In the opposite limit of large redshifts, in contrast, \eq{eq:fit} predicts that $d_L^{\,\rm gw}/d_L^{\,\rm em}$ approaches a constant value $\Xi_0$. This is motivated by the fact that,
in  typical  DE models, the deviations from GR only appear in the recent cosmological epoch, so $\delta(z)$ goes to zero at large redshift, and therefore the integral in \eq{dLgwdLem}
saturates to a constant. The parametrization (\ref{eq:fit}), that in terms of scale factor reads, even more simply, $\dgw(a)/\dem(a)=\Xi_0+ (1-\Xi_0)a^n$,
 interpolates between these two limiting behaviors, with a power-law determined by $n$.

This simple parametrization turns out to work remarkably well for most modified gravity model. It was first proposed in \cite{Belgacem:2018lbp}, in the context of a non-local modification of gravity that we will further discuss in sect~\ref{sect:RT}, where it perfectly reproduced the exact prediction of the model. As shown in 
ref.~\cite{Belgacem:2019pkk}, the same happens for most of the other best-studied modified gravity models, including   several examples of
Horndeski and DHOST theories (with the only exception of bigravity, where $\dgw(z)/\dem(z)$ displays non-trivial  oscillations due to the interaction between the two metrics). In fact, when used over a broad range of redshifts, it does a much better job at reproducing the behavior of $\dgw(z)/\dem(z)$, compared to
how the parametrization (\ref{w0wa}) works for $\wde(z)$. This is due to the fact that \eq{w0wa}
is nothing but a Taylor expansion of $w_{\rm DE}(a)$ around $a=1$, truncated to  linear order. Rather than expanding around $a=1$, it can be improved by expanding around a `pivot' point $a_*$, or the corresponding pivot redshift $z_*$,  where  a given experiment is more sensitive. In any case, in the comparison with the actual prediction of a specific model, it will typically go astray far from the pivot point, as we can expect for a Taylor expansion truncated to linear order. In contrast, \eq{eq:fit} catches the correct limiting behaviors at both small and large $z$, and, if in between the ratio $\dgw(z)/\dem(z)$  is smooth, it typically works very well for all redshifts.

From \eq{eq:fit} we can obtain a corresponding parametrization for $\delta(z)$. Observing that \eq{dLgwdLem} can be inverted as
\be
\delta(z)=-(1+z)\frac{d}{dz}\log\[ \frac{d_L^{\,\rm gw}(z)}{d_L^{\,\rm em}(z)} \]
\, ,
\ee
and using \eq{eq:fit} on the right-hand side, we get~\cite{Belgacem:2018lbp}
\be\label{paramdeltaz}
\delta(z)=\frac{n  (1-\Xi_0)}{1-\Xi_0+ \Xi_0 (1+z)^n}
\, .
\ee
In the literature, for phenomenological studies, the function $\delta(z)$ has sometimes been approximated by a constant (see e.g. \cite{Nishizawa:2017nef}). This is probably not well justified physically, since in typical explicit modified gravity models, $\delta(z)$ rather goes to zero for large redshifts. In any case, a constant $\delta(z)=\delta_0$ is a special case of the parametrization (\ref{eq:fit}), with $\Xi_0=0$ and $n=\delta_0$, as we see from \eq{paramdeltaz}, and leads to $\dgw(z)/\dem(z)=1/(1+z)^n$. 

Just as, for the DE equation of state, there is a hierarchy in the importance of the parameters, with $w_0$ being more significant than $w_a$, similarly for modified GW propagation the crucial parameter is $\Xi_0$, which fixes the asymptotic value of $\dgw(z)/\dem(z)$ at large redshift, while $n$ only determines the precise shape of the function that interpolates between $\dgw(z)/\dem(z)=1$ at $z=0$ and $\dgw(z)/\dem(z)\simeq \Xi_0$ at large $z$. In the following, we will therefore focus on $\Xi_0$, fixing $n$ to some typical value of some particularly interesting model, see below.

\subsection{Perspectives for modified GW propagation at 2G detectors}\label{sect:persp}

As we have recalled above, at the level of background evolution the signature of a dynamical DE is given by a non-trivial DE equation of state. However, 
forecasts for the measurement of the DE equation of state at 2G detectors~\cite{Belgacem:2019tbw} indicate that, even for a LIGO/Virgo/KAGRA network at target sensitivity, and  even combining the GW observations with other cosmological datasets, the accuracy on $w_0$ will not   really improve significantly, compared to  the  $(4-5) \%$  accuracy that is already  obtained from current CMB+BAO+SNe data.
\red{Another way to exploit BBHs as dark sirens is to make use of the mass scale imprinted on the mass distribution of BHs  by the pulsational pair instability supernova (PISN) process~\cite{Farr:2019twy}. For 2G detectors, after five years of operations, this could provide a  measurement of $H_0$ at the $2.9\%$ level or, assuming an independent measurement of  $H_0$ at the $1\%$ level, to a measurement of $w_0$ at the $12\%$ level~\cite{Farr:2019twy}.}
 To obtain more significant constraints on $(w_0,w_a)$ from standard sirens, one must wait for
third-generation (3G) ground-based detectors such as the Einstein Telescope (ET)~\cite{Punturo:2010zz,Maggiore:2019uih} and Cosmic Explorer~\cite{Reitze:2019iox}, but even in this case the improvement will be minor. The most recent forecasts for standard sirens with counterpart (subject, however, to significant  uncertainties on the network of electromagnetic observatories that will operate at the time of ET, and to the telescope time that they will devote to the follow up of GW signals)
indicate that, from  ET data only, one  could reach an accuracy of about $10\%$ on $w_0$, i.e. $\Delta w_0\simeq 0.1$, to be compared with the accuracy of about $(4-5)\%$ already obtained from  CMB+BAO+SNe observation; combining ET data with (current) CMB+BAO+SNe data one could shrink the error to about $2\%$~\cite{Belgacem:2019tbw} (see also  \cite{Sathyaprakash:2009xt,Zhao:2010sz,Belgacem:2018lbp} for earlier work). Even if useful, this does not look like a spectacular improvement on the current knowledge.
For  the LISA space interferometer~\cite{Audley:2017drz}, using as standard sirens the coalescence of supermassive BH binaries (which are expected to have an electromagnetic counterpart) and state-of-the-art models for the coalescence rate depending on the seed BHs,
one finds  again that the improvement on $w_0$ is  marginal compared to current CMB+BAO+SNe observations~\cite{Belgacem:2019pkk}.

The situation is quite different for modified GW propagation, to the extent that significant results on $\Xi_0$ could be obtained even  at 2G detectors~\cite{Belgacem:2018lbp,Belgacem:2019tbw,Belgacem:2019lwx,Belgacem:2020pdz}.
There are three main reasons for this:

\begin{enumerate}

\item  By itself, the accuracy that can be reached on $\Xi_0$ from standard sirens is in general better than the corresponding accuracy on $w_0$. This can be traced to the fact that the effect of $w_0$ on the electromagnetic luminosity distance (\ref{dLemmod}) is masked  by partial degeneracies with $H_0$ and $\oma$  in the electromagnetic luminosity distance (see the discussion in~\cite{Belgacem:2018lbp}). 
As a result, a variation of $\Xi_0$ by, say, $5\%$ with respect to the GR values $\Xi_0=1$, has a  more significant effect on the luminosity distance than a $5\%$ variation of $w_0$ with respect to the $\Lambda$CDM value $w_0=-1$. This  has been  confirmed by the explicit Markov Chain Monte Carlo (MCMC) analysis performed in \cite{Belgacem:2018lbp} for ET and in \cite{Belgacem:2019pkk} for LISA, that show that the accuracy that can be obtained on $\Xi_0$ is significantly better than the accuracy that can be obtained on $w_0$.

\item From electromagnetic cosmological observations, such as CMB, BAO, SNe and structure formation, we  already know that deviations from $\Lambda$CDM  cannot exceed a few percent, both in the background evolution and in the scalar perturbation sector   (at least, in the regime of linear perturbations). Precise numbers depend on the details of the model considered but, for instance, in a simple $w$CDM model [i.e.,  in a phenomenological  extension of $\Lambda$CDM where $\wde(z)$ is taken to be a constant $w_0$ which is allowed to be different from $-1$, and scalar perturbations are taken to be the same as in $\Lambda$CDM], combining  CMB, BAO, SNe and DES data, one finds that $w_0$ cannot differ from $-1$ by more than $5\%$~\cite{Abbott:2018xao}. 
Therefore, if with standard sirens we cannot improve on this accuracy, it is hard to expect surprises (even if a measurement with completely different systematics, such as that obtained with GWs, would anyhow be valuable).
In contrast, the sector of tensor perturbations is an uncharted territory, that we are beginning to explore thanks to GW observations, and any information on it would be worthwhile, even without the need of reaching percent level accuracy.\label{ref:degen}

\item  One could  have expected that, if a modified gravity model complies with existing observational bounds, and therefore does not differ from $\Lambda$CDM by more than a few percent in the background evolution and in the scalar perturbations, then even in the tensor perturbation sector it should display deviations of, at most, the same order. However, the study of explicit models shows that this is not necessarily the case, and the deviations in the tensor sector can be much larger. A particularly striking example is given  by the so-called RT non-local gravity model, where $\Xi_0$ can be as large as $1.80$
\cite{Belgacem:2019lwx,Belgacem:2020pdz}, corresponding to a $80\%$ deviation from the GR value $\Xi_0=1$, despite the fact that, in the background and in the scalar perturbations, the model is very close to 
$\Lambda$CDM, and indeed it
fits CMB, BAO, SNe and structure formation data  at a level statistically equivalent to $\Lambda$CDM.

\end{enumerate}

The RT model will be briefly discussed in sect.~\ref{sect:RT}, and its predictions will be our reference example in this paper, although most of our analysis will be completely general. 
First, let us give an orientative discussion of the effects of modified GW propagation  that can be expected at the redshifts accessible to 2G detectors.

The first observational bound on modified GW propagation was  obtained in \cite{Belgacem:2018lbp} using GW170817 and its electromagnetic counterpart.  Since GW170817 is at a very small redshift, $z\simeq 0.01$, we can expand \eq{dLgwdLem} as
\be\label{eq:fitlowz}
\frac{d_L^{\,\rm gw}(z)}{d_L^{\,\rm em}(z)}=1-z \delta(0)+{\cal O}(z^2)\, ,
\ee
so at very low redshift we are actually sensitive to $\delta(0)\equiv \delta(z=0)$. In  the $(\Xi_0,n)$ parametrization  $\delta(0)=n(1-\Xi_0)$, but in fact the analysis for sources at such a low redshift can be carried out independently of any parametrization of the function $\delta(z)$ or of $\dgw(z)/\dem(z)$.

For GW170817,  we can compare
the GW luminosity distance obtained from the GW observation, $d^{\,\rm gw}_L=43.8^{+2.9}_{-6.9}\,\, {\rm Mpc}$ (at one sigma, i.e.  68.3\% c.l.), to the electromagnetic luminosity distance of the galaxy hosting the counterpart. The latter has been  obtained in \cite{Cantiello:2018ffy}  from surface brightness fluctuations, which gives
$d^{\,\rm em}_L=40.7 \pm 1.4 \, {\rm (stat)} \pm 1.9\,  {\rm (sys)} \, \, {\rm Mpc}$. Note that, using the value of $\dem$ obtained directly from surface brightness fluctuations,  we do not even need to assume a cosmological model and a value of $H_0$, contrary to what happens if  we reconstruct the electromagnetic luminosity distance from the redshift of the source.

Interpreting the difference from one of the resulting ratio $\dgw(z)/\dem(z)$ as an effect of modified GW propagation, we get a measurement of $\delta(0)$~\cite{Belgacem:2018lbp}, 
\be\label{limitdelta0}
\delta(0)=-7.8^{+9.7}_{-18.4}\, ,
\ee
which is of course consistent with the GR prediction $\delta(z)=0$. 
Setting for illustration $n=1.91$ (which is the value predicted by the RT model in the same limit in which $\Xi_0=1.80$, see sect.~\ref{sect:RT} below) and using $\delta(0)=n(1-\Xi_0)$, this can be translated into 
$\Xi_0=5.1^{+9.1}_{-5.1}$, and therefore into an upper bound
\be
\Xi_0<14.2\, ,
\ee
at 68.3\% c.l. This  is not a stringent bound, as a consequence of the  fact that GW170817 is at very small redshift, and the effect of modified GW propagation vanishes as $z\ra 0$. Similar limits on modified GW propagation from GW170817 have also been later found in \cite{Lagos:2019kds}. The authors use a parametrization of modified GW propagation of the form $\alpha_M(z)=c_M\ode(z)/\ode(0)$ (where $\alpha_M(z)=-2\delta(z)$ in our notation), which is often  used in the context of Horndeski theories.\footnote{In should be observed that this parametrization is just a simple ansatz, which was  first introduced  in \cite{Bellini:2014fua} to study the scalar sector of Horndeski theories (and, in Horndeski theories, the same function turns out to enter also in the modification of the tensor sector, in the form of a running Planck mass). However, as shown in \cite{Belgacem:2019pkk}, even for Horndeski theories the effect of modified GW propagation is actually very well reproduced by the parametrization (\ref{eq:fit}).}   The result found in ref.~\cite{Lagos:2019kds}, combining the GW luminosity distance with the redshift of the galaxy and using a {\em Planck} prior on $H_0$, is  $c_M=-9^{+21}_{-28}$, corresponding to a constraint somewhat broader, but  comparable to \eq{limitdelta0}.   Another application  has been developed in \cite{Pardo:2018ipy}, which studied the implication of GW170817 on a parametrization of modified GW propagation inspired by the idea that GWs could leak into extra dimensions.\footnote{It should also be observed that 
the propagation of electromagnetic and gravitational waves is affected, in the same way, by the presence of inhomogeneities in the Universe. The GW luminosity distance obtained from standard sirens automatically includes this effect. If the
electromagnetic luminosity distance is inferred from the redshift of the source, assuming propagation through a homogeneous FRW Universe without correcting for the effect of inhomogeneities along the actual propagation, this would introduce a bias, partially degenerate with modified GW propagation. On linear scales, detailed computations of Doppler, lensing and ISW effects \cite{Bertacca:2017vod} show that inhomogeneities induces a relative error $\Delta d_L/d_L$ on the luminosity distance at a level below $1\%$ for all redshifts $z<5$ (see also Fig.~12 of \cite{Maggiore:2019uih}). See~\cite{Kalomenopoulos:2020klp} for a modeling, at an effective level, of the effect of non-linear scales.}

\begin{figure}[t]
\centering
\includegraphics[width=0.65\textwidth]{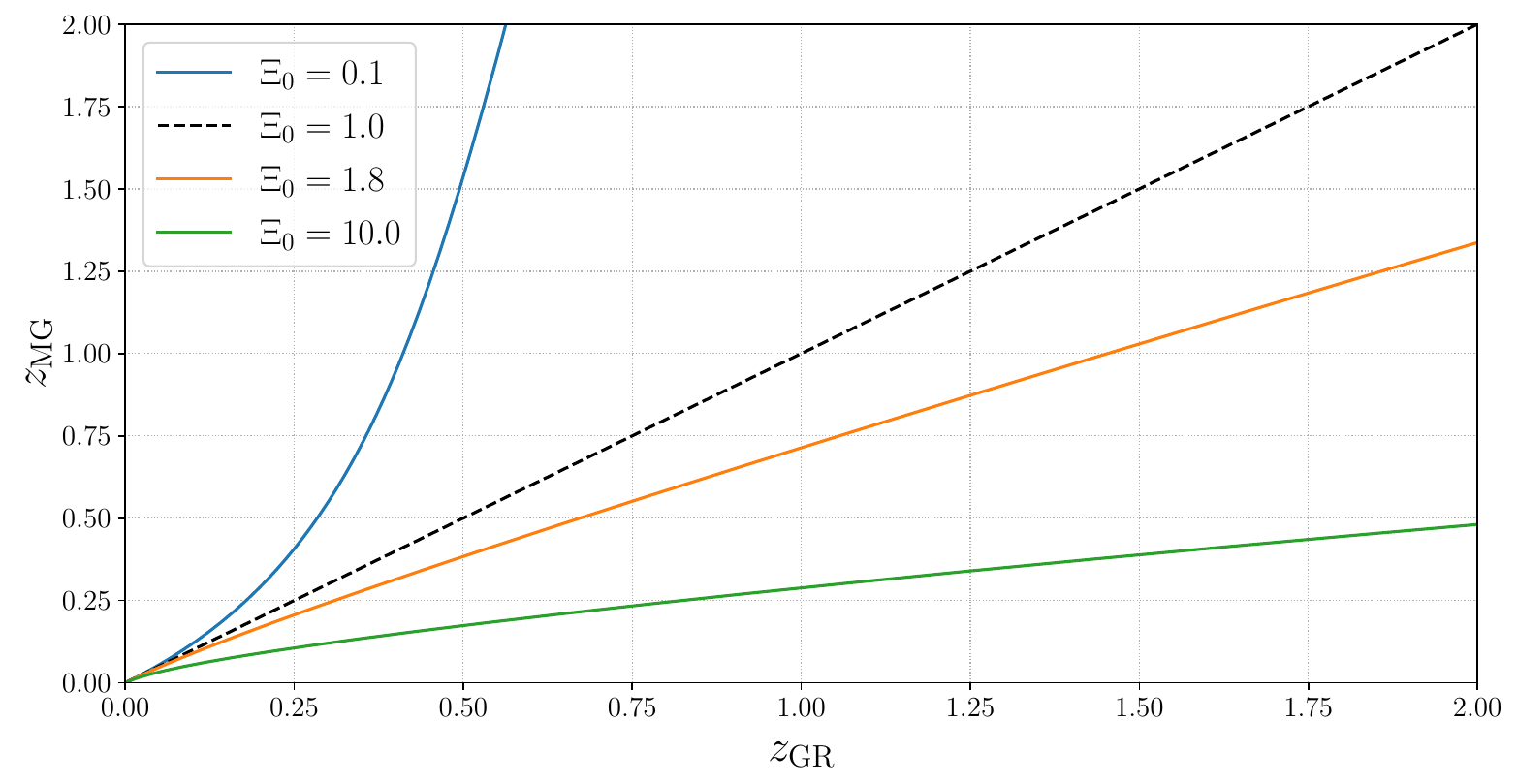}
\caption{
The  redshift $z_{\rm MG}$ of a source inferred from a measurement  of $\dgw$ in a modified  gravity  model, for different values of $\Xi_0$,   as a function of the value $z_{\rm GR}$ that would be inferred in GR
(we  set for definiteness $n=1.91$).}
\label{fig:zMG_vs_zGR}
\end{figure}

The very small redshift of GW170817, $z\simeq 0.01$, is the main limiting factor of these studies.
The situation, however, can change significantly for  standard sirens at higher redshifts, although still in the range accessible to 2G detectors. This can be appreciated from Fig.~\ref{fig:zMG_vs_zGR}. In this plot
 $z_{\rm MG}$ is defined as  the  redshift  of the source that would be inferred from a measurement of $\dgw$ obtained by GW observations, using a modified  gravity  model with a given value of $\Xi_0$. We show it as a function of   $z_{\rm GR}$, which is the redshift that would be inferred using GR (and $\Lambda$CDM). To produce this plot we have assumed that the corresponding electromagnetic luminosity distances are the same, in order to single out the effect of $\Xi_0$ [for definiteness, we set $n=1.91$, as above, but the results vary little with $n$, for $n={\cal O}(1)$]. In any case, as we discussed above, for viable models the electromagnetic luminosity distance cannot differ from that of $\Lambda$CDM by more than a few percent, which is very small
with respect to the effect induced by the values of $\Xi_0$ shown in the plot. 
From Fig.~\ref{fig:zMG_vs_zGR} (that, for illustration, includes also very high and very low values of $\Xi_0$, such as $\Xi_0=0.1$ or $\Xi_0=10$, well beyond even those obtained in the RT model or in any other known viable model, but which can be considered at the level of a first phenomenological analysis) we see that, if the actual theory describing Nature is a modified gravity model with $\Xi_0>1$, the actual redshift of the source would be smaller than that inferred from the GW data using $\Lambda$CDM and, vice versa, for $\Xi_0<1$ it would be larger. Even without using extreme values such as $\Xi_0=10$ or $\Xi_0=0.1$, the effect can be quite significant, as long as the sources are at sufficiently large redshift.

As an example, consider the event GW190814, resulting from the coalescence of  $23 \msun$ BH with a compact object with a mass $2.6\msun$~\cite{Abbott:2020khf}. Its measured (GW) luminosity distance is
$\dgw=241^{+41}_{-45}$~Mpc. In $\Lambda$CDM, 
using $H_0=67.9$ and $\oma=0.30$, this gives
\be
z=0.052^{+0.009}_{-0.010}\, .
\ee
Given that this redshift is still quite small, in this case the effect  of modified GW propagation will not be large. Using again $H_0=67.9$ and $\oma=0.30$, but setting now   $\Xi_0=1.80$, we  get 
\be
z=0.049^{+0.008}_{-0.008}\, ,
\ee
which is not distinguishable from  the GR prediction, within the error. Consider, however, a source at a larger distance, say $\dgw=2920$~Mpc. In $\Lambda$CDM, with the above values of $H_0$ and $\oma$, this corresponds to $z=0.50$. In contrast, in a model with  the same values of $H_0$ and $\oma$ but $\Xi_0=1.80$ (as the RT model with a large  value of a parameter $\Delta N$, discussed in sect.~\ref{sect:RT} below),  the prediction for  the redshift of the source would rather be $z=0.38$, a rather significant difference; similarly, for a source at $\dgw=6812$~Mpc, for which GR would predict $z=1.0$, the same modified gravity model would predict $z=0.71$. \red{Of course, increasing the distance to the source, the average error on the measurement of the luminosity distance also increases, in a way that depends on the GW detector network under consideration, and this must be taken into account in the above considerations. In any case, in particular for the sources with the best distance reconstruction,
looking for the electromagnetic counterpart by assuming a priori the validity of $\Lambda$CDM could  lead to missing a possible electromagnetic counterpart.}\footnote{For instance, the criterion on the redshift of the counterpart  used in \cite{Morgan:2020wvf} for
GW190814 was $|z_{\rm GR}-z_{\rm em}|<3\sigma$, where $z_{\rm GR}$ is the redshift inferred from the GW observation assuming GR and $\Lambda$CDM, $z_{\rm em}$ the observed redshift of the electromagnetic counterpart, and $\sigma$ combines in quadrature the error on the redshift from the GW and electromagnetic observations.}
Given that the most important outcome of these observations would be a possible falsification of $\Lambda$CDM and of GR, it is important to extend the search for a counterpart to a range of redshifts not limited to that predicted by  $\Lambda$CDM.

The  predictions for the redshift of the source, which  can  discriminate between  GR and modified gravity theories, can be tested either using standard sirens with counterpart, or using dark sirens. On the counterpart side, we have seen above that GW170817 is too close to give stringent results. For the BBH GW190521 a tentative electromagnetic counterpart, ZTF19abanrhr, has been reported in \cite{Graham:2020gwr}, although,  using $\Lambda$CDM as the underlying cosmological model, the probability of a real association based on volume localization is not statistically sufficient for a confident identification of ZTF19abanrhr as the counterpart~\cite{Ashton:2020kyr}. Furthermore, uncertainties exist about the physical mechanism providing a flare from the coalescence of the BBH, believed to have taken place in a AGN disk. However, this  event is the furthest detection reported by LIGO/Virgo in the O3a run, at $\dgw\sim (4-5)$~Gpc, and it is therefore very interesting from the point of view of modified GW propagation. We will examine it in 
sect.~\ref{sect:GW190521} (see also the related analysis in \cite{Mastrogiovanni:2020mvm}),  where  we will also see that modified GW propagation could remove a significant objection to the interpretation of ZTF19abanrhr as a counterpart, namely the limited spatial localization overlap of the GW and electromagnetic signals. Most of our results, however, will be focused on the measurement that can be obtained with current dark sirens observations, both for $H_0$ and for $\Xi_0$.

\subsection{An explicit example: the RT nonlocal model}\label{sect:RT}

Our analysis of modified GW propagation, in this paper, will be of purely phenomenological nature: we will  assume a modified gravity model with a  prediction  for the `electromagnetic' luminosity distance very close to that of  $\Lambda$CDM (say, at a few percent or better, so that, at the level of the present analysis, we will neglect this difference) and, on top of it, a much larger effect due to modified GW propagation, parametrized in terms of $(\Xi_0,n)$ with a value of $\Xi_0$ significantly different from 1. To support the potential interest of this phenomenological scenario, it is useful to have in mind an explicit model that realizes it, even if, in the end, our analysis will be more general. 

An explicit realization of this scenario is provided by 
the `RT' nonlocal gravity  model. This model was originally proposed in \cite{Maggiore:2013mea}, and has been much explored by our group in recent years. A complete and updated review of its conceptual and phenomenological aspects  can be found in \cite{Belgacem:2020pdz} (which updates the review
\cite{Maggiore:2016gpx}), whom we refer the reader for more details.  In short,
the underlying idea is that, because of infrared effects in spacetimes of cosmological interest, the conformal mode of the metric becomes massive, and this is described by the generation of a suitable non-local term in the quantum effective action.\footnote{Recall, from standard quantum field theory, that, while the fundamental action of a theory must be local, the corresponding quantum effective action develops unavoidably nonlocal terms whenever the theory contains massless particles, such as the graviton in GR. See 
\cite{Maggiore:2016gpx} for a pedagogical discussion in this context.}
Various studies of this model and of variants of the same underlying physical idea~\cite{Foffa:2013vma,Kehagias:2014sda,Maggiore:2014sia,Nesseris:2014mea,Dirian:2014ara,Barreira:2014kra,Dirian:2014bma,Dirian:2016puz,Maggiore:2016gpx,Dirian:2017pwp,Belgacem:2017cqo,Belgacem:2018wtb}
have shown that the model passes all phenomenological constraints: it generates dynamically a  dark energy density that drives  accelerated expansion in the recent epoch; it has stable cosmological perturbations in the scalar and  tensor sectors (a nontrivial condition that ruled out many modified gravity models even before reaching the stage of comparison with data);   the model fits CMB, BAO, SNe and structure formation  at a level statistically equivalent to $\Lambda$CDM; and it reproduces the successes of GR at solar system scale, including  bounds on the time variation of Newton's constant, and laboratory scales (without the need of a screening mechanism since, contrary to all other modifications of GR, it does not introduce extra degrees of freedom).  Studies of  possible variations of the idea have singled out the RT model as the only known nonlocal model that passes all these  tests.\footnote{The RT model has also  been selected as one of the priority models for further studies and development of 
dedicated pipelines by the Dark Energy Science Collaboration (DESC) of the Vera Rubin Observatory (formerly LSST)~\cite{Ishak:2019aay}.}

The fact that the RT model passes all phenomenological tests is partly related to the fact that 
it differs from $\Lambda$CDM by less that $0.5\%$ at the level of background evolution and by a few percent to below percent level, depending on the wavenumber of the Fourier modes, for the scalar perturbations~(see e.g. Fig.~2 and Figs.~5-14 of \cite{Belgacem:2020pdz}). In the tensor sector, one finds that GWs propagate at the speed of light, therefore complying with the bound from GW170817, and the model displays the phenomenon  of modified GW propagation. Quite surprisingly, despite the fact that the model is so close to $\Lambda$CDM at the background level and in the scalar perturbations, the deviations from GR in the tensor sector, as parametrized by $\Xi_0$, can be very large~\cite{Belgacem:2019lwx,Belgacem:2020pdz}. More precisely, the model has a free parameter $\Delta N$, related to the choice of initial conditions (defined by starting the evolution during a phase of primordial inflation, $\Delta N$ e-folds before the end of inflation) and, for large $\Delta N$, the prediction for $\Xi_0$  of the RT model saturates to a value $\Xi_0\simeq 1.80$,  corresponding to a $80\%$ deviation from the GR value $\Xi_0=1$. In contrast, if the model is started with initial conditions of order one during radiation dominance, one rather finds $\Xi_0\simeq 0.93$, a $7\%$ deviation from GR.\footnote{In more detail, in the RT model with large values of the parameter $\Delta N$, fitting the cosmological parameters to {\em Planck 2018} CMB data, the Pantheon SNe dataset and a selection of BAO measurements (and including  also the sum of neutrino masses among the free cosmological parameters, as appropriate when dealing with modified gravity~\cite{Dirian:2017pwp}) one finds, for $H_0$ and $\oma$, the mean values $H_0=67.88\pm 0.48$ and $\oma=0.3076\pm 0.0060$; these are practically indistinguishable from the values $H_0=67.89\pm 0.47$ and $\oma=0.3085\pm 0.0060$ obtained in  $\Lambda$CDM, using the same observational datasets and letting again free the sum of neutrino masses~\cite{Belgacem:2020pdz}. The DE equation of state $\wde(z)$ is also very close to $-1$ at all $z$, with $w_0=-1.06$, and as a result the relative difference in $H(z)$ between the RT model with large $\Delta N$ and $\Lambda$CDM is below $0.5\%$ at all redshifts (see Table~2 and Fig.~2 in \cite{Belgacem:2020pdz}). At the same time, the RT model with large $\Delta N$ predicts values of $\Xi_0$ that deviates strongly from $\Xi_0=1$ (e.g. $\Xi_0\simeq 1.64$ for a number of e-fold $\Delta N=64$, that saturates to $\Xi_0\simeq 1.80$ increasing $\Delta N$ further, corresponding to deviations at the $60\%-80\%$ level), so the effect from modified GW propagation dominates completely.\label{foot:H0RT}}

Apart from its specific virtues as a viable dynamical dark energy model,  the RT  model therefore provides an explicit example of a phenomenologically viable modification of GR that produces very large deviations from GR in the sector of tensor perturbations.
This shows how the study of GWs on cosmological scales is a genuinely new window, accessible only to GW detectors, that has the potential of providing significant surprises.

\section{Methodology for correlation between dark sirens and galaxy catalogs}\label{sect:Meth}

We now move to the study of standard sirens without electromagnetic counterpart (`dark sirens'), using a statistical method based on the correlation with  a galaxy catalog. In this section, we will give a detailed discussion of the methodology used, and we suggest various  improvements with respect to the current state of the art.

\subsection{Hierarchical Bayesian framework}\label{sect:HBf}

We use a hierarchical Bayesian statistical framework, elaborating on the formalism discussed in 
refs.~\cite{Mandel:2018mve,Chen:2017rfc}. This formalism has been recently used for extracting $H_0$ from various LIGO/Virgo detections discussed in the Introduction  (in \cite{Fishbach:2018gjp}
for GW170817 without using the known electromagnetic counterpart, 
in \cite{Soares-Santos:2019irc} for GW170814  and in~\cite{Palmese:2020aof} for  GW190814). A related  formalism is discussed in \cite{Gray:2019ksv} and has been used  by the LIGO/Virgo collaboration for extracting a value of  $H_0$ from the detections of the O1 and O2 observation runs~\cite{Abbott:2019yzh},  and
in \cite{Abbott:2020khf,Vasylyev:2020hgb} for GW190814. See also \cite{Loredo:2004nn,DelPozzo:2011yh,Adams:2012qw} for earlier work on applications of the hierarchical Bayesian method to astrophysics, and \cite{Thrane:2018qnx,Vitale:2020aaz} for reviews.

We consider an ensemble of binary coalescences, each characterized by a set of parameters $\theta$ that determine the GW signal. Among them, we separate the position of the source from the other parameters. The position can be described  by its luminosity distance $d_L$ and by its direction, identified by a unit vector $\hatO$, so  we write
\be\label{vectheta}
\theta=\{d_L,\hatO,\theta'\}\, .
\ee
Equivalently, given a cosmology, we can use 
the redshift $z$ instead of the luminosity distance $d_L$, so $\theta=\{z,\hatO,\theta'\}$. Note also that, in the context of modified gravity, the quantity directly measured by the GW signal is the GW luminosity distance $\dgw$, rather than the electromagnetic luminosity distance $\dem\equiv d_L$. In any case, the actual position of the source is given by the electromagnetic luminosity distance $d_L$.
The set $\theta'$ contains all other source parameters that affect the GW signal. Among them, the most important are the chirp mass and the total mass of the system, and the inclination of the orbit, but at higher and higher level of accuracy other parameters will enter, such as the spins of the compact bodies, tidal deformability in the case of a NS, eccentricity of the orbit, etc. 

We collectively denote by $\lambda$ the set of parameters on which we want to draw an inference from the set of measurements of binary coalescences. These could be properties of the underlying astrophysical population, such as their rate of coalescence or their mass distribution, as well as parameters of the underlying cosmological model, such as $H_0$ or, in the context of modified gravity, $\Xi_0$. 
As in ref.~\cite{Mandel:2018mve}, we separate 
$\lambda$ into an overall normalization for the event rate $N$ (that will be treated separately, see sect.~\ref{sect:rate}) and the remaining parameters, denoted by $\lambda'$, so $\lambda=\{N,\lambda'\}$. The distribution 
$dN/d\theta$
of the   events (in our case, the GW events due to binary coalescences) is written as
\be\label{dNdtheta}
\frac{dN}{d\theta} (\lambda)=Np_0(\theta|\lambda')\, .
\ee
When we study the Hubble parameter in the context of $\Lambda$CDM  \red{we fix all parameters related to the underlying astrophysical population  to typical values suggested by astrophysical models (we will then explore the effect of different choices)} and we keep all other cosmological parameters fixed to the mean values obtained from CMB+BAO+SNe, so $\lambda'=\{H_0\}$. This is the strategy that has been used also  in \cite{Fishbach:2018gjp,Soares-Santos:2019irc}. In principle, we should also  allow  $\oma$, which is the other relevant cosmological parameter that enters in \eq{dLemLCDM}, to vary within a prior fixed by CMB+BAO+SNe data. However, the latter fix $\oma$ to an accuracy of order $2\%$,  which, as we shall see, is much better than what we will obtain for $H_0$ using only standard sirens. Then, to the level of accuracy that can be currently reached,   it is  sensible  to keep $\oma$ fixed.\footnote{\red{Note that CMB+BAO+SNe data also fix $H_0$ to a sub-percent precision. However, the tension between early and late Universe measurements of $H_0$ makes significant an independent measurement of $H_0$ from standard sirens, while a measurement of $\oma$ from standard sirens is less compelling. Furthermore, an estimate of $\oma$ from standard sirens would also be much  less precise, given that, at low redshift, the dependence of $\oma$ disappears in the Hubble law, so the precision on $\oma$ from standard sirens alone is about comparable to that obtained for $w_0$. Conversely, this means that the precise value at which we fix $\oma$ has little impact on our bounds on $H_0$ and $\Xi_0$.}}$^{,}$\footnote{Of course,  $\ora$ is fixed to much higher precision by the CMB temperature measurement and, given its smallness, is totally irrelevant at the redshift of standard sirens; given $\oma$ and $\ora$,  within flat $\Lambda$CDM, $\ola$ is then fixed by the flatness condition.} \red{More generally, one could perform a global fit inferring both the cosmology and the population properties. This is particularly useful when some feature of the population property, such as the narrowness of the neutron star mass distribution~\cite{Taylor:2011fs}, or the mass scale that can be imprinted on the BH mass distribution by the  pulsational pair-instability supernova mechanism~\cite{Farr:2019twy}, can break the degeneracy between source frame masses and redshifts, thereby providing   independent statistical information of the redshift (see also \cite{Mastrogiovanni:2021wsd} for a recent discussion of the interplay between population property and cosmological parameters inference). In this paper we will limit ourselves to inference of the cosmological parameters, for fixed choices of astrophysical population properties, and we will examine how the results depends on the assumed population properties.}

Similarly, when
we study  modified GW propagation, we  keep $\Xi_0$ as the only free parameter and we fix all other cosmological parameters, including $H_0$, to the mean values obtained from CMB+BAO+SNe, so  
$\lambda'=\{\Xi_0\}$.  Once again, rather than fixing them, in principle one should  vary all other cosmological parameters within a prior fixed by CMB+BAO+SNe data. However, given that current CMB+BAO+SNe data fix $H_0$ to an accuracy of about 
$0.7\%$~\cite{Aghanim:2018eyx}, while the bounds that we will find on $\Xi_0$ from dark sirens will turn out to be about two orders of magnitude  less constraining, with current GW data keeping $H_0$ fixed to the mean value obtained from CMB+BAO+SNe is again an adequate approximation. As the amount and quality of GW data will improve, this will no longer be appropriate. Actually,  at the level of 3G detectors, when are expected millions of BBH detections per year~\cite{Maggiore:2019uih}, it would not even be appropriate to  separate 
electromagnetic from GW data, using only the former to determine the prior on $H_0$ and the other cosmological parameters, and then using GW data to study $\Xi_0$, with the other parameters allowed to vary within these priors. Rather, the best approach will be  to run full MCMCs on the whole electromagnetic+GW datasets, as done in \cite{Belgacem:2018lbp} with simulated ET data 
and in \cite{Belgacem:2019pkk} with simulated LISA data. Eventually, especially in view of the $H_0$ tension between early- and late-Universe probes, one would also like to have  a joint measurement of $(H_0,\Xi_0)$ from standard sirens only, without using electromagnetic data at all. However, to achieve this with sufficient accuracy  will be challenging even for a 3G detector such as ET, see Fig.~13 of \cite{Belgacem:2018lbp}.
At present, given the limited constraining power on the cosmological parameters of current GW data, the approach that we take in this work, of fixing all parameters except $H_0$ when we study $H_0$ within $\Lambda$CDM, and all parameters (including $H_0$) except $\Xi_0$ when we study modified GW propagation, is adequate.

The distribution (\ref{dNdtheta}) is sampled with  a set $\{{\cal D}_{\rm GW}\}$ of $N_{\rm obs}$ observations of binary coalescences, each one providing GW data
${\cal D}_{\rm GW}^i$, $i=1, \ldots, N_{\rm obs}$, and we assume that there is 
no electromagnetic counterpart, i.e. no corresponding electromagnetic data. As in  \cite{Mandel:2018mve}, we begin by focusing on the possibility of extracting information on the parameters $\lambda'$ (we will later specify our notation to $\lambda'=\{H_0\}$ or to $\lambda'=\{\Xi_0\}$, depending on the context.
In this section, however, we still use the more general notation  $\lambda'$). The 
posterior for $\lambda'$, given the GW data, is obtained from
\be\label{posteriorXi0}
p(\lambda'|\{{\cal D}_{\rm GW}\})=
\frac{p(\{{\cal D}_{\rm GW}\}|\lambda')\, \pi(\lambda')}{p(\{{\cal D}_{\rm GW}\})}\, ,
\ee
where $p(\{{\cal D}_{\rm GW}\}|\lambda')$ is the likelihood
of the data given $\lambda'$, and $\pi(\lambda')$ is the prior on $\lambda'$. Both for $\lambda'=\{H_0\}$ and for $\lambda'=\{\Xi_0\}$ we will use a flat prior  within broad limits, see below. The probability of the data
$p(\{{\cal D}_{\rm GW}\})$ (the `evidence') ensures the normalization
$\int d\lambda'\, p(\lambda'|\{{\cal D}_{\rm GW}\})=1$, so
\be
p(\{{\cal D}_{\rm GW}\}) =\int d\lambda'\, p(\{{\cal D}_{\rm GW}\}|\lambda')\, \pi(\lambda')\, ,
\ee
and is therefore independent of 
$\lambda'$.
For a set $\{{\cal D}_{\rm GW}\}$ of $N_{\rm obs}$ independent detections (neglecting for the moment the information on the overall rate, see sect.~\ref{sect:rate}), the total likelihood is given by
\be\label{pDl}
p(\{{\cal D}_{\rm GW}\}|\lambda')=\prod_{i=1}^{N_{\rm obs}} p({\cal D}^i_{\rm GW}|\lambda')\, .
\ee
The likelihood $p({\cal D}^i_{\rm GW}|\lambda')$ of the $i$-th GW event can be written as~\cite{Mandel:2018mve}
\be\label{pDi}
p({\cal D}^i_{\rm GW}|\lambda')=\frac{1}{\beta(\lambda')}
\int d\theta\, p({\cal D}^i_{\rm GW}|\theta,\lambda')\, p_0(\theta|\lambda')\, ,
\ee
where $ p({\cal D}^i_{\rm GW}|\theta,\lambda')$ is the likelihood of the data ${\cal D}^i_{\rm GW}$ given the value $\theta$ of the parameters, and $p_0(\theta|\lambda')$, defined by \eq{dNdtheta}, is the prior on the parameters $\theta$. Both are in principle  conditioned on $\lambda'$, although this depends on the choice of the variables $\theta$, as we will see below.
The qualification `hierarchical' Bayesian  inference refers to the fact that, through the data, we do not have access directly to $\lambda'$ but rather to the posterior of the parameters $\theta$ given the data, and therefore, through Bayes' theorem, to the
likelihood $p({\cal D}^i_{\rm GW}|\theta,\lambda')$. The desired likelihood with respect to the parameters $\lambda'$ is then obtained using \eq{pDi} [and the corresponding posterior from \eq{posteriorXi0}].  In this context the parameters $\lambda'$ (or, more generally, $\lambda$, including the overall rate) are often called hyper-parameters, and $p(\{{\cal D}_{\rm GW}\}|\lambda)$ is  called the hyper-likelihood.

In our problem, when $\theta=\{d_L,\hatO\}$, the prior $p_0(\theta|\lambda')$ will be provided by the galaxy catalog, with the precise connection discussed in sect.~\ref{sect:cat}. We will then discuss how to include the prior on the remaining parameters $\theta'$.

The function $\beta(\lambda')$ is a normalization factor, that ensures that the integral of $p({\cal D}^i_{\rm GW}|\lambda')$ over all data above the detection threshold is equal to one,
so~\cite{Mandel:2018mve}
\be\label{defalpha}
\beta(\lambda')=\int_{{\cal D}_{\rm GW}>{\rm threshold}} d{\cal D}_{\rm GW}\,
\int d\theta\, p({\cal D}_{\rm GW}|\theta,\lambda')\, p_0(\theta|\lambda')\, .
\ee
We can rewrite this expression as 
\be\label{alpha}
\beta(\lambda')=\int d\theta\, P_{\rm det}(\theta,\lambda')\, p_0(\theta|\lambda')\, ,
\ee
where 
\be\label{Pdet}
P_{\rm det}(\theta,\lambda')\equiv \int_{{\cal D}_{\rm GW}>{\rm threshold}} 
d{\cal D}_{\rm GW}\,
p({\cal D}_{\rm GW}|\theta,\lambda')\, 
\ee
is the detection probability for a \red{source} characterized by the parameters $\theta$, for the given $\lambda'$.
The normalization factor $\beta(\lambda')$ represents the fraction of events that can be detected, in a population characterized by the distribution 
$p_0(\theta|\lambda')$.
As stressed in \cite{Mandel:2018mve,Chen:2017rfc,Fishbach:2018gjp}, since this normalization factor depends on $\lambda'$, we must include its dependence  in order to obtain the correct  likelihood and indeed, as we will see, it is quite crucial to compute it accurately to obtain meaningful results. 

Putting together \eqs{pDl}{pDi}, in the approximation in which we neglect the information on the total rate,  we have
\be\label{likeprod}
p(\{{\cal D}_{\rm GW}\}|\lambda')=\, \frac{1}{[\beta(\lambda')]^{N_{\rm obs}}}\,
\prod_{i=1}^{N_{\rm obs}} \int d\theta\, p({\cal D}^i_{\rm GW}|\theta,\lambda')\, p_0(\theta|\lambda')\, ,
\ee
with $\beta(\lambda')$ given by \eqs{alpha}{Pdet}. 

The hierarchical  Bayesian framework has been first developed to study the hyper-parameters $\lambda$ that characterize the underlying astrophysical population, see e.g. \cite{Adams:2012qw,Mandel:2018mve,Thrane:2018qnx,Vitale:2020aaz}. In that case (with a natural choice for the parameters $\theta$, see below), the hyper-parameters $\lambda'$ (or, more, generally, $\lambda$) only affect $p_0(\theta|\lambda')$,  which encodes the information on the population, and do not enter in $p({\cal D}^i_{\rm GW}|\theta,\lambda')$, which can therefore be written simply as 
$p({\cal D}^i_{\rm GW}|\theta)$. As an example, the hyper-parameters could describe the shape of the mass function of the  BH  masses in a BH-BH binary. These parameters enter in the description of the population, i.e. in $p_0(\theta|\lambda')$; however, the probability of detection is uniquely fixed in terms of the parameters $\theta$, e.g. the distance, masses, etc. of the specific event.
The formalism, however,  can be extended to the case where $\lambda'$ include also cosmological parameters such as $H_0$. In this case, in general, $\lambda'$ also enters in $p({\cal D}^i_{\rm GW}|\theta,\lambda')$. More precisely, the conditioning on
$\lambda'$, in \eq{likeprod}, depends on the choice of the variables  $\theta$. 
Indeed,  suppose that we start from variables $\theta$  such  that $p({\cal D}_{\rm GW}|\theta,\lambda')$ is independent of
$\lambda'$, so
\be\label{pDinolambda}
p({\cal D}^i_{\rm GW}|\lambda')=\frac{1}{\beta(\lambda')}
\int d\theta\, p({\cal D}^i_{\rm GW}|\theta)\, p_0(\theta|\lambda')\, .
\ee
One might then wish to make a transformation to new variables 
\be\label{ththtilde}
\theta\ra \tilde{\theta}(\theta,\lambda')\, ,
\ee 
which involves the hyper-parameters $\lambda'$. Such a transformation would be possible, in principle, even  when $\lambda'$ only describe the astrophysical population, but in that case it would be very unnatural.
In contrast, we will see on some explicit examples below that it  can be very natural when the hyper-parameters describe the cosmological model. 
Under this transformation
\bees 
p_0(\theta|\lambda')&\ra& \tilde{p}_0(\tilde{\theta}|\lambda')=p_0(\theta(\tilde{\theta},\lambda') |\lambda') \frac{d\theta}{d\tilde{\theta}}\, ,\label{p0tilde}\\
p({\cal D}_{\rm GW}|\theta)&\ra&
\tilde{p}({\cal D}_{\rm GW}|\tilde{\theta},\lambda')= p({\cal D}_{\rm GW}|\theta(\tilde{\theta},\lambda') )\, .\label{pDpDtilde}
\ees
These are the correct transformation laws for these quantities, since $p_0(\theta|\lambda')$ is a probability distribution with respect to $\theta$, so $\tilde{p}_0(\tilde{\theta}|\lambda')d\tilde{\theta}=
p_0(\theta |\lambda') d\theta$,
while $p({\cal D}_{\rm GW}|\theta,\lambda')$ is rather a probability distribution with respect to ${\cal D}_{\rm GW}$ and  is a `scalar' with respect to transformations of the $\theta$ variables. This transformation therefore induces a dependence on the hyper-parameters in $\tilde{p}({\cal D}_{\rm GW}|\tilde{\theta},\lambda')$, even if it was absent in $p({\cal D}_{\rm GW}|\theta)$.

As an example, consider  the case where we want to infer the value of $H_0$ within $\Lambda$CDM, so that $\lambda'=\{H_0\}$, and  we chose as variables
$\theta=\{d_L,\hatO,\theta'\}$, where $d_L$ is the standard luminosity distance in GR. Then
\eq{pDi} reads
\be\label{pDidLO}
p({\cal D}^i_{\rm GW}|H_0)=\frac{1}{\beta(H_0)}
\int d(d_L)d\Omega d\theta'\, p({\cal D}^i_{\rm GW}|d_L,\hatO,\theta')\, p_0(d_L,\hatO,\theta'|H_0)\, ,
\ee
and 
\be\label{alphadLO}
\beta(H_0)=\int d(d_L)d\Omega d\theta'\, P_{\rm det}(d_L,\hatO,\theta')\, p_0(d_L,\hatO,\theta'|H_0)\, .
\ee
Note that the GW measurement is sensitive directly to the luminosity distance $d_L$ and the direction $\hatO$ of the source, so $p({\cal D}^i_{\rm GW}|d_L,\hatO,\theta')$ is independent of $H_0$.\footnote{\red{Actually, a dependence on $H_0$ re-enters from the fact that the GW signal depends not on the `source frame' masses $m_i$, which, physically, are the most natural quantities to include among the $\theta'$ parameters, but on the `detector frame' masses $m_i^{(d)}=m_i(1+z)$ and, when using $d_L$ as independent variable, $z=z(d_L,H_0)$. Similarly, 
in \eq{pDizO}  there is a small explicit $z$ dependence entering in the form $m_i(1+z)$,
not contained in $d_L(z,H_0)$.\label{foot:H0_from_mi}}}
In contrast,  the  galaxy radial positions are first obtained by measuring their redshifts, which can then be translated into  values for $d_L$ by making use of $H_0$. Therefore, the term  $p_0(d_L,\hatO,\theta'|H_0)$ depends on $H_0$. The choice
$\theta=\{d_L,\hatO,\theta'\}$ is therefore an example of a choice of variables such that  the dependence on the hyper-parameters is only in $p_0(\theta|\lambda')$.

We could, however, equally well choose  as variables $\theta=\{z,\hatO,\theta'\}$, so that both the radial position of the GW event and of the  galaxies are expressed in terms of the redshift $z$ (this is indeed the choice made in \cite{Chen:2017rfc,Fishbach:2018gjp,Soares-Santos:2019irc}). In this case, \eq{pDidLO} is replaced by
\be\label{pDizO}
p({\cal D}^i_{\rm GW}|H_0)=\frac{1}{\beta(H_0)}
\int dzd\Omega d\theta'\, p({\cal D}^i_{\rm GW}|d_L(z,H_0),\hatO,\theta')\, p_0(z,\hatO,\theta')\, .
\ee
Note that now, in \eq{pDizO}, the dependence on $H_0$ has been entirely moved from the prior  to the likelihood. In fact, the galaxy catalog is obtained directly in redshift space, so the probability density for finding a galaxy in the catalog at a given $z$, expressed by the prior $p_0(z,\hatO,\theta')$,  is independent of $H_0$.
In contrast, the probability of the GW data ${\cal D}^i_{\rm GW}$ depends on the luminosity distance of the source, and can be expressed as a function of redshift through the function $d_L(z,H_0)$,
that carries the dependence on $H_0$.\footnote{Note that the formalism of ref.~\cite{Gray:2019ksv} also belongs to the hierarchical Bayesian framework, even if the authors do not use this terminology explicitly, as can be seen for instance from 
their eq.~(A5). There, the conditioning on $H_0$ is correctly written both in the prior and in the likelihood for the data.}

Another option, for example,  could be to use comoving coordinates $\vx$, with $|\vx|\equiv \dcom$. In that case, both $p({\cal D}^i_{\rm GW})$ and $p_0$ would depend on $H_0$, since $H_0$ (and, more generally, the cosmology)  enters both in the transformation from $z$ to $\dcom$ and in the transformation from $d_L$ to $\dcom$  (although, for close sources, the latter dependence disappears and $d_L\simeq \dcom$).

In the following, to make easier the comparison with refs.~\cite{Chen:2017rfc,Fishbach:2018gjp,Soares-Santos:2019irc}, we will use $\{z,\hatO\}$.
Observe also that the normalization constant $\beta(\lambda')$ is invariant under the transformation
(\ref{ththtilde})--(\ref{pDpDtilde}), as we see inserting \eqs{p0tilde}{pDpDtilde} 
into \eqs{alpha}{Pdet}.

From the above discussion, we see that there are three main ingredients in the formalism: the first is the probability of the GW data given the parameters of the source and the hyper-parameters $\lambda'$, $p({\cal D}^i_{\rm GW}|\theta,\lambda')$. 
For the study of $H_0$ or of $\Xi_0$  we are interested mostly in the distance and direction to the GW event. In this case a useful approximation is obtained assuming a
gaussian likelihood with direction-dependent parameters~\cite{Singer:2016eax,Singer:2016erz},
\be\label{likelihoodSinger}
{\cal L}_i(d_L,\hatO)\equiv p({\cal D}^i_{\rm GW}|d_L, \hatO)=\rho_i(\hatO)\, 
\frac{{\cal N}_i (\hatO)}{\sqrt{2\pi}\,\sigma_i(\hatO)}\, \exp\left\{-\frac{[d_L-\mu_i(\hatO)]^2}{2\sigma_i^2(\hatO)}\right\}\, .
\ee
The quantities $\rho_i(\hatO)$, ${\cal N}_i(\hatO)$, $\sigma_i(\hatO)$ and $\mu_i(\hatO)$ for the $i$-th GW event
are provided by the LIGO/Virgo skymaps, for a  discrete set  of values of $\hatO$ corresponding to  the HEALPix pixelization~\cite{Gorski:2004by}.\footnote{We observe that, for BBH coalescences, the public LIGO/Virgo posteriors and skymaps are provided assuming a uniform prior in a bounded region of the  plane $(m^{(d)}_1,m^{(d)}_2)$, where $m^{(d)}_1,m^{(d)}_2$ are  the detector frame masses of the two BHs. This is an approximation, for two reasons; first, a more physical prior should be expressed in terms of source frame masses $m_1,m_2$. The factor $(1+z)$ needed to pass from source frame to detector frame masses then introduces a dependence on the cosmology, since $z$ is obtained from the measured (GW) luminosity distance, and therefore depends on $H_0$ (and on $\Xi_0$). Second, the actual mass distribution of BH in binaries is now determined by the GW data to have quite a different shape, see \eq{brokenpowerlaw} below, and, for instance, becomes singular as the heavier mass $m_1$ approaches the assumed minimum value, so is quite different from a uniform distribution in the $(m_1,m_2)$ plane. \label{foot:reweight}}
The corresponding posterior distribution of  $d_L$, conditional on the direction $\hatO$, is  given by
\be\label{posterior_gau}
p(d_L|\hatO)d(d_L)=\frac{ {\cal N}_i (\hatO)}{\sqrt{2\pi}\,\sigma_i(\hatO)}\, 
\exp\left\{-\frac{[d_L-\mu_i(\hatO)]^2}{2\sigma_i^2(\hatO)}\right\}\, \pi(d_L)d(d_L)\, .
\ee
The term $\pi(d_L)$ is the prior chosen by the LIGO/Virgo collaboration  to obtain the posterior.
For our purposes  we  only need  the likelihood (\ref{likelihoodSinger}), which we obtain from the publicly  available posteriors by removing the prior $\pi(d_L)\propto d_L^2$ that has been used by  the LIGO/Virgo collaboration. The factor ${\cal N}_i (\hatO)$ is a normalization constant, given for convenience,  fixed by the condition
\be
\int_0^{\infty} d(d_L)\, p(d_L|\hatO)=1\, ,
\ee
so that $p(d_L|\hatO)$ is a normalized pdf. The remaining overall factor $\rho_i(\hatO)$ in \eq{likelihoodSinger} then gives the probability that the event is in the direction $\hatO$. As discussed in \cite{Singer:2016eax,Singer:2016erz},
\eq{posterior_gau} provides a good approximation to the full posterior (conditional on $\hatO$) obtained from the MCMC. 

For the case of $H_0$, the likelihood can be written as a function of  $z$ as  
\be\label{likelihoodSingerH0}
{\cal L}_i(z,\hatO,H_0)=\rho_i(\hatO)\, 
\frac{{\cal N}_i(\hatO)}{\sqrt{2\pi}\,\sigma_i(\hatO)}\, \exp\left\{-\frac{[d_L(z,H_0)-\mu_i(\hatO)]^2}{2\sigma_i^2(\hatO)}\right\}   \, .
\ee
Similarly, for modified gravity,
\be\label{likelihoodSingerXi0}
{\cal L}_i(z,\hatO,\Xi_0)\equiv p({\cal D}^i_{\rm GW}|\dgw(z,\Xi_0), \hatO)=
\rho_i(\hatO)\,
\frac{{\cal N}_i (\hatO)}{\sqrt{2\pi}\,\sigma_i(\hatO)}\, \exp\left\{-\frac{[\dgw(z,\Xi_0)-\mu_i(\hatO)]^2}{2\sigma_i^2(\hatO)}\right\}\, .
\ee
 In this work, we will always use the publicly available gaussian likelihoods (\ref{likelihoodSinger}).\footnote{See \url{https://dcc.ligo.org/LIGO-P1800381/public} and  \url{https://dcc.ligo.org/LIGO-P2000223/public}.}

The second ingredient is the prior $p_0(z,\hatO,\theta')$. For $\{z,\hatO\}$, this will be obtained from a galaxy catalog. This entails the physical assumption that the spatial location of BH-BH coalescences is correlated with the position of galaxies. This is a natural assumption for BHs of astrophysical origin, but might not be true if most of the BHs in the observed coalescences were of primordial, rather than astrophysical, origin.\footnote{Indeed, the techniques discussed here could also be used to study whether there is a statistically significant correlation between galaxy positions and binary BH  coalescences, providing important clues on their stellar versus primordial origin.} As stressed in \cite{Chen:2017rfc,Fishbach:2018gjp,Gray:2019ksv}, the issue of completeness of the catalog introduces some delicate aspects, that we will discuss in sect.~\ref{sect:cat}.

Finally, the third ingredient is the computation of the normalization factor, $\beta(H_0)$ or $\beta(\Xi_0)$. 
When using $z$ as  variable, the expression for $\beta(H_0)$ is given by 
\be\label{betadL}
\beta(H_0)=\int dzd\Omega d\theta'\, P_{\rm det}[d_L(z,H_0),\hatO,\theta']\, p_0(z,\hatO,\theta')\, ,
\ee
where, again, we  used  the fact that the, in  $ P_{\rm det}$, the dependence on $z$ and $H_0$ enters through $d_L(z,H_0)$ \red{(see, however, footnote~\ref{foot:H0_from_mi})}.
When we study modified GW propagation, we rather have
\be\label{betaXi}
\beta(\Xi_0)=\int dzd\Omega d\theta'\, P_{\rm det}[\dgw(z,\Xi_0),\hatO,\theta']\, p_0(z,\hatO,\theta')\, .
\ee 
The computation of these normalization factors therefore involves a model for the detection process, and 
will be discussed in sect.~\ref{sect:beta}, where we will elaborate and improve on the computation in refs.~\cite{Chen:2017rfc,Fishbach:2018gjp}.

\subsection{The $p_0(z,\hatO)$ prior: galaxy catalogs and completeness}\label{sect:cat}

We now assume that the prior $p_0(z,\hatO,\theta')$ factorizes as
\be\label{factorp0}
p_0(z,\hatO,\theta')= p_0(z,\hatO) \tilde{p}_0(\theta')\, ,
\ee 
and we separately normalize
\be\label{normp0zO}
\int dz d\Omega\, p_0(z,\hatO)=1\, ,
\ee
and
\be\label{normptheta}
\int d\theta'\, \tilde{p}_0(\theta')=1\, .
\ee
At some level of accuracy this factorization assumption will have to be improved since, for instance, the distribution of masses in a binary black hole (BBH) might be correlated with redshift \red{(see \cite{Fishbach:2021yvy} for a recent discussion in the context of the GWTC-2 catalog)}. However, this factorization is a sensible starting point. \red{We will later improve  this prior by including a parametrization of the redshift dependence  of the BBH merger rate, see \eq{factorp0lambda} below}.

To determine the prior $p_0(z,\hatO)$ we use  the public GLADE galaxy catalog (v2.4) \cite{Dalya:2018cnd} (\url{http://glade.elte.hu}) because of its full sky coverage.\footnote{We have also explored the possibility of using the public DES~Y1 catalog (\url{https://des.ncsa.illinois.edu/releases/y1a1}). However, because of the tiny footprint of DES~Y1,  only one event in O2 falls in it while, in O3a, there is again only one event whose localization  volume completely  falls in it, and a few others that only partially fall in it.  Despite the higher completeness of DES we find that, because of the relatively large redshift errors,  the posteriors obtained from these events
using DES~Y1 do not improve significantly with respect to the posteriors of the same events  that we obtain with GLADE. The situation could improve  with the DES~Y3 catalog, which is not yet public (and which should also have improved redshift estimations). In any case, the posterior that we obtain for GW190814 using GLADE is  comparable  to that obtained in \cite{Soares-Santos:2019irc} using DES~Y3 (and in sect.~6.4 of \cite{Abbott:2020khf} using only GLADE). This  could  be traced  to the facts that, for GLADE, the use of luminosity weighting with a lower cut on luminosity improves significantly the completeness (see the discussion in app.~\ref{sect:GLADE}), and, furthermore, the multiplicative completion that we introduce below appears to work quite well.
We also considered the  GWENS catalog (\url{https://astro.ru.nl/catalogs/sdss_gwgalcat}), but we found it difficult to  make use of it in the absence, apparently, of any accompanying documentation.
\label{foot:catalogs}}
When using a  galaxy catalog,  completeness is  a crucial issue. In an ideal complete catalog, the prior would simply be given by $p_0(z,\hatO)=p_{\rm cat}(z,\hatO)$, where 
\be\label{pcat}
p_{\rm cat}(z,\hatO)=\frac{\sum_{\alpha=1}^{N_{\rm cat}} w_\alpha \delta(z-z_{\alpha})\delta^{(2)}(\hatO-\hatO_{\alpha})}{\sum_{\alpha=1}^{N_{\rm cat}} w_\alpha} \, .
\ee
Here $N_{\rm cat}$ is the number of galaxies in the catalog,
$z_{\alpha}$ are the cosmological redshifts of the galaxies (obtained correcting the measured redshift for peculiar velocities) and $\hatO_{\alpha}$ their directions. Note that we use the index $i$ to label the GW events, and $\alpha$ to label the galaxies. Here we have assumed that the errors on $z_{\alpha},\hatO_{\alpha}$ are negligible with respect to the  corresponding errors of the GW events. This is an excellent approximation for $\hatO_{\alpha}$, compared to the current and future angular resolution of GW detector networks.  For the redshift this is not always a good approximation, particularly for galaxies with photometric redshift determination, and
can be improved by replacing, for each galaxy, the Dirac delta  by a gaussian approximation to the  $z_{\alpha}$ posterior, as in  \cite{Fishbach:2018gjp},
so that
\be\label{pcatN}
p_{\rm cat}(z,\hatO)=\frac{\sum_{\alpha=1}^{N_{\rm cat}} w_\alpha 
{\cal N}(z;z_{\alpha},\sigma_{\alpha})\delta^{(2)}(\hatO-\hatO_{\alpha})}{\sum_{\alpha=1}^{N_{\rm cat}} w_\alpha} \, ,
\ee
where ${\cal N}(z;z_{\alpha},\sigma_{\alpha})$ is a gaussian pdf in the variable $z$, with mean $z_{\alpha}$ and variance $\sigma_{\alpha}$. In particular,  for the GLADE catalog, the likelihoods are taken to be  gaussian with a width  
$\Delta z\simeq 1.5\times 10^{-4}$ for galaxies with spectroscopic redshift and 
$\Delta z\simeq 1.5\times 10^{-2}$ for galaxies with photometric redshift~\cite{Dalya:2018cnd}. \red{More generally, one could actually use the full posterior  of the redshift distribution of each galaxy.
This is obtained multiplying the likelihood (that need not be gaussian) by a prior, which is naturally chosen as a volume prior  $dV_c/dz$, where $V_c$ is the comoving volume (followed by normalization). For narrow, gaussian likelihoods, such as in GLADE, these posteriors are well fitted by another gaussian, that, eventually, is the quantity that we denote by  ${\cal N}(z;z_{\alpha},\sigma_{\alpha})$
in \eq{pcatN}.}
However, our code can treat the general case, see also footnote~\ref{foot:metalog} on page \pageref{foot:metalog}.\footnote{A slightly different formalism is used in
\cite{Soares-Santos:2019irc}, which also uses the gaussian likelihoods, and in
\cite{Palmese:2020aof}, that use the full non-gaussian likelihood for the redshifts.}
In the formulas below, for notational simplicity, we simply write 
$\delta(z-z_{\alpha})$, but all these formulas are trivially generalized replacing 
$\delta(z-z_{\alpha})$ by the full posterior for the redshifts.

In \eq{pcatN}
we have inserted the normalization factor
$\sum_{\alpha=1}^{N_{\rm gal}} w_\alpha$ at the denominator, so that 
$p_{\rm cat}(z,\hatO)$ is a properly normalized pdf,
\be\label{pcatzOnorm}
\int dz d\Omega\, p_{\rm cat}(z,\hatO)=1\, .
\ee
The factors $w_{\alpha}$ allow for the possibility of weighting differently the probability that a galaxy is the host of a given GW signal. As in \cite{Fishbach:2018gjp}, we will examine three different choices. The simplest option is to  set all $w_{\alpha}=1$, corresponding to equal probability for each galaxy to host the source.  Alternatively, we can weight the galaxies by their B-band luminosity $L_B$, which is  a proxy for their star formation rate, or by their K-band luminosity $L_K$, which is a proxy for their total stellar mass. 
As we will discuss in app.~\ref{sect:GLADE}, for the GLADE catalog the most appropriate choice is to use luminosity weighting, either in the B or in the  K band.

In general, however, galaxy catalogs are incomplete, especially  all-sky catalogs that reach large distances. 
In app.~\ref{sect:GLADE},  expanding on the results presented in \cite{Dalya:2018cnd,Fishbach:2018gjp},  we will study the completeness of the GLADE catalog according to the measures of completeness that we develop in this section.
The incompleteness of the catalog means that the simple prior $p_0(z,\hatO)=p_{\rm cat}(z,\hatO)$, with
$p_{\rm cat}(z,\hatO)$ given by \eq{pcat}, must be modified. To this purpose we  consider a  region ${\cal R}$, sufficiently large to include all the GW events potentially detectable by the GW detector network under consideration; typically this will be a spherical ball in redshift space, with a  sufficiently large radius $z_{\cal R}$, but for the moment we keep the shape of ${\cal R}$ generic. 
We denote by 
$N_{\rm gal}({\cal R})$ the total number of galaxies in this region, 
by $N_{\rm cat}({\cal R})$ the number of galaxies in ${\cal R}$ that are actually present in the catalog,  and by   $N_{\rm miss}({\cal R})= N_{\rm gal}({\cal R})-N_{\rm cat}({\cal R})$ the number of galaxies that have been missed.
The natural prior would be given by the sum over all  galaxies in ${\cal R}$,
\be\label{p0Ntot}
p_0(z,\hatO;{\cal R})=\frac{\sum_{\alpha=1}^{N_{\rm gal}({\cal R})} w_\alpha \delta(z-z_{\alpha})\delta^{(2)}(\hatO-\hatO_{\alpha})}{\sum_{\alpha=1}^{N_{\rm gal}({\cal R})} w_\alpha} \, .
\ee
(In this section we will use the notation $p_0(z,\hatO;{\cal R})$, where the region ${\cal R}$ is written explicitly).
We can split the sum  over the galaxies in  ${\cal R}$ into a sum over those that are the catalog and over those that have been  missed, 
\bees
p_0(z,\hatO;{\cal R})&=&\frac{1}{\sum_{\alpha=1}^{N_{\rm gal}({\cal R})} w_\alpha }\, 
\label{p0assum}\\
&&\times \[ \sum_{\alpha_1=1}^{N_{\rm cat}({\cal R})} w_{\alpha_1} \delta(z-z_{\alpha_1})\delta^{(2)}(\hatO-\hatO_{\alpha_1})
+\sum_{\alpha_2=1}^{N_{\rm miss}({\cal R})} w_{\alpha_2}\delta(z-z_{\alpha_2})\delta^{(2)}(\hatO-\hatO_{\alpha_2})
\] \, ,\nn
\ees
where, for clarity, we have used  the dummy index $\alpha$  for the sum over all galaxies, $\alpha_1$
for the sum over the galaxies in the catalog and  $\alpha_2$ for the sum over  missed galaxies. We  define the normalized pdf's
\bees
p_{\rm cat}(z,\hatO;{\cal R})&=&\frac{\sum_{\alpha_1=1}^{N_{\rm cat}({\cal R})} w_{\alpha_1} \delta(z-z_{\alpha_1})\delta^{(2)}(\hatO-\hatO_{\alpha_1})}{\sum_{\alpha_1=1}^{N_{\rm cat}({\cal R})} w_{\alpha_1}} \, ,\label{pcatdelta}\\
p_{\rm miss}(z,\hatO;{\cal R})&=&\frac{\sum_{\alpha_2=1}^{N_{\rm miss}({\cal R})} w_{\alpha_2} \delta(z-z_{\alpha_2})\delta^{(2)}(\hatO-\hatO_{\alpha_2})}{\sum_{\alpha_2=1}^{N_{\rm miss}({\cal R})} w_{\alpha_2}}\, .\label{pmissdelta}
\ees
From these definitions it follows that
\be\label{p0pcatpmiss}
p_0(z,\hatO;{\cal R})=f_{\cal R}\,  p_{\rm cat}(z,\hatO;{\cal R})+(1-f_{\cal R}) p_{\rm miss}(z,\hatO;{\cal R})\, ,
\ee
where
\be\label{fR}
f_{\cal R}=\frac{ \sum_{\alpha_1=1}^{N_{\rm cat}({\cal R})} w_{\alpha_1}}{  \sum_{\alpha=1}^{N_{\rm gal}({\cal R})} w_{\alpha}  }\, .
\ee
In the case of weights $w_{\alpha}=1$, $f_{\cal R}=N_{\rm cat}({\cal R})/N_{\rm gal}({\cal R})$ is just the fraction of galaxies that have been detected in the region under consideration, while, more generally, it is a weighted fraction of detected galaxies.

In order to define these quantities, it has been necessary to introduce a region ${\cal R}$, and
it is important to understand the dependence on the choice of this region. Observe that $f_{\cal R}$ does depend on  ${\cal R}$.   Extending the region ${\cal R}$  toward very large distances, where the catalog will be highly incomplete, has the effect of decreasing $f_{\cal R}$, because, including highly incomplete regions, the denominator in \eq{fR} increases much more than the numerator. As a limiting case, including a region that is beyond the reach of the galaxy surveys would not increase at all the numerator, while the denominator would still increase significantly. Thus, increasing ${\cal R}$ so to include very incomplete regions, the weight of $p_{\rm cat}(z,\hatO;{\cal R})$ in \eq{p0pcatpmiss} is diminished and the weight of the unknown component 
$p_{\rm miss}(z,\hatO;{\cal R})$ is increased.
One might then fear that we are downplaying the importance of the information contained in $p_{\rm cat}(z,\hatO;{\cal R})$. This would not be meaningful because, after all, the only relevant aspect, for performing the GW/galaxy correlation for a given GW event, is how complete the catalog is in the localization region of the GW event. Further reflection, however, shows that this is not the case, as long as ${\cal R}$  is larger than the region where we can observe GW events; indeed,  extending ${\cal R}$ further, not only $f_{\cal R}$ changes, but also $p_{\rm miss}(z,\hatO;{\cal R})$, and precisely in such a way that the effect on $p_0(z,\hatO;{\cal R})$ is just an overall multiplicative factor, that cancels between numerator and denominator in \eq{pDizO}. We prove this in app.~\ref{app:R}.

We next want to  understand how to estimate $p_{\rm miss}(z,\hatO)$. To this purpose,
in the case $w_{\alpha}=1$, we first need the  relation between $p_0(z,\hatO)$, $p_{\rm cat}(z,\hatO)$, $p_{\rm miss}(z,\hatO)$ and the corresponding comoving number densities 
$n_{\rm gal}(z,\hatO)$, $n_{\rm cat}(z,\hatO)$ and
$n_{\rm miss}(z,\hatO)$ (we will generalize in sect.~\ref{sect:compl_lum_w} to the case of luminosity weighting). We 
write the comoving volume element $dV_c$ as 
$dV_c=\dcom^2 d(\dcom) d\Omega$, where $\dcom(z)$ is the comoving distance. Using 
$d(\dcom)/dz= c/H(z)$, we have
\be\label{dVcdzdO1}
dV_c=\frac{dV_c}{dzd\Omega}\, 
dz d\Omega\, ,
\ee
where
\be\label{dVcdzdO2}
\frac{dV_c}{dzd\Omega}=\dcom^2(z) \frac{c}{H(z)} \, .
\ee
As the region ${\cal R}$ that enters the definition of
 $p_0(z,\hatO;{\cal R})$, $p_{\rm cat}(z,\hatO;{\cal R})$ and
$p_{\rm miss}(z,\hatO;{\cal R})$ we take  a spherical ball extending up to a maximum redshift $z_{\cal R}$, sufficiently large to include the localization region of all  GW events detectable by a given detector network, for the whole range of priors on $H_0$ or $\Xi_0$. We then observe that, in terms of $n_{\rm gal}(z,\hatO)$, the number of galaxies in an infinitesimal  comoving volume $dV_c$ is $n_{\rm gal}(z,\hatO)dV_c$, while, in terms of 
$p_0(z,\hatO;{\cal R})$, which is a density with respect to $dz$ and $d\Omega$, it is given by $N_{\rm gal}({\cal R})p_0(z,\hatO;{\cal R})dzd\Omega$, where the factor $N_{\rm gal}({\cal R})$ compensates the denominator in the definition (\ref{p0Ntot}) (given  that we are considering the case $w_{\alpha}=1$).
Therefore
\be\label{pbarvsp}
p_0(z,\hatO;{\cal R})=\frac{n_{\rm gal}(z,\hatO)}{N_{\rm gal}({\cal R})}\, 
\frac{dV_c}{dzd\Omega}\, .
\ee
Similarly,
\be\label{pcatncat}
p_{\rm cat}(z,\hatO;{\cal R})=\frac{n_{\rm cat}(z,\hatO)}{N_{\rm cat}({\cal R})}\, 
 \frac{dV_c}{dzd\Omega}\, ,
\ee
\be\label{pmissnmiss}
p_{\rm miss}(z,\hatO;{\cal R})=\frac{n_{\rm miss}(z,\hatO)}{N_{\rm miss}({\cal R})} \,   
\frac{dV_c}{dzd\Omega}\, .
\ee
To estimate   $p_{\rm miss}(z,\hatO;{\cal R})$ (again, considering first 
the weighting $w_{\alpha}=1$) we observe that  the number density of all galaxies per comoving volume  can be inferred from the observed ones by extrapolating their distribution with a Schechter mass or luminosity function. The result is that,  after  averaging over a sufficiently large  region to smooth out local inhomogeneities,   the number density of all galaxies per comoving volume  is  approximately constant in redshift, at a value $\bar{n}_{\rm gal}$ of order $(0.1-0.2) \, {\rm Mpc}^{-3}$,
up to $z\sim 1-2$ (see e.g.~\cite{Conselice2016}).\footnote{\red{Actually, the number density of galaxies depends also on the lower mass limit assumed for galaxies, which in ref.~\cite{Conselice2016} is taken to be $10^6\msun$. Most systems at mass lower than this are star clusters within galaxies, or debris from galaxy interactions.  In any case, as we will see below, in the end we will only use luminosity weighting, applying  the same lower cut on the luminosity on the Schechter function and on the galaxies in the GLADE catalog.
In principle, these is also a dependence on the cosmology, and the values in ref.~\cite{Conselice2016} are obtained setting  $H_0=70~{\rm km}\, {\rm s}^{-1}\, {\rm Mpc}^{-1}$ and $\oma=0.3$.  However, it will be clear from the discussion in sect.~\ref{sect:conemask} that completeness is  an intrinsically approximate notion, that can be defined in different `quasi-local' ways (or using different weights such as B or K band luminosity weighting). It is useful  in order to associate a  weight to the information contained in different regions of a catalog, but the intrinsic uncertainty associated to its very definition, that we will explore below, is certainly much larger than its dependence on cosmology}.}
We will discuss in the next sections how to exploit this information (and its even more relevant generalization to luminosity weighting).
 
\red{A further  refinement in the choice of prior can be obtained observing that the prior should reflect the probability of a binary BH merger in a given galaxy. This depends not only on the `size' of the galaxy (as measured for instance by its luminosity in different bands), but also on its redshift, since the  rate of BBH mergers is redshift dependent. Following \cite{Fishbach:2018edt,Callister:2020arv}, for  redshifts well below the peak of the star formation rate,  we describe this with a prior
\be\label{dNdVclambda}
 p(z|\lambda)\propto \frac{dV_c}{dz} \,  (1+z)^{\lambda -1}\, .
\ee
The case $\lambda=0$ corresponds to a merger rate density uniform in comoving volume and source-frame time (the remaining factor $(1+z)^{-1}$ transforms the rate from source-frame time to observer time).
\Eq{dNdVclambda} is  motivated by the fact that, if the merger rate follows the star formation rate (SFR), then we should expect that 
\be\label{dNdVpsi}
 p(z|\lambda)\propto \frac{dV_c}{dz} (1+z)^{-1} \,  \psi(z)\, ,
\ee
where $\psi(z)$ is the SFR, for which a common parametrization is given by  the Madau-Dickinson SFR~\cite{Madau:2014bja,Madau:2016jbv}
\be\label{MadauSFR} 
\psi(z)\propto 
\frac{(1+z)^{\alpha}}{1+\(\frac{1+z}{1+z_p}\)^{\alpha+\beta}}\, . 
\ee
In this expression,  $z_p$ is the peak of the star formation rate (which is in the range $z_p\sim 2-3$) and  $\alpha$ and $\beta$ are constants. This functional form  interpolates between a power-like behavior $\psi(z)\propto (1+z)^{\alpha}$ for redshifts well below $z_p$, and $\psi(z)\propto (1+z)^{-\beta}$ at $z\gg z_p$. 
For $z\ll 1$, \eq{dNdVpsi} reduces to \eq{dNdVclambda} with $\lambda=\alpha$, while, for intermediate values $0.1\,\lsim\, z\, \lsim\, 1$, \eqs{dNdVpsi}{MadauSFR} are still well approximated by \eq{dNdVclambda}, with a value of $\lambda$ close to $\alpha$~\cite{Fishbach:2018edt}.}  If the BBH merger rate indeed follows the star formation rate (SFR), one expects  $\lambda\sim 3$. More realistically, one should  convolve the SFR with a distribution of time delays between binary formation and merger, and this  affect significantly  the estimate of $\lambda$. Formation channels with short delays of order 50~Myr, typical of field formation, lower $\lambda$ from the value $\lambda\sim 3$ but still give   $\lambda >0$, while formation channels with very long delays, of several Gyr, such as the chemical homogeneous formation channel, drive 
$\lambda$ toward negative values, $-6\,\lsim\, \lambda \,\lsim\, -4$. Even longer delays, $\sim 10$~Gyr, are expected for binaries that form dynamically
inside clusters but are ejected prior to merger, resulting in $\lambda\sim -10$ (see \cite{Fishbach:2018edt} for discussion and references).  \red{The current observational limit on $\lambda$ from GW observation themselves, using the data from the GWTC-2 catalog,  is
$\lambda=1.8^{+2.1}_{-2.2}$ at $90\%$ c.l.~\cite{Abbott:2020gyp} (for the broken power law model of the BBH mass distribution that we will use below, see sect.~\ref{sect:MC}). Observe that this limit assumes the validity of GR since, as we will discuss in sect.~\ref{sect:rate}, the effect of modified GW propagation is partially degenerate with the redshift dependence of the rate.} 

\red{In the following, we will  first use as reference value $\lambda=1$, which amounts to neglecting any redshift dependence in the  prior induced by the BBH merger rate. We will then investigate the effect of  $\lambda$, replacing the prior in \eq{factorp0} by
\be\label{factorp0lambda}
p_0(z,\hatO,\theta',\lambda)= p_0(z,\hatO) \tilde{p}_0(\theta') c_{\lambda}(1+z)^{\lambda-1}\, ,
\ee 
where the normalization constant $c_{\lambda}$ ensures that
\be\label{pcatzOnormclambda}
c_{\lambda} \int dz d\Omega\, p_0(z,\hatO)\, (1+z)^{\lambda-1}=1\, ,
\ee
while $\tilde{p}_0(\theta')$ is separately normalized as in \eq{normptheta}.\footnote{Note that the factor $dV_c/dz$ in \eq{dNdVpsi} is already included in 
$p_0(z,\hatO;{\cal R})$, see \eq{pbarvsp}, so for $\lambda\neq 1$ we only need to add the factor
$(1+z)^{\lambda-1}$.}
We will then study  how the results change fixing  $\lambda$ to other values within  a plausible  range, that we will chose to be 
$\lambda\in [0,3]$.
Eventually, in particular when sufficient statistics will be available,  the best approach would be to perform a simultaneous inference on both the cosmological and astrophysical parameters, including therefore $\lambda$ (or, more generally, $\alpha,\beta$ and $z_p$) in the inference procedure, together with $H_0$ and $\Xi_0$, see also the discussion in sect.~\ref{sect:rate} below. This would be much more demanding computationally and, given the current limited statistics, we will limit ourselves to the simpler approach of inferring $\Xi_0$ or $H_0$ for different, fixed values of the parameters that describe the astrophysical population.}

\subsubsection{Cone completeness and mask completeness}\label{sect:conemask}

Given an expected value of the galaxy number density, a natural measure of the completeness of a catalog is obtained by comparing the number of galaxies in the catalog within some volume, to the number that should be expected for a complete catalog. A crucial issue here is the choice of such a volume.
In previous studies of galaxy completeness, for a catalog such as GLADE, the volume has been chosen as a $4\pi$ spherical ball, extending up to a redshift $z$, and completeness has been studied as a function of $z$~\cite{Dalya:2018cnd,Fishbach:2018gjp}.
The limitation of this approach is that it can mix up regions with  very different completeness level. Indeed, even after cutting out the galactic plane, which is obscured by dust, in a $4\pi$ catalog such as GLADE that combines different surveys, different region on the sky can have wildly different completeness level, as it is apparent for instance from Fig.~\ref{fig:Mollview_z02} in app.~\ref{sect:GLADE} (which is already on a logarithmic scale). Furthermore, also in the radial direction the completeness can vary quite strongly, especially beyond some redshift, and a measure of completeness which integrates over all redshifts from $z=0$ up to a given $z$ is not really representative of what happens in a shell around $z$.
Given a GW event with its localization region, when we correlate it with a galaxy catalog what we need to know is the level of completeness of the catalog within this localization region. We do not want to weight the information provided by the catalog according to how good, or how bad, the catalog completeness is elsewhere. We therefore need a `quasi-local' notion of completeness. In this section we develop this notion in two different variants, that we will denote as `cone completeness' and `mask completeness'.

The qualification `quasi-local' refers to the fact that 
completeness is necessarily a coarse-grained notion, defined over a region ${\cal S}$  sufficiently large to contain a statistically significant sample of galaxies, since it is only over such a region that we can use the expected number density of galaxies in a complete sample, $\bar{n}_{\rm gal}$ (or the expected luminosity density, see sect.~\ref{sect:compl_lum_w}), as a measure of completeness.
If  ${\cal S}$ is chosen too small, completing the catalog is equivalent to smoothing out true structures, and in this case we would loose valuable  information for the correlation with GW events. For instance, if, for a range of values of $H_0$, the measured luminosity distance of a GW event would correspond to redshifts such that the GW event happens to fall in a cosmic void, such range of values of $H_0$ would be statistically disfavored by such an event. However, if we artificially deem that void region as incomplete, and `top it up' with galaxies, we would just throw away the important  information.
On the other hand, taking  ${\cal S}$  too large we loose the `quasi-local' notion of completeness with the risk that, in the localization volume of a GW event, we would add  `missing' galaxies on the basis of how galaxy surveys covered very far away regions. A good compromise between these two effects must therefore be found: the region ${\cal S}$   must be sufficiently  large that the notion of average number of galaxies in it makes sense, and can be compared with the expected number for a complete sample, but sufficiently small so that we can still consider the completeness as a function of `coarse-grained' variables $z$ and $\hatO$.\footnote{Recall that, in contrast, ${\cal R}$ was defined as a large region covering all detectable GW events, typically a spherical ball extending up to a high redshift $z_{\cal R}$. It has nothing to do with the small region ${\cal S}$ that we are  using to coarse-grain the number density of galaxies. In a field-theoretical jargon, ${\cal S}$ plays the role of a short-distance, or ultraviolet, regulator and ${\cal R}$ of a long-distance, or infrared, regulator.}
In the following we will introduce two different quasi-local notions of completeness, that we will denote as cone completeness and mask completeness, respectively,  and, in section~\ref{sect:results} and in app.~\ref{sect:dependence_results}, we will compare the results obtained with these two prescriptions.

Let us denote by $V_c({\cal S})$ the comoving volume of the region ${\cal S}$.
The information that we have on the number of missing galaxies can  be written as
\be\label{infopmissw1}
\frac{1}{V_c({\cal S})}\, 
\int_{\cal S} dV'_c\[ n_{\rm cat}(z',\hatO')+n_{\rm miss}(z',\hatO')\]=\bar{n}_{\rm gal}\, ,
\ee
where $ dV'_c$ can be expressed in terms of $dz'd\Omega'$ using \eqs{dVcdzdO1}{dVcdzdO2}. Note that we use a prime on the integration variables, to distinguish them from the values $z,\hatO$ around which the region ${\cal S}$ is centered.
The number of galaxies in the catalog in the region ${\cal S}$ is
\be
N_{\rm cat}({\cal S})=\int_{\cal S} dV'_c\,  n_{\rm cat}(z',\hatO')\, ,
\ee
and we  define
the {\em completeness fraction} $P_{\rm compl}({\cal S})$ as
\be\label{Pcompletedef}
P_{\rm compl}({\cal S})\equiv  
\frac{N_{\rm cat}({\cal S})}{\bar{n}_{\rm gal}V_c({\cal S}) }\, .
\ee
Cone completeness is defined by choosing, for the  region  ${\cal S}$, 
a section of a cone  with axis  $\hatO$ and  half-opening angle $\theta_c$
[so that it covers a solid angle $\Delta\Omega=2\pi (1-\cos\theta_c)$] restricted, in the radial direction, to redshifts between $z-(\Delta z/2)$ and $z+(\Delta z/2)$. 
We will denote the corresponding region by ${\cal S}(z,\hatO;\Delta z,\Delta\Omega)$ and the corresponding 
completeness fraction by $P_{\rm compl}(z,\hatO;\Delta z,\Delta\Omega)$. For cone completeness, and the GLADE catalog, our default choice will be  to use  $\Delta z=0.05$, 
and  $\theta_c=5^{\circ}$. These choices will be justified in app.~\ref{sect:GLADE}.

\begin{figure}[t]
\centering
\includegraphics[width=0.45\textwidth]{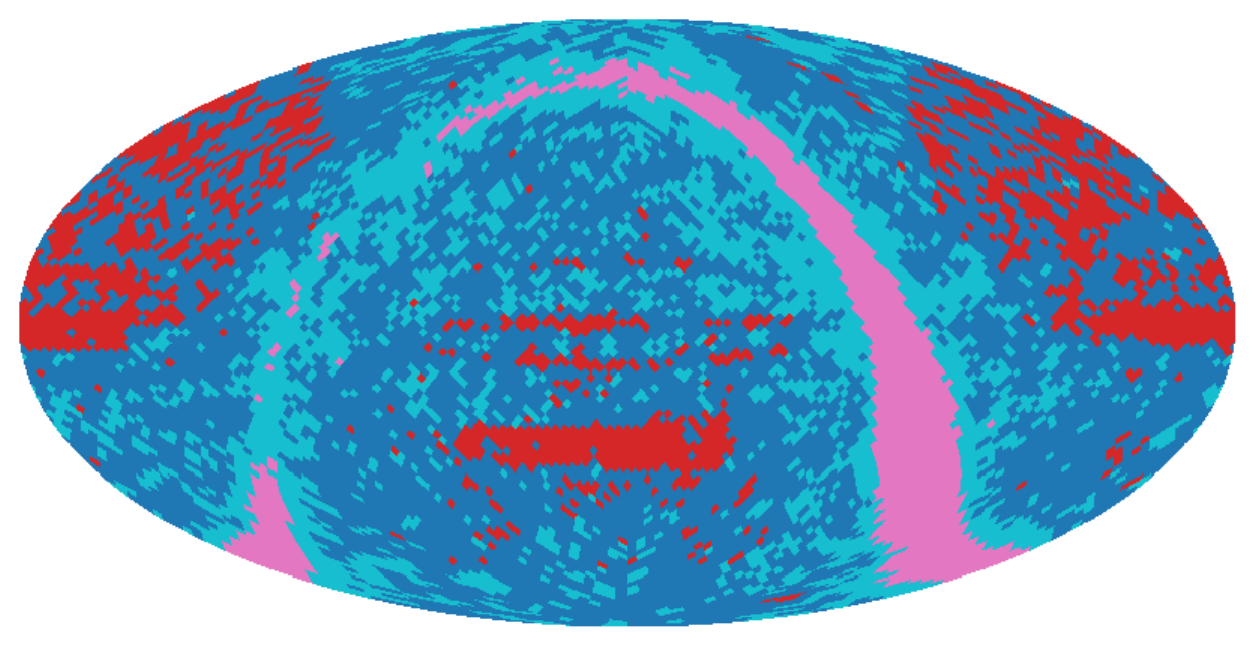}
\includegraphics[width=0.45\textwidth]{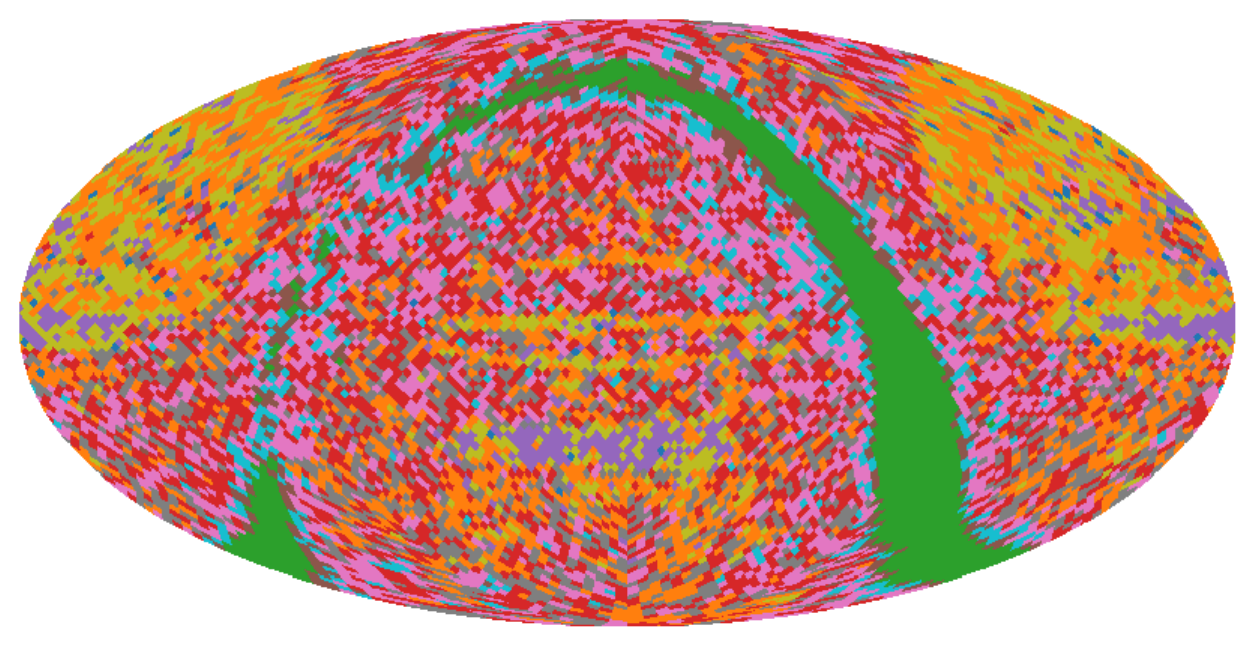}
\caption{The masks that emerge when applying to GLADE the pattern recognition algorithm described in the text, with 4 masks (left) and 10 masks (right). The different colors, chosen as non-sequential for visual clarity, define the masks.}
\label{fig:masks}
\end{figure}

The choice $\Delta z=0.05$  provides relatively large bins in the radial directions. This allows us to have a sufficiently large volume ${\cal S}$, in the sense discussed above, even restricting to a rather narrow cone, as obtained with $\theta_c=5^{\circ}$. In this sense, cone completeness is a good quasi-local property from the angular point of view, but might imply too much coarse graining in the radial direction, especially
considering that the completeness of GLADE drops relatively fast with redshift. A complementary option is to reduce the binning in the radial direction
to about $\Delta z \lesssim 0.01$. This  reduces the volume of the region ${\cal S}$, with the result that the true number of galaxies inside it might no longer be representative of the expected average value. To compensate, we attempt to include  into one single region ${\cal S}$ all the areas in which the sky has been observed with similar telescope time and instrument characteristics, and in which the masking due to dust is also similar. Instead of increasing the cone opening angle, which would smooth out changes of these features with angle, that can be seen to be sharp transitions in  Fig.~\ref{fig:Mollview_z02} in app.~\ref{sect:GLADE}, we introduce a mask. It assigns each direction on the sky to one of a few regions, making the completeness at fixed redshift a piecewise constant function on the sphere. The mask is generated using a basic machine learning technique. We start by splitting the sky into a set of small cones, or rather HEALPix~\cite{Gorski:2004by} pixels in practice, and then group similar pixels together by solving a clustering problem. The most relevant feature of the small cones for this purpose is  the total number of galaxies within them.\footnote{We  apply the logarithm, as this nonlinear transformation compresses the galaxy count feature into a more linear range, improving the results of the clustering. Note that Fig.~\ref{fig:Mollview_z02} also uses a logarithmic scale for improved visualization.}
We found agglomerative clustering applied to this single feature to work sufficiently well.
 The number of clusters is chosen by hand, and we find that should not be less than about 4-5 for GLADE. The use of a single cluster reduces to the  completeness defined over a $4\pi$ spherical shell with radial size between $z-\Delta z/2$ and $z+\Delta z/2$ (this is still more refined than the completeness used in~\cite{Dalya:2018cnd,Fishbach:2018gjp}, which uses a $4\pi$ ball extending in redshift from zero to $z$),  while two clusters can handle footprints of surveys like DES and GWENS, or exclude the galactic plane for GLADE. We find no evidence that more than 10 clusters are needed for the considered catalogs.
An example of the masks that emerge by this procedure, when applied to the GLADE catalog, is shown in Fig.~\ref{fig:masks}, with 4 masks (left panel) and 10 masks (right panel). Note that spatial proximity is not a feature we use for the clustering, but the masks are nevertheless mostly connected if there are only a few of them. This changes for larger mask number. Eventually, for increasing mask number, mask completeness converges toward cone completeness with small independent cones of the opening angle of a pixel of the HEALPix pixelization we choose.
Using masks allows us to adapt the redshift bin size  $\Delta z $ as follows. First, we define  the depth  of  a  mask  as the redshift $z_{\rm depth}$  that contains  $90\%$ of the galaxies in the mask. We expect the completeness to be a function decaying significantly on that scale. 
We then subdivide the mask into 30 bins from $z=0$ to $z=z_{\rm depth}$. Typically,
in the densest cluster we get in this way bins of size  $\Delta z = 0.014$, while in the faintest regions (apart from the galactic plane), the bin size is about  $\Delta z = 0.008$. We then compute the completeness, according to \eq{Pcompletedef}, for the regions ${\cal S}$ defined in this way. We finally apply a smoothing filter to the $30$ values of completeness obtained in each mask, to further restrict sensitivity to intrinsic fluctuations of the galaxy structure.

\subsubsection{Homogeneous and multiplicative completion}\label{sect:hom_multi_compl}

In the previous section we have introduced  two  quasi-local notions of completeness. Once assessed the level of completeness of a region ${\cal S}$, i.e. `how many galaxies are missing', we still have to specify how exactly to complete it, i.e. how to distribute these galaxies within ${\cal S}$.
Indeed, whether we use cone completeness or mask completeness,  \eq{infopmissw1} is an integrated relation (with the choice of cone vs. mask completeness reflected in the choice of the integration region ${\cal S}$) and,
to exploit the information contained in it,
we must still make some assumption on how the missing galaxies are distributed within the region ${\cal S}$. Below, we examine two natural possibilities. In the following, for definiteness, we use the notation appropriate to cone completeness, with the region ${\cal S}$ denoted as ${\cal S}(z,\hatO;\Delta z,\Delta\Omega)$, but the considerations below are straightforwardly adapted to mask completeness,  with an index $I$ labeling the region ${\cal S}_I$ corresponding to the $I$-th mask.

\vspace{2mm}\noindent
{\bf Homogeneous completion}.
The simplest hypothesis is to assume that, within the   region 
${\cal S}(z,\hatO;\Delta z,\Delta\Omega)$,  
$n_{\rm miss}(z',\hatO')$ is distributed uniformly, i.e. has a value that depends only on the `central' values of $z,\hatO$ that enter in the definition of  the region ${\cal S}(z,\hatO;\Delta z,\Delta\Omega)$, but not on the `microscopic' variables  $z',\hatO'$ that are used to span  that region. In that case, 
from \eq{infopmissw1},
\bees\label{nmiss_additive}
n_{\rm miss}(z,\hatO;\Delta z,\Delta\Omega)&=&\bar{n}_{\rm gal}-
\frac{1}{V_c({\cal S})}\, 
\int_{\cal S} dV'_c\, n_{\rm cat}(z',\hatO')\nn\\
&=&\bar{n}_{\rm gal}-\frac{N_{\rm cat}({\cal S})}{V_c({\cal S})}
\, .
\ees
Note that  the dependence of $n_{\rm miss}$
on $(z,\hatO;\Delta z,\Delta\Omega)$  enters through ${\cal S}$.   In terms of
$P_{\rm compl}({\cal S})$, defined in \eq{Pcompletedef},
\be\label{nmiss_additivePc}
n_{\rm miss}(z,\hatO;\Delta z,\Delta\Omega)=\bar{n}_{\rm gal}\, 
\[ 1-P_{\rm compl}(z,\hatO;\Delta z,\Delta\Omega)\]\, .
\ee
We will refer to this procedure as  `homogeneous completion'.
In the following, to make the notation lighter, we will not write explicitly the arguments
$\Delta z,\Delta\Omega$. It is, however, always understood that quantities such as 
$n_{\rm miss}(z,\hatO)$ are coarse-grained quantities, whose definition requires the introduction of a binning defined by $\Delta z, \Delta\Omega$, in the simple case of the section of a cone, or, more generally, the introduction of a region ${\cal S}$.

For  homogeneous completion, from \eqs{pmissnmiss}{nmiss_additivePc} we  find
\bees
p_{\rm miss}(z,\hatO;{\cal R})&=&\frac{1}{N_{\rm miss}({\cal R})} \,   
\bar{n}_{\rm gal}\, 
\[ 1-P_{\rm compl}(z,\hatO)\]
\frac{dV_c}{dzd\Omega}\nn\\
&=&\frac{N_{\rm gal}({\cal R})}{N_{\rm miss}({\cal R})} \,  
\[ 1-P_{\rm compl}(z,\hatO)\]
\frac{1}{ V_c({\cal R}) }
\, \frac{dV_c}{dzd\Omega}\, ,
\ees
where, in the second line, we used
$\bar{n}_{\rm gal}=N_{\rm gal}({\cal R})/V_c({\cal R})$.
From \eq{fR} (with $w_{\alpha}=1$) we have 
$1-f_{\cal R}=N_{\rm miss}({\cal R})/N_{\rm gal}({\cal R})$,
and therefore, from \eq{p0pcatpmiss}, the prior $p_0(z,\hatO;{\cal R})$ in the case of homogeneous completion is given by
\be\label{p0fRPcompl}
p^{\rm hom}_0(z,\hatO;{\cal R})=f_{\cal R}\,  p_{\rm cat}(z,\hatO;{\cal R})+
\[ 1-P_{\rm compl}(z,\hatO)\]\, \frac{1}{V_c({\cal R})}\, \frac{dV_c}{dzd\Omega}
\, .
\ee
We next express $f_{\cal R}$ in terms of $P_{\rm compl}(z,\hatO)$, as follows. 
We first define
\be
\bar{n}_{\rm cat}(z,\hatO)=\frac{N_{\rm cat}({\cal S})}{V_c({\cal S})}\, .
\ee
This is the average density of the galaxies present in the catalog, within the region 
${\cal S}={\cal S}(z,\hatO;\Delta z,\Delta\Omega)$ (as before, we have not written explicitly the bin sizes $\Delta z,\Delta\Omega$ among the arguments of $\bar{n}_{\rm cat}$).
\Eq{Pcompletedef} can then be written as 
\be\label{Pcompletebarngal}
P_{\rm compl}(z,\hatO)= 
\frac{\bar{n}_{\rm cat}(z,\hatO)}{\bar{n}_{\rm gal}}\, .
\ee
Integrating this equation over the whole region ${\cal R}$ used to define $f_{\cal R}$, and dividing by
$V_c({\cal R})$, we get
\be\label{NcatsuNgalPc}
\frac{1}{V_c({\cal R})}\, \int_{\cal R}dV'_c\,  P_{\rm compl}(z',\hatO')=
\frac{N_{\rm cat}({\cal R})}{N_{\rm gal}({\cal R})}\, .
\ee
The right-hand side is nothing but $f_{\cal R}$ (for the weights $w_{\alpha}=1$ that we are considering here), so
\be\label{fRPcomplete}
f_{\cal R}=\frac{1}{V_c({\cal R})}\, \int_{\cal R}dV'_c\,  P_{\rm compl}(z',\hatO')\, .
\ee
Inserting this in \eq{p0fRPcompl} we get the  expression for the prior in the case of homogeneous completion, in the form
\be\label{p0isocompl}
p^{\rm hom}_0(z,\hatO;{\cal R})=\frac{1}{V_c({\cal R})}
\left\{
\[  \int_{\cal R}dV'_c\,  P_{\rm compl}(z',\hatO')\] 
p_{\rm cat}(z,\hatO;{\cal R}) +
\[ 1-P_{\rm compl}(z,\hatO)\]\,\, \frac{dV_c}{dzd\Omega} \right\}
\, .
\ee
For $P_{\rm compl}(z,\hatO)=1$ we recover 
$p_0(z,\hatO;{\cal R})=p_{\rm cat}(z,\hatO;{\cal R})$, as it should, while in the absence of a catalog, i.e.  for  $P_{\rm compl}(z,\hatO)=0$, we get back the  prior uniform in comoving volume. The above equations agree with those given in \cite{Fishbach:2018gjp}, while providing, in our view,  a more clear conceptual basis.

In \eq{p0isocompl} it is convenient to extract the  factors of $H_0$ that are implicit in $dV_c$ and $dV_c/dzd\Omega$. We write the Hubble parameter  as
\be\label{HzH0Ez}
H(z)=H_0E(z)\, ,
\ee 
so, for instance, in a  spatially flat cosmological model with 
matter density $\oma$, radiation density $\ora$ and
dark energy density $\rde(z)$, we have
$E(z)=[\oma (1+z)^3 +\ora (1+z)^4 + \rde(z)/\rho_0]^{1/2}$. Similarly, we write
\be\label{defdcom}
\dcom(z)\equiv c\int_0^z\, \frac{d\tilde{z}}{H(\tilde{z})}
= \frac{c}{H_0}\int_0^z\, \frac{d\tilde{z}}{E(\tilde{z})}\equiv \frac{c}{H_0}\, u(z)\, .
\ee
Then, from \eq{dVcdzdO2},
\be\label{dVzdOdexplicit}
\frac{dV_c}{dzd\Omega}= \(\frac{c}{H_0}\)^3\,  \frac{u^2(z)}{E(z)}\equiv 
\(\frac{c}{H_0}\)^3\,  j(z)\, .
\ee
We also define the dimensionless comoving volume ${\cal V}_c$ of the region ${\cal R}$ from
\be\label{Vcj}
V_c({\cal R})=\(\frac{c}{H_0}\)^3\, {\cal V}_c({\cal R})\, .
\ee
Notice that  $d{\cal V}_c=j(z)dzd\Omega$, i.e. $j(z)$ is the  Jacobian of the transformation from the dimensionless volume element 
$d{\cal V}_c$ to $dzd\Omega$.
We also introduce the angular average of the completeness,
\be\label{defPcomplz}
P_{\rm compl}(z)\equiv  \int \frac{d\Omega}{4\pi} \, P_{\rm compl}(z,\hatO)\, .
\ee
Then (assuming henceforth that the region ${\cal R}$ is a spherical region whose radius, in redshift space, is $z_{\cal R}$),
\be
\int_{\cal R}dV'_c\,  P_{\rm compl}(z',\hatO')= \(\frac{c}{H_0}\)^3\,  4\pi\,
\int_0^{z_{\cal R}}dz'\, j(z')P_{\rm compl}(z')\, ,
\ee
and
\eq{p0isocompl} can be rewritten as
\be\label{p0isocompldimensionless}
p^{\rm hom}_0(z,\hatO;{\cal R})=\frac{1}{{\cal V}_c(z_{\cal R})}
\left\{ \[4\pi\int_0^{z_{\cal R}}dz'\, j(z')P_{\rm compl}(z')\]
p_{\rm cat}(z,\hatO;{\cal R}) +
\[ 1-P_{\rm compl}(z,\hatO)\]\,j(z) \right\}
\, .
\ee
where, for the spherical region ${\cal R}$,  we have written ${\cal V}_c({\cal R})={\cal V}_c(z_{\cal R})$, and 
\be\label{Vczintj}
{\cal V}_c(z_{\cal R})=4\pi \int_0^{z_{\cal R}}dz\, j(z)\, .
\ee
We finally observe that (when using luminosity weighting, see below), in the local Universe, $P_{\rm compl}(z,\hatO)$ can be larger than one because of actual over-densities, see in particular Fig.~\ref{fig:Bcompleteness} in app.~\ref{sect:GLADE}, and similarly it can be below one just because of local under-densities. For each direction $\hatO$, we will therefore perform the completion only for $z>z_*(\hatO)$, where $z_*(\hatO)$ is
the largest value of $z$ for which  $P_{\rm compl}(z,\hatO)=1$. In other words, for $z<z_*(\hatO)$ the difference of $ P_{\rm compl}(z,\hatO)$ with respect to 1 will be attributed to actual over-densities and under-densities, while for $z>z_*(\hatO)$ we attribute them to the lack of completeness. In particular this means that, in the regions where $P_{\rm compl}(z,\hatO)>1$, we set
$n_{\rm miss}(\vx)=0$, rather than giving it a meaningless negative value, as would follow from a blind use of \eq{nmiss_additivePc}.
So, for $z<z_*$, we simply set $p_0(z,\hatO;{\cal R})=p_{\rm cat}(z,\hatO;{\cal R})$, which is formally equivalent to setting $P_{\rm compl}(z,\hatO)=1$,
while for 
$z\geq z_*(\hatO)$ we use the actual value of $P_{\rm compl}(z,\hatO)$ obtained from the catalog, and we 
use as prior $p_0(z,\hatO;{\cal R})$ the expression given in \eq{p0isocompldimensionless}.

\vspace{2mm}\noindent
{\bf Multiplicative completion}.
We now come back to the assumption that the missing galaxies are uniformly distributed 
within ${\cal S}$. This  is not the only natural possibility. If, for instance, within ${\cal S}$ there are  clusters, filaments and voids,  and the galaxy surveys have explored this volume uniformly, it could be more natural to assume that the galaxies that have been missed reside preferentially near high-density regions rather than, say, in the middle of a  void. In its most extreme form, this amounts to assuming that $n_{\rm miss}(z,\hatO)$ is proportional to $n_{\rm cat}(z,\hatO)$, i.e.  
\be\label{nmiss_multiplicative}
n_{\rm miss}(z',\hatO')=b({\cal S})\, n_{\rm cat}(z',\hatO')\, ,
\ee
where $z',\hatO'$ are the `microscopic' integration variables within ${\cal S}$ that appears in \eq{infopmissw1} and $b({\cal S})$ is
a multiplicative  `bias factor' that depends on the region ${\cal S}$, i.e. on the average redshift and direction of the region.
This factor is then determined from \eq{infopmissw1}, which gives
\be\label{1piub}
1+b({\cal S})=\frac{\bar{n}_{\rm gal}V_c({\cal S}) }{N_{\rm cat}({\cal S})}
=\frac{1}{P_{\rm compl}({\cal S})}\, .
\ee
We will refer to \eq{nmiss_multiplicative} as the `multiplicative' completion. Multiplicative completion can be seen as a quantitative formulation of the idea  \cite{Nair:2018ign,Bera:2020jhx} that, to perform the correlation of GW events with galaxy catalogs, it is not necessary to have a complete catalog, but it is enough to have a catalog that reproduces the clustering structure of the galaxies. 
An intermediate option, which might be  the best  justified physically, could be to distribute the missing galaxies with respect to the observed ones, according to the correlation length of the spatial distribution of the observed galaxies. In this work, however, we will limit ourselves to homogeneous and multiplicative completions.

Inserting \eq{nmiss_multiplicative} into \eqs{pmissnmiss}{p0pcatpmiss} we get, for the prior
$p_0(z,\hatO;{\cal R})$ in the case of multiplicative completion,
\be
p^{\rm multi}_0(z,\hatO;{\cal R})=\frac{N_{\rm cat}({\cal R})}{N_{\rm gal}({\cal R})}\, [1+b(z)]
p_{\rm cat}(z,\hatO;{\cal R})\, .
\ee
Using \eqs{NcatsuNgalPc}{1piub}, we finally get 
\be\label{p0multipcompl1}
p^{\rm multi}_0(z,\hatO;{\cal R})=
\frac{1}{V_c({\cal R})}\,  
\[\int_{\cal R}dV'_c\,  P_{\rm compl}(z',\hatO')\]\,   \frac{1}{P_{\rm compl}(z,\hatO)} 
p_{\rm cat}(z,\hatO;{\cal R})\, .
\ee
Observe that \eq{p0multipcompl1} is singular at the points  $(z,\hatO)$ where $P_{\rm compl}(z,\hatO)=0$ since, there, both the numerator  $p_{\rm cat}(z,\hatO;{\cal R})$ and the denominator $P_{\rm compl}(z,\hatO)$ vanish. In fact, multiplicative completion is only meaningful in regions where $P_{\rm compl}(z,\hatO)$ is above a minimum value, otherwise one would unreasonably amplify the effect of random detections of galaxies in regions that have been very under-covered by the surveys. For instance, it would not be meaningful to apply it to regions such as those obscured by the galactic plane, where the distribution of the relatively few galaxies observed might reflect, for instance, the existence of directions with less absorption in our Galaxy, such as the Baade window, and has nothing to do with the actual distribution of galaxies.  In contrast, homogeneous completion provides a neutral, uninformative prior that can be applied even in regions of low completeness (of course, in practice, when the completeness is too low the correlation of the GW events with the galaxy catalog will anyhow become uninformative).
When using multiplicative completion, we will therefore interpolate it with homogeneous completion, depending on the completeness of the region. A simple way to do so is to use, as a prior,
\bees
p^{\rm multi/hom}_0(z,\hatO;{\cal R})&\equiv& 
\theta\[ P_{\rm compl}(z,\hatO)-P_c^{\rm min}\]\, p^{\rm multi}_0(z,\hatO;{\cal R})\nn\\
&&+\theta\[P_c^{\rm min}- P_{\rm compl}(z,\hatO)\]\, p^{\rm hom}_0(z,\hatO;{\cal R})\, ,
\label{p0multihom}
\ees
where $\theta$ is the Heaviside theta function,  and
$P_c^{\rm min}$ is a threshold on the completeness, so that regions with 
$P_{\rm compl}(z,\hatO)>P_c^{\rm min}$ are completed multiplicatively, while
in regions with 
$P_{\rm compl}(z,\hatO)<P_c^{\rm min}$ we apply homogeneous completion.\footnote{In practice, we make two modifications to this formula: first, we interpolate smoothly, replacing the theta function by a  smoothed step function (chosen so that it raises from 0 to 1 around a value $P_{\rm compl}\simeq 0.1$ of the completeness, reaching a value 0.9 around $P_{\rm compl}\simeq 0.3$). Second, for simplicity,  in the occurrence of
$p_0^{\rm hom}$ that enters in \eq{p0multihom}, rather than using the full expression (\ref{p0fRPcompl}), we use
$p^{\rm hom}_0(z,\hatO;{\cal R})=V^{-1}_c({\cal R})\, dV_c/dzd\Omega$, which is the 
$P_{\rm compl}(z,\hatO)\ra 0$ limit of \eq{p0fRPcompl}, that is, we drop the contribution from catalog, which is small once the term $p^{\rm hom}_0(z,\hatO;{\cal R})$ takes over in \eq{p0multihom}.
\label{foot:smoothedtheta}}

\subsubsection{Completion with luminosity weighting}\label{sect:compl_lum_w}

We can  proceed similarly when  the weights $w_{\alpha}$ are equal to the B- or  K-band luminosity. 
The computation is the same as before, with the number of all galaxies per comoving volume, $n_{\rm gal}(\vx)$, replaced by the luminosity per comoving volume of all galaxies, that we denote by $l_{\rm gal}(\vx)$;
$n_{\rm cat}(\vx)$ is replaced  by
the luminosity per comoving volume of the  galaxies in the catalog, $l_{\rm cat}(\vx)$, and
$n_{\rm miss}(\vx)$ is replaced  by
the luminosity per comoving volume of the missed galaxies, $l_{\rm miss}(\vx)$, all  restricted to the desired band (and, possibly, with a lower cut on the luminosity of individual galaxies, see below).  
For all these quantities,  we define $L({\cal R})=\int_{\cal R} d^3x\, l(\vx)$. Then \eqst{pbarvsp}{pmissnmiss} are replaced by
\bees
p_0(z,\hatO;{\cal R})&=&\frac{l_{\rm gal}(\vx)}{L_{\rm gal}({\cal R})} 
\,   \frac{dV_c}{dzd\Omega}\, ,\label{pbarvspL}\\
p_{\rm cat}(z,\hatO;{\cal R})&=&\frac{l_{\rm cat}(\vx)}{L_{\rm cat}({\cal R})} 
\,   \frac{dV_c}{dzd\Omega}\, ,\label{pcatncatL}\\
p_{\rm miss}(z,\hatO;{\cal R})&=&\frac{l_{\rm miss}(\vx)}{L_{\rm miss}({\cal R})} 
\,   \frac{dV_c}{dzd\Omega}\, .\label{pmissnmissL}
\ees
The condition (\ref{infopmissw1}) is replaced by
\be\label{NgaldiR}
\frac{1}{V_c({\cal S})}\, 
\int_{\cal S} dV'_c\[ l_{\rm cat}(z',\hatO')+l_{\rm miss}(z',\hatO')\]=\bar{l}_{\rm gal}\, ,
\ee
where $\bar{l}_{\rm gal}$ is the average luminosity per comoving volume of a complete sample. This is  obtained  assuming that the luminosity of a complete sample follows 
the  Schechter luminosity function, 
\be\label{Schfun}
\Phi(L) dL =\phi^*\, \(\frac{L}{L^*}\)^{\alpha}\, e^{-L/L^*}\, \frac{dL}{L^*}\, ,
\ee
where $\phi^*$, $\alpha$ and $L^*$ are the Schechter parameters in the corresponding band. 
We  restrict to galaxies with luminosity above a given minimum value $L_{\rm cut}$.
Then, the average luminosity density of a sample of galaxies with $L>L_{\rm cut}$ is
\bees\label{intSch}
\bar{l}_{\rm gal}(L_{\rm cut})&\equiv& \int_{L_{\rm cut}}^{\infty} dL\, L\phi(L)  \nn\\
&=&
\phi^*L^*\int_{L_{\rm cut}/L^*}^{\infty} dx\, x^{\alpha+1} e^{-x}\nn\\
&=&\phi^*L^*\Gamma(\alpha+2,L_{\rm cut}/L^*)\, ,
\ees
where $\Gamma(\alpha,x)$ is the incomplete Gamma function. We will discuss in app.~\ref{sect:GLADE} the choice of Schechter parameters in the B and K bands, and the quality of the fit to the Schechter function of the galaxies in GLADE. The best choice of $L_{\rm cut}$ is determined by a compromise between the fact that, increasing this cut, we have less galaxies overall, and the fact that the remaining sample is more and more complete. As we discuss in app.~\ref{sect:GLADE}, for GLADE a good compromise is provided by  $L_{\rm cut}=0.6L^*$ (both in B-band and in K-band), and we will henceforth use this as a default choice.

The definition (\ref{Pcompletedef}) is then generalized to 
\be\label{PcompletedefLum}
P_{\rm compl}({\cal S};L_{\rm cut})\equiv  
\frac{L_{\rm cat}({\cal S};L_{\rm cut})}{\bar{l}_{\rm gal}(L_{\rm cut})V_c({\cal S}) }\, ,
\ee
where $L_{\rm cat}({\cal S};L_{\rm cut})$ is the total luminosity of the galaxy in the catalog, in the region ${\cal S}$, and with luminosity above $L_{\rm cut}$, and
the notation stresses also the dependence of the various quantities on the chosen lower cutoff on the luminosity. Using  cone completeness, the region ${\cal S}$ is written more explicitly as ${\cal S}(z,\hatO;\Delta z,\Delta\Omega;L_{\rm cut})$, and, correspondingly,  
$P_{\rm compl}({\cal S};L_{\rm cut})$ can be written as
$P_{\rm compl}(z,\hatO;\Delta z,\Delta\Omega;L_{\rm cut})$. However, again, to keep the notation lighter, we will often simply write it as $P_{\rm compl}(z,\hatO)$; for mask completeness we use again the notation ${\cal S}_I$.

In terms of these variables, 
the computation of $p_0(z,\hatO;{\cal R})$ is identical to  that performed for $w_{\alpha}=1$.  So, in the case of isotropic completion, we get again \eq{p0isocompldimensionless}, 
while, assuming   $l_{\rm miss}(z',\hatO')=b({\cal S})\, l_{\rm cat}(z',\hatO')$,
 we get again
\eq{p0multipcompl1}, where, in both cases, 
 $p_{\rm cat}(z,\hatO;{\cal R})$ now includes the weights $w_{\alpha}$ for the corresponding band, see \eq{pcatdelta}, and 
$P_{\rm compl}(z)$ is now given by \eq{PcompletedefLum}. Again, in the case of multiplicative completion, we will then interpolate it with homogeneous completion as in \eq{p0multihom}.

\red{Observe that the assignment of absolute luminosities to galaxies assumes a cosmology, in particular a value of $H_0$. However, changing $H_0$ only produces an overall 
multiplicative shift of the weights $w_{\alpha}$,  which  cancels among the numerators and denominators  in  \eqss{p0Ntot}{pcatdelta}{pmissdelta}, and therefore does not affect the corresponding priors. There are also possible systematics in assuming a fixed luminosity function, since the latter 
could have a redshift-dependent evolution. Nevertheless, the error that this assumption may introduce is negligible at the current level of precision.}

\subsection{The  normalization factors  $\beta(H_0)$ and $\beta(\Xi_0)$}\label{sect:beta}

We now turn to the normalization factor, treating in parallel   the cases  $\lambda'=\{H_0\}$ and $\lambda'=\{\Xi_0\}$. We use \eqs{betadL}{betaXi}, that we rewrite, respectively, as
\be\label{betadLR}
\beta(H_0)=\int_0^{z_{\cal R}}dz \int d\Omega \, P_{\rm det}[d_L(z,H_0),\hatO]\, p_0(z,\hatO)\, ,
\ee
and
\be\label{betaXiR}
\beta(\Xi_0)=\int_0^{z_{\cal R}}dz \int d\Omega \, P_{\rm det}[\dgw(z,\Xi_0),\hatO]\, p_0(z,\hatO)\, ,
\ee
where we have used the factorization (\ref{factorp0}) and we defined a detection probability averaged over the $\theta'$ parameters,\footnote{Note that, because of the normalization (\ref{normptheta}), the integral over $d\theta'$ of  $\tilde{p}_0(\theta')$ actually corresponds to {\em averaging} over the $\theta'$ parameters. For instance, if $\theta'=\cos\iota$ is the inclination angle of the orbit, and we choose a prior flat in 
$\cos\iota$, then \eq{normptheta} gives
$\tilde{p}_0(\cos\iota)=1/2$, and
$\int d\theta' \, \tilde{p}_0(\theta')$ becomes 
$(1/2)\int_{-1}^{1}d\cos\iota$.\label{foot:normtheta}}
\be\label{pdetthetaH0}
P_{\rm det}[d_L(z,H_0),\hatO]=\int d\theta'\, P_{\rm det}[d_L(z,H_0),\hatO,\theta']\tilde{p}_0(\theta')\, ,
\ee
and, respectively,
\be\label{pdetthetaXi0}
P_{\rm det}[\dgw(z,\Xi_0),\hatO]=\int d\theta'\, P_{\rm det}[\dgw(z,\Xi_0),\hatO,\theta']\tilde{p}_0(\theta')\, .
\ee
As discussed in sect.~\ref{sect:cat}, $p_0(z,\hatO)$ actually depends also on the region ${\cal R}$, which must be specified in order to define it. However, from the discussion in sect.~\ref{sect:cat} and App.~\ref{app:R}, it follows that the final result for the posterior of $H_0$ (or $\Xi_0$) will not depend on the choice of ${\cal R}$, as long as it contains the localization region of all the detectable GW events (in the whole range of priors considered for $H_0$ or $\Xi_0$). In \eqs{betadLR}{betaXiR}, to keep the notation lighter, we have not written explicitly
the argument ${\cal R}$  in $p_0(z,\hatO;{\cal R})$ and in $\beta(H_0;{\cal R})$,
$\beta(\Xi_0;{\cal R})$. Note also that the integral over $dz$ in \eqs{betadLR}{betaXiR} extends up to the maximum redshift $z_{\cal R}$ of the region ${\cal R}$
which has been used to define $p_0(z,\hatO;{\cal R})$.   \red{Note that, in \eq{pdetthetaH0}, if among the parameters $\theta'$ we include the source frame masses $m_i$, there is a further  dependence on the redshift, that we have not written explicitly, entering from the fact that the detection probability depends on the detector frame masses
$m_i^{(d)}=m_i(1+z)$, see also footnote~\ref{foot:H0_from_mi} on page~\pageref{foot:H0_from_mi}.}

Before entering \label{page:geom_effect}
in the details of the computations, it is interesting to understand physically the meaning of the normalization factor $\beta(H_0)$ [or, similarly, of $\beta(\Xi_0)$]. Its role is to offset a dependence on $H_0$ in the numerator of \eq{pDizO}, which is of purely geometric nature, and that is present  even when the GW events are completely uncorrelated to the galaxy distribution. Indeed, changing $H_0$ has the effect of changing the redshift of the GW event (whose position has been determined through a measurement of $d_L$) with respect to the positions of the galaxies in the catalog, that, in contrast, have been determined by a direct measurement of their  redshift, independently of $H_0$. Thus, for instance,  for a given measured value of $d_L$, increasing  $H_0$  has the effect of increasing the redshift inferred for the  GW event.  Increasing the redshift, also the prior $p_0(z)$ increases; for instance,   in a distribution  of galaxies  uniform in comoving volume,  where $n_{\rm gal}(z,\hatO)$ is independent of $z$, $p_0(z)$ increases
as $\dcom^2(z)/H(z)$, see \eqs{dVcdzdO2}{pbarvsp},  and therefore as $z^2$ at low $z$. As a result,  the correlation between the GW probability $p({\cal D}^i_{\rm GW}|z, \hatO,H_0)$ and the  prior $p_0(z,\hatO)$ necessarily increases  if we increase $H_0$, since we are moving the events (in redshift) toward a region where the prior is higher  (and, conversely, the correlation  decreases as we decrease $H_0$). This is a geometric effect that  takes place even if the GW event is totally uncorrelated with the galaxy distribution: for instance, it would take place even if we would randomly generate fictitious GW events, or for sources, such as primordial BHs, that might not be correlated with the distribution of luminous matter. The role of the normalization factor $\beta(H_0)$  is to compensate for such `trivial' geometric effects. Note that this volume effect could give highly non-trivial information when it comes to comparing the overall observed rate of events with the theoretical predictions as a function of $H_0$ or $\Xi_0$. This will be discussed in sect.~\ref{sect:rate}, where we will see that, unfortunately, current astrophysical uncertainties on the rate make it difficult to exploit this information.

In the following subsections we will discuss the computation of $\beta(H_0)$  [or $\beta(\Xi_0)$] with increasing levels of sophistication. In the simplest approximation, one assumes a homogeneous distribution of galaxies and a monochromatic population of sources \red{(i.e., binaries all with the same  source-frame masses)}. The corresponding computation, performed in
\cite{Chen:2017rfc}, is reviewed in sect.~\ref{sect:betaChenetal}.
To go beyond this result, we  first consider
the effect of the approximation that the distribution of galaxies is uniform in comoving volume. In principle,  once made a choice for the prior $p_0(z,\hatO;{\cal R})$, the same  choice  should  be consistently used both in the  numerator  of 
\eq{pDizO}  and in the denominator [i.e. in $\beta(H_0)$].  In the numerator of
\eq{pDizO} we certainly cannot use a uniform density of galaxies: the meaning of the whole procedure is precisely to correlate the localization of the GW event, expressed by $p({\cal D}^i_{\rm GW}|d_L(z,H_0),\hatO)$,  with the inhomogeneities of the galaxy distribution (i.e. with the actual positions of galaxies) expressed by $p_0(z,\hatO)$, to see which values of $H_0$ (or of $\Xi_0$, for modified gravity) maximize this correlation. If we take a uniform distribution of galaxies in the numerator, there is nothing to correlate with.  For the denominator, however, the approximation of uniform distribution is a meaningful  first
approximation, since the correlation only appears at the numerator.
Still, it is necessary to quantify the effect  of this approximation, especially in view of the sub-percent accuracy that is the final aim of studies of $H_0$. In sect.~\ref{sect:betafromcat} we will see that the computation
of sect.~\ref{sect:betaChenetal}
 admits a simple and elegant generalization when   the prior is taken to be that given by the full expression (\ref{p0isocompldimensionless}) for homogeneous completion, or by (\ref{p0multipcompl1}) for multiplicative completion.

In sect.~\ref{sect:MC} we will then include the effect of a more realistic detection model, for a population of BBHs with a distribution of masses. To some extent, this can be done with a 
semi-analytic approach that is quite instructive, and that we will discuss  in app.~\ref{app:betafit}. Eventually, however, we will see that, in order to obtain reliable results, it will be necessary to restrict to GW events that fall in regions of sufficiently high completeness of the catalog used. This introduces a further selection effect, that must be taken into account in the computation of $\beta(H_0)$ or $\beta(\Xi_0)$, and that can only be reliably included with a full Monte Carlo (MC) computation. \red{Furthermore, while the semi-analytic method that we will present uses only  the inspiral part of the waveform, the use of full inspiral-merger-ringdown waveforms is necessary for BBHs with a large total mass (which is the case   in  particular for many detections of the O3a run),  since these  have only few inspiral cycles in the detector bandwidth. Our MC, which includes both the selection effect due to completeness and full inspiral-merger-ringdown waveforms, is described in app.~\ref{sect:ourMC}, and our final results, discussed in sect.~\ref{sect:results} only make use of the MC computation of $\beta(H_0)$ and $\beta(\Xi_0)$.}

\subsubsection{Computation of $\beta$ assuming a homogeneous galaxy distribution}\label{sect:betaChenetal}

The first computation of $\beta(H_0)$ was performed in \cite{Chen:2017rfc}. In that computation the prior 
$p_0(z,\hatO)$ that enters in $\beta(H_0)$, see \eq{betadLR}, has been taken to be that obtained from a  distribution of galaxies uniform in comoving volume, i.e.
\be\label{p0Fish}
p_0(z,\hatO;{\cal R})=p_0(z;{\cal R})=\frac{1}{V_c({\cal R})}
\, \frac{dV_c}{dzd\Omega} 
\, ,
\ee
where $dV_c/dzd\Omega$ is given in \eq{dVcdzdO2}.
This is formally equivalent to setting $P_{\rm compl}(z,\hatO)=0$ in \eq{p0isocompl}. 
Using \eqss{dVzdOdexplicit}{Vcj}{Vczintj}, and taking as usual for ${\cal R}$ a sphere in redshift space with radius $z_{\cal R}$, we can rewrite  \eq{p0Fish} more explicitly as
\be
p_0(z;z_{\cal R})=\frac{j(z)}{4\pi \int_0^{z_{\cal R}}dz' j(z')}\, ,
\ee
with $j(z)$ defined by \eqst{HzH0Ez}{dVzdOdexplicit}. 
The computation  performed in \cite{Chen:2017rfc} is also  restricted to a monochromatic population of sources, e.g. BNS with $(1.4+1.4)\msun$. In this case one can use a simplified model  for detection, 
\be\label{pdetTheta}
P_{\rm det}(d_L)\equiv \int d\Omega \, 
P_{\rm det}(d_L,\hatO)=
\theta\( d_{\rm max}-d_L\)\, ,
\ee
where  $\theta$ is the Heaviside theta function and
$d_{\rm max}$ is the  value of the luminosity distance  (averaged over the solid angle and over the $\theta'$ parameters, so in particular over the source inclination)  to which the  given monochromatic population of source could be detected, given the sensitivity of the detector network.\footnote{Observe that our  normalization (\ref{pcatzOnorm}) of $p_0(z,\hatO;{\cal R})$ is such that the integral of
$p_0(z,\hatO;{\cal R})$ over $d\Omega$ already corresponds to an average (rather than a sum) over the solid angle, since a flat  prior over the angles (for all $z$) corresponds to $p_0(z,\hatO)=
p_0(z)\times [1/(4\pi)]$; compare also with footnote~\ref{foot:normtheta}.}
 Equivalently, in terms of redshift, we can write
\be\label{pdetzmaxH0}
P_{\rm det}(z;H_0)=\theta\[z_{\max}(H_0;d_{\rm max})-z\]\, ,
\ee
where $z_{\max}(H_0;d_{\rm max})$ is the solution of $d_L(z;H_0)= d_{\rm max}$.
With these assumptions, one finds~\cite{Chen:2017rfc}
\bees
\beta_0(H_0;\dmax)&=&\int_0^{z_{\cal R}}dz \int d\Omega \, P_{\rm det}[d_L(z,H_0),\hatO]\, p_0(z,\hatO)\nn\\
&=&\frac{1}{V_c({\cal R})}\,  \int_0^{z_{\cal R}}dz\int   d\Omega \, \frac{dV_c}{dzd\Omega} 
\theta\[z_{\max}(H_0;d_{\rm max})-z\]\nn\\
&=&\frac{V_c[H_0,z_{\max}(H_0;d_{\rm max})]}{V_c(H_0,z_{\cal R})}\, .
\label{Chenbeta}
\ees
The notation $\beta_0(H_0;\dmax)$ in this expression stresses that, with this simplified detection model, $\beta(H_0)$ inherits a dependence on the choice of $\dmax$, and we also added a suffix `zero' to distinguish the result of this `zero-th' order approximation from the more elaborate results that we will present below.
In the above expression,
$V_c(H_0,z_{\max})$ is the comoving volume of a sphere reaching the redshift 
$z_{\max}$, and we have similarly written $V_c({\cal R})$ as $V_c(H_0,z_{\cal R})$.
The notation $V_c[H_0,z_{\max}(H_0;d_{\rm max})]$ emphasizes that  this quantity depends on $H_0$ both explicitly, through the overall $(c/H_0)^3$ factor in \eq{Vcj}, and through $z_{\max}(H_0;d_{\rm max})$. However, the explicit $(c/H_0)^{3}$ factors cancel between $V_c[H_0,z_{\max}(H_0;d_{\rm max})]$
and  $V_c(H_0,z_{\cal R})$, so in the end the dependence on $H_0$ is only through $z_{\max}(H_0;d_{\rm max})$. We can make this explicit by rewriting the result
in terms of the dimensionless comoving volumes ${\cal V}_c$ defined 
in \eq{Vcj},
\bees
\beta_0(H_0;\dmax)&=&\frac{{\cal V}_c[z_{\max}(H_0;d_{\rm max})]}{{\cal V}_c(z_{\cal R})}\nn\\
&=&\frac{ \int_0^{z_{\max}(H_0;d_{\rm max})}dz\, j(z)}{ \int_0^{z_{\cal R}}dz\, j(z)}
\, .\label{betaHChen}
\ees
For small values of $z_{\rm max}$, such that the Hubble law is valid, from \eqs{defdcom}{dVzdOdexplicit} we have $u(z)\simeq z$ and $E(z)\simeq 1$  so $j(z)\simeq z^2$, and  $z_{\rm max}(H_0;d_{\rm max})\simeq H_0d_{\rm max}/c$; then \eq{betaHChen} gives
\be
\beta_0(H_0;\dmax)\simeq \frac{z^3_{\rm max}(H_0;d_{\rm max})}{3 \int_0^{z_{\cal R}}dz\, j(z)}
\simeq \frac{(H_0d_{\rm max}/c)^3}{3 \int_0^{z_{\cal R}}dz\, j(z)}\, ,
\ee
so $\beta_0(H_0;\dmax)\propto H_0^3$~\cite{Chen:2017rfc}. However, the limit of small redshift  is not appropriate in most situations, so the full expression (\ref{betaHChen}) has rather been used in the literature.

\begin{figure}[t]
\centering
\includegraphics[width=0.49\textwidth]{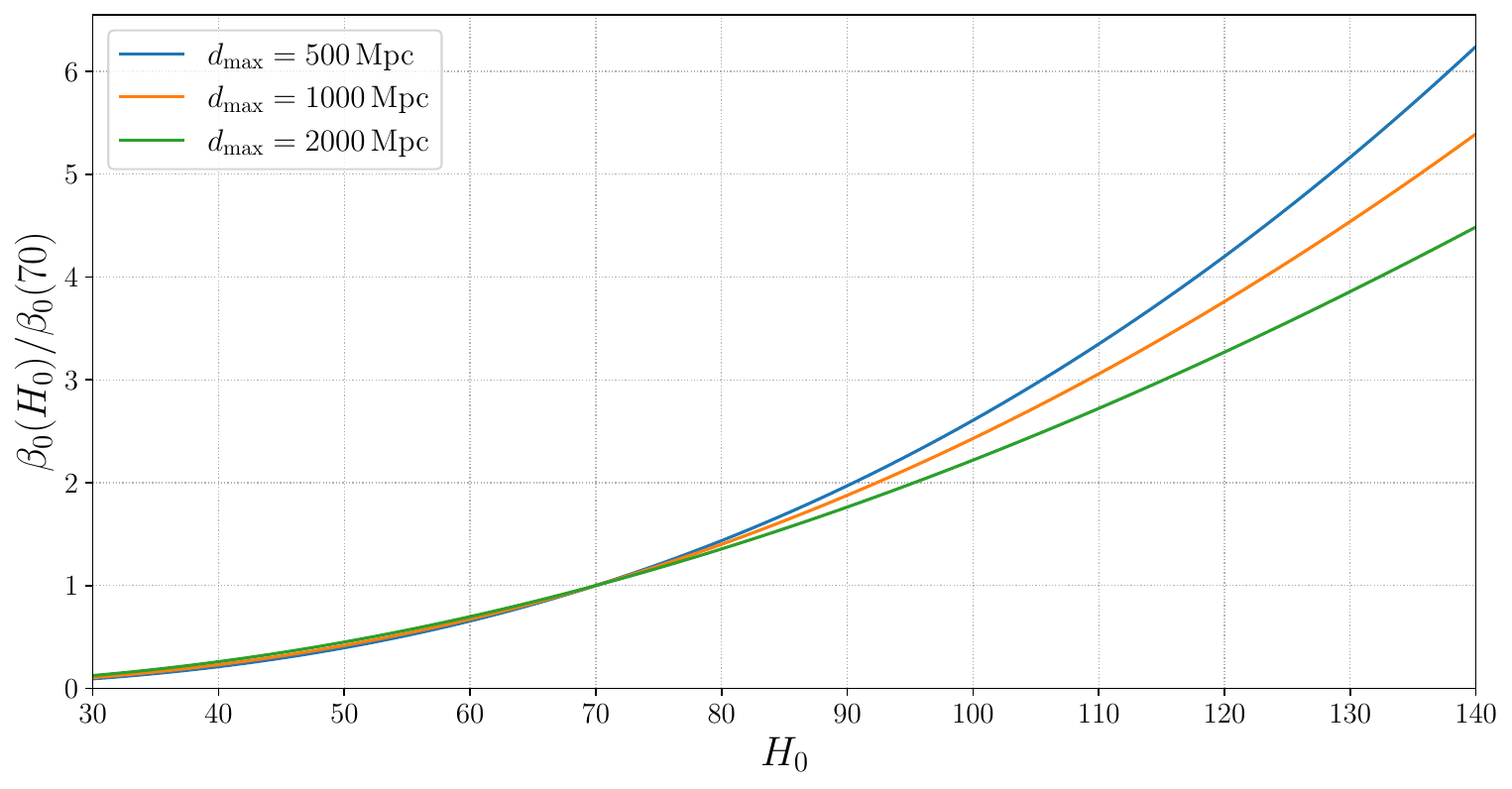}
\includegraphics[width=0.49\textwidth]{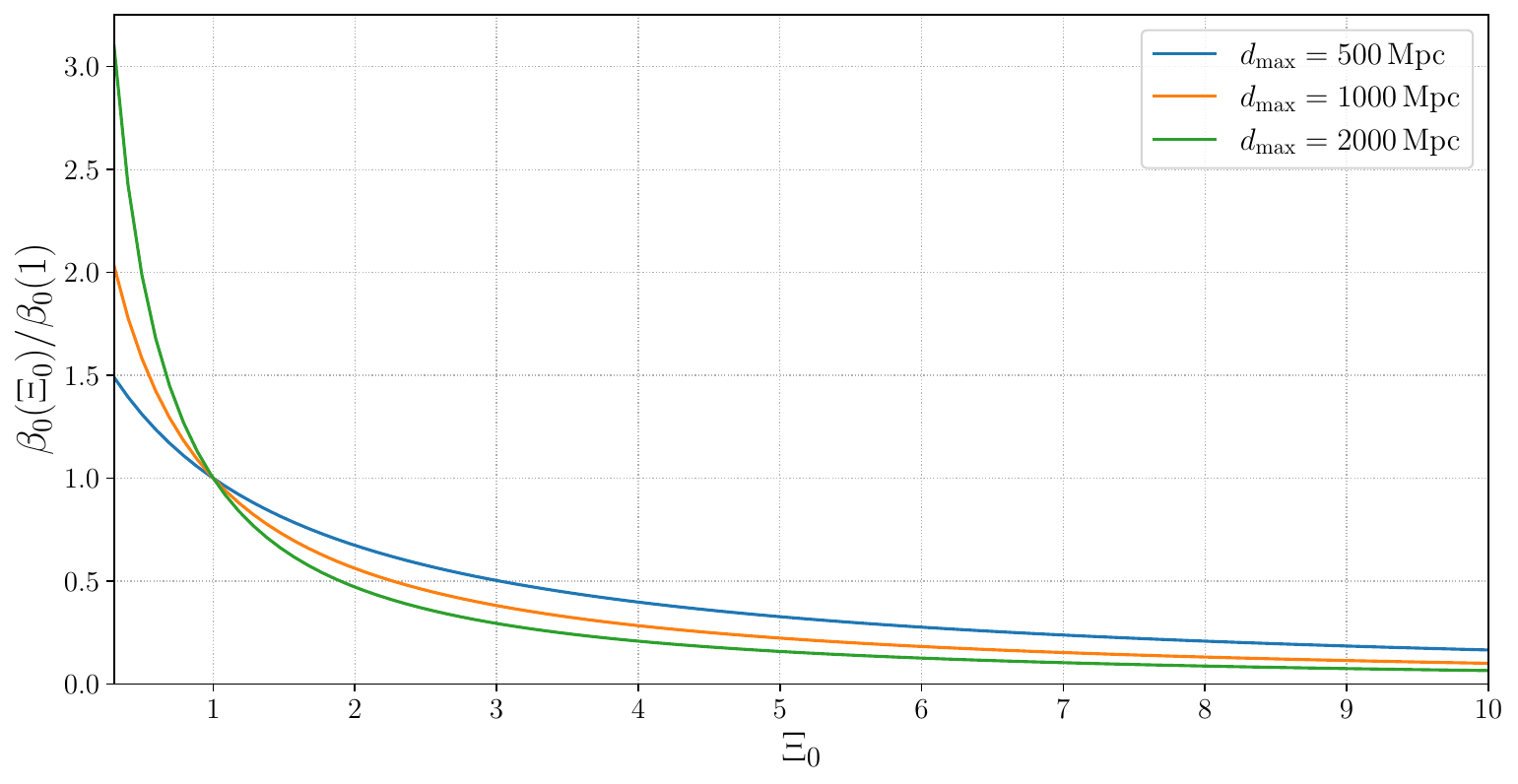}
\caption{The functions $\beta_0(H_0)$ (left panel) and $\beta_0(\Xi_0)$ (right panel)
computed as in \eqs{betaHChen}{betaXiChen}, respectively. In order to show the results for different $d_{\rm max}$ on the same scale, we plot $\beta_0(H_0)$ normalized to its value at $H_0=70~{\rm km}\, {\rm s}^{-1}\, {\rm Mpc}^{-1}$, and
$\beta_0(\Xi_0)$ normalized to its value at $\Xi_0=1$. In the left panel, as well as in all subsequent figures involving $H_0$, the units of the horizontal axis are ${\rm km}\, {\rm s}^{-1}\, {\rm Mpc}^{-1}$.}
\label{fig:beta_Fish}
\end{figure}

Note also that, in \eq{betaHChen}, we made use of the fact that, by definition, $z_{\cal R}$ is chosen larger than $z_{\rm max}$ otherwise, formally, 
the integral over $z$  would give
$\beta_0(H_0;\dmax)={\rm min}(1, {\cal V}_c[z_{\max}(H_0;d_{\rm max})]/{\cal V}_c(z_{\cal R}))$.
However, the condition $z_{\cal R}>z_{\rm max}$ is necessary for the consistency of the formalism, as we see from the derivation in app.~\ref{app:R}, and the apparent saturation of $\beta$ is just an artifact due to the fact that, if we take $z_{\cal R}<z_{\rm max}$, we miss some events in the count.

The corresponding result for $\beta_0(\Xi_0;\dmax)$ is trivially obtained replacing  
$z_{\max}(H_0;d_{\rm max})$ by  $z_{\max}(\Xi_0;d_{\rm max})$, defined as  the solution of $\dgw(z;\Xi_0)= d_{\rm max}$, so
\be\label{betaXiChen}
\beta_0(\Xi_0;\dmax)=\frac{ \int_0^{z_{\max}(\Xi_0;d_{\rm max})}dz\, j(z)}{ \int_0^{z_{\cal R}}dz\, j(z)}
\, .
\ee
The result, for a few choices of $ d_{\rm max}$, are shown in 
Fig.~\ref{fig:beta_Fish}, for $\beta_0(H_0)$ (left panel) and $\beta_0(\Xi_0)$ (right panel).

\subsubsection{Computation of  $\beta$  including catalog completion}\label{sect:betafromcat}

\vspace{2mm}\noindent
{\em Homogeneous completion}.
We now generalize this computation to the full expression of the prior which includes the completion, still keeping for the moment the simplified detection model (\ref{pdetzmaxH0}). 
We begin with homogeneous completion, so that the prior is given by  \eq{p0isocompldimensionless}.
Inserting \eq{p0isocompldimensionless} into 
\eq{betadLR}, and using the detection model (\ref{pdetzmaxH0}),
we obtain [setting $z_{\cal R}$ sufficiently large, so that  $z_{\rm max}(H_0;d_{\rm max})<z_{\cal R}$ for the range of $H_0$ considered]
\bees
\beta^{\rm hom}(H_0;\dmax)&=&\frac{\int_0^{z_{\cal R}}dz'\, j(z')P_{\rm compl}(z')}{\int_0^{z_{\cal R}}dz'\, j(z')} 
\int_0^{z_{\rm max}(H_0;d_{\rm max})}dz \int d\Omega \, p_{\rm cat}(z,\hatO;{\cal R})\nn\\
&&+\frac{1}{\int_0^{z_{\cal R}}dz'\, j(z')} \,  \int_0^{z_{\rm max}(H_0;d_{\rm max})}dz\, j(z) \,
\[1-P_{\rm compl}(z)\]\, ,
\ees
where we also used \eqs{defPcomplz}{Vczintj}.
Inserting  the expression (\ref{pcatdelta}) for $p_{\rm cat}(z,\hatO)$, we get our final result,
\bees
\beta^{\rm hom}(H_0;\dmax)&=&\(\frac{\int_0^{z_{\cal R}}dz'\, j(z')P_{\rm compl}(z')}{\int_0^{z_{\cal R}}dz'\, j(z')} \)\, 
\frac{\sum_{\alpha: z_{\alpha}<z_{\rm max}(H_0;d_{\rm max})}w_{\alpha}}
{\sum_{\alpha=1}^{N_{\rm cat}({\cal R})} w_{\alpha}}
\nn\\
&&+\frac{1}{\int_0^{z_{\cal R}}dz'\, j(z')} \,  \int_0^{z_{\rm max}(H_0;d_{\rm max})}dz\, j(z) \,
\[1-P_{\rm compl}(z)\]\, ,
\label{betaH0simpledetectionmodel}
\ees
where, in the first term, the sum in the numerator is restricted to the galaxies with redshift $z_{\alpha}<z_{\rm max}(H_0;d_{\rm max})$, while that in the denominator is over all the galaxies of the catalog  in the region ${\cal R}$.
If one rather uses \eq{pcatN} instead of the Dirac delta's, we remain with a convolution in redshift involving the likelihoods for the $z_{\alpha}$.

In the limit $P_{\rm compl}(z,\hatO)\ra 0$, we recover \eq{betaHChen}. In the opposite  limit
$P_{\rm compl}(z,\hatO)\ra 1$ we get
\be
\beta^{\rm hom}(H_0;\dmax)= \frac{\sum_{\alpha: z_{\alpha}<z_{\rm max}(H_0;d_{\rm max})}w_{\alpha}}
{\sum_{\alpha=1}^{N_{\rm cat}({\cal R})} w_{\alpha}}\, .
\ee
For $w_{\alpha}=1$, this reduces to 
\be
\beta^{\rm hom}(H_0;\dmax)=\frac{N_{\rm cat}(z<z_{\rm max})}{N_{\rm cat}(z<z_{\cal R})}\, ,
\ee 
where
$N_{\rm cat}(z<z_{\rm max})$ is the number of galaxies  in the catalog with $z<z_{\rm max}(H_0)$, and
$N_{\rm cat}(z<z_{\cal R})$ is the number of galaxies  in the catalog with $z<z_{\cal R}$. If the galaxies in the catalog are distributed uniformly in comoving volume, this is again the same as the ratio
$V_c[z_{\max}(H_0)]/V_c(z_{\cal R})$, so we expect that \eq{betaHChen} should be a reasonable zero-th order approximation.

It is  straightforward to repeat these computations for $\beta(\Xi_0)$. The only modification is that  
$z_{\rm max}(H_0;d_{\rm max})$ is replaced by 
$z_{\rm max}(\Xi_0;d_{\rm max})$, which  is defined as the solution of
\be\label{dgwdmaxXi0}
\dgw ( z_{\rm max},\Xi_0 )=d_{\rm max}\, .
\ee
Therefore
\bees
\beta^{\rm hom}(\Xi_0;\dmax)&=&\(\frac{\int_0^{z_{\cal R}}dz'\, j(z')P_{\rm compl}(z')}{\int_0^{z_{\cal R}}dz'\, j(z')} \)\, 
\frac{\sum_{\alpha: z_{\alpha}<z_{\rm max}(\Xi_0;d_{\rm max})}w_{\alpha}}
{\sum_{\alpha=1}^{N_{\rm cat}({\cal R})} w_{\alpha}}
\nn\\
&&+\frac{1}{\int_0^{z_{\cal R}}dz'\, j(z')} \,  \int_0^{z_{\rm max}(\Xi_0;d_{\rm max})}dz\, j(z) \,
\[1-P_{\rm compl}(z)\]\, .
\label{betaXi0simpledetectionmodel}
\ees


\vspace{2mm}\noindent
{\em Multiplicative completion}. We now consider multiplicative completion (at first without interpolation with homogeneous completion), so that the prior is given by  \eq{p0multipcompl1}.  Repeating the same steps as above, \eq{betaH0simpledetectionmodel} is replaced by
\be\label{betamultiplicative2}
\beta^{\rm multi}(H_0;\dmax)=
\(\frac{\int_0^{z_{\cal R}}dz'\, j(z')P_{\rm compl}(z')}{\int_0^{z_{\cal R}}dz'\, j(z')} \)\, 
\frac{\sum_{\alpha: z_{\alpha}<z_{\rm max}(H_0;d_{\rm max})}w_{\alpha}/P_{\alpha}}{\sum_{\alpha=1}^{N_{\rm cat}({\cal R})}w_{\alpha}}\, .
\ee
where we have defined
\be\label{defPalpha}
P_{\alpha}\equiv P_{\rm compl}(z_{\alpha},\hatO_{\alpha})\, .
\ee
The factor $1/P_{\alpha}$ in \eq{betamultiplicative2} reflects the fact that, within multiplicative completion, in regions where the completeness is smaller than one each observed galaxy is assumed to be representative of many more galaxies, and is therefore weighted with an extra factor $1/P_{\alpha}$. 
Once we interpolate between multiplicative and homogeneous completion, using \eq{p0multihom} as a prior, we get

\bees
\beta^{\rm multi/hom}(H_0;\dmax)&=&\(\frac{\int_0^{z_{\cal R}}dz'\, j(z')P_{\rm compl}(z')}{\int_0^{z_{\cal R}}dz'\, j(z')} \)\, \label{betaH0multihomsdm}\\
&&\hspace*{-30mm}\times\frac{\sum_{\alpha: z_{\alpha}<z_{\rm max}(H_0;d_{\rm max})} 
\left\{\theta[ P_{\rm compl}(z_{\alpha},\hatO_{\alpha})-P_c^{\rm min}] w_{\alpha}/P_{\alpha}+ 
\theta[ P_c^{\rm min}-P_{\rm compl}(z_{\alpha},\hatO_{\alpha})] w_{\alpha} \right\}}
{\sum_{\alpha=1}^{N_{\rm cat}({\cal R})} w_{\alpha}}
\nn\\
&&\hspace*{-30mm}
+\frac{1}{\int_0^{z_{\cal R}}dz'\, j(z')} \,  \int_0^{z_{\rm max}(H_0;d_{\rm max})}dz \, j(z)\,
\int \frac{d\Omega}{4\pi}\, 
\theta[ P_c^{\rm min}-P_{\rm compl}(z,\hatO)] \,
\[1-P_{\rm compl}(z,\Omega)\]\, .\nn
\ees
The corresponding expressions for $\beta(\Xi_0;\dmax)$ are obtained again replacing
$z_{\rm max}(H_0;d_{\rm max})$ by $z_{\rm max}(\Xi_0;d_{\rm max})$.

\subsubsection{Inclusion of the distribution of source parameters and  more realistic detection model}\label{sect:MC}

The  computation of $\beta(H_0)$ and $\beta(\Xi_0)$ discussed above still has a  main limitation, particularly when applied to an ensemble of $N_{\rm obs}$ observed events, due to the simplified detection model (\ref{pdetTheta}). 
For an ensemble of events with different parameters, such as masses, inclination, etc., there is no  notion of maximum detection distance.
The distance to which an event can be detected depends strongly on its parameters, in particular on the masses of the components of the binary system, and on the inclination of the orbit. 

The most direct way of including these effects is to perform a Monte Carlo (MC) evaluation of $\beta$  (the technical detail sof our MC are discussed in app.~\ref{sect:ourMC}) although  one can get several insights also with a semi-analytic approximation, that we will discuss in app.~\ref{app:betafit}. Both for the semi-analytic approximation and for the MC evaluation,
we consider  BBHs with a  distribution of component masses $\tilde{p}_0(m_1,m_2)$ (we use a tilde to distinguish it from the various other probability distributions that entered above). The formalism can be developed for an arbitrary choice of $\tilde{p}_0(m_1,m_2)$, but, for presenting the numerical results, we will  use a broken power-law distribution, which is one of the models  proposed in \cite{Abbott:2020niy}, where it is shown that it provides a good fit to the ensemble of BBHs observed in the O1,O2 and O3a LIGO/Virgo runs. This is given by
\be
\tilde{p}_0(m_1,m_2)dm_1dm_2=\tilde{p}_0(m_1)\tilde{p}_0(m_2|m_1)dm_1dm_2\, ,
\ee
where $m_2<m_1$. The function $\tilde{p}_0(m_1)$ is chosen to follow a broken power law, $\tilde{p}_0(m_1)\propto m_1^{-\gamma_1}$ for
$m_{\rm min}<m_1<m_{\rm break}$ and $\tilde{p}_0(m_1)\propto m_1^{-\gamma_2}$ for
$m_{\rm break}<m_1<m_{\rm max}$, while the conditional distribution of the lighter mass, $\tilde{p}_0(m_2|m_1)$, is taken to be proportional to $(m_2/m_1)^{b_q}$, for some parameters $\gamma_1,\gamma_2,b_q$.\footnote{The parameters $\gamma_1$, $\gamma_2$, $b_q$ are denoted by $\alpha_1$, $\alpha_2$  and $\beta_q$ in
\cite{Abbott:2020niy}. We change notation to avoid clashes with our use of $\alpha$ to label the galaxies and of $\beta$ for the functions $\beta(H_0)$, $\beta(\Xi_0)$.}
The distribution is taken to vanish outside the interval $[m_{\rm min},m_{\rm max}]$. More precisely, for $\tilde{p}_0(m_2|m_1)$ (writing explicitly its normalization factor, that depends on $m_1$)  we have 
\be
\tilde{p}_0(m_2|m_1)=\frac{1+b_q}{m_1}\, 
\frac{(m_2/m_1)^{b_q}}{1-(m_{\rm min}/m_1)^{1+b_q} }\, .
\ee
For $b_q=0$, corresponding to a conditional distribution flat in $m_2$, this reduces to
$\tilde{p}_0(m_2|m_1)=1/(m_1-m_{\rm min})$~\cite{LIGOScientific:2018mvr}. Then
\be\label{brokenpowerlaw}
\tilde{p}_0(m_1,m_2)=  A\,
\frac{ (m_2/m_1)^{b_q}}{m_1 [1-(m_{\rm min}/m_1)^{1+b_q}] }\, \times
\left\{\begin{array}{lc}
(m_1/m_{\rm break})^{-\gamma_1} & \hspace{10mm} (m_{\rm min}<m_1<m_{\rm break})\\
(m_1/m_{\rm break})^{-\gamma_2} &\hspace{10mm}  (m_{\rm break}<m_1<m_{\rm max})
\end{array}\right. \, ,
\ee
where $A$ is an overall normalization constant independent of $m_1,m_2$, and we have enforced continuity at $m_1=m_{\rm break}$. The values of the various parameters that enter the distribution are determined by a fit to the masses of the observed BBHs  and depend on several details such as, for instance, on whether 
GW190814, with its very  low secondary mass, is considered a binary BH or not. \red{The values obtained from the fit to the GWTC-2 data are}\footnote{These values, though not explicitly reported in the
 LVC population analysis paper \cite{Abbott:2020gyp},
are obtained from the associated
data release, available at \url{https://dcc.ligo.org/LIGO-P2000434/public}. These are the values obtained including also GW190814 in the fit.}
\be\label{mminbmax}
m_{\rm min}=2.2^{+0.27}_{-0.20}\, \msun\, ,\qquad m_{\rm max}=86.16^{+12.37}_{-13.65}\, \msun\, ,
\ee
\be\label{gamma1gamma2}
\gamma_1=1.05^{+0.68}_{-1.08}\, ,\qquad \gamma_2=5.17^{+3.22}_{-2.31}\, ,\qquad
b_q=0.28^{+1.28}_{-1.04}\, ,
\ee
and 
\be\label{bbreak}
b_{\rm break}=0.41^{+0.16}_{-0.14}\, ,
\ee
where $b_{\rm break}$ determines $m_{\rm break}$ from 
$m_{\rm break}=m_{\rm min}+b_{\rm break}(m_{\rm max} -m_{\rm min})$, so the mean value of $m_{\rm break}$ is $36.76\,\msun$.
To have a quantitative idea of the dependence of the results  on these choices, in sect.~\ref{sect:results} we   will vary the parameter  $\gamma_1$, \red{as well as the mass scales $m_{\rm break}$ and $m_{\rm max}$}, first over the typical range of values 
which corresponds to their errors in the fit. It should be observed, however,  that the fit to the mass function in ref.~\cite{Abbott:2020gyp} is performed under the assumption of a fixed cosmology. Varying also the cosmology would broaden the allowed range of the parameters that enter in the mass function. We will therefore examine also the effect of larger variations of these parameters.

We first briefly summarize the results of the semi-analytic approximation, relegating the (rather long) technical details to App.~\ref{app:betafit}.  For $\beta(H_0)$, using homogeneous completion, we find
\bees
\beta^{\rm hom}(H_0)&=&\(\frac{\int_0^{z_{\cal R}}dz'\, j(z')P_{\rm compl}(z')}{\int_0^{z_{\cal R}}dz'\, j(z')} \)\, 
\frac{\sum_{\alpha=1}^{N_{\rm cat}({\cal R})} w_{\alpha} B[d_L(z_{\alpha}, H_0) ]}
{\sum_{\alpha=1}^{N_{\rm cat}({\cal R})} w_{\alpha}}\nn\\
&&+ 
 \frac{1}{\int_0^{z_{\cal R}}dz'\, j(z')}\, 
\int_0^{z_{\cal R}} dz\,j(z)  \, \[ 1-P_{\rm compl}(z)\]\, B[d_L(z, H_0)]\, ,
\label{betaH0homfinal_text}
\ees
while, for multiplicative completion,
\be\label{betaH0multfinal_text}
\beta^{\rm multi}(H_0)=\frac{\sum_{\alpha=1}^{N_{\rm cat}({\cal R})}w_{\alpha}B[d_L(z_{\alpha}, H_0) ]/P_{\alpha}}{\sum_{\alpha=1}^{N_{\rm cat}({\cal R})}w_{\alpha}}\, .
\ee
The function $B[d_L(z, H_0)]$, defined in \eq{defofB}, is obtained from an integration over the inclination of the orbits (with a flat prior) and over the distribution of  the component BH masses.
Comparing with  \eqs{betaH0simpledetectionmodel}{betamultiplicative2}, we see that the function $B[d_L(z,H_0)]$ plays the role of a  smooth cutoff function, that replaces the  theta function $\theta[z_{\rm max}(H_0;d_{\rm max})-z]$.  

\begin{figure}[t]
\centering
\includegraphics[width=0.48\textwidth]{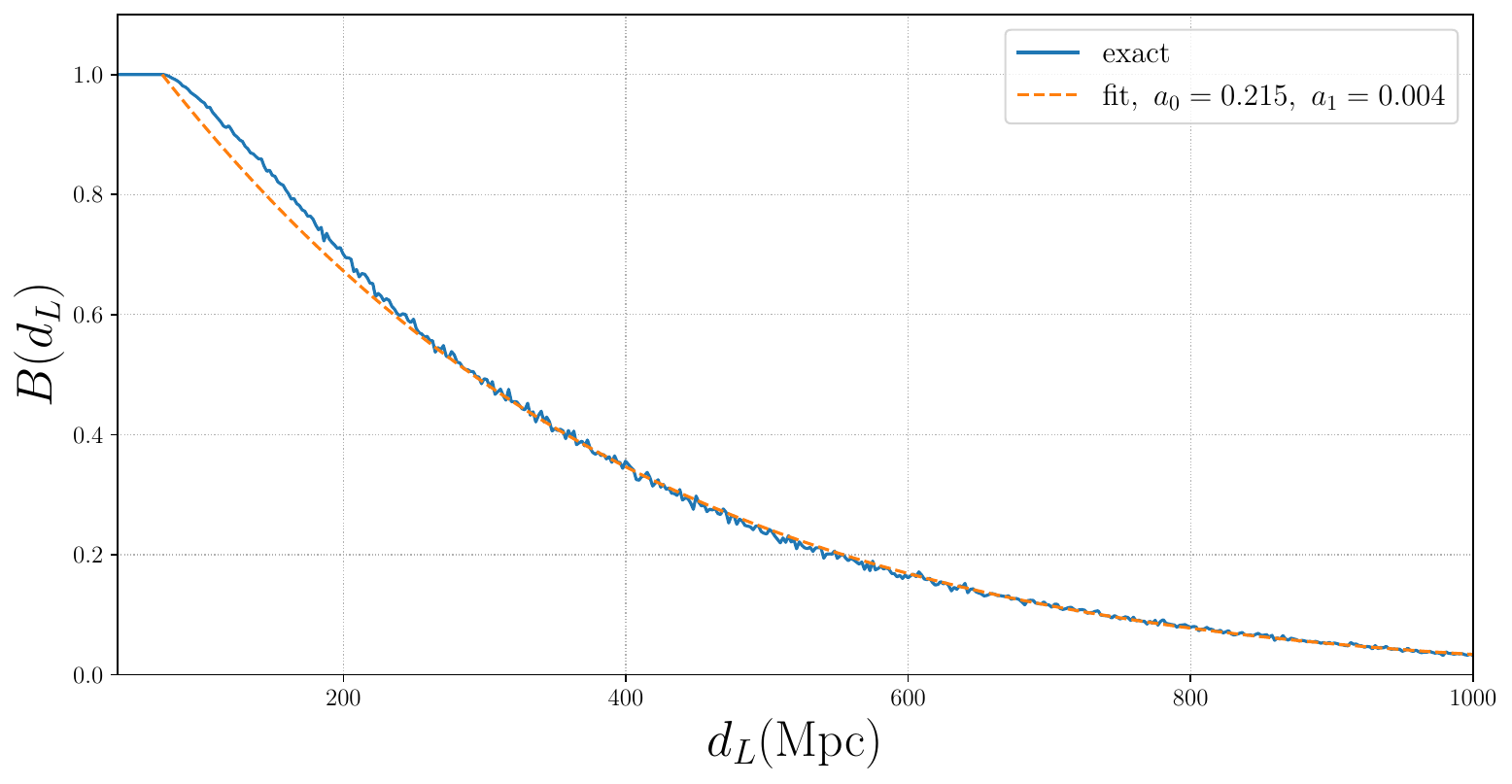}
\includegraphics[width=0.48\textwidth]{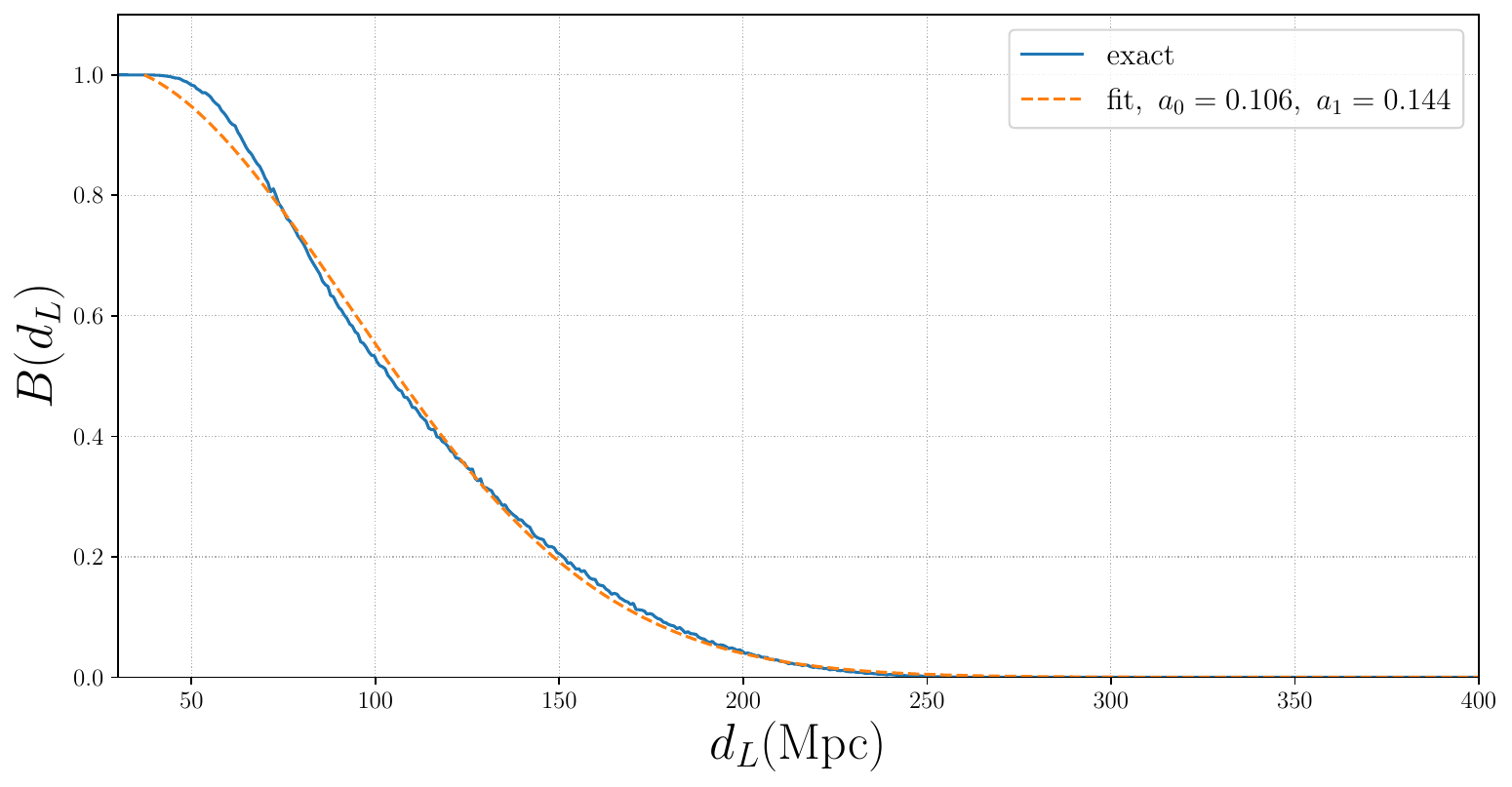}
\caption{Left panel: The function $B[d_L;z(d_L,H_0)]$ for BBHs, using the best fit values given in \eqst{mminbmax}{bbreak}
in the BBH mass distribution (\ref{brokenpowerlaw}) and  the O2 strain sensitivity, and
setting $H_0=70~{\rm km}\, {\rm s}^{-1}\, {\rm Mpc}^{-1}$. The blue solid curve is the numerical result
and the dashed orange curve is the fit (\ref{BdL2}).
Right panel: the function $B[d_L;z(d_L,H_0)]$ for a BNS mass distribution,   using a distribution for the source-frame masses flat in the range $[1-3] \msun$.  
}
\label{fig:B_vs_dL}
\end{figure}

\begin{figure}[t]
\centering 
\includegraphics[width=0.48\textwidth]{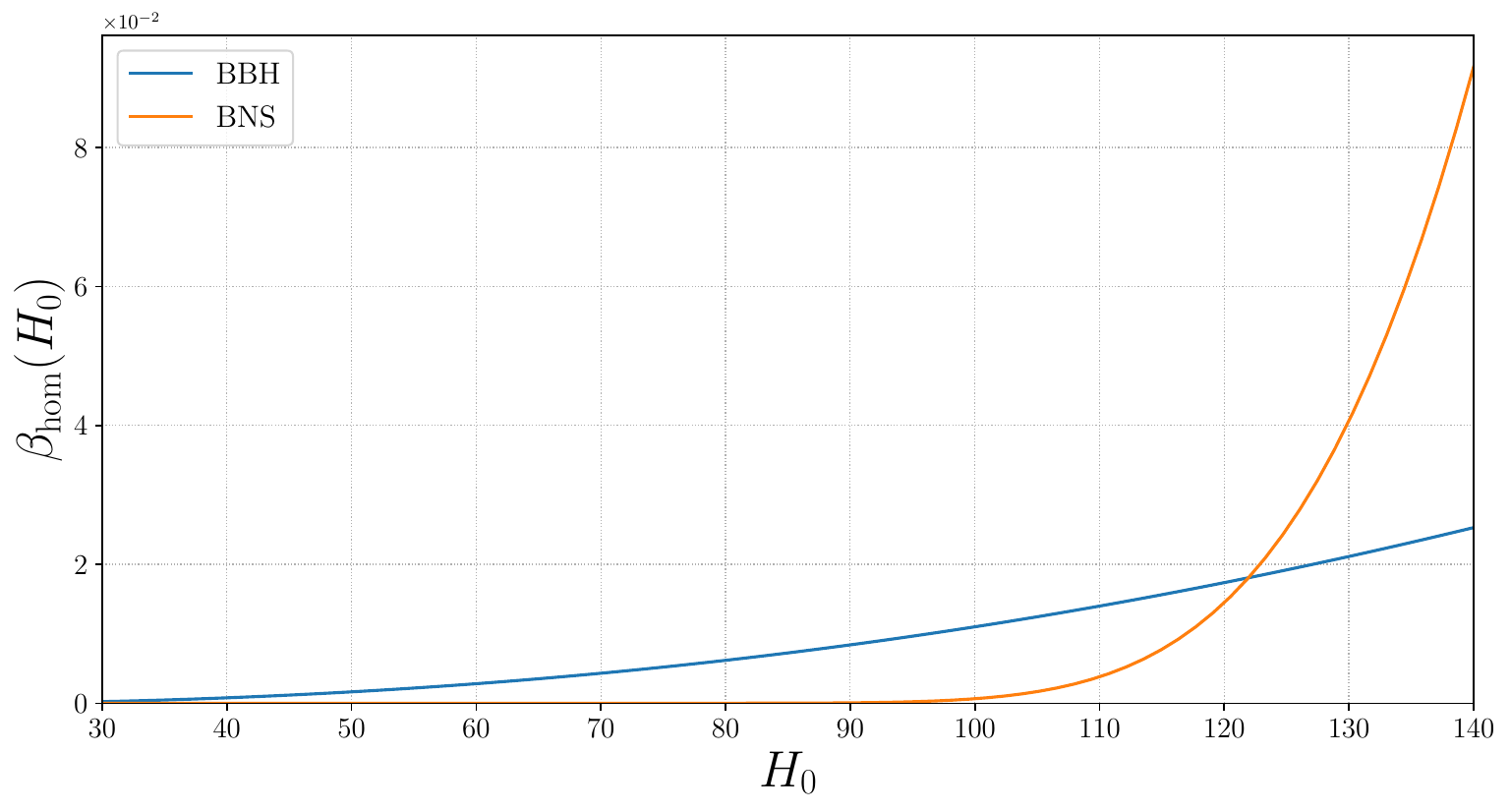}
\includegraphics[width=0.48\textwidth]{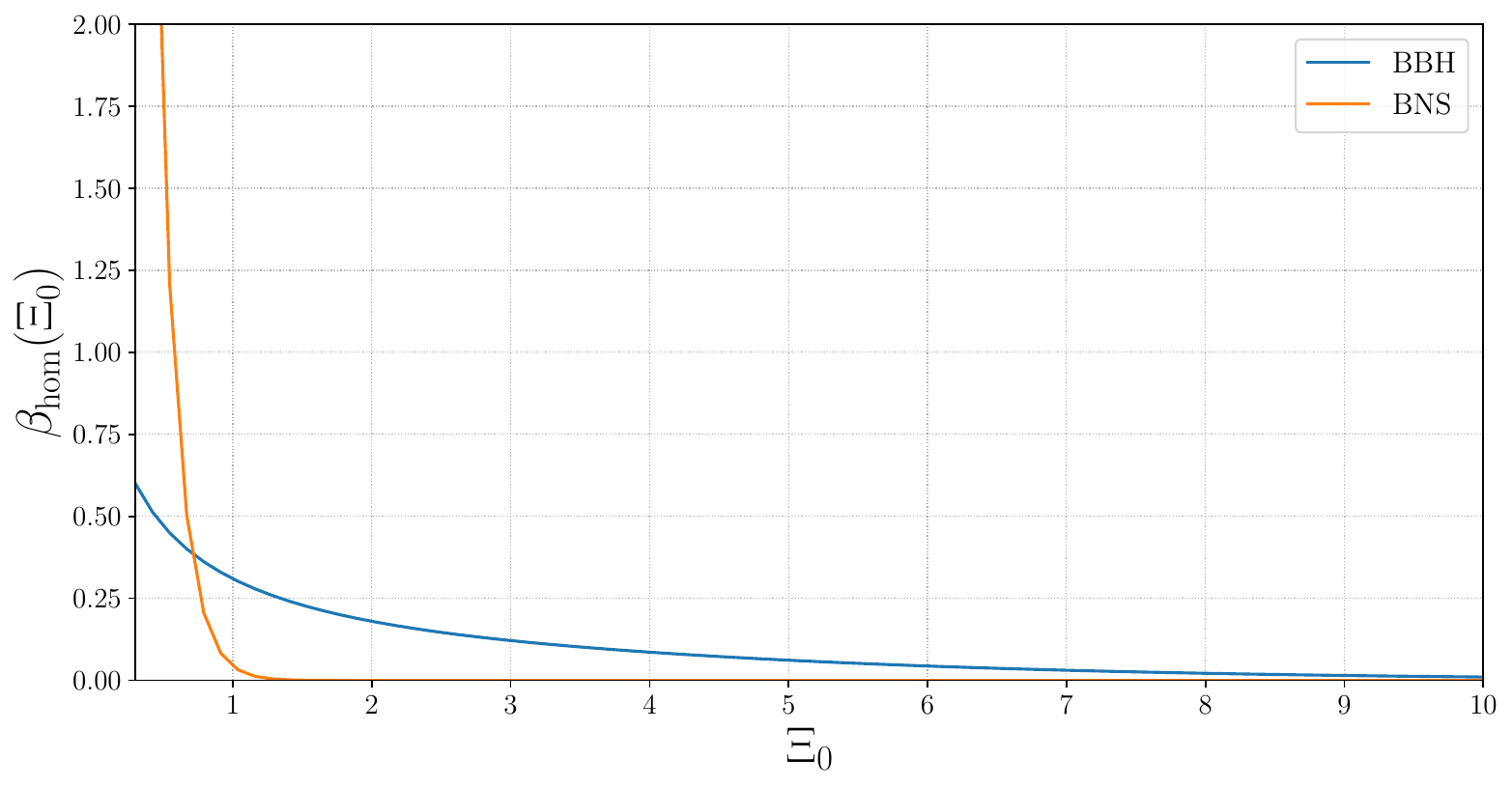}
\caption{The functions $\beta_{\rm hom}(H_0)$ (left) and $\beta_{\rm hom}(\Xi_0)$  (right)
for BBHs and for BNS, obtained using  \eq{betaH0homfinal_text},
using the functions $B[d_L;z(d_L,H_0)]$ shown in Fig.~\ref{fig:B_vs_dL}.}
\label{fig:beta_semianalytic_several_gamma}
\end{figure}

A plot of the function $B(d_L)$ is shown in  the left panel of Fig.~\ref{fig:B_vs_dL}, using  the BBH mass function (\ref{brokenpowerlaw}) with the parameters set to  the best fit values given in \eqst{mminbmax}{bbreak}. In the right panel we show the result for a (source-frame) mass function appropriate to neutron stars in BNS, taken to be flat in the interval  $[1-3] \msun$.
The corresponding results for $\beta^{\rm hom}(H_0)$, for  the BBH and BNS  mass function, are shown in Fig.~\ref{fig:beta_semianalytic_several_gamma}.
Similar results are obtained for $\beta(\Xi_0)$, in which simply $\dgw(z, \Xi_0)$ replaces $d_L(z, H_0)$ in the argument of $B[d]$, and are shown in the right panel of Fig.~\ref{fig:beta_semianalytic_several_gamma}, again for BBHs and BNSs.

A further issue that deserves consideration is that, for a catalog such as GLADE, most of the events of the O2 and O3 LIGO/Virgo runs fall in regions of low or very low completeness; this is true  in particular for all events at large redshift, given that, as we discuss in app.~\ref{sect:GLADE}, even with luminosity weighting and optimal choices of the cuts, GLADE becomes very incomplete beyond redshift $z\sim 0.2$. Events in region of very low completeness are in practice uninformative for our purpose of performing correlations between the GW events and the galaxy distribution  and, if correctly treated, should give an essentially  flat contribution to the posterior (at least after averaging over several such events). However, this flat contribution comes from a very delicate compensation of potentially large effects between the numerator and denominator of the posterior (\ref{pDinolambda}), and would require a very tight control over the function $\beta(H_0)$ [or $\beta(\Xi_0)$], including the precise BBH mass function, details of the detector sensitivities at time of detection, etc.\footnote{In particular, we find that events falling in very incomplete regions give contributions to the numerator in \eq{pDinolambda} that are peaked at very small values of $\Xi_0$. This can be traced to the fact that, in the absence of a significant GW/galaxy correlation due to the incompleteness of the catalog in the localization region, the correlation tends to push the GW events toward larger redshift, because of the geometric effect described on page~\pageref{page:geom_effect}. In principle, this should be compensated by the strong rise of $\beta(\Xi_0)$ near $\Xi_0=0$, well visible, for instance, in
the right panels of Fig.~\ref{fig:beta_Fish} and  Fig.~\ref{fig:beta_semianalytic_several_gamma}. However, in practice, an accurate compensation between these two large effects is impossible to achieve, given the approximations anyhow implicit in the computation of $\beta$ (approximate BBH mass distribution, approximate modelization of the detector noise, etc.).\label{foot:compens}}
A much more sound strategy, \label{page:sound} therefore, consists in simply excluding from the analysis the GW events that fall in regions of low completeness. This, however, introduces a selection effect that must be included also in the computation of $\beta$. In practice, this can only be done with a Monte Carlo (MC), which we perform to provide our final results. 

\red{Another element of the MC computation of $\beta(H_0)$ [or of $\beta(\Xi_0)$] is the choice of the waveform used for computing the SNR of a given event.  The use of a waveform including only the inspiral part (that we have used in the semi-analytic technique, as described in app.~\ref{app:betafit}) could still be a reasonably good approximation for the O2 run. However, 
for the detections in the O3 run this is no longer the case, given the large total mass of many of the detected BBHs, that implies that they only have  few inspiral cycles in the detector bandwidth, and a significant portion of the SNR comes from the merger (and possibly ringdown) phase. We  therefore use full inspiral-merger-ringdown waveforms. In our MC, we use in particular the IMRPhenomXAS waveforms~\cite{Pratten:2020fqn}.}\footnote{This is an important improvement compared to the v1 version of this paper, that only used inspiral waveforms and (together with the use of a higher completeness threshold) is the main element that affects  the results shown in sect.~\ref{sect:results}. } \red{Note that this is still an approximation to the actual detection process, since it is based on a threshold on the SNR, as opposed to injecting mock signals into the actual detection pipelines. }

\subsection{Information from the overall rate}\label{sect:rate}

Finally, we consider the information that could be extracted from the  total number of events detected in a given run, and from their redshift distribution. For definiteness, we use $\Xi_0$ in the discussion below since it can potentially give larger effects,
but the equations apply to the case of $H_0$ with the replacement $\Xi_0\ra H_0$. 

Including the information on the rate, for an ensemble of $N_{\rm det}$ GW detections, the posterior (\ref{posteriorXi0})  for $\Xi_0$ becomes~\cite{Gray:2019ksv} 
\be\label{fullPost}
p(\Xi_0|\{{\cal D}_{\rm GW}\}) =p(N_{\rm det}|\Xi_0)\,
 \, \frac{\pi(\Xi_0) \prod_{i=1}^{ N_{\rm det} } p({\cal D}^i_{\rm GW}|\Xi_0)}
 {p(\{{\cal D}_{\rm GW}\})}\, ,
\ee
where $p(N_{\rm det}|\Xi_0)$ is the likelihood of detecting $N_{\rm det}$ events in the observation run, given $\Xi_0$. Since $\Xi_0$ effectively changes the horizon distance of the GW events, it can significantly affect the total number of detections. For instance, at $z=1$,  for  $\Xi_0=1.8$ we have $\dgw/\dem\simeq 1.6$, so the actual electromagnetic luminosity distance probed is smaller by a factor 1.6 compared to GR. For these redshifts, a reduction factor of about the same order takes place for the comoving distance,
corresponding to  a reduction by a factor $(1.6)^3\sim 4$ in the comoving volume explored. In principle, this is a  large effect. However, to exploit  it we need to have a comparably accurate estimate of the rate of BBH detections and of their distribution in redshift, in the absence of modified GW propagation.   Unfortunately, this introduces large astrophysical uncertainties that, currently, seem to make difficult  of obtaining convincing information on modified GW propagation from the the study of the overall  number of detections and of their redshift distribution. This is apparent, for instance, from the clear  discussion in~\cite{Fishbach:2018edt}. The relevant quantity,
to compute the number of detected events and their redshift distribution, is  the differential mass-redshift distribution of BBH,
\be
\frac{dN}{dm_1dm_2dz}=N_{\rm BBH}\, \tilde{p}_0(m_1,m_2,z)\, ,
\ee
where  $N_{\rm BBH}$ is the total number of BBHs across all masses and redshifts, and 
$\tilde{p}_0(m_1,m_2,z)$ is the (normalized) mass distribution, for which in sect.~\ref{sect:MC} we assumed an expression independent of redshift, given by  \eq{brokenpowerlaw}. Here, however,  it is crucial  to include also the redshift dependence, since this is  partially degenerate with the effect of $\Xi_0$ on the rate.
Note that, when computing the posterior $p({\cal D}^i_{\rm GW}|\Xi_0)$ that expresses the GW-galaxy correlation,  we are anyhow limited to small redshifts by the incompleteness of the  galaxy catalog. For the overall rate we do not have this limitation, since it does not involve a correlation with galaxies, and indeed the effect become large only going to sufficiently large redshift. However, this also makes more important to understand the redshift dependence in $\tilde{p}_0(m_1,m_2,z)$.

Ref.~\cite{Fishbach:2018edt}  discusses several typical assumptions on 
$\tilde{p}_0(m_1,m_2,z)$, and their effect on the predictions. For instance,
one in general starts from the simplifying assumption that the distribution factorizes,
$\tilde{p}_0(m_1,m_2,z)=\tilde{p}_0(m_1,m_2)\tilde{p}_0(z)$, i.e. that the mass distribution of BBH does not evolve with $z$, an assumption that most probably breaks down at large redshift. One then proposes a parametrization for $\tilde{p}_0(m_1,m_2)$, involving some phenomenological and relatively poorly constrained parameters, such as \eq{brokenpowerlaw}. For the redshift dependence, as we saw in \eq{dNdVclambda}, \red{for redshifts $z\,\lsim \, 1$}    
a common parametrization  is
\be\label{pzlambda}
\tilde{p}_0(z)\propto \frac{dV_c}{dz}\, (1+z)^{\lambda -1}\, ,
\ee
where $V_c$ is  comoving volume and $\lambda$ a free parameter, with $\lambda=0$ corresponding  to a merger rate density uniform in comoving volume and source-frame time.
\red{However, as discussed below \eq{dNdVclambda}, if the BBH merger rate follows the star formation rate (SFR), at low redshifts $z\, \lsim \, 1$ one rather  has $\lambda\sim 3$ while, convolving  the SFR with a distribution of time delays between binary formation and merger can lead to estimates of $\lambda$ over a very broad range, from $\lambda\sim -10$ to $\lambda$ of order a few (see \cite{Fishbach:2018edt} for discussion and references). }

With all these uncertainties on the mass distribution and redshift dependence, it is difficult to convincingly disentangle any extra redshift dependence on the BBH coalescence rate induced by modified GW propagation from the unknown redshift dependence of astrophysical origin (see  fig.~2 of \cite{Fishbach:2018edt}, where it is shown that the modification from a selected modified gravity model can be easily mimicked by changes in the astrophysical assumptions). The same limitations apply to  the idea of exploiting the statistics of the signal-to-noise ratio from an ensemble of events, as attempted in refs.~\cite{Chen:2014yla,Calabrese:2016bnu,Garcia-Bellido:2016zmj}.\footnote{\red{One could try to break some of these degeneracies by making stronger astrophysical assumptions, for instance, assuming that there is just a single population of BHs, whose mass distribution has a PISN mass gap \cite{Farr:2019twy}, and no contamination from primordial BHs or  `second generation' BHs that are themselves the result of a merger 
(see \cite{Mastrogiovanni:2021wsd,MariaEzquiaga:2021lli} for recent work in this direction). In that case, however, the results may depend crucially on such astrophysical assumptions. In our case, in contrast, the bulk of the information comes from the correlation with the galaxy catalog. Even if in the sample of GW detections there was, for instance, a population of primordial BHs that does not correlate with luminous matter, still this would simply give a contribution to the posterior that, averaged over a sufficiently large statistics, would be basically flat, but should not induce a strong bias.}}

One could attempt to use the information on the rate only for  events with $z\ll 1$, where one can hope to neglect the effect of source evolution. This could possibly be of use for the statistical measurement of $H_0$, but not for $\Xi_0$,   because in the limit $z\ll 1$ also the effect of modified GW propagation disappears. Both modified GW propagation and astrophysical effects such as those parametrized in \eq{pzlambda}  give corrections $1+{\cal O}(z)$ at low $z$, that, again, are very difficult to disentangle, at least with the current astrophysical uncertainties. 

For this reason, in this work we will only consider the correlation with galaxy catalogs, and we will neglect the information potentially contained in the term $p(N_{\rm det}|\Xi_0)$ in \eq{fullPost}. We will  therefore set $p(N_{\rm det}|\Xi_0)$ to a constant independent of $\Xi_0$  (and similarly for $H_0$). As shown in
\cite{Fishbach:2018gjp,Mandel:2018mve}, this is formally equivalent to marginalizing \eq{fullPost} over the event rate $R$, assuming a prior $\pi(R)dR=dR/R$ i.e. a prior flat with respect to $\log R$, which is the typical prior that one use when one has no previous knowledge on a quantity, here $R$, that can vary over several orders of magnitudes.

\subsection{Summary}

We can now summarize the results of this long section as follows.
From \eq{likeprod}, in the approximation in which we neglect the information on the total rate, the total likelihood for $H_0$ is
\be
p(\{{\cal D}_{\rm GW}\}|H_0)=
\prod_{i=1}^{N_{\rm obs}}p(\{{\cal D}^i_{\rm GW}\}|H_0)\, 
\ee
where $p(\{{\cal D}^i_{\rm GW}\}|H_0)$ is conveniently written as a ratio,
\be\label{pinbeta}
p(\{{\cal D}^i_{\rm GW}\}|H_0)=\frac{n_i(H_0)}{\beta(H_0)}\, .
\ee
The expressions of the function $\beta(H_0)$ and of the numerator $n_i(H_0)$ depend on the prior $p_0(z,\hatO)$ and therefore on how we complete the catalog. We consider first the case in which the parameter $\lambda$ in \eq{factorp0lambda} is set to $\lambda=1$, so there is no redshift dependence in the prior induced by the merger rate. In this case, for homogeneous completion, inserting
\eqs{pcatdelta}{p0isocompldimensionless} into \eq{pDizO} and using the likelihood (\ref{likelihoodSinger}) that depends  on $z$ and $H_0$ [through $d_L(z,H_0)$] and on $\hatO$, we get
\bees
n_i(H_0) 
&=&\left[
\frac{\int_0^{z_R}dz'\, j(z')P_{\rm compl}(z')}{\int_0^{z_R}dz'\, j(z')}\right]\,\label{niH0}\\
&&\times \left\{ 
\frac{\sum_{\alpha=1}^{N_{\rm cat}({\cal R})}w_{\alpha}{\cal L}_i(z_{\alpha},\Omega_{\alpha};H_0)}{\sum_{\alpha=1}^{N_{\rm cat}({\cal R})}w_{\alpha} }
+\frac{\int_0^{z_R} dz j(z)\int d\Omega [ 1-P_{\rm compl}(z,\hat{\Omega})]\,
\, {\cal L}_i(z,\hat{\Omega};H_0)}{4\pi\int_0^{z_R}dz\, j(z)P_{\rm compl}(z)}\right\}\, ,\nn
\ees
where the jacobian $j(z)$ is defined in \eq{dVzdOdexplicit}; $P_{\rm compl}(z,\hat{\Omega})$ describes the completeness of the catalog, and has been defined in eqs.~(\ref{Pcompletedef}) or (\ref{Pcompletebarngal}) for the case $w_{\alpha}=1$, and generalized in \eq{PcompletedefLum} for the case of luminosity weighting.  
The  structure in braces nicely shows how the catalog term is an average of the likelihood with respect to the weights $w_{\alpha}$, and the completion term is an average over the comoving volume, with the numerator weighted by
$ 1-P_{\rm compl}(z,\hat{\Omega})$ and the denominator by $P_{\rm compl}(z)$.
As we discussed below \eq{pcatN}
actually, rather than
using a Dirac delta in  redshift as in \eq{pcatdelta}, we use a full galaxy 
posterior.\footnote{\red{More precisely, we use a distribution of the class of so-called metalog distributions \cite{doi:10.1287/deca.2016.0338}, which are quantile-parametrized  distributions that, in principle,  allow our code to handle non-gaussian redshift distributions for galaxy catalogs, when this information is available. For GLADE, we effectively fit the quantile distribution to a gaussian. For catalogs such as DES, this more general formalism can, however, be useful.\label{foot:metalog}}} 
Then, the first term in braces  ${\cal L}_i(z_{\alpha},\Omega_{\alpha};H_0)$, which has been obtained by carrying out the integral over the redshift using  a Dirac delta $\delta(z-z_{\alpha})$, is replaced by the full convolution in redshift. 

\red{We next include the effect of a generic parameter $\lambda$ in the merger rate. In this case, using \eq{factorp0lambda} [with $\tilde{p}_0(\theta')=1$]
we get 
\bees
n_i(H_0) &=&c_{\lambda}
\[\frac{\int_0^{z_R}dz'\, j(z')P_{\rm compl}(z')}{\int_0^{z_R}dz'\, j(z')}\]\,
\frac{\sum_{\alpha=1}^{N_{\rm cat}({\cal R})}w_{\alpha}(1+z_{\alpha})^{\lambda -1}
{\cal L}_i(z_{\alpha},\Omega_{\alpha};H_0)}
{\sum_{\alpha=1}^{N_{\rm cat}({\cal R})}w_{\alpha} }\nn\\
&&+c_{\lambda}
\frac{\int_0^{z_R} dz j(z)(1+z)^{\lambda-1}\int d\Omega [ 1-P_{\rm compl}(z,\hat{\Omega})]\,
\, {\cal L}_i(z,\hat{\Omega};H_0)}{4\pi\int_0^{z_R}dz\, j(z)}\, .
\label{niH0lambda}
\ees
The constant $c_{\lambda}$ is determined from \eq{pcatzOnormclambda} and is independent of $H_0$, so it can be omitted. In any case, it is given by
\bees
c_{\lambda}^{-1}&=&\[\frac{\int_0^{z_R}dz'\, j(z')P_{\rm compl}(z')}{\int_0^{z_R}dz'\, j(z')}\]\,
\frac{\sum_{\alpha=1}^{N_{\rm cat}({\cal R})}w_{\alpha}(1+z_{\alpha})^{\lambda -1}}
{\sum_{\alpha=1}^{N_{\rm cat}({\cal R})}w_{\alpha} }\nn\\
&&+
\frac{\int_0^{z_R} dz j(z)(1+z)^{\lambda-1} [ 1-P_{\rm compl}(z)]}{\int_0^{z_R}dz\, j(z)}\, .
\label{clambda}
\ees
Observe that, for a complete catalog, $P_{\rm compl}(z,\hat{\Omega})=1$, \eqs{niH0lambda}{clambda} reduce to 
\be
n_i(H_0) =
\frac{\sum_{\alpha=1}^{N_{\rm cat}({\cal R})}w_{\alpha}(1+z_{\alpha})^{\lambda -1}
{\cal L}_i(z_{\alpha},\Omega_{\alpha};H_0)}
{\sum_{\alpha=1}^{N_{\rm cat}({\cal R})}w_{\alpha} (1+z_{\alpha})^{\lambda -1}}\, ,
\ee
which amounts to using $w_{\alpha}(1+z_{\alpha})^{\lambda -1}$ as weights. Similarly, in the opposite limit $P_{\rm compl}(z,\hat{\Omega})=0$,  \eqs{niH0lambda}{clambda} become
\be\label{nilambdalast}
n_i(H_0) =\frac{\int_0^{z_R} dz j(z)(1+z)^{\lambda-1}\,
\, {\cal L}_i(z,\hat{\Omega};H_0)}{\int_0^{z_R}dz\, j(z) (1+z)^{\lambda-1}}\, ,
\ee
which corresponds to using a weight  $j(z) (1+z)^{\lambda-1}$. However, in the presence of both the catalog term and the completion term, the result is more complicated, since the normalization constant involves both. We similarly compute the function $\beta(H_0)$ by performing a Monte Carlo (MC) integration  using \eq{factorp0lambda} as the prior.}

\red{Repeating the computation for purely multiplicative completion, we get
\be
n_i(H_0)=
\frac{\sum_{\alpha} P_{\alpha}^{-1} w_{\alpha}  (1+z_{\alpha})^{\lambda -1}
{\cal L}_i(z_{\alpha},\Omega_{\alpha};H_0)}{\sum_{\alpha} P_{\alpha}^{-1} w_{\alpha}  (1+z_{\alpha})^{\lambda -1}}\, ,
\ee
where $P_{\alpha}$ was defined in \eq{defPalpha}. So, for pure multiplicative completion, each galaxy is weighted with a factor $P_{\alpha}^{-1} w_{\alpha}  (1+z_{\alpha})^{\lambda -1}$.}
For  the interpolation between multiplicative completion and homogeneous completion, in the case $\lambda=1$, inserting 
\eqss{pcatdelta}{p0multipcompl1}{p0multihom} into \eq{pDizO},  and writing again the result in the form
(\ref{pinbeta}), 
we get 
\be\label{niH0multi}
n_i(H_0) =
 n_i^{(1)}(H_0)+n_i^{(2)}(H_0)\, ,
\ee
where (apart from a normalization factor independent of $H_0$)
\bees
n_i^{(1)}(H_0)&=&\frac{1}{\sum_{\alpha=1}^{N_{\rm cat}({\cal R})}w_{\alpha} }\\
&&\hspace*{-15mm}\times\sum_{\alpha=1}^{N_{\rm cat}({\cal R})} 
\left\{ P_{\alpha}^{-1} w_{\alpha} (1+z_{\alpha})^{\lambda -1}\, \theta[ P_{\rm compl}(z_{\alpha},\hatO_{\alpha})-P_c^{\rm min}]\right.\nn\\
&&
\left. +w_{\alpha}    (1+z_{\alpha})^{\lambda -1}     \theta[ P_c^{\rm min}-P_{\rm compl}(z_{\alpha},\hatO_{\alpha})]  \right\}
{\cal L}_i(z_{\alpha},\Omega_{\alpha};H_0)\nn
\ees
and
\be
n_i^{(2)}(H_0)=
\frac{\int_0^{z_R} dz j(z)\, (1+z)^{\lambda -1}\int d\Omega \,
 \theta[ P_c^{\rm min}-P_{\rm compl}(z_{\alpha},\hatO_{\alpha})] \, 
[ 1-P_{\rm compl}(z,\hat{\Omega})]\,
\, {\cal L}_i(z,\hat{\Omega};H_0)}{4\pi\int_0^{z_R}dz\, j(z)P_{\rm compl}(z)}\, ,
\ee
where  the theta functions are smoothed as in footnote~\ref{foot:smoothedtheta} on page~\pageref{foot:smoothedtheta}.

The denominator
$\beta(H_0)$ is computed through a MC evaluation, as discussed in sect.~\ref{sect:MC}, using full inspiral-merger-ringdown waveforms. Details of our (publicly available) MC are given in app.~\ref{sect:ourMC}. The analysis is restricted to events that pass a cut in term of completeness (see footnote~\ref{foot:evcompl} on page~\pageref{foot:evcompl} for details). The same cut is applied in the MC evaluation of $\beta(H_0)$.
The results for $\Xi_0$ are simply obtained replacing $H_0$ with $\Xi_0$ in the above expressions.

\section{Results}\label{sect:results}

We now present our results for the posteriors of $H_0$ and of $\Xi_0$, using the published data from the O1, O2 and O3a LIGO/Virgo runs. With current data, we cannot yet expect  very stringent constraints. Our emphasis, however, will also be  on the methodological aspects and on the dependence of the results on the various assumptions and approximations used, with the aim of understanding the systematic errors that,  with the expected increase in the amount and quality of the GW data, will eventually become the limiting factor of the statistical method. In particular, 

\begin{itemize}

\item we will  compare the results obtained with different ways of quantifying the completeness of the catalog, i.e. mask vs. cone completeness, introduced in sect.~\ref{sect:conemask}. We will limit our main analysis to the GLADE catalog (see footnote~\ref{foot:catalogs} on page~\pageref{foot:catalogs}). 

\item For each of these two ways of quantifying the quasi-local completeness of the catalog, we will consider two  different ways of distributing the missing galaxies, the homogeneous completion and the interpolation between multiplicative and homogeneous completion, introduced in sect.~\ref{sect:hom_multi_compl}. For brevity, we will refer to the interpolation between multiplicative and homogeneous completion simply as `multiplicative completion'.

\item As discussed at the end of  sect.~\ref{sect:MC},  GW events that fall in a region of low completeness  should  in principle give almost flat contributions to the posteriors. However, this comes from a delicate cancellation between the numerator and denominator of \eq{pinbeta} which, in practice, cannot be achieved unless one is able to compute the `exact' expression for $\beta$,  including the correct BBH mass function, the details of the detector sensitivities and their evolution during the run, etc. A better strategy is therefore to define the completeness of a GW event (relative to a given catalog) in terms of the completeness $P_{\rm comp}$ of the catalog in the region where it falls, and 
include only events whose completeness  exceeds a threshold value
$P_{\rm th}$. We will then have to  account for this selection effect in the MC computation of the $\beta$ function.\footnote{There is of course some freedom in the exact definition of  completeness of an event, since, for instance, the localization region where an event falls depends  on the values of $H_0$, $\Xi_0$, and one could use an average over the localization region, or the best fit or mean values for the position of the event. We will use a simple definition in which  the completeness of an event (relative to a given catalog) is defined as the completeness of the catalog evaluated at the nominal position of the event, obtained from the mean value of its luminosity distance and of its direction, using fiducial values $H_0=70 \,\, {\rm km}\, {\rm s}^{-1}\, {\rm Mpc}^{-1}$  and $\Xi_0=1$. We then require this to exceed a threshold value
$P_{\rm th}$. Actually, the details of the definition have limited importance, and different variants of this definition simply have the effect of removing or adding some GW event from the selected subset. Since already the completeness obtained  with mask or cone completeness, are somewhat different (and also depend on whether one uses B band or K band luminosity weighting), and again lead to a slightly different subset of GW events that pass the threshold, the comparison of these two cases, and  also the comparison of results for different  thresholds
$P_{\rm th}$, 
already gives an estimate of the systematic effect involved; what is more important is that the same selection criterion is used in the MC computation of $\beta(H_0)$ [or, respectively, $\beta(\Xi_0)$].
\label{foot:evcompl}} \red{We will explore the effect of different choices for 
$P_{\rm th}$, in the range $0.1-0.7$.}

\item We will always use luminosity weighting, since, as discussed in app.~\ref{sect:GLADE}, for the GLADE catalog the weights $w_{\alpha}=1$ result in a level of completeness too low for the correlation with standard sirens. As discussed again in  app.~\ref{sect:GLADE}, we adopt a lower cut on the luminosity $L_{\rm cut}=0.6L_B^*$ in B-band, and $L_{\rm cut}=0.6L_K^*$ in K-band. We will then 
compare the results obtained with B-band  and with K-band luminosity weighting.

\item \red{We will study how the posterior of $H_0$ or $\Xi_0$ are affected by different choices for the values of the parameters that describe the astrophysical population. In particular, we will discuss the effect of changes in
 the parameters $m_{\rm break}$, $m_{\rm max}$ and $\gamma_1$ that characterize the BBH mass function (\ref{brokenpowerlaw}), and of the parameter $\lambda$ that characterize the BBH rate (at low redshifts) according to \eq{dNdVclambda}.}
 
\end{itemize}

\red{In order to avoid a proliferation of plots, in the main text we will only show the results for our `baseline' set of choices, which is defined as follows: we use mask completeness (with 9 masks, but the results with, say, 5 masks, are very similar) to quantify the quasi-local completeness level of the catalog;  we use multiplicative completion (more precisely,  the interpolation between multiplicative and homogeneous completion) to distribute the missing galaxies; we use K-band luminosity weighting; and, finally, we set a rather high value of the completeness threshold, 
$P_{\rm th}=0.7$, for the selection of the GW events. In App.~\ref{sect:dependence_results} we will justify these choices, and we will study how the results depend on them. As we will discuss there, mask completeness gives results consistent with cone completeness, and, similarly, homogeneous completion gives results consistent with multiplicative completions, within our (rather large) statistical uncertainty. In contrast, the choice of 
$P_{\rm th}$ is quite important, and, as we will discuss in App.~\ref{sect:dependence_results}, we make the conservative choice of  setting it to a rather high value, in order to minimize potential biases that could otherwise be induced by the GW events that fall in rather incomplete regions of the catalog. B-band is also consistent with K-band, but would require a higher completeness threshold to give similar results.}

A consequence of the current limited statistics will be the dependence of the results on the 
 range of prior chosen for $H_0$, and particularly on the upper limit.  This is due to the fact that, for a given measured luminosity distance of the GW event, increasing $H_0$ corresponds to increasing the inferred redshift of the source and, as long as there are galaxies in the catalog at large $z$ (and as long as we complete the catalog, adding by hand `missing' galaxies) we will always find a non-zero correlation. 
Therefore, the posterior on $H_0$ obtained from each  individual GW event can extend to arbitrarily large values of $H_0$ (unless we use values of $H_0$ so high that  the redshifts of the GW events are pushed into the dark epoch where no more galaxies are present, and we  stop `completing' the catalog), and is simply cut off by the choice of prior. As the mock data challenge  in \cite{Gray:2019ksv} shows, eventually, with a sufficiently large number of dark sirens, the total posterior on $H_0$ will take a well-behaved gaussian-like shape, with tails that  become  suppressed as we increase the number of events, effectively removing the dependence on the prior  (see, in particular, Fig.~6 of \cite{Gray:2019ksv}). However, with the current limited statistics of real data, we have to live with a dependence on the prior range chosen for $H_0$.\footnote{For $\Xi_0$ the problem is less severe. Indeed, the redshift of the galaxies now increases if $\Xi_0$ decreases. However, $\Xi_0$ is naturally bounded to be greater than zero, since it represents a ratio of distances and, while in the limit $H_0\ra\infty$ the redshift of a source, for given measured luminosity distance,  is formally pushed to infinity, in the limit $\Xi_0\ra 0$ the ratio $\dgw/\dem$ in \eq{eq:fit} stays finite, so the  inferred redshift also stays finite.}
In the following, we will use a flat prior in the range $[30,140] \,\,{\rm km}\, {\rm s}^{-1}\, {\rm Mpc}^{-1}$, similar to the prior $[20,140] \,\,{\rm km}\, {\rm s}^{-1}\, {\rm Mpc}^{-1}$ used in \cite{Soares-Santos:2019irc,Palmese:2020aof}. \red{When a posterior does not go to zero as we approach the upper limit of the prior range  chosen, the notion of median value loses meaning, since the integral of the posterior distribution still gets significant extra contributions if we increase the upper limit of the prior range. In that cases, if the posterior shows a well defined peak, rather than the median value we quote the value at the maximum, which is independent of the prior range (as long as it includes the maximum), and  the error that we quote corresponds to the $68\%$ highest density interval, i.e. the smallest interval which includes the maximum and contains $68\%$ of the area of the posterior; this quantity depends again on the prior range.}

\subsection{Hubble constant}\label{sect:H0results}

\subsubsection{Results using only dark sirens}\label{sect:H0darkonly}

We now present our results, using  the BBH detections of the O1-O2 runs and of the O3a run. 
In Table~\ref{tab:O2events} we list the O1-O2 BBH events, including some relevant data such as distance and masses, and the completeness at their nominal location (defined as in footnote~\ref{foot:evcompl}), computed using cone completion (in B band and in K band), and using mask completion (again, in B band and K band). Observe that mask completeness gives higher values than cone completeness. This can be traced to the fact that cone completeness extends the region ${\cal S}$  toward larger redshifts compared to mask completeness, and increasing the redshift the  catalog becomes less complete. In this sense, mask completeness seems a more faithful measure of the local completeness. 
In any case, we see that only few events have a high completeness. For instance, using mask completeness, both in B and K band there are 4 events that exceed a threshold  $P_{\rm th}=0.5$, that reduce  to 3 events if we raise the threshold to $P_{\rm th}=0.7$. We also observe that, for the majority of events, K band completeness is higher than B band completeness (the same happens for O3a data). We will discuss this further in app.~\ref{sect:dependence_results}, and this will lead us to chose K-band in our baseline model.
Observe also that events at comparable redshifts can have very different completeness level, according to our quasi-local notions of completeness (compare, for instance, GW170809 with GW170818). This shows the importance of using a quasi-local notion of completeness such as one  of those that we have introduced, rather than  a notion of completeness obtained from a $4\pi$ average.

\begin{table}[t]
	\centering
	\resizebox{\columnwidth}{!}{
		\begin{tabular}{|c|c|c|c|c|c|c|c|c|}
			\toprule
			name & $d_L$ (Mpc) & $z$ & $m_1 (\msun)$ & $m_2 (\msun)$ 
			& $P^{\rm cone}_{\rm comp, B}$ &$P^{\rm cone}_{\rm comp, K}$ 
			& $P^{\rm mask}_{\rm comp, B}$ &$P^{\rm mask}_{\rm comp, K}$\\ \midrule
			GW150914 &440&0.09  &35.6 &30.6 &0.75 &1.00&1.00&1.00\\
			GW151012 &1080&0.21&23.2 &13.6 &0.00 &0.05&0.04&0.07\\
			GW151226 &450&0.09  &13.7   &7.7&0.72 &1.00&0.95&1.00\\
			GW170104 &990&0.20  &30.8 &20.0&0.10 &0.41&0.27&0.41\\
			GW170608 &320&0.07  &7.6   &14.9&0.62 &0.77&1.00&0.85\\
			GW170729 &2840&0.49&50.2 &34.0&0.00 &0.00&0.00&0.00\\
			GW170809 &1030&0.20&35.0 &23.8&0.23 &0.19&0.28&0.28\\
			GW170814 &600&0.12  &30.6 &25.2&0.36 &0.69&0.52&0.57\\
			GW170818 &1060&0.21&35.4 &26.7&0.01 &0.06&0.03&0.08\\
			GW170823 &1940&0.35&39.5 &29.0&0.00 &0.00&0.00&0.00\\
			\bottomrule
		\end{tabular}%
	} 
	\caption{Values of luminosity distance (or GW luminosity distance, in the context of modified gravity), redshift (obtained from $d_L$ assuming $\Lambda$CDM with {\em Planck} cosmology) and source frame masses, for all BBH events detected in O1-O2 (from 
		\url{https://www.gw-openscience.org/eventapi/html/GWTC-1-confident/}). In the last four columns we add the values of the completeness of GLADE at the nominal position of the event, computed using cone completeness (in B band and in K band), and mask completeness (again in B band and in K band).  \label{tab:O2events}}
\end{table}

We next consider the data from the O3a run~\cite{Abbott:2020niy}. The data release corresponds to 39 candidate GW events; given the threshold in the false alarm rate it is expected, statistically, that about $3$ of them will be false alarms. Despite this significantly larger number of events, compared to O2, we find that most of them fall into regions of very low completeness of GLADE, and are therefore not useful for correlation with this catalog.  This is due to the fact that the increase in the number of detections comes from  the increase  of the  reach of the detectors between O2 and O3, and so there are many more events at large 
redshift. However, these are precisely the events that are more likely to  fall in regions where GLADE has low completeness.
Table~\ref{tab:O3aevents} lists all O3a events  whose nominal position (defined as in footnote~\ref{foot:evcompl} on page~\pageref{foot:evcompl}) falls into a region with GLADE completeness higher than $0.1$, in at least one among B~band and K~band (restricting now, for definiteness, to mask completeness). We observe, from the value of the masses, that they all corresponds to BBH candidates, except for GW190426\_152155, which has a component with mass $1.5\msun$, and it is rather a potential BH-NS binary. Its inclusion would therefore require a different computation of the beta function, with the mass function appropriate to  BH-NS binaries, which is currently not really constrained. Furthermore, this is one of the three events  in O3a (and the only one that makes it to  Table~\ref{tab:O3aevents}) that, based on the false alarm rate, are more likely to be noise, rather than  real GW events~\cite{Abbott:2020niy}. For these reasons, we will simply discard this event, and focus on the contribution from the BBH binaries  in Table~\ref{tab:O3aevents}, which are all highly likely detections.

\begin{table}[t]
	\centering
	\begin{tabular}{|c|c|c|c|c|c|c|}
		\toprule
		name & $d_L$ (Gpc) & $z$ & $m_1 (\msun)$ & $m_2 (\msun)$ 
		& $P^{\rm mask}_{\rm comp, B}$& $P^{\rm mask}_{\rm comp, K}$ \\ \midrule
		GW190412               &0.74&0.15 &30.0 &8.3 &0.74 &0.65\\
		\cellcolor{gray!25}GW190426\_152155 &\cellcolor{gray!25}0.38&\cellcolor{gray!25}0.08 &\cellcolor{gray!25}5.7 &\cellcolor{gray!25}1.5 &\cellcolor{gray!25}0.87&\cellcolor{gray!25}1.00\\
		GW190527\_092055 &3.10&0.53  &36.2   &22.8&0.31&0.48\\
		GW190630\_185205 &0.93&0.19  &35.0 &23.6&0.22&0.18\\
		GW190707\_093326 &0.80&0.16  &11.5   &8.4&0.23&0.38\\
		GW190708\_232457 &0.90&0.18&17.5 &13.1&0.34&0.51\\
		GW190814               &0.24&0.05&23.2 &2.59&1.00&1.00\\
		GW190924\_021846 &0.57&0.12  &8.8 &5.0&0.99&1.00\\
		GW190930\_133541 &0.78&0.16&12.3 &7.8&0.20&0.24\\
		\bottomrule
	\end{tabular}
	\caption{Median values of luminosity distance (or GW luminosity distance, in the context of modified gravity), redshift (obtained from $d_L$ assuming $\Lambda$CDM with {\em Planck} cosmology) and source frame masses, for the events in the O3a run (from \cite{Abbott:2020niy}), restricting to the events whose mask completeness, either  in B-band or in K-band (shown in the last two columns), is higher than $0.1$. As discussed in the text, we exclude GW190426\_152155 (gray entry) from further analysis, since its masses do not correspond to a BBH and its false alarm probability is high.
		\label{tab:O3aevents}}
\end{table}

\begin{figure}[t]
\centering
\includegraphics[width=\textwidth]{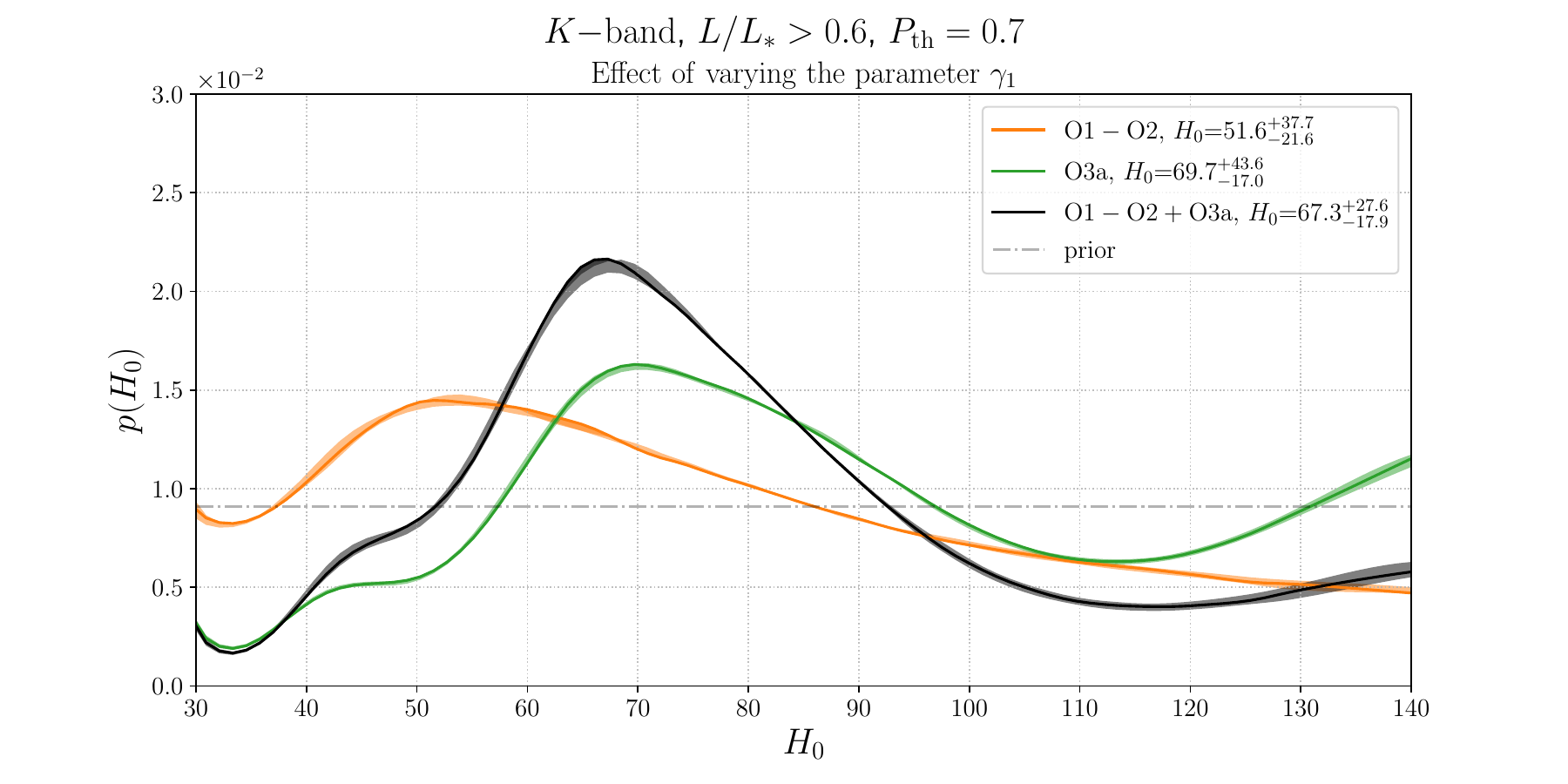}
\caption{The separate posterior for $H_0$ from the O1+O2 data and from the O3a data, and their combined posterior, using mask completeness, multiplicative completion,  K-band luminosity weighting, and a completeness threshold $P_{\rm th}=0.7$. The shaded bands give the effect of changing $\gamma_1$ within the range $\gamma_1=1.05^{+0.68}_{-1.08}$, with the central lines corresponding to our reference value $\gamma_1=1.05$.}
\label{fig:varyingGamma_H0}
\end{figure}

In Fig.~\ref{fig:varyingGamma_H0} we show the separate posterior for $H_0$ from the O1+O2 data and from the O3a data, and their combined posterior (the central lines corresponds to our reference value $\gamma_1=1.05$, while the shaded bands
band give the effect of changing $\gamma_1$ within the range $\gamma_1=1.05^{+0.68}_{-1.08}$, as we will discuss in more detail below).

\begin{figure}[t]
\centering
\includegraphics[width=\textwidth]{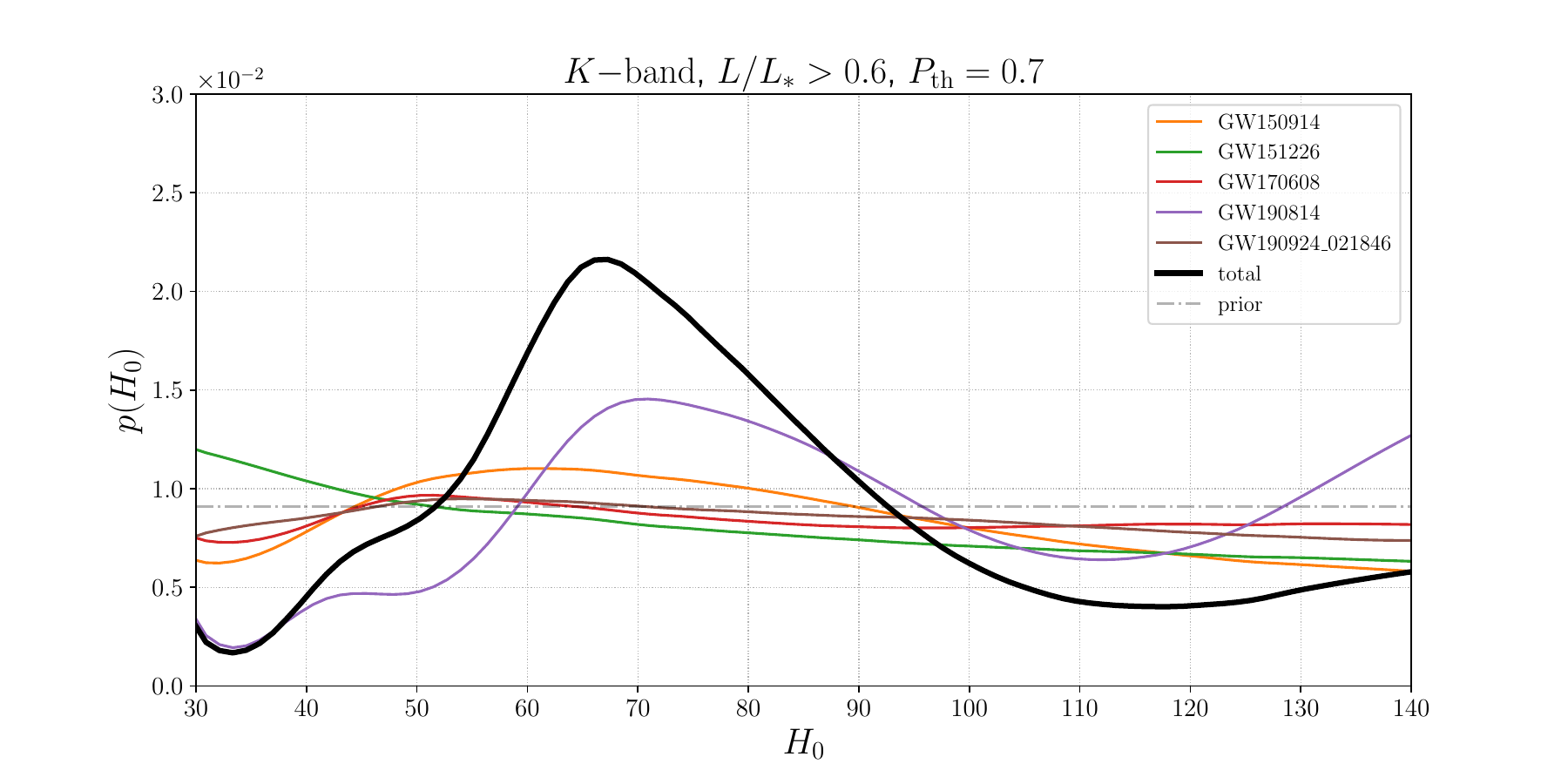}
\caption{The separate  contributions to the posteriors of $H_0$, from   the individual events  in O1+O2+O3a that pass the completeness threshold $P_{\rm th}=0.7$. The black line is the total posterior from O1+O2+O3a, already  shown as the  line inside the black band
in Fig.~\ref{fig:varyingGamma_H0}.}
\label{fig:separate_H0}
\end{figure}

Fig.~\ref{fig:separate_H0} shows the separate contributions to the posterior of $H_0$, from   the  five events  in O1+O2+O3a that pass the completeness threshold $P_{\rm th}=0.7$. We see that the single most important contribution comes from GW190814, followed  by GW150914. Note, from Tables~\ref{tab:O2events} and \ref{tab:O3aevents}, that these  two events have a completeness $P^{\rm mask}_{\rm comp}=1$. It is, however, interesting to remark that even if, individually, the other posteriors seem quite uninformative, still their combination produces a peak in the total posterior much more significant than that due to any single event. This nicely illustrates the power of the statistical method.

\red{We observe that, with current statistics,  the posterior for $H_0$ does not go to zero at the upper limit of the prior range considered (and in fact, for O3a, it even grows near the upper limit). Therefore, as discussed at the beginning of this section,
the notion of median value loses meaning, since  in this case it depends strongly on the prior range chosen. To condense  the information contained in the posteriors shown in Fig.~\ref{fig:varyingGamma_H0} into a single value of $H_0$ and a corresponding  error, it is more meaningful to provide  the value of the maximum of the posterior, with an error that corresponds to the $68\%$ highest density interval, i.e. the smallest interval which includes the maximum and contains $68\%$ of the area of the posterior. With this procedure, for the combined O1+O2+O3a data, we get}
\be\label{H0KbandPth07}
H_0=67.3^{+27.6}_{-17.9}\,\,\,\, {\rm km}\, {\rm s}^{-1}\, {\rm Mpc}^{-1}\, .
\ee
The maximum a posteriori values and the corresponding highest density intervals obtained separately from O1+O2 and from O3a are given in the inset of Fig.~\ref{fig:varyingGamma_H0}.
\Eq{H0KbandPth07} corresponds to a measurement of $H_0$ at about $34\%$ accuracy from dark sirens only. This improves on the result
$H_0=75^{+40}_{-32}\,\, {\rm km}\, {\rm s}^{-1}\, {\rm Mpc}^{-1}$
obtained in  \cite{Soares-Santos:2019irc} 
by correlating the dark siren GW170814 with the DES galaxy catalog,  and also on the result $H_0=66^{+55}_{-18}\,\, {\rm km}\, {\rm s}^{-1}\, {\rm Mpc}^{-1}$ obtained in \cite{Palmese:2020aof} for the dark siren GW190814 (which is the event that gives the dominant contribution in our case), correlated with DES. Of course, the interest of these results is at a level  of proof-of-principle of the statistical method, and much higher statistics will be needed to obtain an interesting accuracy.

We next investigate the dependence of the results on the astrophysical population parameters. 
We  first vary the parameters $\gamma_1$, $m_{\rm break}$, $m_{\rm max}$ that enter in the broken power-law mass function (\ref{brokenpowerlaw}) within the $1\sigma$  range found by the LVC fit, and we also vary the parameter $\lambda$ in the parametrization (\ref{pzlambda}) of the merger rate, within the astrophysically plausible range $\lambda\in [0,3]$, which is also of the order of the
range  $\lambda=1.8^{+2.1}_{-2.2}$ ($90\%$ c.l.) obtained using the data from the GWTC-2 catalog~\cite{Abbott:2020gyp}, see the discussion below \eq{MadauSFR}.
We present the results as  an estimate of the corresponding systematic error. 
As already remarked, however,   the mass distribution parameters, such as those given in \eqs{mminbmax}{gamma1gamma2}, have been obtained from fits to the mass distribution, under the assumption of a fixed cosmology. The correct first-principle approach would be to  turn this systematic error into a statistical error, by including the parameters that enter in the mass function into the set of parameters  on which we wish to draw an inference, perform such a global inference, and finally present results for $H_0$ or $\Xi_0$ marginalized over the parameters that enter in the mass function. The same should be done for $\lambda$.
Such a first-principle approach, however, would be  numerically quite demanding. Given that anyhow the current dark sirens data do not yet allow us to get very stringent bounds on $H_0$ or $\Xi_0$, in this paper we limit ourselves to the simplest approach of fixing the parameters in the mass function to a reference value, and then vary this reference values within a plausible range, to obtain a first estimate of their effect, which, at this level, is treated as a  systematic error. We begin by showing the results obtained varying the parameters in the mass function within the $1\sigma$ range found by the LVC fit, and $\lambda$ in the  the  range $\lambda\in [0,3]$. We will then show how the results change with more extreme values of  these parameters, that might possibly emerge from changes in the underlying cosmology.

\begin{figure}[t]
\centering
\resizebox{1.04\columnwidth}{!}{
\includegraphics[width=0.52\textwidth]{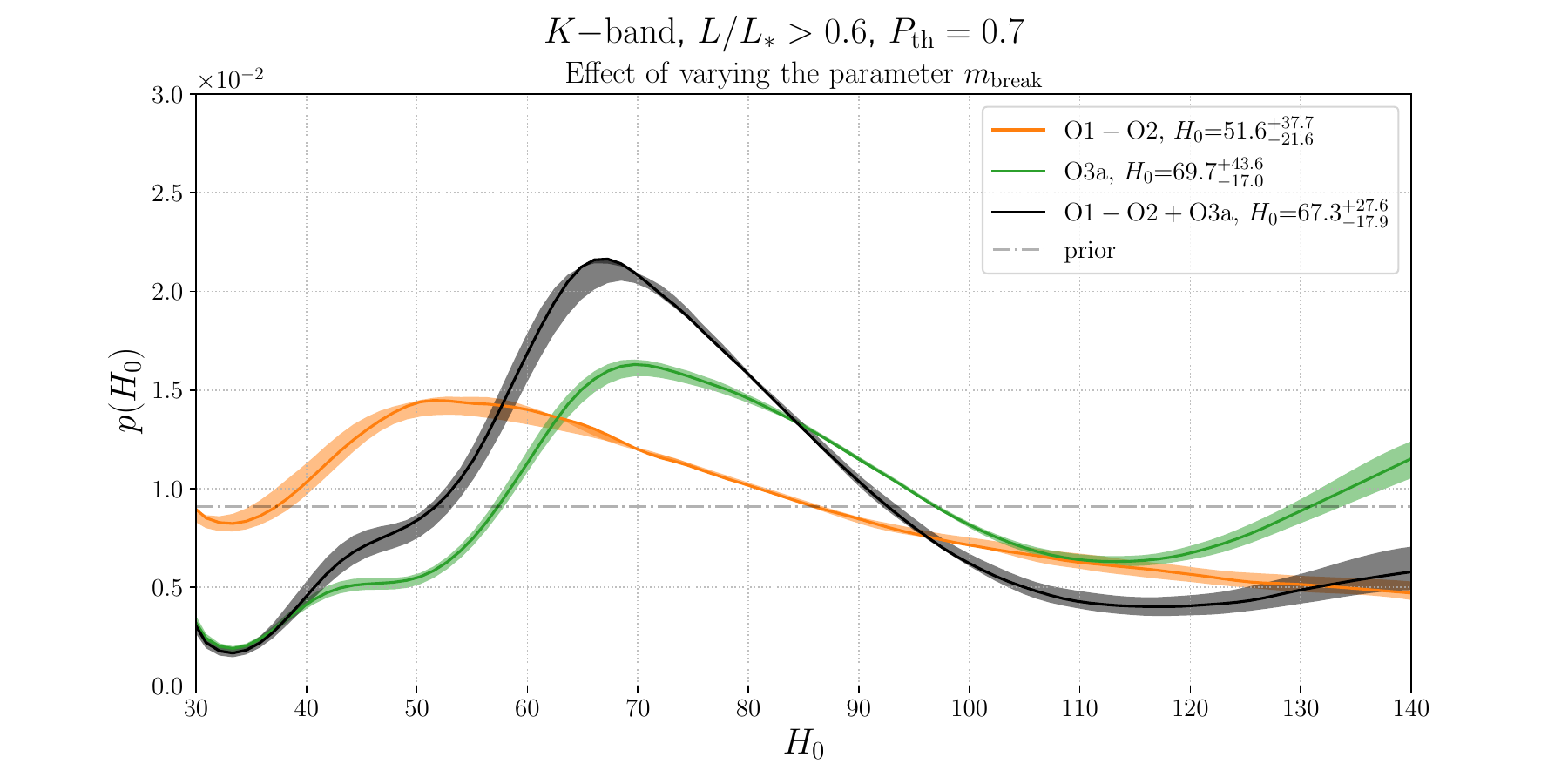}
\includegraphics[width=0.52\textwidth]{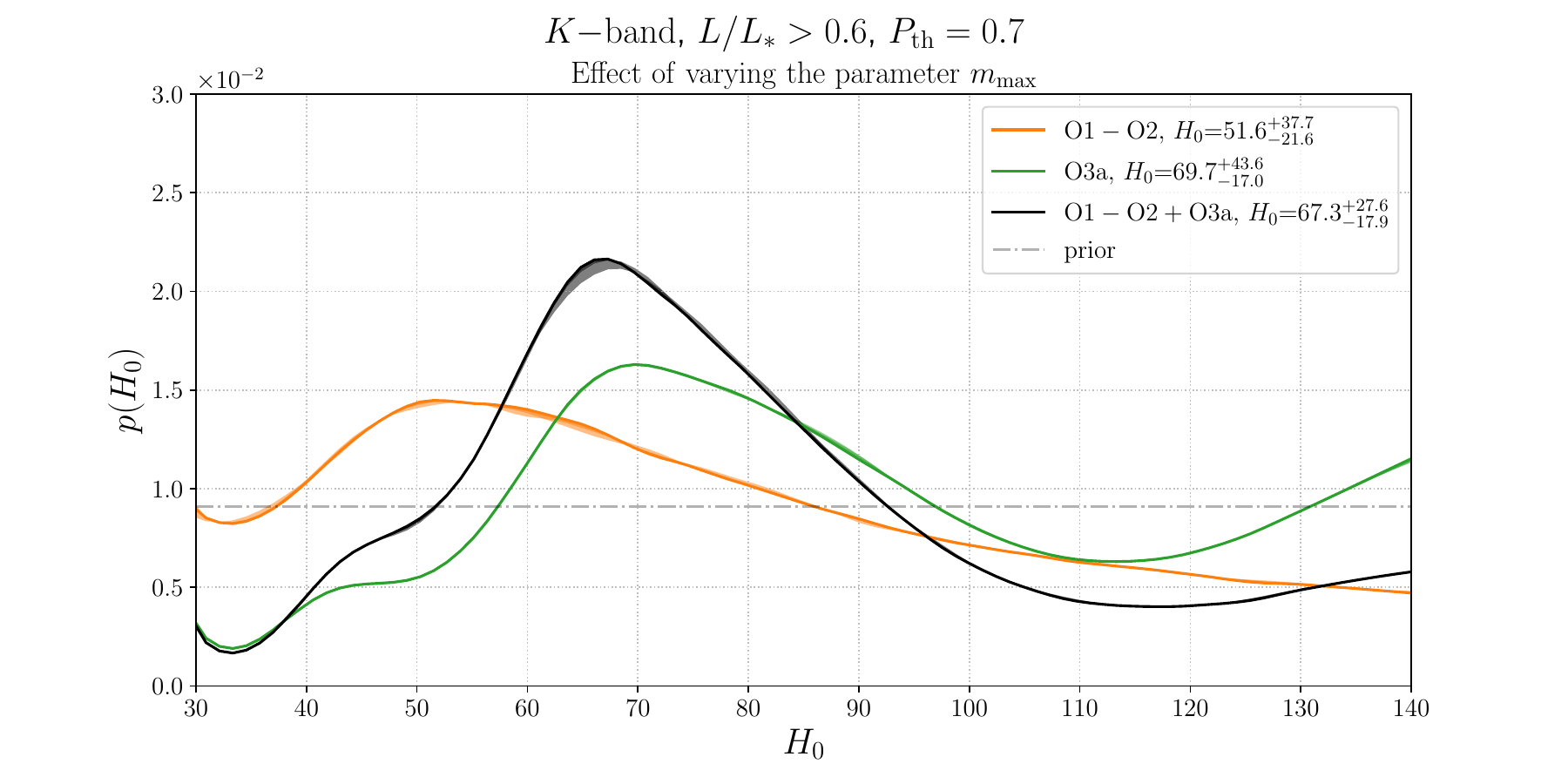}
}
\includegraphics[width=0.52\textwidth]{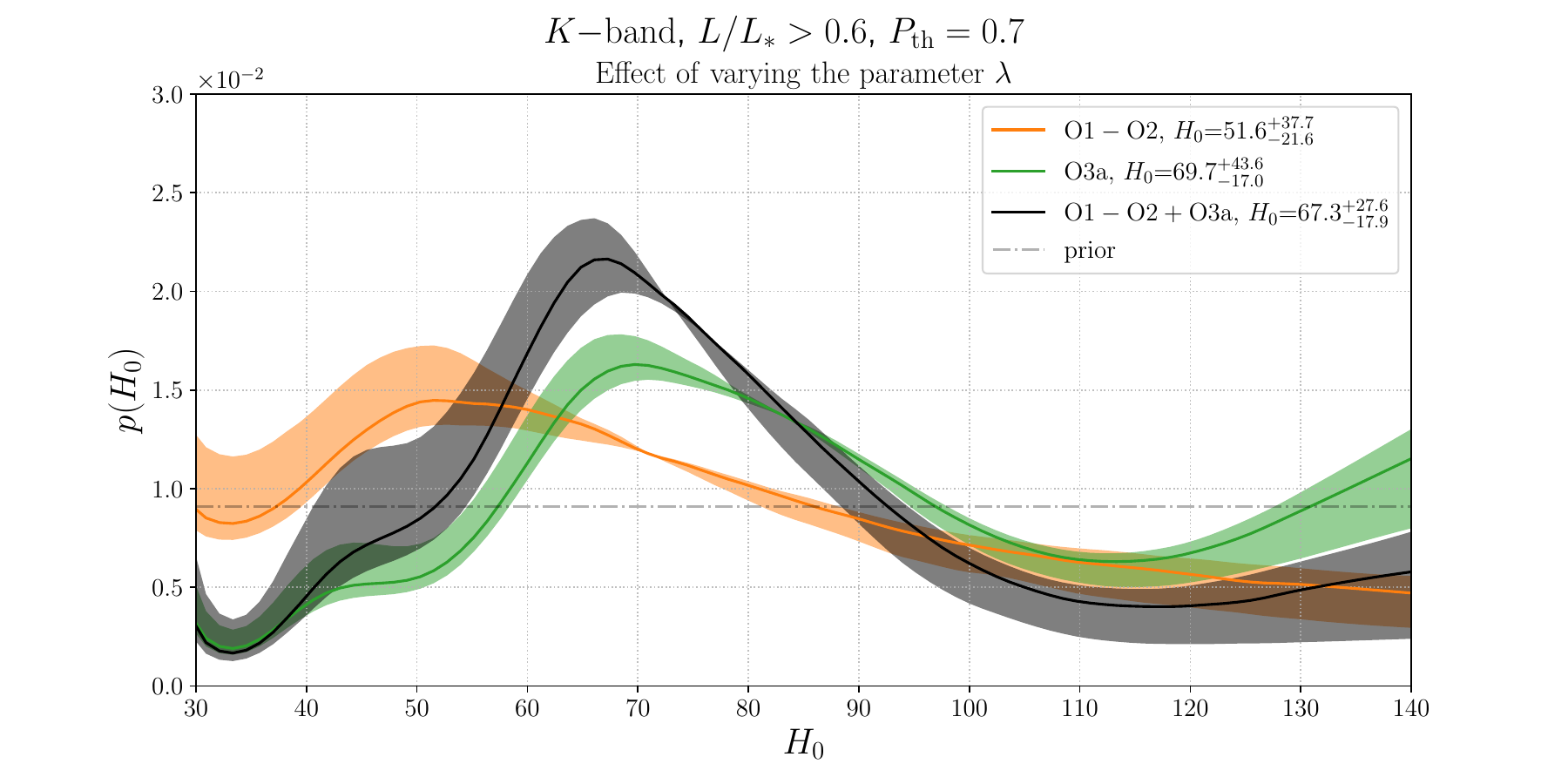}
\caption{Upper row: the effect of  changing $m_{\rm break}$ within the range $m_{\rm break}=36.76^{+11.37}_{-12.48}\, \msun$ (left panel) and the effect of  changing $m_{\rm max}$ within the range $m_{\rm max}=86.16^{+12.37}_{-13.65} \, \msun$ (right panel). Lower panel: the effect of changing $\lambda$ in \eq{pzlambda} within the range $[0,3]$, with the central lines corresponding to our default value $\lambda=1$. We use mask completeness, multiplicative completion,  K~band, and a completeness threshold $P_{\rm th}=0.7$.}
\label{fig:varyingMbreakMmaxLambda_H0}
\end{figure}

The effect of 
varying the parameter $\gamma_1$ within $1\sigma$ has already been shown in Fig.~\ref{fig:varyingGamma_H0}, where the shaded bands are obtained varying $\gamma_1$ within the range $\gamma_1=1.05^{+0.68}_{-1.08}$, which corresponds to its  error in the fit to the observed BBH mass 
distribution~\cite{Abbott:2020niy}.\footnote{\red{In these plots, and in the followings, the bands are just visualization tools, obtained varying a given parameter within a chosen range, and do not have a statistical meaning.  As discussed above, in principle the correct procedure would be to include these parameters in  a global inference, and  marginalize over them.}} 
\red{In Fig.~\ref{fig:varyingMbreakMmaxLambda_H0} the upper row (left panel) shows the effect of changing 
$m_{\rm break}$ within the range $m_{\rm break}=36.76^{+11.37}_{-12.48}\, \msun$, (corresponding to the $1\sigma$ variation of the LVC fit), with all other parameters of the mass function fixed at the reference values (\ref{mminbmax})--(\ref{bbreak}). The right panel shows the effect of changing  $m_{\rm max}$ (with the other parameters fixed) in the interval 
$m_{\rm max}=86.16^{+12.37}_{-13.65} \, \msun$. We observe that changing $m_{\rm max}$ (at fixed $m_{\rm break}$)  has only a minor effect. This can be understood from the fact that this only affects the tails of the mass distribution. The effects of changing $\gamma_1$ or $m_{\rm break}$ are more important, and somewhat comparable.} 

\red{A much more important effect is obtained changing the parameter $\lambda$ that enters in the low-redshift limits of the BBH merger rate, \eq{dNdVclambda}. The results are shown in the lower panel of Fig.~\ref{fig:varyingMbreakMmaxLambda_H0}, where the bands give the effect of changing $\lambda$ in \eq{pzlambda} within the astrophysically plausible range $[0,3]$, around  our reference value $\lambda=1$. The effect becomes particularly large at small values of $H_0$, say $H_0< 50\, \,\,{\rm km}\, {\rm s}^{-1}\, {\rm Mpc}^{-1}$. In principle,  one could cut out this region with a narrower choice of prior range, given what we already know on $H_0$ from cosmological observation. Still, even in the most interesting range, say $H_0\in  [60-80]\, \,\,{\rm km}\, {\rm s}^{-1}\, {\rm Mpc}^{-1}$, the effect of $\lambda$ on the shape of the posterior is quite significant, and appears to produce the dominant astrophysical uncertainty, at least among the astrophysical parameters that we have tested.}

\red{However, even if the shaded bands due to $\lambda$, in Fig.~\ref{fig:varyingMbreakMmaxLambda_H0}, are quite thick, the effect on the maximum a posteriori value and on the corresponding highest density intervals is not so large: for instance, for $\lambda=0$,  we get $H_0=68.6^{+31.3}_{-18.1}$, while,
for $\lambda=3$, we find $H_0=65.9^{+18.1}_{-21.6}$, to be compared with \eq{H0KbandPth07} for $\lambda=1$.  Of course, these results are all consistent, within their large statistical errors.}

\begin{figure}[t]
\centering
\includegraphics[width=\textwidth]{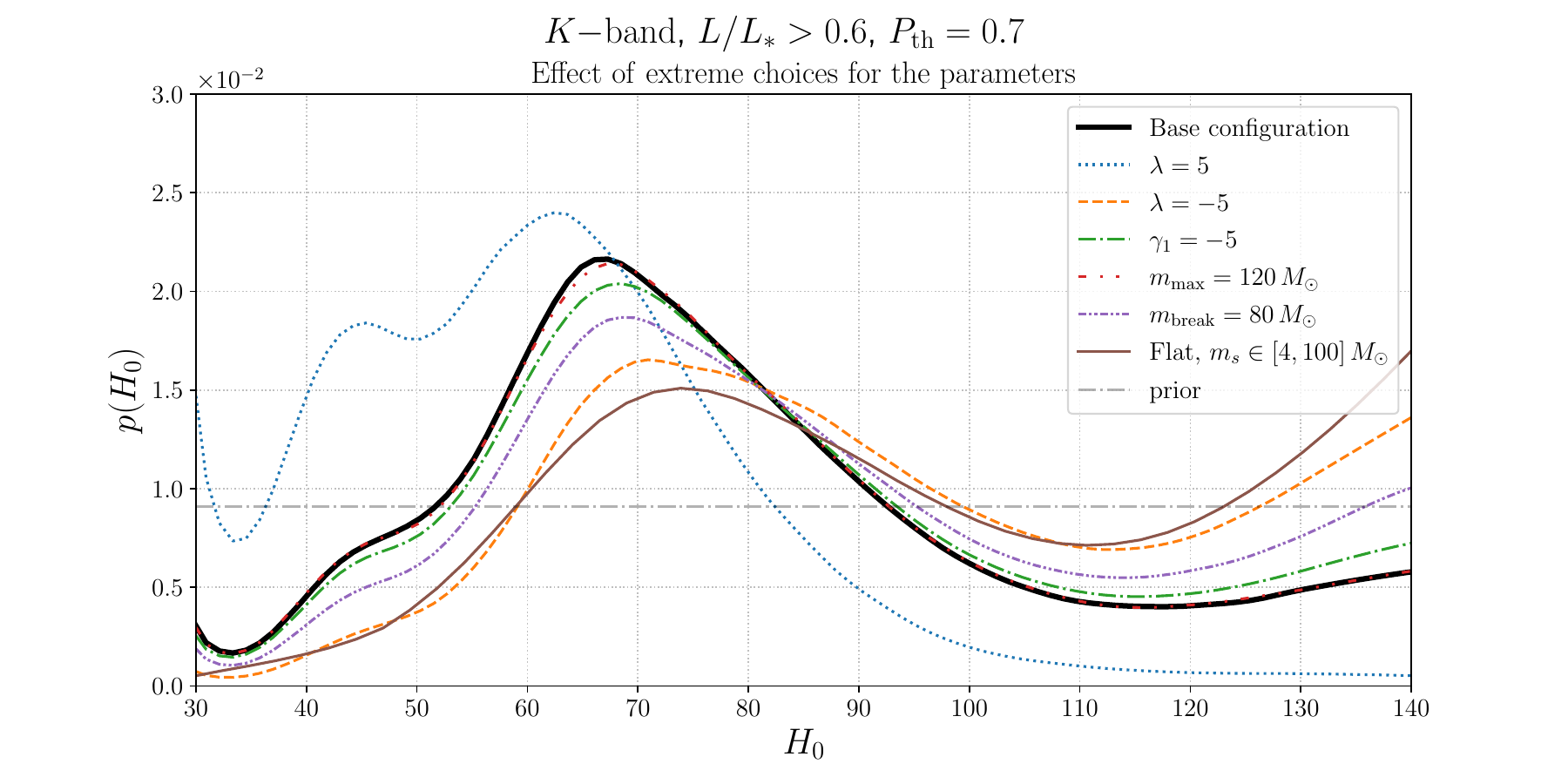}
\caption{The baseline result compared with that obtained with some extreme choices for the values of the parameters in the mass function (\ref{brokenpowerlaw}), as well as with a (source-frame) BBH mass distribution  flat in the range $[4,100] \msun$.}
\label{fig:Extreme_var_all_H0}
\end{figure}

In Fig.~\ref{fig:Extreme_var_all_H0} we show how our baseline result changes with some more extreme choices for the values of these parameters. In particular (fixing each time all other parameters at our baseline configuration), we test the values $\lambda=5$ and $\lambda=-5$, as well as the value $m_{\rm max}=120\msun$, and the value $m_{\rm break}=80\msun$. We even consider, as an extreme case, a source-frame BBH mass distribution  flat between $4\msun$ and $100\msun$.  While, as expected, with such large changes the 
posterior for $H_0$ is appreciably modified, still we see that a clear peak always emerges, showing that the results are coming from the galaxy catalog and not from the assumed mass distribution.

\red{In app.~\ref{sect:dependence_results} we will study how the results presented here change with the completeness threshold, we will compare the results obtained for mask completeness with those obtained adopting cone completeness; for each of these two choices  we will  compare multiplicative with homogeneous completion, and we will compare K-band with B-band luminosity weighting.}

\subsubsection{Dark sirens combined with GW170817}\label{sect:GW170817}

Finally, it is interesting to combine the results obtained from dark sirens with the posterior for $H_0$ obtained from the BNS GW170817 and its electromagnetic counterpart. 
For GW170817, the likelihood (\ref{pDizO})  must be supplemented by the EM data about the position of the host galaxy NGC4993, and reads
\be\label{pDizOEM}
p({\cal D}_{\rm GW}, {\cal D}_{\rm EM}|H_0)=\frac{1}{\beta(H_0)}
\int dzd\Omega d\theta'\, p({\cal D}_{\rm EM}|z) p({\cal D}_{\rm GW}|d_L(z,H_0),\hatO,\theta')\, p_0(z,\hatO,\theta')\, ,
\ee
where now $\beta(H_0)$ is the normalization factor appropriate to BNS.
The likelihood for the EM measurement is given by
\be\label{EMlikGW170817}
p({\cal D}_{\rm EM}|z) = \mathcal{N}(z; z_{\rm EM}, \sigma_{\rm EM}) \times \delta\left( \hatO -\hatO_{\rm EM}\right) \, , 
\ee
where $z_{\rm EM}$, $\sigma_{\rm EM}$ denote the mean and standard deviation of the redshift of the EM counterpart, and $\hatO_{\rm EM}$ its angular position.
For the mean and standard deviation of the redshift, it is important to take into account peculiar velocity corrections. The uncorrected value of the recession velocity of the host galaxy NGC4993 is $v=3327 \pm 72\, \rm{km \; s^{-1}}$ \cite{Crook:2006sw, Abbott:2017xzu}. 
In principle, to fully take into account peculiar velocity corrections, one should add to \eq{pDizOEM} a marginalisation over the peculiar velocity distribution. In Ref.~\cite{Abbott:2017xzu} this is taken to be a gaussian with mean and standard deviation $v_{\rm p}=310 \pm 150 \, \rm{km \; s^{-1}}$. In this case the correction amounts to  using in \eq{EMlikGW170817} a gaussian with mean obtained by the recession velocity of NGC4993 corrected for the peculiar velocity, and standard deviation obtained by summing the errors in quadrature, which gives $v= 3017\pm 166\, \rm{km \; s^{-1}}$. Ref.~\cite{Mukherjee:2019qmm} uses a more refined estimate of the peculiar velocity PDF, resulting in $v_{\rm p}=373 \pm 130 \, \rm{km \; s^{-1}}$. The full PDF is non-gaussian, but in absence of its exact shape we will use a gaussian approximation with this mean and standard deviation. The resulting velocity of NGC4993 in this approximation is $v= 2954\pm149  \, \rm{km \; s^{-1}}$. We shall use this value in the following, which corresponds to using $z_{\rm EM}=0.0098 $, $\sigma_{\rm EM}=0.0004$ in~\eq{pDizOEM}.\footnote{We checked that, when using the value of Ref.~\cite{Abbott:2017xzu}, our result fully agree with that obtained with the publicly available code  $\tt{gwcosmo}$ (\url{https://git.ligo.org/lscsoft/gwcosmo}) used by the LVC.}
 
 As for the prior $p_0(z,\hatO,\theta')$, we take $\theta'=\{m_1, m_2\}$,  where $m_1, m_2$ are the source-frame masses, and we write the prior as $p_0(z,\hatO,\theta')=\tilde p_0(z) \tilde p_0(m_1, m_2)$, as in sect.~\ref{sect:rate}. Now, however, $\tilde p_0(m_1, m_2)$ is the distribution of (source-frame) NS masses in a BNS, and we take it to be   flat  between 1 and 3 solar masses. We choose $\tilde p_0(z)$ as in \eq{pzlambda}, setting for definiteness   $\lambda=1$. In any case, given the smallness of the redshift of GW170817, the choice of $\lambda$ is here irrelevant.
 
The term $p({\cal D}_{\rm GW}|d_L(z,H_0),\hatO,\theta')$ in \eq{pDizOEM} is the GW likelihood, marginalised over all variables except $d_L(z,H_0),\hatO$, and $\theta' =\{m_1, m_2\}$.
Using Bayes' theorem and \eq{EMlikGW170817}, and integrating over $\hatO$, we can rewrite \eq{pDizOEM} as
\bees 
p({\cal D}_{\rm GW}, {\cal D}_{\rm EM}|H_0)&=&\frac{1}{\beta(H_0)}
\int dz \, \mathcal{N}(z; z_{\rm EM}, \sigma_{\rm EM})\, \tilde p_0(z) \nn\\
&\times& \int d m_1 d m_2\, \frac{p(d_L(z,H_0),\hatO_{\rm EM}, m_1, m_2 | {\cal D}_{\rm GW})}{\pi(m_1, m_2) \pi(d_L(z,H_0))}\, \tilde p_0(m_1, m_2)\, .\label{pDizOEM1}
\ees
where $\pi(m_1, m_2)$, $\pi(d_L)$ are the prior used in the LVC analysis.
The LVC provides posterior samples for GW170817 restricted to the sky location of the host galaxy~\cite{LIGOScientific:2018mvr}, with a prior on luminosity distance  $\pi(d_L)\propto d_L^2$ and a flat prior in detector frame masses.\footnote{Available at \url{https://dcc.ligo.org/public/0157/P1800370/005/GW170817_GWTC-1.hdf5}.} Taking into account the $(1+z)$ factor  between source frame and detector frame masses, the LVC prior is equivalent to $\pi(m_1, m_2) \propto (1+z)^2$, where $m_1, m_2$ are the source-frame masses.  \red{To get the likelihood, we must  divide the  posteriors provided by the LVC by the priors that they use.  Finally, we must multiply by the probability distribution 
$\tilde p_0(m_1, m_2) $, which corresponds exactly to use the same prior as in $\beta(H_0)$. 
In order to evaluate the integral over the masses in \eq{pDizOEM1},
for given $H_0$ we then re-weight the LVC posterior samples by a factor}
\be\label{weights}
w \propto \frac{\tilde p_0(m_1, m_2)}{ [1+z(d_L, H_0)]^2}\, .
\ee
The resulting samples correspond to draws from the distribution 
\be
p(d_L(z,H_0),\hatO_{\rm EM},m_1, m_2 |  {\cal D}_{\rm GW}) \frac{\tilde p_0(m_1, m_2)} {\pi(m_1, m_2)} \, .
\ee
Then, the integral over masses in~\eq{pDizOEM1} corresponds to a marginalisation yielding the marginal likelihood $p({\cal D}_{\rm GW}|d_L(z,H_0),\hatO_{\rm EM} ) $. We can evaluate it by interpolating the $d_L$ samples and further removing the  $d_L^2$ prior.\footnote{For the interpolation, we use a Kernel Density Estimator (KDE)  with unit bandwidth in cartesian coordinates $(x_1, x_2, x_3) = \allowbreak (d_L \cos{\alpha_{\rm EM}} \cos{\delta_{\rm EM}}, \allowbreak
 d_L \sin{\alpha_{\rm EM}} \cos{\delta_{\rm EM}}, \allowbreak d_L \sin{\delta_{\rm EM}})$ where $\alpha_{\rm EM}$, $\delta_{\rm EM}$ are the right ascension and declination of the counterpart. One might  wonder why using this coordinate system if we are provided with posterior samples in the direction of the counterpart only. Actually, in this case the result is equivalent to a simple KDE on the luminosity distance; however, in more general cases, as it will be for GW190521, this is not the case, and we find that using cartesian coordinates gives better results (see also~\cite{Chen:2020gek}). Note that one must take care when using algorithms to interpolate a probability distribution, since these yield normalised pdfs, thus loosing the information contained in the normalization given by the weights~(\ref{weights}). This factor must then tracked and used to multiply the result of the interpolation. We thank Simone Mastrogiovanni for discussion about this point.\label{foot:KDE}} 
The remaining integral in redshift is evaluated by MC integration sampling from the gaussian distribution in redshift.
Finally,  for GW170817 the selection effect $\beta(H_0)$ is computed from our MC (as discussed above,   
in this case for $\tilde p_0(m_1, m_2)$ we use a flat mass distribution in the range $[1-3] \msun$ for the source-frame masses of the neutron stars) in the  absence of the additional selection effects given by the completeness of the catalog, described at the end of sect.~\ref{sect:MC}. This also amounts to assuming that the EM counterpart for this event could always be detected, hence there are no selection effects depending on the EM part. We use the O2 strain sensitivity.  

\begin{figure}[t]
\centering
\includegraphics[width=\textwidth]{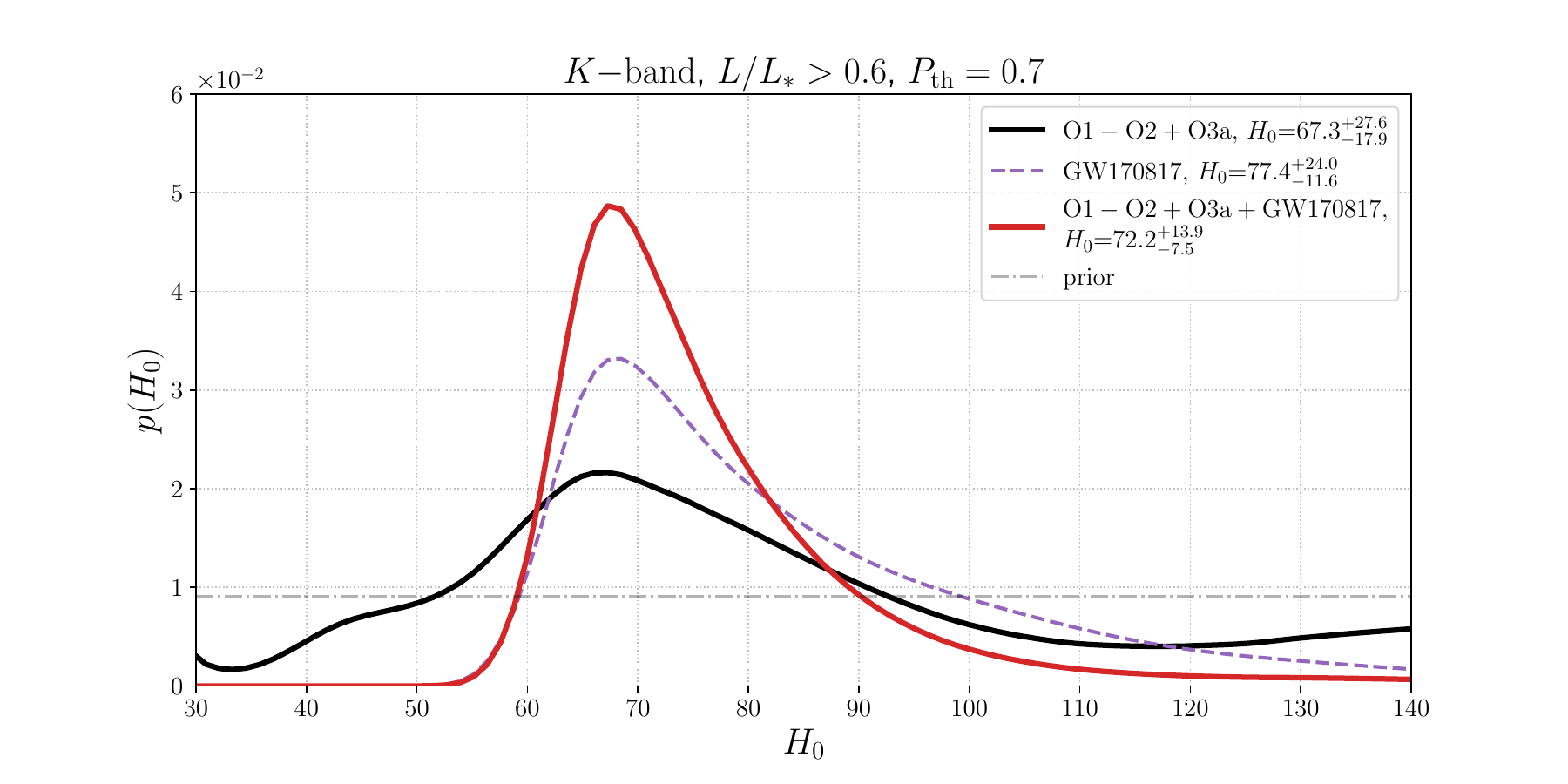}
\caption{The  posterior of $H_0$ obtained from GW170817 with its electromagnetic counterpart (violet dashed line), the posterior obtained from dark sirens in the O1+O2+O3a run obtained using mask completeness, multiplicative completion, K band, and $P_{\rm th}=0.7$ (black line, the same as the black lines in Figs.~\ref{fig:varyingGamma_H0} and \ref{fig:separate_H0}), and the combination of GW170817 and dark sirens (red solid line).
}
\label{fig:dark_and GW1708}
\end{figure}

The results obtained combining O1+O2+O3a dark sirens with GW170817+counterpart
are shown in  Fig.~\ref{fig:dark_and GW1708}, where we show, as before,  the results for dark sirens is obtained  using  mask completeness, multiplicative completion,  K~band luminosity weighting, and a threshold on completeness
$P_{\rm th}=0.7$. We see that the posterior of  GW170817+counterpart falls to zero well before the upper limit of the prior range, and induces a similar behavior on the combined posterior. It therefore makes sense to quote the median of the posterior distribution. 
Using GW170817 alone, for  the  median and symmetric $68.3\%$ credible interval we find
\be\label{H0GW17only}
H_0= \red{77.4 ^{+24.0}_{-11.6}}\,\,\,  {\rm km}\, {\rm s}^{-1}\, {\rm Mpc}^{-1}\, ,  \quad\quad 
\mbox{\small (GW170817 only)} \, .\\
\ee
This is consistent with the value $H_0=74^{+16}_{-8}\,  {\rm km}\, {\rm s}^{-1}\, {\rm Mpc}^{-1}$ (again median and symmetric $68.3\%$ credible interval)
obtained in  ref.~\cite{Abbott:2017xzu} using a prior proportional to $d_L^2$, a flat-in-log prior on $H_0$,
and the value of the recession velocity of NGC4993 discussed above. 

Combining the posterior from GW170817 with that from dark sirens, we get
\be\label{H0GW17anddark}
H_0=\red{72.2^{+13.9}_{-7.5}}\,\,  {\rm km}\, {\rm s}^{-1}\, {\rm Mpc}^{-1}\, , \quad\quad \mbox{\small(GW170817+dark sirens) }\, .
\ee
\red{We see that the addition of dark sirens provides a non-negligible improvement on the accuracy of the measurement of $H_0$. \Eq{H0GW17anddark} corresponds to a  relative error on $H_0$ of $15\%$, to be compared with the relative error of $23\%$, obtained from \eq{H0GW17only}, using  GW170817 only. \Eq{H0GW17anddark} also  improves  on  the  result  $H_0=70^{+20}_{-9} \,{\rm km}\, {\rm s}^{-1}\, {\rm Mpc}^{-1}$,  which
corresponding to  $21\%$ error,
obtained in~\cite{Abbott:2019yzh}  from GW170817 and dark sirens from O1-O2 (using again a flat prior on $H_0$, as in our case). 
To compare with  \cite{Abbott:2019yzh}, we also report  the result that we obtain from GW170817+dark sirens using  a uniform in log prior on $H_0$,
$H_0=70.9^{+12.2}_{-6.8}
\,\,{\rm km}\, {\rm s}^{-1}\, {\rm Mpc}^{-1}$, corresponding to a $13\%$ error. This should be compared with  the value
$H_0=69^{+16}_{-8}
\,\,{\rm km}\, {\rm s}^{-1}\, {\rm Mpc}^{-1}$ obtained in \cite{Abbott:2019yzh}. However, for $H_0$, our prior information from cosmological observations  does not span several orders of magnitudes, so a prior uniform in $H_0$ is more justified than a prior uniform in $\log H_0$.}

The above results were obtained without any modelization of the jet emission from GW170817, and rely only on the fact that standard sirens provide an absolute measurement of the luminosity distance. Therefore, they are purely `gravitational' results.
However, Very Long Baseline Interferometry (VLBI) constrains the emission angle of the jet associated to the EM counterpart of GW170817~\cite{Mooley:2018qfh}. In particular, one can constrain the viewing angle $\Theta = \min(\theta_{\rm {JN}}, \pi-\theta_{\rm {JN}})$, where $\theta_{\rm {JN}}$ is the angle between the line of sight and the total angular momentum of the binary~\cite{TheLIGOScientific:2017qsa}, 
as
\be\label{VLBI}
0.25\, {\rm rad} \leq  \left(  \frac{d_L}{41\,  \rm{Mpc}} \right) \Theta \, \leq0.45\, {\rm rad} \, .
\ee
For small spins of the individual binaries and negligible precession of the orbital plane, as in the case of GW170817, $\theta_{\rm {JN}} \simeq \iota$ where $\iota$ is the inclination angle,  so this information can be used to break the degeneracy between the distance and the inclination of the orbit. 
One then obtains a significantly tighter measurement of the Hubble parameter, $H_0=68.3^{+4.6}_{-4.5}\,  {\rm km}\, {\rm s}^{-1}\, {\rm Mpc}^{-1}$~\cite{Mukherjee:2019qmm}. Of course, one must stress that   this result is based on some jet modeling, which could potentially introduce extra  uncertainties and systematic errors. 

\begin{figure}[t]
\centering
\includegraphics[width=\textwidth]{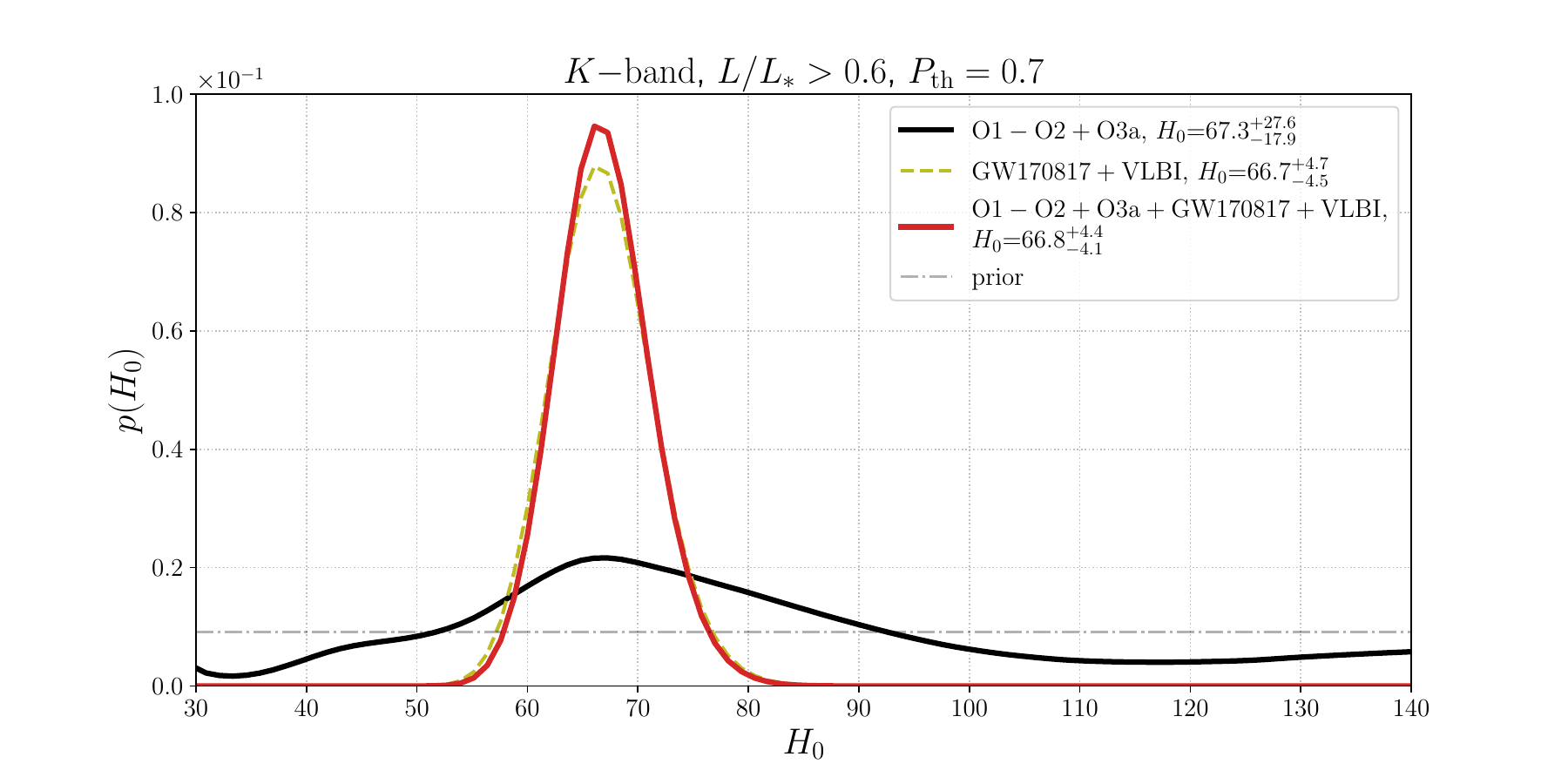}
\caption{As in Fig.~\ref{fig:dark_and GW1708}, adding the VLBI information on GW170817.}
\label{fig:dark_and GW1708_VLBI}
\end{figure}

In any case, it is obviously interesting to combine this result with dark sirens. First of all, we have repeated  
 the analysis for GW170817 using the VLBI constraint, by selecting only the posterior samples in the  interval~(\ref{VLBI}), finding $H_0=66.7^{+4.7}_{-4.5}\, {\rm km}\, {\rm s}^{-1}\, {\rm Mpc}^{-1}$, consistent with~\cite{Mukherjee:2019qmm}.\footnote{The small relative difference could be due to a slightly different re-weighting of the samples, a different interpolation of the marginal likelihood, or to our simplified assumption of a gaussian pdf for the peculiar velocity.} We have then combined it with the dark sirens posterior. 
The results are shown in  Fig.~\ref{fig:dark_and GW1708_VLBI}, 
using again  mask completeness, multiplicative completion, K-band, and $P_{\rm th}=0.7$.
We see that, in this case, given that the posterior from GW170817 alone becomes much more narrow, combining it with the posteriors from current dark sirens gives only a very marginal improvement.\footnote{Note, however, that events like GW170817, where the constraint on the inclination can give such an  improvement, are not expected to be the majority~\cite{Mastrogiovanni:2020ppa}.} Our result, from GW170817  with VLBI  plus dark sirens, is
\be\label{H0VLBIdark}
H_0=66.8^{+4.4}_{-4.1}\, \,{\rm km}\, {\rm s}^{-1}\, {\rm Mpc}^{-1} \quad 
\mbox{\small  (GW170817+VLBI+dark\,  sirens) },
\ee
corresponding to a measurement of $H_0$ at the $6.3\%$ level. By comparison, the {\em Planck} value  is $H_0=67.4^{+0.5}_{-0.5}\, {\rm km}\, {\rm s}^{-1}\, {\rm Mpc}^{-1}$~\cite{Aghanim:2018eyx} and  the SH0ES value is 
$H_0=74.0^{+1.4}_{-1.4}\, {\rm km}\, {\rm s}^{-1}\, {\rm Mpc}^{-1}$~\cite{Riess:2019cxk}
(which becomes $H_0=73.8^{+1.1}_{-1.1}\, {\rm km}\, {\rm s}^{-1}\, {\rm Mpc}^{-1}$ when combined with H0LiCOW~\cite{Wong:2019kwg}).\footnote{See, however, ref.~\cite{Khetan:2020hmh} for a result closer to {\em Planck} obtained by calibrating type Ia SNe with surface brightness fluctuations.}
 The result (\ref{H0VLBIdark}) slightly favors the {\em Planck} result. However,  the error  is still quite large (with both the  SH0ES  and the combined SH0ES+H0LiCOW results still within $1.5\sigma$), and one should also keep in mind the potential uncertainty due to possible systematics in the jet modelization, so that no definite conclusion can yet be drawn.

\subsection{Modified GW propagation}\label{sect:Xi0results}

We now repeat the analysis for modified GW propagation. As discussed in sect.~\ref{sect:HBf} (see also footnote~\ref{foot:H0RT} on page~\pageref{foot:H0RT}), 
in this case we will fix $H_0$ to the value $H_0=67.88\,\, {\rm km}\, {\rm s}^{-1}\, {\rm Mpc}^{-1}$, which is the mean value obtained from a fit to CMB+BAO+SNa data, both in the $\Lambda$CDM and in the RT nonlocal model  with large $\Delta N$ (and letting the sum of neutrino masses vary freely within the observational limits).  In principle, a more correct approach would be to let both
$H_0$ and $\Xi_0$ free, with a prior on $H_0$ fixed by the  CMB+BAO+SNa data. However, CMB+BAO+SNa data fix $H_0$ to a precision of about $0.7\%$  while, as we will see, the bounds that we will obtain on $\Xi_0$  will rather be in the range $\Xi_0<{\cal O}(5)$, i.e. a  deviation of order $500\%$  from the GR value $\Xi_0=1$. With current data, therefore, the degeneracy with $H_0$ can be neglected.\footnote{As already remarked in sect.~\ref{sect:HBf}, eventually one might also want to extract $H_0$ and $\Xi_0$ from standard sirens only, without using at all priors from electromagnetic data, but this is  beyond what can be done with current data, at least to any useful accuracy.}
We will also fix $n$ in \eq{eq:fit} to the value $n=1.91$, which is the value predicted by the RT model in the  RT nonlocal model  with large $\Delta N$, but this choice has a limited impact on our analysis, and similar results are obtained as long as $n={\cal O}(1)$.

\begin{figure}[t]
\centering
\includegraphics[width=\textwidth]{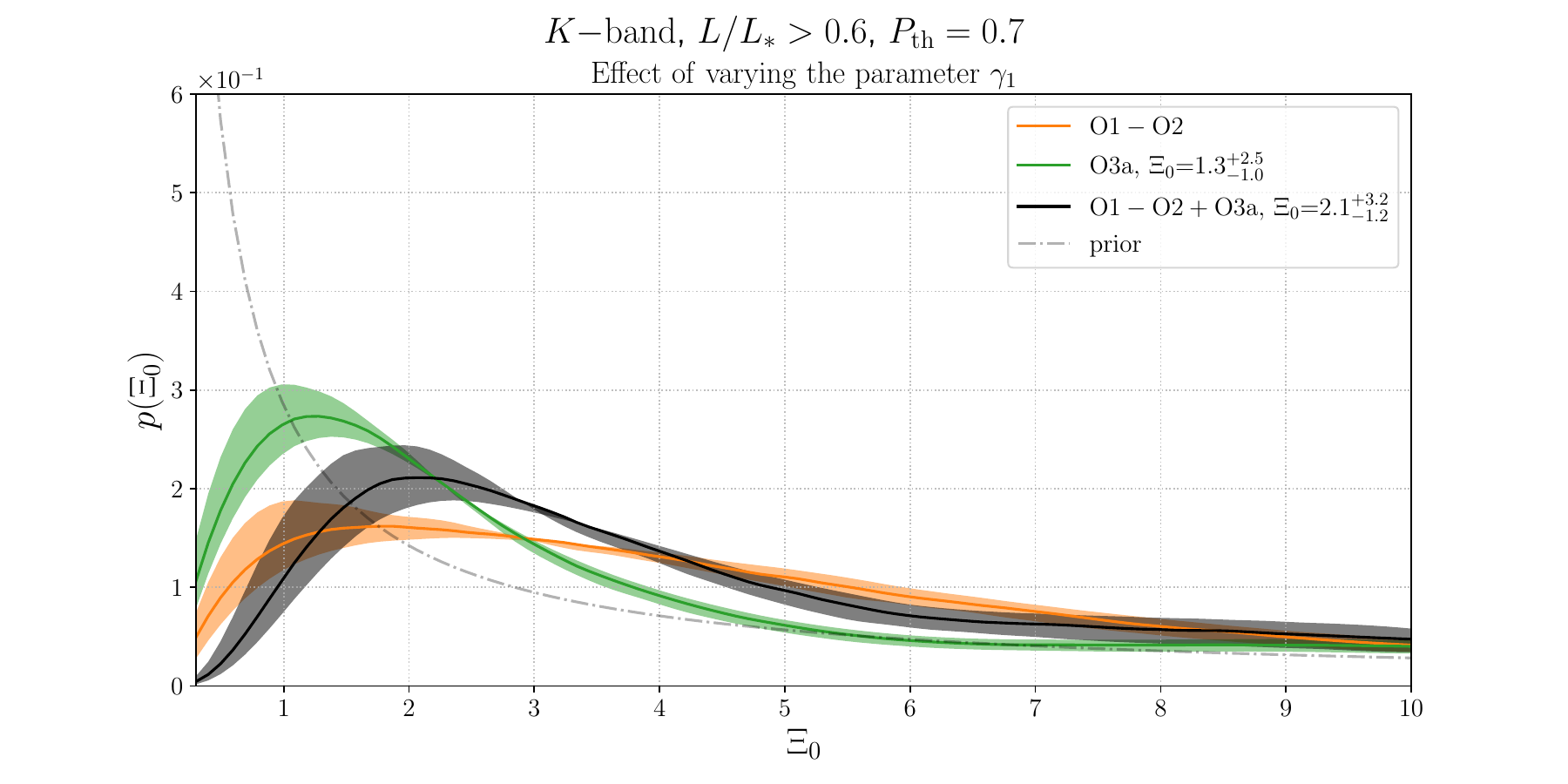}
\caption{The separate posterior for $\Xi_0$ from the O1+O2 data and from the O3a data, and their combined posterior (using mask completeness, multiplicative,  completion, $P_{\rm th}=0.7$, K-band luminosity weighting, and a prior uniform in $\log\Xi_0$). 
The shaded bands give the effect of changing $\gamma_1$ within the range $\gamma_1=1.05^{+0.68}_{-1.08}$.}
\label{fig:varyingGamma_Xi0}
\end{figure}

\begin{figure}[th]
\centering
\resizebox{1.04\columnwidth}{!}{
\includegraphics[width=0.52\textwidth]{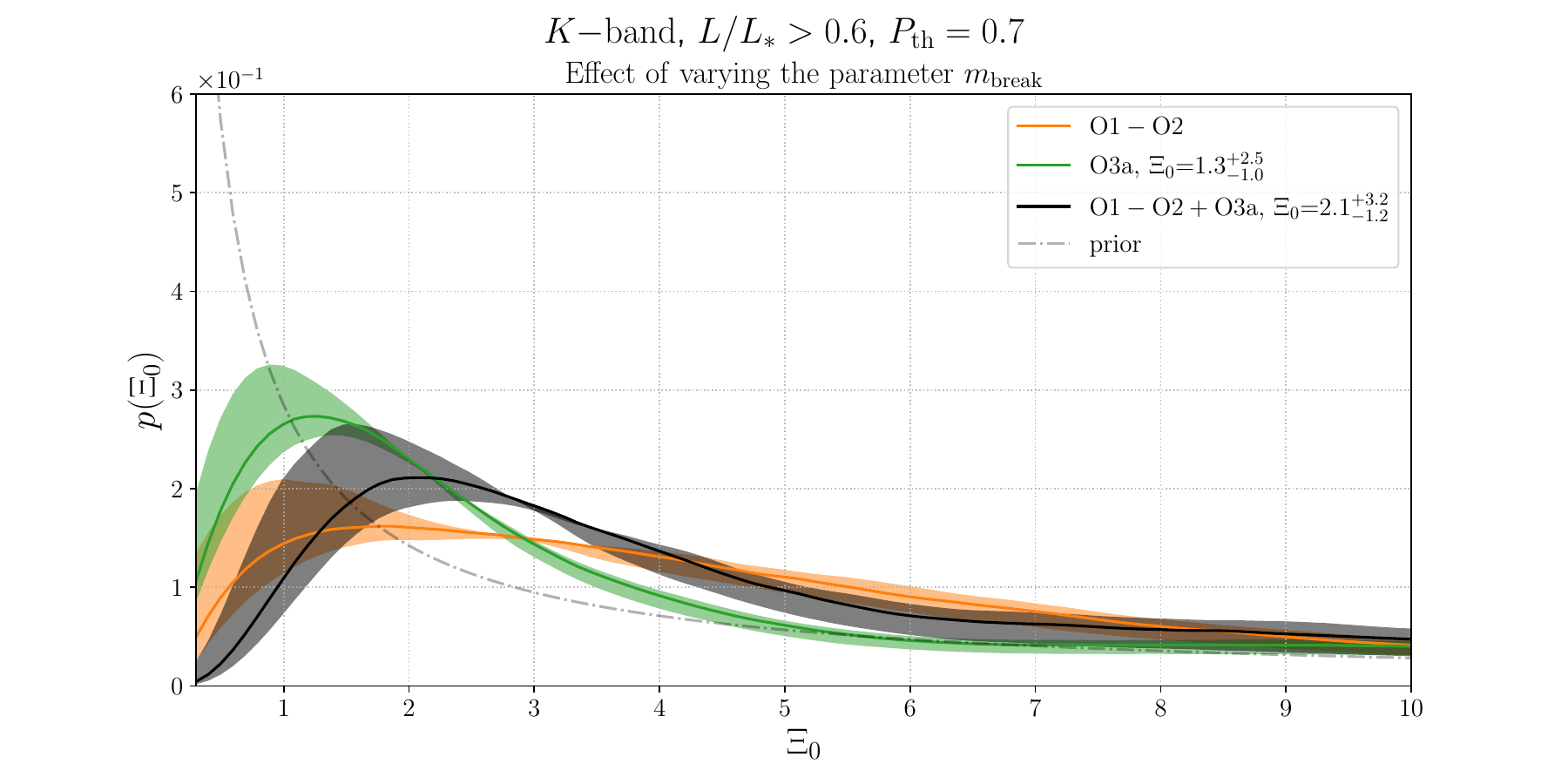}
\includegraphics[width=0.52\textwidth]{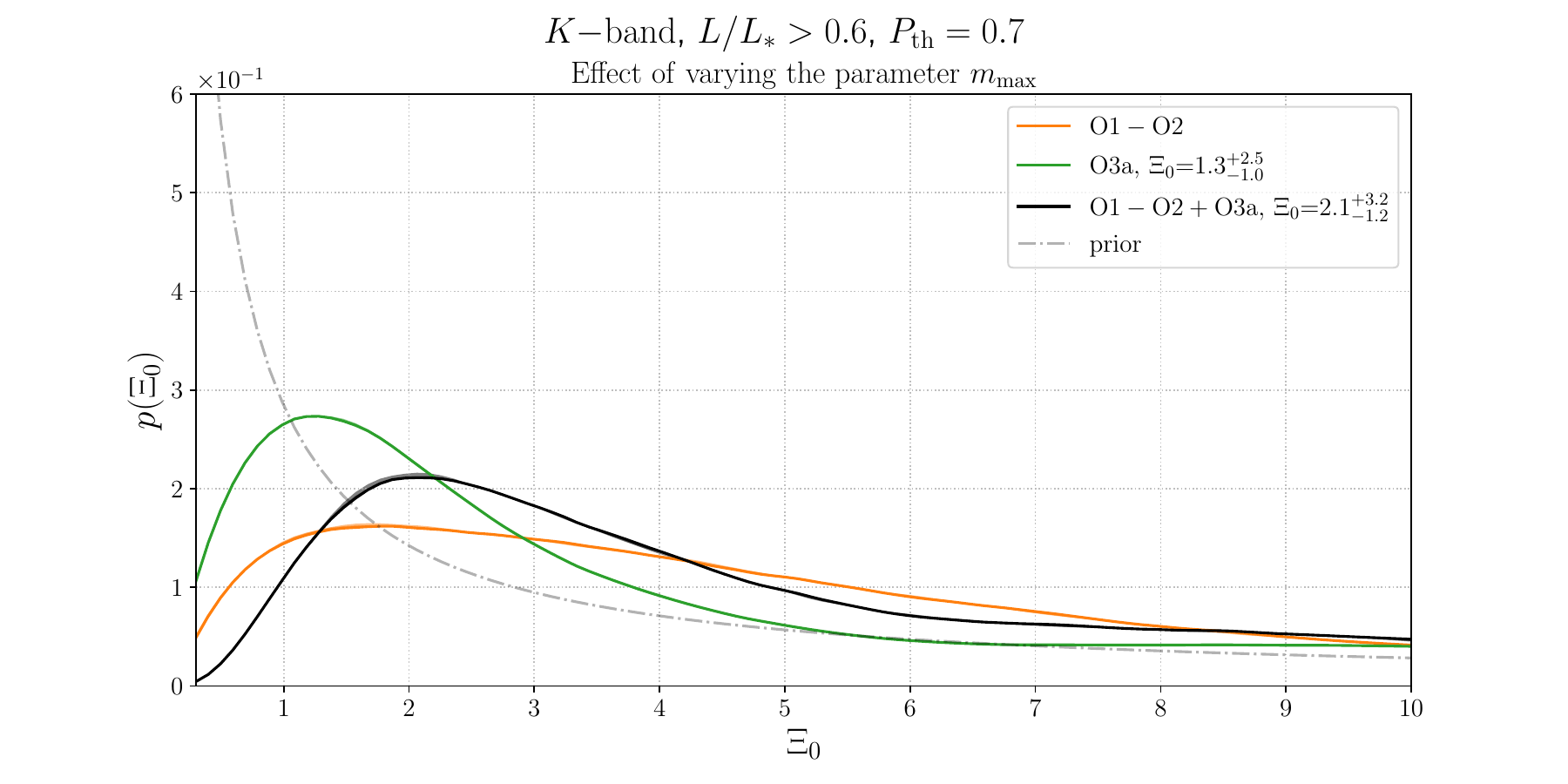}
}
\includegraphics[width=0.52\textwidth]{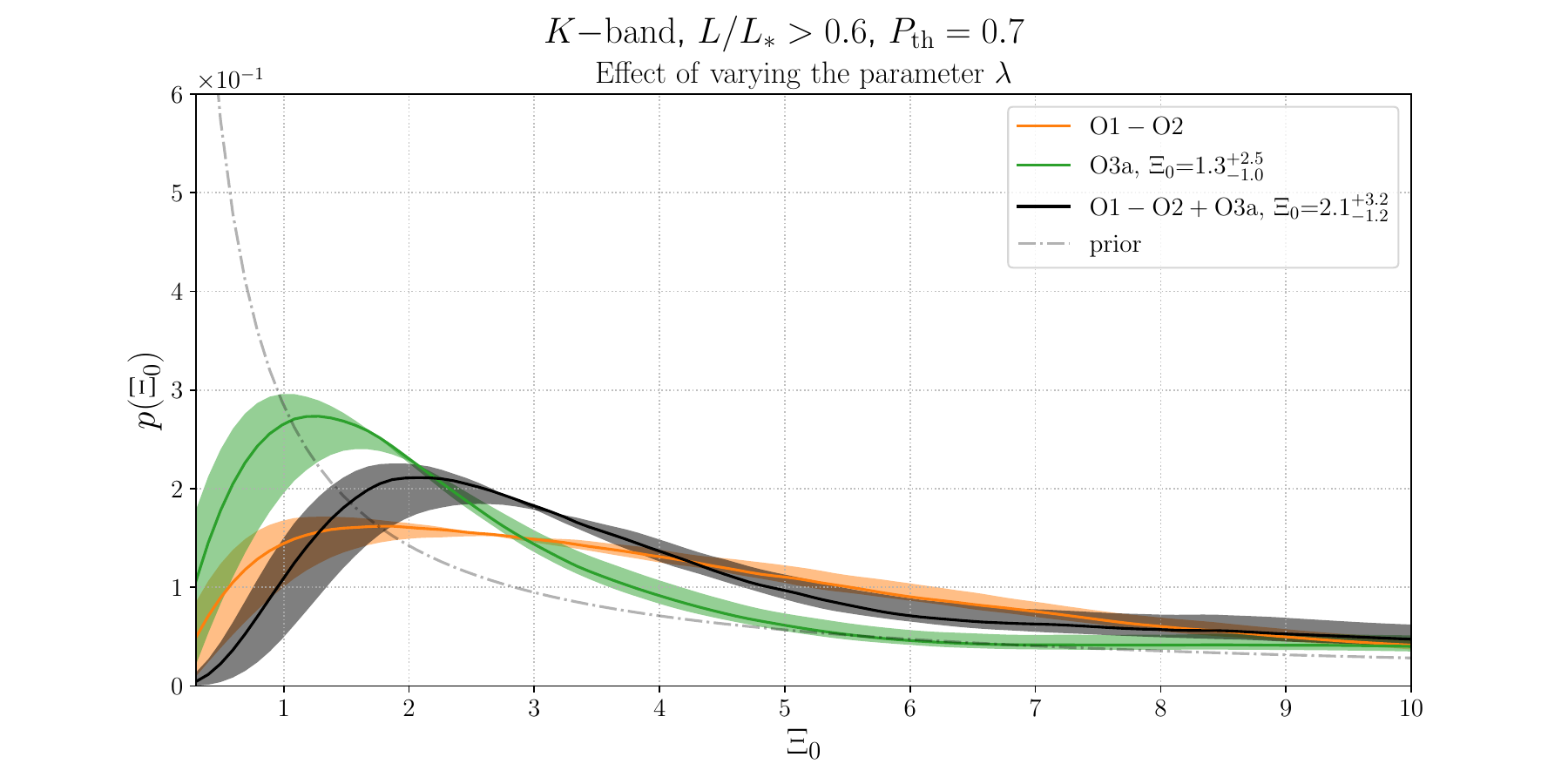}
\caption{Upper row: the effect of  changing $m_{\rm break}$ within the range $m_{\rm break}=36.76^{+11.37}_{-12.48}\, \msun$ (left panel) and the effect of  changing $m_{\rm max}$ within the range $m_{\rm max}=86.16^{+12.37}_{-13.65} \, \msun$ (right panel). Lower panel: the effect of changing $\lambda$ in \eq{pzlambda} within the range $[0,3]$, with the central lines corresponding to our default value $\lambda=1$. We use mask completeness, multiplicative completion,  K~band, and a completeness threshold $P_{\rm th}=0.7$.}
\label{fig:varyingMbreakMmaxLambda_Xi0}
\end{figure}

\red{For the Hubble constant, we have used a prior uniform in $H_0$. This is the most natural choice, given that electromagnetic observations already constrain $H_0$ significantly. In contrast, as we reviewed in sect.~\ref{sect:persp}, current limits on $\Xi_0$ are very broad. We will therefore use a correspondingly broad prior range 
$\Xi_0\in [0.3,10]$. In such a range, that extends over almost two orders of magnitude, a more natural choice is a prior uniform in $\log\Xi_0$, rather than in $\Xi_0$. In the following, we will therefore always use a uniform-in-log prior for $\Xi_0$. Since $d\log\Xi_0=d\, \Xi_0/\Xi_0$, with respect to a prior uniform in $\Xi_0$ this has the effect of suppressing the tails of the posterior at  large $\Xi_0$. As discussed in App.~\ref{sect:dependence_results}, the more stringent prior knowledge that we have on $H_0$ allowed us
to determine the best settings for our baseline model. We will then use exactly the same settings for $\Xi_0$, namely mask completeness, multiplicative completion, $P_{\rm th}=0.7$ and K-band luminosity weighting.
}
 
In Fig.~\ref{fig:varyingGamma_Xi0} (which is the analogous of Fig.~\ref{fig:varyingGamma_H0} for $H_0$) we  show the posteriors for $\Xi_0$  from the O1-O2 data, the O3a data, and the total posterior obtained by combining the  O1+O2 data with  the O3a data. As in Fig.~\ref{fig:varyingGamma_H0},
the central lines in the shaded bands give the posteriors for $\gamma_1=1.05$, while the shaded bands give the effect of varying $\gamma_1$ in the interval $\gamma_1=1.05^{+0.68}_{-1.08}$.
For the maximum a posteriori value and its corresponding highest density interval, from the combined O1+O2+O3a data, we get
\be\label{Xi0limitdarkK02}
\Xi_0=2.1^{+3.2}_{-1.2} \, .
\ee
The effect of changing  $\gamma_1$ in the BBH mass function, within the limits $\gamma_1=1.05^{+0.68}_{-1.08}$,  can be appreciated from the shaded bands in Fig.~\ref{fig:varyingGamma_Xi0}.
In Fig.~\ref{fig:varyingMbreakMmaxLambda_Xi0} we show the effect of changing $m_{\rm break}$, 
$m_{\rm max}$ and $\lambda$, similarly to the plots made for $H_0$ in Fig.~\ref{fig:varyingMbreakMmaxLambda_H0}. We observe that changing $m_{\rm max}$  (at fixed $m_{\rm break}$) has almost no effect. The effect of changes in $\lambda$ on the posterior of $\Xi_0$ is less pronounced, compared to its effect on the posterior of $H_0$, and about comparable to that of changes in $m_{\rm break}$ and $\gamma_1$.

\begin{figure}[t]
\centering
\includegraphics[width=\textwidth]{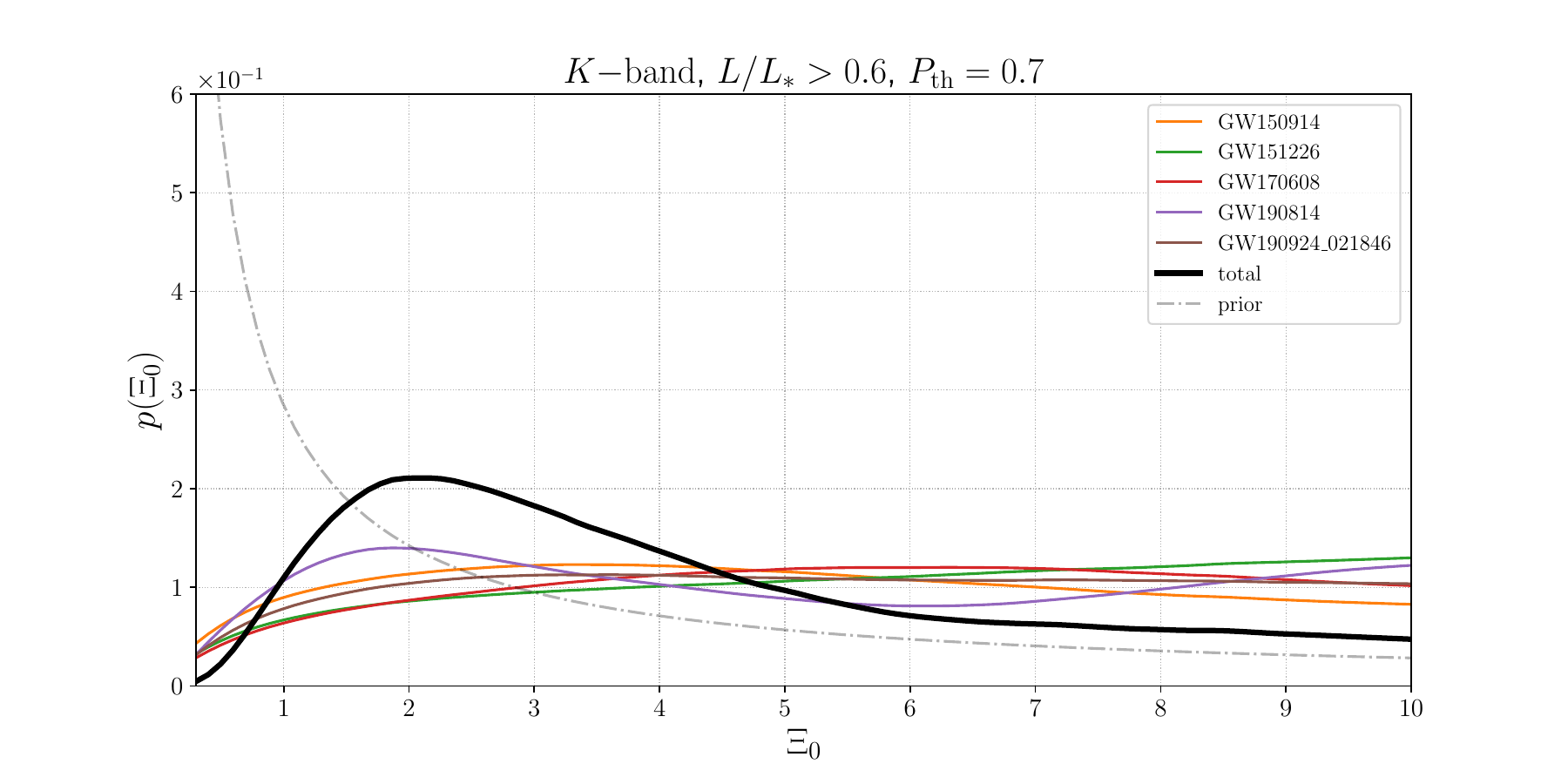}
\caption{The separate  contributions to the posteriors of $\Xi_0$, from   the individual events  in O1+O2+O3a that pass the completeness threshold $P_{\rm th}=0.7$. The black line is the total posterior from O1+O2+O3a, already  shown as the  line inside the black band
in Fig.~\ref{fig:varyingGamma_Xi0}.}
\label{fig:separate_Xi0}
\end{figure}

The individual contributions from the five  events  in O1+O2+O3a that pass the completeness threshold $P_{\rm th}=0.7$ are shown in Fig.~\ref{fig:separate_Xi0}. Similarly to what we found for $H_0$, the single most significant contribution comes from GW190814, but all other events contribute to the total posterior.

\subsection{GW190521 and its candidate counterpart}\label{sect:GW190521}

One of the most interesting detections in the O3a run is the BBH GW190521~\cite{Abbott:2020tfl,Abbott:2020mjq}. Assuming a quasi-circular orbit, it is interpreted as the coalescence of two BHs with masses $85^{+21}_{-14}$ and $66^{+17}_{-18}\msun$. One reason that makes this event remarkable is that, because of the   pulsational pair-instability supernova mechanism,  it is  believed that no BH should form from stellar collapse  in the approximate mass range $\sim (65-135)\msun$. The primary BH mass in GW190521 lies in this  mass gap, with only $0.32\%$ probability of being below $65\msun$. The inferred mass of the final BH is $142^{+28}_{-16}\msun$, which classifies it as the first observation of an  intermediate mass BH.
Another remarkable property of this event is its luminosity distance, estimated to be 
$d_L\simeq 5.3^{+2.4}_{-2.6}$~Gpc, making it currently the farthest  GW detection.\footnote{This value actually refers to the `NRSurPHM' waveform model; see below for more details on the posterior on $d_L$ and the dependence on the waveform model used.} In $\Lambda$CDM, and assuming a {\em Planck} cosmology, the corresponding redshift is $z=0.82^{+0.28}_{-0.34}$.  However, a main theme of this paper is that in modified gravity this luminosity distance becomes a GW luminosity distance and the redshift will be different, and we will elaborate on this below.

Ref.~\cite{Graham:2020gwr} has  proposed a potential electromagnetic counterpart to this GW detection, consisting of an electromagnetic flare detected by the Zwicky Transient Facility,  labeled ZTF19abanrhr. The flare is associated with an AGN at $z=0.438$ and could be interpreted as due to the remnant BH, which has been  kicked by the merger and is moving at high speed through the accretion disk of the AGN, according to a model previously proposed in ref.~\cite{McKernan:2019hqs}. As we already mentioned in sect.~\ref{sect:persp}, doubts have been raised on this association. One aspect of the problem  is that it is not currently clear how plausible is the model for an AGN flare from a kicked BH, and on this issue we will have nothing to add. On top of this, in ref.~\cite{Ashton:2020kyr} it  has been argued that the probability of a real association based on volume localization is not statistically sufficient for a confident identification of ZTF19abanrhr as the counterpart. This analysis, however, is based on the redshift determination of the GW event made using GR and $\Lambda$CDM. In modified gravity the situation  is different, and the comparison of the GW posterior on the luminosity distance, that in this case is a GW luminosity distance, with the redshift of the candidate counterpart, can be translated into a posterior for $\Xi_0$. It is therefore interesting to reanalyze the possible association with the counterpart, in the context of modified GW propagation. A similar  analysis has already been done in \cite{Mastrogiovanni:2020mvm}, using various parametrization of modified GW propagation including our \eq{eq:fit}, while in \cite{Chen:2020gek} the event as been studied in the context of $w$CDM, and we will compare below with their results.

We consider again a likelihood of the form  (\ref{pDizOEM}), with the likelihood for the EM measurement  given by \eq{EMlikGW170817}, and we will study both the posterior for $H_0$ in the framework of $\Lambda$CDM, obtained from GW190521 assuming that ZTF19abanrhr is indeed the counterpart, and the corresponding posterior for $\Xi_0$ in the framework modified gravity. As discussed after \eq{dNdtheta}, when
we study the Hubble parameter in the context of $\Lambda$CDM  we fix  all other cosmological parameters to the mean values obtained from CMB+BAO+SNe; similarly, when
we study  modified GW propagation, we  keep $\Xi_0$ as the only free parameter and we fix all other cosmological parameters, including $H_0$, to their mean values obtained from CMB+BAO+SNe. Indeed, we will see below that, also in the case of GW190521+ZTF19abanrhr, just as for dark sirens, the limits that one can currently obtain on $\Xi_0$ are still very broad, allowing variation of about a factor of 3, i.e. of about $300\%$, or more compared to the value $\Xi_0=1$ of GR. Then, letting $H_0$ vary within a prior consistent with 
CMB+BAO+SNe data, which fix it to an accuracy at the $0.7\%$ level, would have  a negligible effect, compared to fixing it to the corresponding mean value; on the other hand, the current data have too little constraining power to provide useful limits if we do not use at all the CMB+BAO+SNe prior, and we attempt a purely gravitational measurement. 

Since the candidate counterpart is at considerably higher redshift than in the case of GW170817, there is no significant effect from peculiar velocities, and we can now neglect the small uncertainty in the electromagnetic determination of the redshift. We therefore fix the position of the counterpart  to 
$z_{\rm EM}=0.438$, 
$(\alpha_{\rm EM}, \delta_{\rm EM}) = (192.42625^{\circ}, 34.82472^{\circ})$~\cite{Graham:2020gwr}. As for the prior on masses, as explained in sect.~\ref{sect:GW170817}, one should in principle re-weight the samples to include the correct mass distribution, which can also induce a dependence on the cosmology through the conversion between source and detector frame masses (see also footnote~\ref{foot:reweight} on page~\pageref{foot:reweight}). However, the choice of a mass function for GW190521 is a subtle point. First, as discussed above, when analysed with a standard flat prior in detector frame mass, this event has a reported primary mass that lies inside the pair instability Supernovae mass gap, and at the edge of all the best-fit primary mass distributions discussed in~\cite{Abbott:2020gyp}, including the broken power-law used in this work, \eq{brokenpowerlaw}, which makes it difficult to efficiently re-sample the LVC posterior samples. Moreover, large uncertainties are still associated to all such mass distributions, which become relevant for such an extreme event. Additionally, it has been argued that the use of a population-informed prior on the secondary mass in the analysis could change substantially the inferred values for the parameters~\cite{Fishbach:2020qag}, and that this event could be a highly eccentric merger, which also gives different values for the component masses~\cite{Gayathri:2020coq}. In conclusion, we make the standard  choice of not re-weighting the LVC samples, which corresponds to the use of a prior flat in detector frame mass in the range  $[30-100]\, \msun $~\cite{Abbott:2020mjq, Chen:2020gek,Mastrogiovanni:2020mvm, Abbott:2020niy}.\footnote{For GW190521, the prior is further restricted so that in the detector frame the total mass does not exceed $200 M{_\odot}$, the chirp mass lies in the interval  $\[70, 150\] M{_\odot}$, and the mass ratio is larger than 0.17~\cite{Abbott:2020niy}.}  It must, however, be kept in mind that the use of a proper mass function can have a large impact on the final result. As  prior for the distance we choose $\tilde p_0(z)$ as in \eq{pzlambda} with $\lambda=0$, 
which corresponds to a merger rate uniform in comoving volume and source-frame time (this is the same prior used in the LVC analysis of O3a data, see sect.~V.C of~\cite{Abbott:2020niy})

\begin{figure}[t]
\centering
\includegraphics[width=0.90\textwidth]{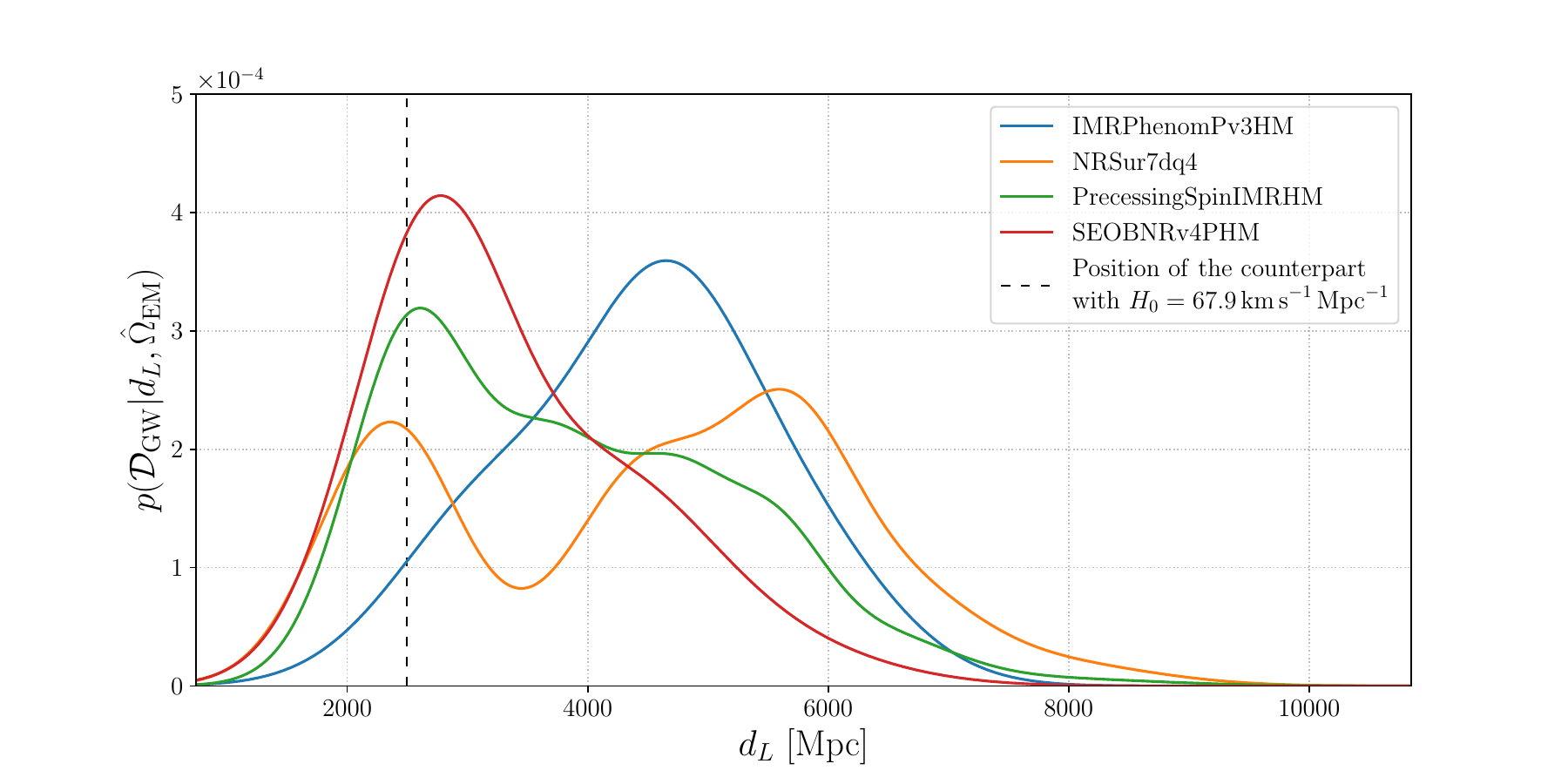}
\caption{The line-of-sight marginal likelihood in the direction of the counterpart, $ p( {\cal D}_{\rm GW} | d_L, \hatO_{\rm EM} )$, for the four different waveform models used by the LVC. The dotted vertical line represents the position of the counterpart for $H_0=67.9 \,  {\rm km}\, {\rm s}^{-1}\, {\rm Mpc}^{-1}$ and $\Xi_0=1$.
}
\label{fig:LOS_dL}
\end{figure}

With the above assumptions, the calculation of the posterior for this event reduces   to the computation of the line-of-sight marginal likelihood  $p({\cal D}_{\rm GW}| d_L,\hatO_{\rm EM} )$.\footnote{For this, we interpolate the LVC posterior samples, 
using a Gaussian Mixture Model (GMM), whose optimal number of components is determined by minimising the Bayesian Information Criterion (BIC), defined as usual by $\text{BIC}=2 k \log(N)-2 \log(\mathcal{L})$, where $k$ is the number of parameters of the model, $N$ the number of data points, and $\mathcal{L}$ is the likelihood of the data given the parameters of the model.
Note, however, that in this case, the LVC posterior samples are not restricted to the line of sight, as in the case of GW170817. 
Thus, as discussed in footnote \ref{foot:KDE} on page \pageref{foot:KDE}, it is convenient to use Cartesian coordinates, since in this case the GMM yields the posterior $p( x_1, x_2, x_3 | {\cal D}_{\rm GW}) = p( d_L, \hatO | {\cal D}_{\rm GW})/d_L^2$ where the $1/d_L^2$ factor comes from the jacobian between Cartesian and spherical coordinates. Since the LVC uses a prior proportional to $d_L^2$ that should eventually be removed, we can directly use the GMM in cartesian coordinates for the marginal likelihood.} The LVC provides posterior samples obtained by using four different waveform models. In Fig.~\ref{fig:LOS_dL} we show our reconstruction of the marginalized  likelihood in the direction of the counterpart, $ p( {\cal D}_{\rm GW} | d_L, \hatO_{\rm EM} ) $, for the four waveforms. 

We  first compute the posterior of $H_0$ in $\Lambda$CDM obtained from GW190521 assuming that ZTF19abanrhr is indeed the counterpart.
To compute $\beta(H_0)$ we use our MC in the absence of additional selection effects given by the incompleteness of the catalog, using the O3 strain sensitivity, the broken power law \eq{brokenpowerlaw} for the mass distribution, and neglecting any EM contribution to the selection effect. 

\begin{figure}[t]
\centering
\includegraphics[width=0.90\textwidth]{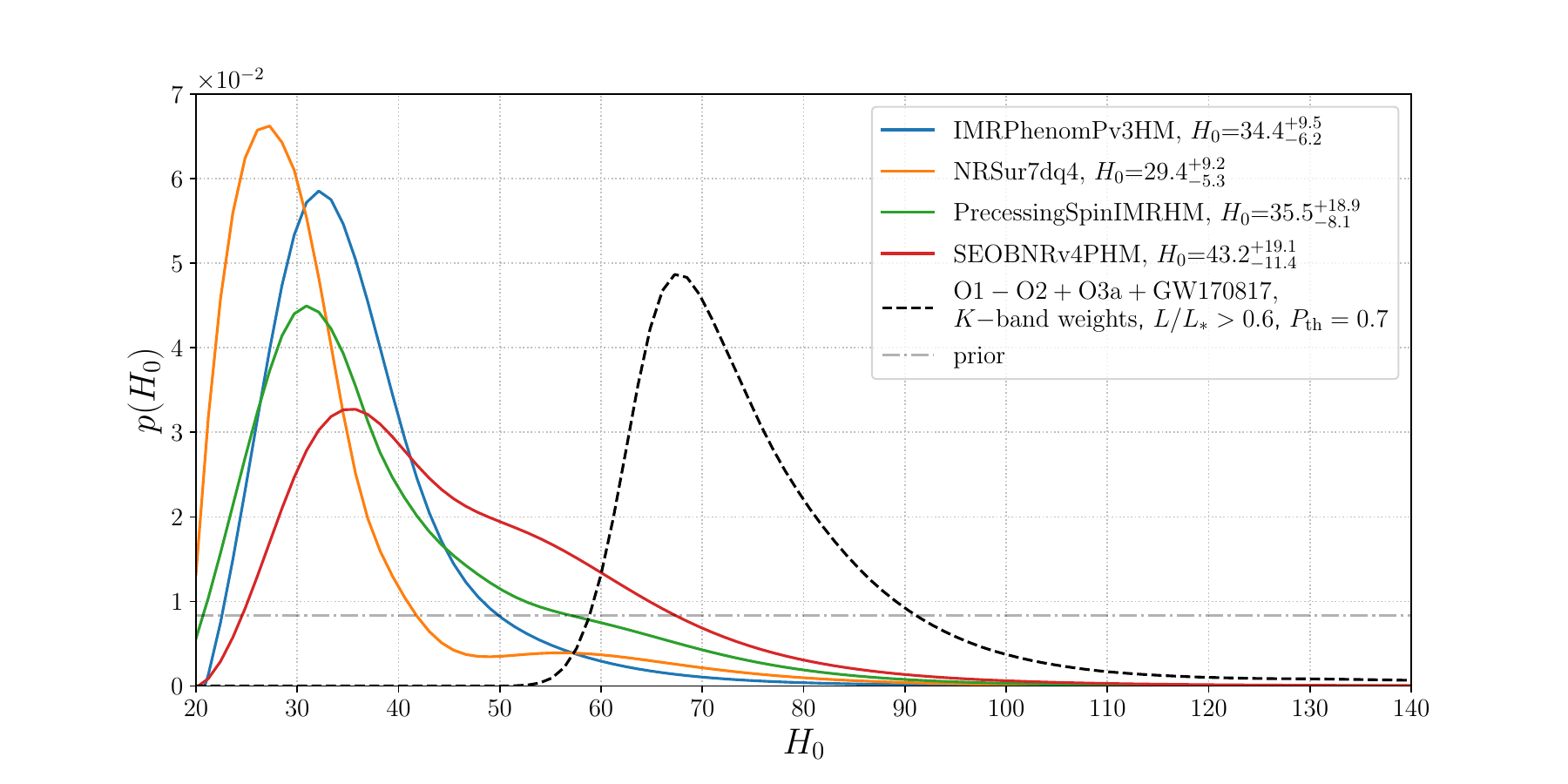}
\caption{Posterior distributions for $H_0$ obtained from GW190521 and its potential EM counterpart for four different waveforms. For comparison, we show also the posterior obtained in the previous section combining GW170817 with dark sirens.
}
\label{fig:H0counterpart}
\end{figure}

The resulting posterior distribution for $H_0$ is shown in Fig.~\ref{fig:H0counterpart}, for the four waveform models provided by the LVC. By comparison, we plot it  together with the posterior that we obtained  by combining GW170817 with dark sirens, already shown as the red line in  Fig.~\ref{fig:dark_and GW1708}. Even if the statistical error is large and the difference can be accounted as a statistical fluctuations, we see that, independently of the waveform model used, the posterior for $H_0$ from  GW190521+counterpart is in tension  with the value of $H_0$ that we already know from electromagnetic observations, whether we use the {\em Planck} value $H_0=67.4^{+0.5}_{-0.5}\, {\rm km}\, {\rm s}^{-1}\, {\rm Mpc}^{-1}$~\cite{Aghanim:2018eyx}, or the even larger SH0ES value 
$H_0=74.0^{+1.4}_{-1.4}\, {\rm km}\, {\rm s}^{-1}\, {\rm Mpc}^{-1}$~\cite{Riess:2019cxk}, 
contrary to the result obtained from dark sirens together with GW170817+counterpart (black dotted line), which is fully consistent with them.\footnote{\red{One might be puzzled by the fact that the posterior for $H_0$ from  GW190521+counterpart, shown in Fig.~\ref{fig:H0counterpart},  is in tension with the {\em Planck} value, despite the fact that (for 3 waveforms out of 4) the likelihood shown in Fig.~\ref{fig:LOS_dL} has a peak at the distance inferred from the redshift by using the 
the {\em Planck} value of $H_0$. There are two effects that concur to this result. First, even if in Fig.~\ref{fig:LOS_dL} the peak of the likelihood, for 3 waveforms, is in the `right' place, the distributions are far from  gaussians, and have very long tails at large $d_L$. The orange curve even has a second peak, higher than the first, at much higher values of $d_L$. Thus,  the median values of the distributions shown in Fig.~\ref{fig:LOS_dL} are all at much higher values.
This has the effect of enhancing the posterior for $H_0$ at low $H_0$ since, for fixed redshift, large $d_L$ corresponds to small $H_0$. Second, the posterior in Fig.~\ref{fig:H0counterpart} is obtained dividing the likelihood by $\beta(H_0)$. Since $\beta(H_0)\sim H_0^3$, and is at the denominator, this further enhances significantly the low-$H_0$ region in the posterior, shifting further the maximum of the posterior   toward the lower values that can be seen in Fig.~\ref{fig:H0counterpart}.}} 

Apart from the possibility of a statistical fluctuation, within $\Lambda$CDM two possible  interpretations are either that ZTF19abanrhr is simply not the counterpart, or that the estimate of the luminosity distance is not correct because the binary had a significant eccentricity. Indeed, as discussed in \cite{Abbott:2020mjq},  the signal only stays in the detector bandwidth for about  four cycles 
(as a consequence of the large mass of the system)
so that, while considered the most plausible, the interpretation in terms of a quasi-circular coalescence is not certain. Using a model in which GW190521 was a highly eccentric BBH with eccentricity $e\sim 0.7$, and assuming that ZTF19abanrhr is  the counterpart,
ref.~\cite{Gayathri:2020coq} finds $H_0=88.6^{+17.1}_{-34.3}\, {\rm km}\, {\rm s}^{-1}\, {\rm Mpc}^{-1}$, consistent with standard cosmology within the very large error.

Another obviously interesting possibility is to analyze the possible association between GW190521 and ZTF19abanrhr by
enlarging the cosmological model, with respect to $\Lambda$CDM. An investigation in this direction has been done in~\cite{Chen:2020gek}, that considers $w$CDM, i.e. the phenomenological extension of $\Lambda$CDM where the DE equation of state is taken to be a constant $w_0$, while perturbations are assumed to be the same as in $\Lambda$CDM. In this case, assuming that ZTF19abanrhr is  the counterpart and varying also $\oma$,   the result of ref.~\cite{Chen:2020gek} is $H_0=48^{+23}_{-10}\, {\rm km}\, {\rm s}^{-1}\, {\rm Mpc}^{-1}$,
$\oma=0.35^{+0.41}_{-0.26}$ and $w_0=-1.31^{+0.61}_{-0.48}$ (median and $68\%$ credible interval). However, as we discussed in sect.~\ref{sect:persp}, current limits from electromagnetic observations 
show  that $w_0$ cannot differ from $-1$ by more than about $5\%$ and, using this as a prior, the result  
for $H_0$  from GW190521+ZTF19abanrhr obtained in $w$CDM would not be significantly different from that  in $\Lambda$CDM. More importantly,
as discussed in  sect.~\ref{sect:dLgw}, any known model that deviates from $\Lambda$CDM at the background level, giving a DE equation of state different from $w=-1$, also predicts modified GW propagation, and the effect of modified GW propagation in general dominates, possibly even by 1-2 orders of magnitude,
 over that induced by the DE equation of state,  because: (1)  in viable models $w_0$ is constrained to be equal to $-1$ within a few percent, while $\Xi_0$ can have large deviations from the GR value $\Xi_0=1$, with known examples where $\Xi_0$ can be as large as $1.80$; and,  (2) the effect of $\Xi_0$ on the luminosity distance is more important than the effect of $w_0$ because of degeneracies among $H_0$, $\oma$ and $w_0$, see the discussion on page~\pageref{ref:degen}.\footnote{For example, we have seen that in the RT non-local model
$\Xi_0$ can deviate from the GR value by as much as $80\%$, and this results in a $80\%$ effect on the luminosity distance. In contrast, in the same model, for the large values of $\Delta N$ for which $\Xi_0\simeq 1.80$, in the relevant range of redshift $\wde(z)$ differ from $-1$ by less than $5\%$. Furthermore, because of the compensating effect of the degeneracy between $w_0$, $H_0$ and $\oma$ mentioned on page~\pageref{ref:degen}, the net effect of this on the electromagnetic luminosity distance is much smaller, and in fact below $0.5\%$, see Fig.~2 of \cite{Belgacem:2020pdz}. In this case, therefore, the effect of a non-trivial DE equation of state is two order of magnitudes smaller than the effect of modified GW propagation.}
For these reasons, opening the parameter space by including $w_0$, while neglecting modified GW propagation, does not appear well justified. Here we will rather pursue the opposite strategy,  investigating the effect of $\Xi_0$ in a model where the electromagnetic luminosity distance differs little from that in $\Lambda$CDM, see also the discussion at the beginning of sect.~\ref{sect:RT}. 

\begin{figure}[t]
\centering
\includegraphics[width=0.90\textwidth]{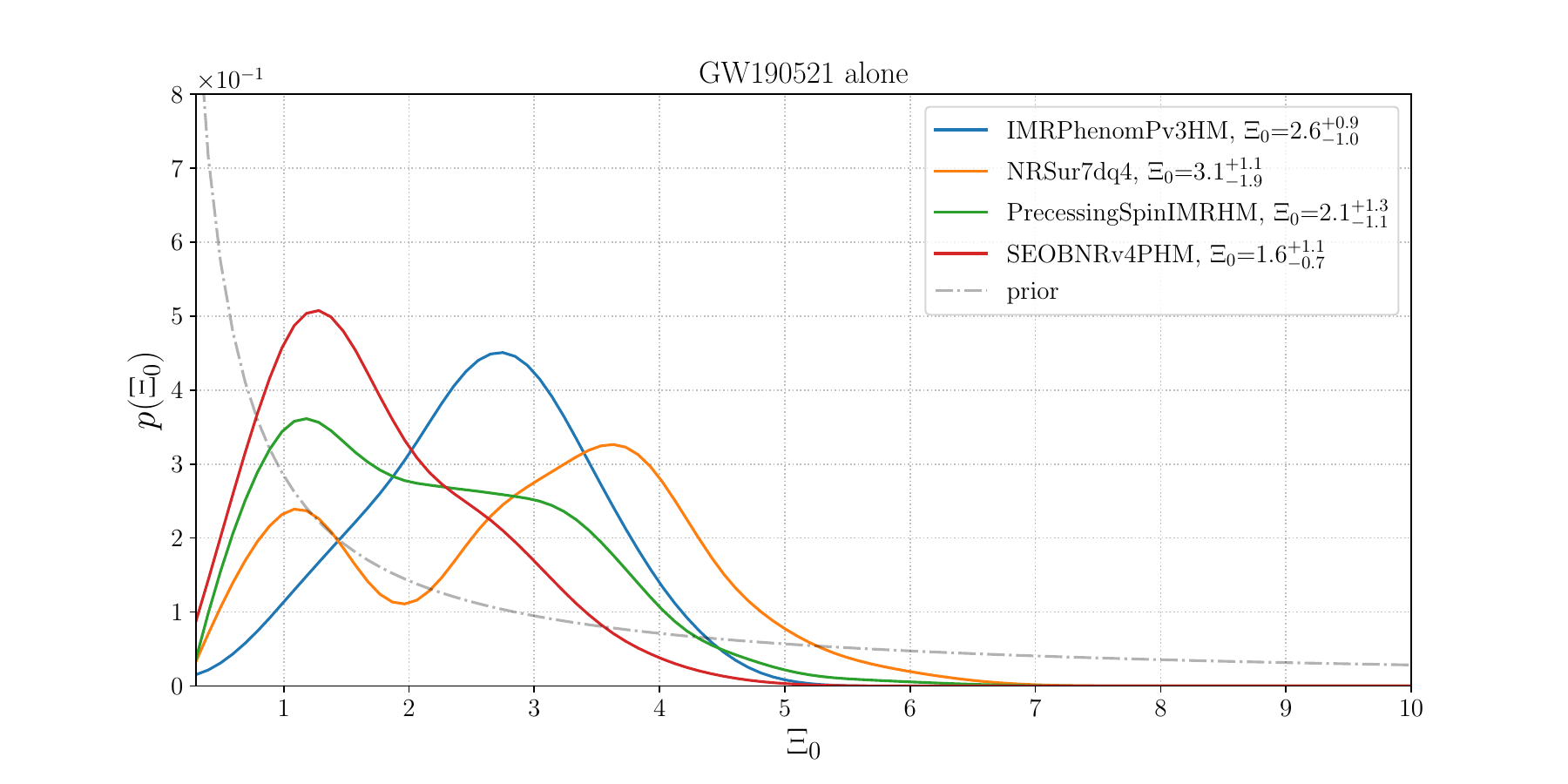}
\caption{ Posterior distributions for $\Xi_0$ obtained from GW190521 and its potential EM counterpart for four different waveform models. 
}
\label{fig:Xi0counterpart1}
\end{figure}

\begin{figure}[t]
\centering
\includegraphics[width=0.90\textwidth]{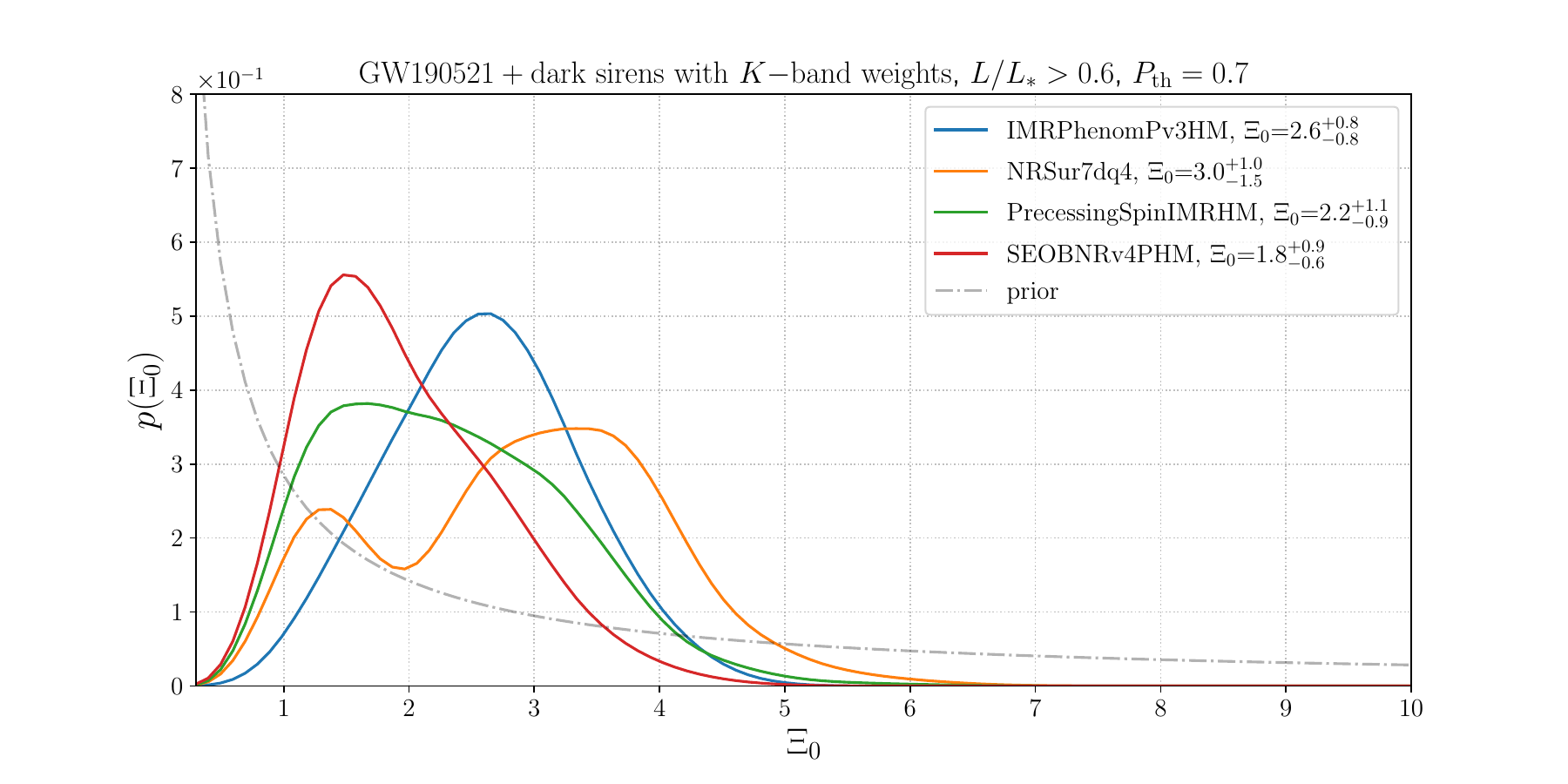}
\caption{Posterior distribution for $\Xi_0$, combining GW190521 and dark sirens using K-band weights, and a threshold probability $P_{\rm th}=0.7$. 
}
\label{fig:Xi0counterpart2}
\end{figure}

Proceeding as we have done for $H_0$, but keeping now $\Xi_0$ as the only free parameter, 
we find the posteriors shown in  Fig.~\ref{fig:Xi0counterpart1}, for the four waveform models.
The corresponding limits on $\Xi_0$ for the four waveform are given in Table~\ref{Tab:Xi0_GW1905only} (median values and limits at $68.3\%$  and $95\%$ credible interval). 

\begin{table}[t]
	\centering
	\begin{tabular}{|l|c|c|c|c|}
		\toprule
		& NRSur                        & IMRPhenom              & SEONBR                      & PrecessingSpin \\ \midrule                                
		\phantom{\Big(}$\Xi_0\, (68.3\%)$  & $3.1^{+1.1}_{-1.9}$ & $2.6^{+0.9}_{-1.0}$ & $1.6^{+1.1}_{-0.7}$   & $2.1^{+1.3}_{-1.1}$\\
		\phantom{\Big(}$\Xi_0\, (95\%)$  & $3.1^{+2.4}_{-2.5}$ & $2.6^{+1.7}_{-1.8}$ & $1.6^{+2.3}_{-1.1}$   & $2.1^{+2.6}_{-1.6}$\\
		\bottomrule
	\end{tabular}
	\caption{Median value and limits on $\Xi_0$  ($68.3\%$  and $95\%$ symmetric credible interval), for the four waveform models, from GW190521+ZTF19abanrhr.
		\label{Tab:Xi0_GW1905only}}
\end{table}

\begin{table}[t]
	\centering
	\begin{tabular}{|l|c|c|c|c|}
		\toprule
		& NRSur                        & IMRPhenom              & SEONBR                      & PrecessingSpin \\ \midrule                                 
		\phantom{\Big(}
		$\Xi_0\, (68.3\%)$  & $3.0^{+1.0}_{-1.5}$ & $2.6^{+0.8}_{-0.8}$ & $1.8^{+0.9}_{-0.6}$   & $2.2^{+1.1}_{-0.9}$\\
		\phantom{\Big(}
		$\Xi_0\, (95\%)$  & $3.0^{+2.1}_{-2.1}$ & $2.6^{+1.6}_{-1.5}$ & $1.8^{+1.9}_{-1.1}$   & $2.2^{+2.2}_{-1.4}$\\
		\bottomrule
	\end{tabular}
	\caption{Median value and limits on $\Xi_0$ ($68.3\%$  and $95\%$ symmetric credible interval), for the four waveform models, 
		from GW190521+ZTF19abanrhr combined with dark sirens (using K-band, with a completeness threshold $P_{\rm th}=0.7$).
		\label{Tab:Xi0_GW1905dark}}
\end{table}

These results can be compared with those in Table~1 of ref.~\cite{Mastrogiovanni:2020mvm}, where the limits are given at $95\%$ credible interval. The results are broadly consistent. However, ref.~\cite{Mastrogiovanni:2020mvm} only obtains upper limits on $\Xi_0$, while the results in our Table~\ref{Tab:Xi0_GW1905only} provide both upper and lower limits on $\Xi_0$.\footnote{Partly, this can be due to differences in the resampling of the LVC posterior in the direction of ZTF19abanrhr; indeed, our Fig.~\ref{fig:LOS_dL} is consistent with, but not identical to Fig.~1 of  ref.~\cite{Mastrogiovanni:2020mvm} (the comparison should be made with the v2 arxiv version of ref.~\cite{Mastrogiovanni:2020mvm}, where an error in the posterior was corrected). The difference at low $\Xi_0$, however, is most likely due to differences in the evaluation of $\beta(\Xi_0)$. In particular, we have computed $\beta(\Xi_0)$ with our MC,
using the mass function (\ref{brokenpowerlaw}), while in ref.~\cite{Mastrogiovanni:2020mvm} a flat mass function has been used in the computation of $\beta(\Xi_0)$. We thank Simone Mastrogiovanni for comparison of our computations.}$^{,}$\footnote{Observe also that $\Xi_0=0$ corresponds to the case
where the function $\delta(z)$ is taken to be constant, see  the comment below \eq{paramdeltaz}. Our results, under the assumption of the GW190521-ZTF19abanrhr association, would
therefore rule out a constant $\delta(z)$, for  values $n={\cal O}(1)$ such as used here [as discussed, we have set $n=1.91$, which is a value  inspired by the RT model, but the results depends very weakly on $n$ as long as $n={\cal O}(1)$]. Note that $\Xi_0=0$ is also ruled out, to good statistical confidence, already from dark sirens only, see \eq{Xi0limitdarkK02}.}

Contrary to what happens for $H_0$ in $\Lambda$CDM, see fig.~\ref{fig:H0counterpart},  the posterior 
for $\Xi_0$ obtained from GW190521+ZTF19abanrhr
is consistent with that obtained from dark sirens found in sect.~\ref{sect:Xi0results}, and therefore it makes sense to combine them.  
In Fig.~\ref{fig:Xi0counterpart2}
we show the total posterior obtained by combining the result from GW190521+ZTF19abanrhr (with the four waveform models) with  the dark sirens posterior that we obtained in sect.~\ref{sect:Xi0results} from O1+O2+O3a data (using K-band weights and a threshold probability $P_{\rm th}=0.7$). The corresponding limits on $\Xi_0$ for the four waveform are given in Table~\ref{Tab:Xi0_GW1905dark}. 
The best bound depends on the waveform used, with the most stringent one being
\be\label{Xi0_GW1905}
\Xi_0=1.8^{+0.9}_{-0.6}  \, ,
\ee
(median and $68.3\%$ credible interval), obtained
from GW190521+ZTF19abanrhr with SEOBNRv4PHM waveform, combined with dark sirens. 

Observe that the `tension' that, in the context of $\Lambda$CDM, exists between the spatial localization of GW190521 and the redshift of  ZTF19abanrhr, and which has led to doubts on the interpretation of 
 ZTF19abanrhr as the counterpart to GW190521 \cite{Graham:2020gwr}, once interpreted in the context of modified gravity translates into a (small) tension between the measured value of $\Xi_0$ in \eq{Xi0_GW1905} and the GR value $\Xi_0=1$. To interprete correctly this result, however, one must  stress that the  association of
ZTF19abanrhr with GW190521 is anyhow tentative.\footnote{In particular, as we already mentioned, there are uncertainties as to how generic and plausible is the mechanism proposed for the emission of a flare from the kicked BH in a AGN disk.  Observe, however, that the model that associates ZTF19abanrhr with GW190521 predicts the possibility of a repeated
flare due to a re-encountering of the remnant BH with the disk, on a timescale of the order of 1.6~yr \cite{Graham:2020gwr}. The confirmation of this prediction would make a  strong case for the association.} Furthermore, the statistical error in \eq{Xi0_GW1905} is large, so that the GR value is still at the border of the $68\%$ credible interval. We have also seen that there is a large uncertainty due to different waveform models, as apparent from Fig.~\ref{fig:Xi0counterpart2},  and  one must also take into account the possibility of explaining  the low value of $H_0$  obtained  from 
GW190521+ZTF19abanrhr within $\Lambda$CDM in terms of a large
eccentricity~\cite{Gayathri:2020coq}. Therefore, the result (\ref{Xi0_GW1905}) should obviously be interpreted with care. It shows, however, the potential of standard sirens for obtaining interesting limits on $\Xi_0$.

\section{Conclusions}\label{sect:Concl}

In this paper we have first of all developed several methodological aspects relevant for performing statistical  correlations of dark sirens with galaxy catalogs, paying special attention to many systematic errors, and to the effect of various choices that are made in the analysis.  This will become more and more important as GW detectors increase their range and collect more and more events at high redshift, and will be even more important for third generation (3G) ground based interferometers such as the Einstein Telescope and Cosmic Explorer, as well as for the LISA space interferometer.\footnote{\red{While current galaxy catalogs are very incomplete for the redshifts that will be reached in the 3G era, significant improvement in the next few years are expected by  surveys such as the Maunakea Spectroscopic Explorer  
(MSE, \url{https://mse.cfht.hawaii.edu/}), the Dark Energy Spectroscopic Instrument (DESI,
\url{https://www.desi.lbl.gov/}, and the 4-metre Multi-Object Spectroscopic Telescope (4MOST, 
\url{https://www.4most.eu/cms/}.
}} 

 We have introduced and compared two  different `quasi-local' ways of quantifying the level of completeness of a galaxy catalog (cone completeness and mask completeness) and, given an assessment of the level of completeness, we have compared the results from two different ways of distributing the missing galaxies, the `homogeneous completion', previously used in the literature, and the `multiplicative completion', that we have proposed here. The latter can be seen as a quantitative realization of the intuitive idea that, for the GW/galaxy correlation, it is sufficient that a galaxy catalog traces the structure of galaxy clusters. 
 
We have  compared the results with different luminosity weightings, and with different astrophysical assumptions, as encoded for instance in the parameters  that enters the BBH mass function (\ref{brokenpowerlaw}) or in the parameter $\lambda$ that parametrizes the redshift dependence of the mass function, \eq{pzlambda}. \red{Eventually, the only way forward
will be to turn some of the systematic uncertainties, at least the ones
regarding the unknown mass and redshift distribution of BBH sources,
into statistical uncertainties, and jointly fit for the population and cosmology.}
 
 We have  paid special attention to the computation of the function $\beta(H_0)$  that enters in \eq{likeprod}, using different levels of refinement. Another delicate point that emerged from our analysis is the need of including a completeness threshold on the events, that must be carefully accounted for in the computation of $\beta(H_0)$. We have seen that there can be a significant dependence on this threshold,  and that large values of this threshold are necessary to avoid the possible introduction of spurious effects.
All these technical details turn out to be quite crucial if one eventually aims at measuring $H_0$ with percent level precision from dark sirens. Current data are still limited by statistics, but the results that we have presented show that, eventually, to reach such a precision,  these  systematic effects must be well understood, and our paper is meant first of all as a contribution in this direction. 

Apart from the methodological aspects, the application to current data is already quite interesting, even if still clearly limited by the statistics and by the fact that, with the current detector network, the localization of GW events is  very broad.  Our  best result from dark sirens only, for $H_0$, is shown in Fig.~\ref{fig:varyingGamma_H0} and in \eq{H0KbandPth07}, and provides
the currently most accurate results obtained from standard sirens only, while 
Fig.~\ref{fig:dark_and GW1708} and \eqs{H0GW17only}{H0GW17anddark}
show how the addition of dark sirens from the O1, O2 and O3a runs improves the constraint on $H_0$ obtained from GW170817 and its counterpart.

While, in the literature,  most of the work on dark sirens  has been currently focused on extracting $H_0$, we have stressed that, eventually, the most interesting contribution that GW observations can give to cosmology might be the observation of modified GW propagation, which is a smoking gun of dark energy and modified gravity, and can only be observed with GW detectors. We have presented our result 
in terms of $\Xi_0$, the main parameter that characterizes modified  GW propagation, and we find that, even with the current limited statistics, dark sirens start to provide interesting limits. Our best results, using dark sirens only, are
given by Fig.~\ref{fig:varyingGamma_Xi0} and \eq{Xi0limitdarkK02}.

Finally, we have considered the limits that can be obtained on modified GW propagation  if one  accepts the interpretation of the flare ZTF19abanrhr as the electromagnetic counterpart of the BBH coalescence GW190521 (assumed to be in a quasi-circular orbit). In this case, combining the result from  GW190521 with the posterior from dark sirens, our most stringent result is given in
\eq{Xi0_GW1905}. Considering that, in nonlocal gravity with initial conditions in inflation, the prediction for $\Xi_0$ is in range $\Xi_0\simeq 1.6-1.8$~\cite{Belgacem:2019lwx,Belgacem:2020pdz}, to be 
compared with $\Xi_0=1$ for GR, the result in \eq{Xi0_GW1905} is suggestive. However, because of the various caveats discussed in sect.~\ref{sect:GW190521}, one cannot yet draw any conclusion, except that, with the expected increase in the amount and quality of the GW data, modified GW propagation can become an extremely interesting observable for GW observations.

\vspace{3mm}\noindent
Together with this paper, we release the publicly available code $\tt{DarkSirensStat}$, which is available under open source license at  \url{https://github.com/CosmoStatGW/DarkSirensStat}.

\vspace{5mm}
{\bf Acknowledgments}. We are grateful to Enis Belgacem for his contributions to the early stages of this work, and for many useful discussions.
We thank  Gergely D\'alya, Archisman Ghosh, Ignacio Hernandez and Simone Mastrogiovanni for very useful discussions, \red{and the referee for an extremely competent, careful and useful report.}
The work  of the authors is supported by the  Swiss National Science Foundation and  by the SwissMap National Center for Competence in Research.

\appendix

\section{Completeness of the GLADE catalog}\label{sect:GLADE}

In this appendix we discuss in some detail the completeness of the GLADE catalog (complementing the discussion in \cite{Dalya:2018cnd} and in \cite{Fishbach:2018gjp}),  both as an application of the tools developed in sect.~\ref{sect:cat}, and for its intrinsic usefulness for the results that we present in the paper. This will also allow us to fix our choices for the parameters $\Delta z$, $\Delta\Omega$ that appear in the cone completeness
$P_{\rm compl}^{\rm cone}(z,\hatO;\Delta z,\Delta\Omega)$ and to test mask completeness, and to select a choice of weights and of luminosity cuts. For definiteness, we will illustrate our results using cone completeness, which is more intuitive, but similar results are obtained from mask completeness (that, as we saw from Tables~\ref{tab:O2events} and \ref{tab:O3aevents} in sect.~\ref{sect:H0darkonly}, produces slightly higher values of the completeness).

\begin{figure}[t]
\centering
\includegraphics[width=0.46\textwidth]{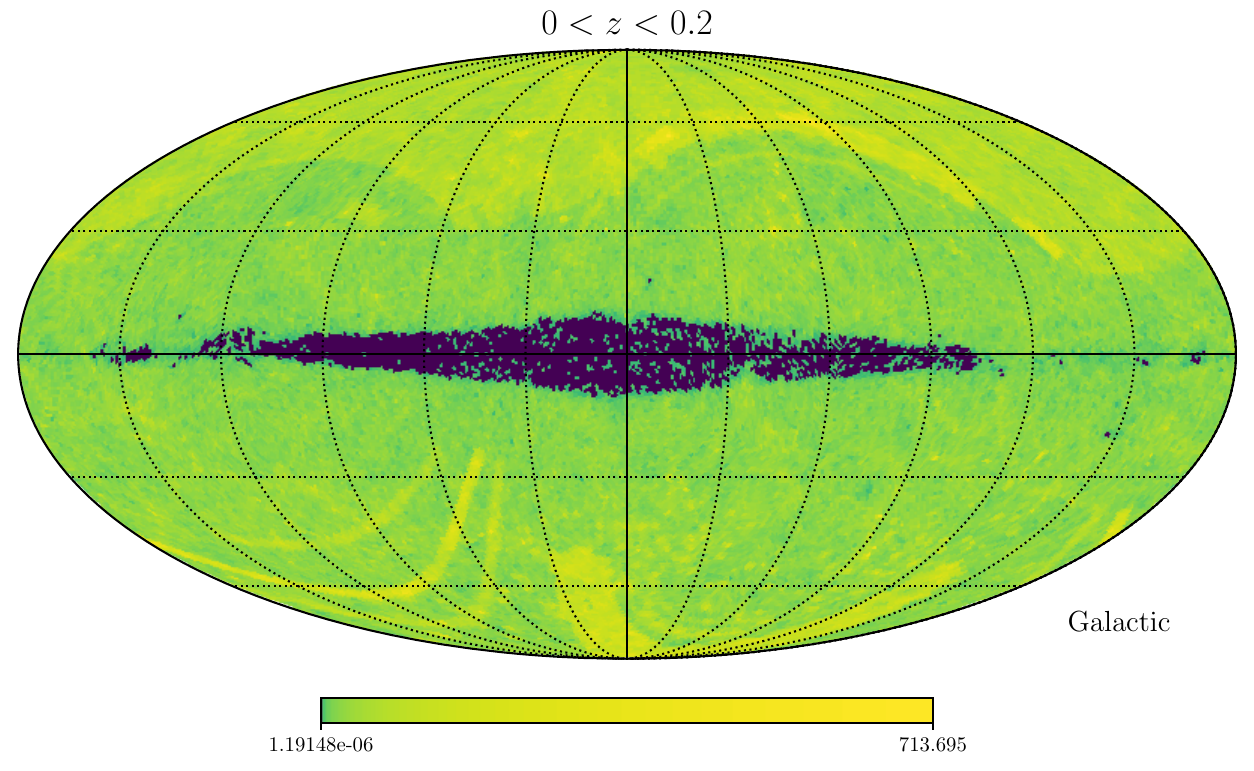}\,\,
\includegraphics[width=0.46\textwidth]{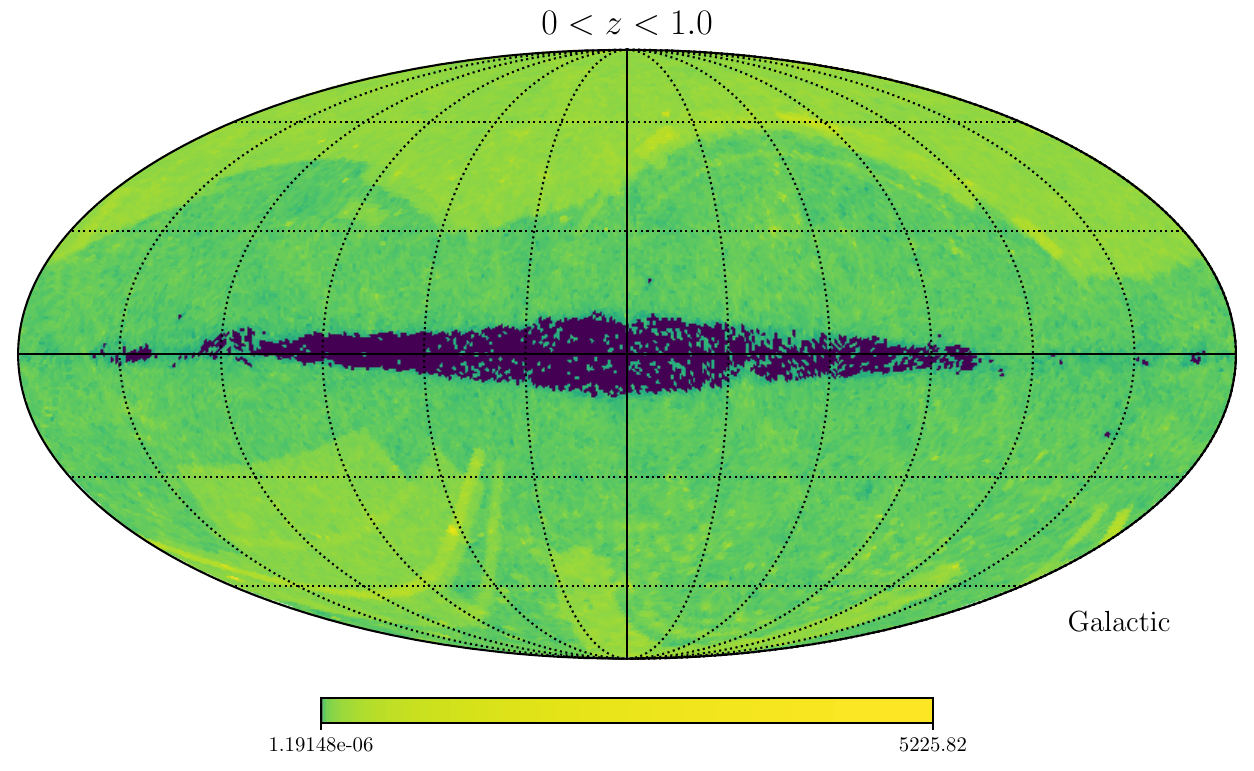}
\includegraphics[width=0.46\textwidth]{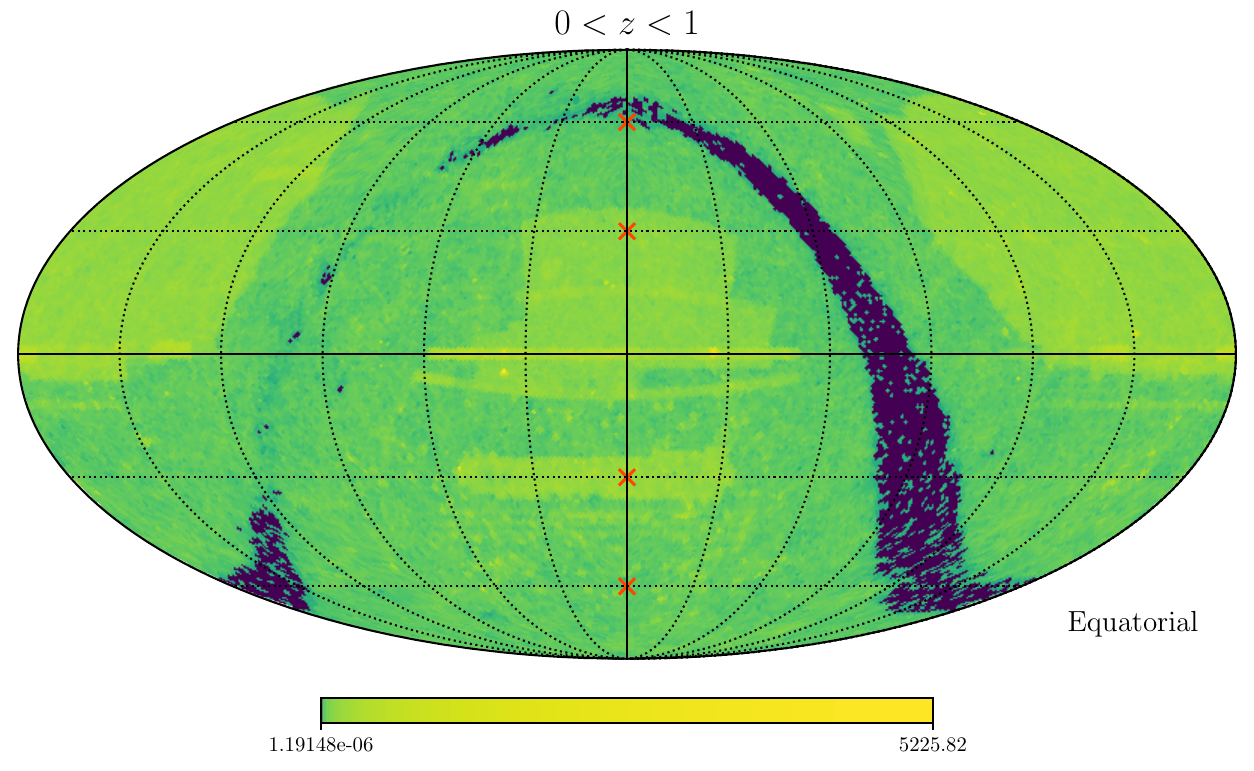}
\caption{Upper panels: the distribution of galaxies  in the GLADE catalog, in galactic coordinates, on a logarithmic scale, using weights $w_{\alpha}=1$.
Left panel: the galaxies with $0<z<0.2$. Right panel: the galaxies with $0<z<1$.
Lower panel: the same as the upper right panel, now in equatorial coordinates, and with the directions with RA=0 and ${\rm dec}=\{60,30,-30,-60\}$  marked by a cross.
\label{fig:Mollview_z02}}
\end{figure}

Completeness strongly depends  on the weights used, and on the direction of the sky.  The upper panels of Fig.~\ref{fig:Mollview_z02} show the distribution of galaxies with $0<z<0.2$ (upper left panel) and with $0<z<1$ (upper right panel) in the GLADE catalog,  assigning a weight $w_{\alpha}=1$ to each galaxy and using galactic coordinates.\footnote{We use a  logarithmic scale. In order to  avoid a divergence of the log in the pixels with no galaxies, we actually plot $\log(\epsilon+n_g)$, where $n_g$ is the number of galaxies in each pixel and 
$\eps=10^{-6}$. Plots such as Fig.~\ref{fig:Mollview_z02} and 
have been obtained using the HEALPix  software~\cite{Gorski:2004by}.} Besides the cut in the Galactic plane, are clearly visible stripes (arising from the HyperLEDA catalog) and regions that have been much better covered than others, corresponding to the footprints of the various surveys that contribute to GLADE. 
From these plots, it is clear that a notion of completeness defined by integrating over $4\pi$, as used until now in the literature, is  very approximate, and the angular-dependent completeness that we have introduced is more appropriate.
 
The lower panel of Fig.~\ref{fig:Mollview_z02} is the same as the upper right panel, except that the plot is shown in equatorial coordinates (RA, dec), and we have marked by a cross the points with RA=0 and ${\rm dec}=\{60,30,-30,-60\}$. We see that the directions with RA=0 and
${\rm dec}=\{30,-30\}$ are examples of well-covered directions, while RA=0 and
${\rm dec}=\{60,-60\}$ are examples of directions with lesser coverage, and we will use them below, to illustrate the behavior of the completeness in different directions.

\begin{figure}[t]
\centering
\includegraphics[width=0.48\textwidth]{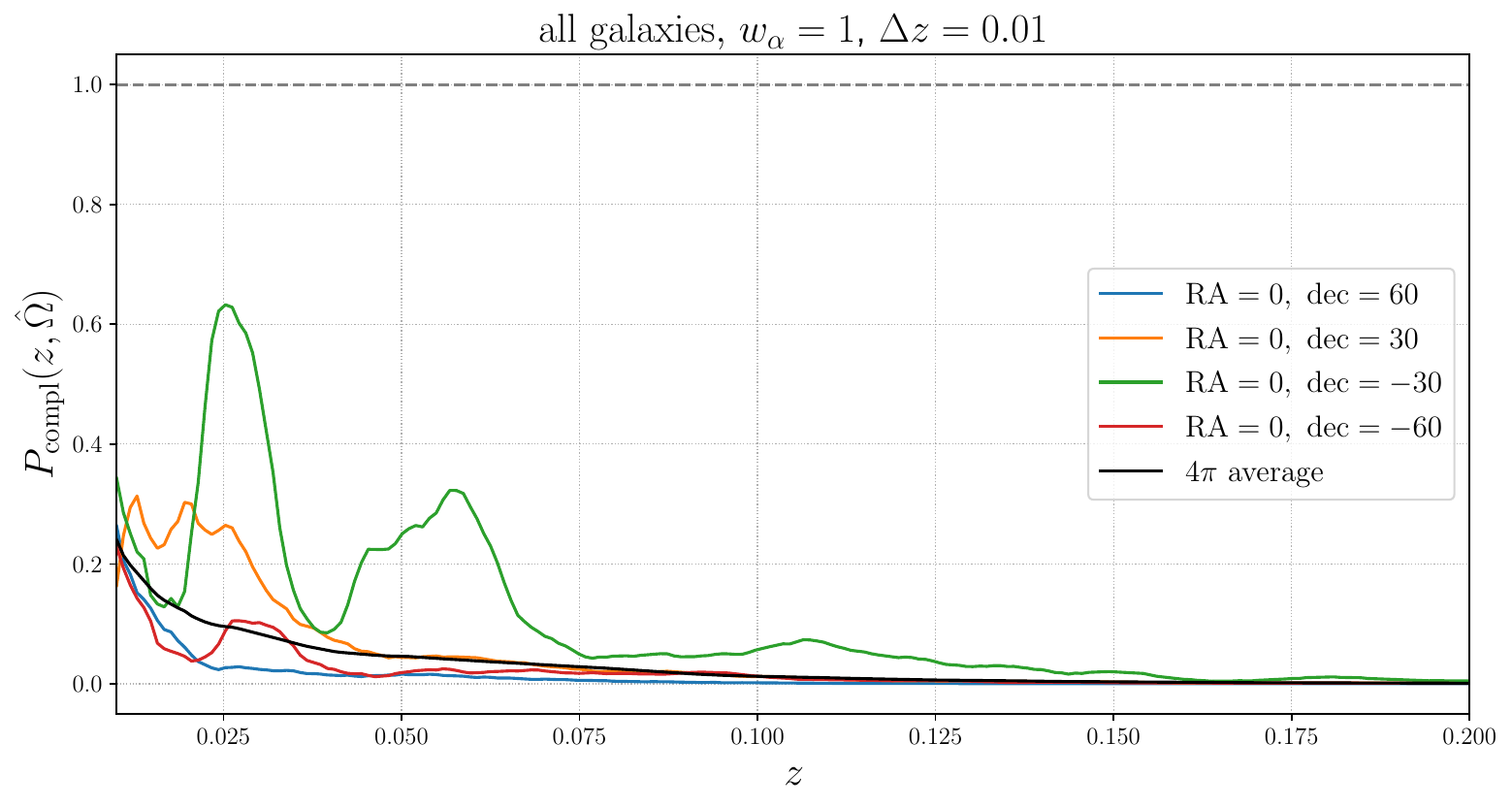}
\includegraphics[width=0.48\textwidth]{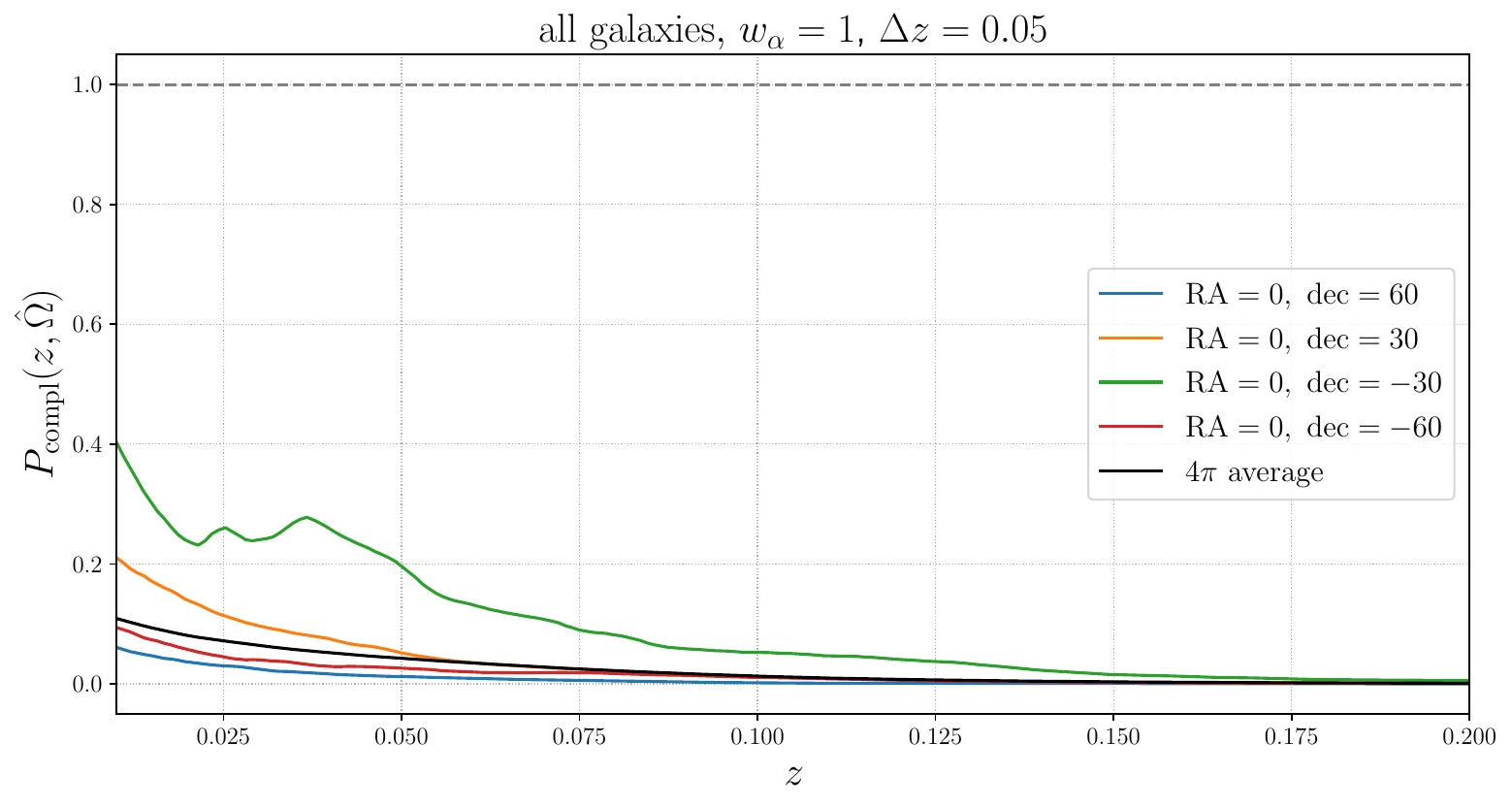}
\caption{The function $P_{\rm compl}(z,\hatO)$ 
using all galaxies in GLADE  and no luminosity weighting, i.e. $w_{\alpha}=1$, for the four directions
$\hatO$ corresponding to the crosses in the lower panel of Fig.~\ref{fig:Mollview_z02} (colored lines) and the all-sky average (black line). The colored line are obtained using sections of cones with half-opening angle $\theta_c=5^{\circ}$. The black line is the $4\pi$ average.  Left panel: using, in the radial direction, a section of a cone with  $\Delta z=0.01$. Right panel: the same, with $\Delta z=0.05$.}
\label{fig:ngal}
\end{figure}

Using  $w_{\alpha}=1$ we can make use of all galaxies in the catalog (GLADE contains about 2.7 millions galaxies, of which 1.6 millions have a known B-band luminosity, and 1.0 millions have a known  K-band luminosity; practically all galaxies with known K-band luminosity also have a known B-band luminosity).\footnote{In GLADE the redshifts of the galaxies  are given in the heliocentric frame and, as in \cite{Fishbach:2018gjp}, we transform them to the CMB rest frame using, for the galactic latitude and longitude of the dipole, in Galactic coordinates, $l_{\rm CMB}=263.99^{\circ}$, $b_{\rm CMB} = 48.26^{\circ}$ and, for the peculiar velocity of the solar system in the CMB rest frame,
$v_{\rm CMB} = 369\, {\rm km/s}$  \cite{Hinshaw:2008kr}.}$^{,}$\footnote{Observe that,
when using the GLADE catalog, some galaxies do not have directly measured redshifts, but rather distances, 
that have been converted to redshift assuming a value for $H_0$ (chosen as $H_0=70\, {\rm km}\, {\rm s}^{-1}\, {\rm Mpc}^{-1}$). These galaxies must therefore be omitted from  applications to the measurement of $H_0$ (but can still be used for $\Xi_0$). However, these galaxies
come from GWGC and are very few (495 objects in GLADE v2.3 and 1528 objects in GLADE v2.4), and  can be found, in GLADE,  by looking for objects with flag2=2 and HyperLEDA\_name=null. 
The bulk of galaxies labeled as `flag2=2', that come from HyperLEDA, have instead a directly measured redshift and can be used also for $H_0$ (we thank Gergely D\'alya for clarifying us this point).}

However, in this case the catalog becomes quickly very incomplete with distance, since we miss most of the low-luminosity galaxies, and, with this weighting, a low-luminosity galaxy counts as much as a high-luminosity one. 
This is clearly visible from Fig.~\ref{fig:ngal} that shows, as a function of $z$, the function 
$P_{\rm compl}(z,\hatO;\Delta z,\Delta\Omega)$ defined in \eq{Pcompletedef}  (setting $\bar{n}_{\rm gal}=0.1 \, {\rm Mpc}^{-3}$), 
with  ${\cal S}(z,\hatO;\Delta z,\Delta\Omega)$ 
given by a section of a cone with axis  $\hatO$ and  half-opening angle 
$\theta_c$ [so $\Delta\Omega=2\pi (1-\cos\theta_c)$], and
restricted, in the radial direction, to redshifts between $z-(\Delta z/2)$ and $z+(\Delta z/2)$. The four colored lines correspond to cones whose axis $\hatO$ is given by the four directions marked with a cross in
the lower panel of Fig.~\ref{fig:Mollview_z02}. We see that the curves corresponding to 
directions that  visually had better coverage, indeed have higher values of $P_{\rm compl}(z,\hatO;\Delta z,\Delta\Omega)$. Figs.~\ref{fig:Mollview_z02} and~\ref{fig:ngal} are complementary: Fig.~\ref{fig:ngal} gives more detailed information on the redshift dependence, which appears only in integrated form in Fig.~\ref{fig:Mollview_z02}, but only for a few selected directions, while
Fig.~\ref{fig:Mollview_z02} shows the full angular dependence.

The colored curves in the left panel of Fig.~\ref{fig:ngal} have been obtained by setting $\theta_c=5^{\circ}$ and $\Delta z=0.01$, while those in the right panel correspond to $\theta_c=5^{\circ}$ and $\Delta z=0.05$. The choice $\theta_c=5^{\circ}$ is suggested by the fact that it is much smaller than the typical angular localization of GW events, and also sufficiently small to resolve the region of the galactic plane which is obscured by dust, where completeness drop dramatically, so that we do not significantly mix regions close to the galactic plane with good covering, with the region near the galactic plane that is plainly obscured. In both panels, the black line is the average of the completeness over the whole sky, i.e. 
the quantity $P_{\rm compl}(z)$ defined by \eq{defPcomplz}.

\begin{figure}[t]
\centering
\includegraphics[width=0.64\textwidth]{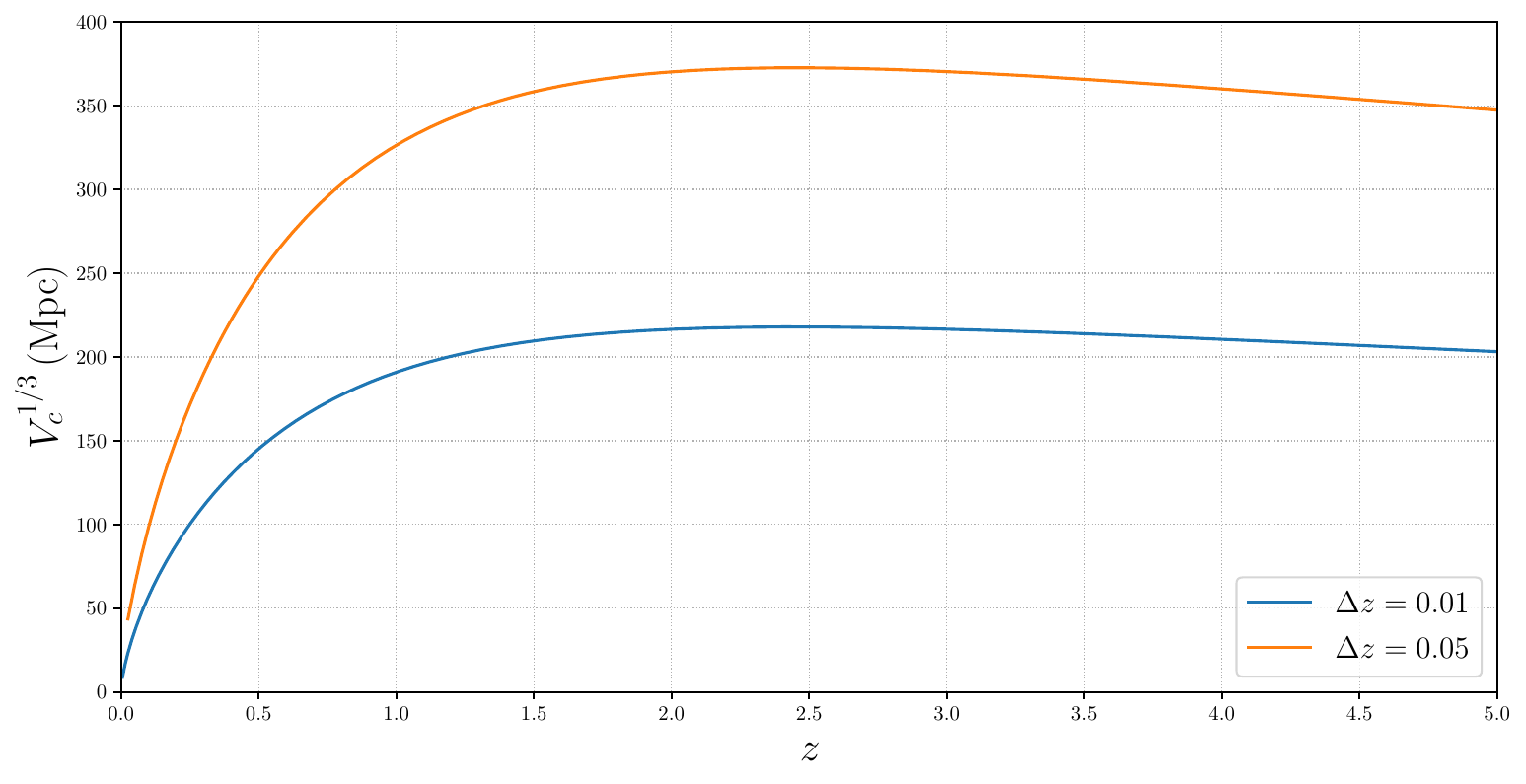}
\caption{The  quantity $V_c^{1/3}$ as a function of $z$,  where $V_c$ is the comoving volume of a truncated cone with  half-opening angle $\theta_c=5^{\circ}$ and redshift between $z-(\Delta z/2)$ and $z+(\Delta z/2)$, for the two choices $\Delta z=0.01$ and $\Delta z=0.05$.}
\label{fig:cone_volume}
\end{figure}

The comparison of these two panels allows us, first of all, to appreciate the effect of the choice of $\Delta z$.
As discussed in sect.~\ref{sect:conemask}, 
the optimal choice of  $\Delta z$ (and of $\theta_c$) must be such that the region ${\cal S}$ is small enough, compared to the typical localization of GW events, to represent a `quasi-local' property; still, it must  be sufficiently large  that the number of galaxies inside ${\cal S}$ (possibly with luminosity weighting) should be representative of the average value  of galaxy number density (or, respectively, luminosity density), rather than being sensitive to  true fluctuations such as  actual voids or actual over-densities. A simple criterion to ensure the latter condition
is that $P_{\rm compl}(z,\hatO;\Delta z,\Delta\Omega)$ should be a smoothly  decreasing function  of $z$, since the effect of  having missed some galaxies  because of observational limitations should  produce a decreasing function of redshift, while actual overdensities and voids will be responsible for fluctuations over this smooth behavior. In the left panel, we see  significant fluctuations superimposed to the smooth behavior. These clearly correspond to actual structures, i.e. under-densities and over-densities. 
As we see from the right panel, taking a larger $\Delta z$ smooths out these short-scale fluctuations.
When we complete the catalog, we want to compensate for galaxies that have been missed because of observational limitations, but we certainly do not what to smooth out actual structures (that contains precisely the information that we need, in order to perform the correlation with the localization of the GW events). Therefore a binning that is too small, such as $\Delta z=0.01$, and that allows us to resolve true fluctuations in the galactic field, is not appropriate in the definition of $P_{\rm compl}(z,\hatO;\Delta z,\Delta\Omega)$. Using such an expression for $P_{\rm compl}$
would lead us to compensate these fluctuations, which represent actual over- or under-densities, as if they where due to short-scale variation (with respect to $z$) in the survey's  capability of detecting galaxies. 
It is only what remains after averaging over these fluctuations, by taking $\Delta z$ sufficiently large, that gives a measure of the fraction of galaxies that have been lost by the surveys.
Thus, the binning must be sufficiently large so that 
$P_{\rm compl}(z,\hatO;\Delta z,\Delta\Omega)$  is a smooth and monotonically decreasing function of $z$. From the right panel we see that, for $w_{\alpha}=1$,  this is the case for $\Delta z=0.05$ (except for the direction corresponding to the green curve, so one should actually take an even larger value of $\Delta z$).\footnote{Notice also that the completeness in general decreases as $\Delta z$ increases. This is a volume effect due to the fact that, within a bin of size $\Delta z$, there are more galaxies  near the upper end of the bin at $z+\Delta z/2$, than near the lower end at $z-\Delta z/2$. Thus, increasing the bin size, we increase the average distance of the galaxies within the bin, and completeness decreases. The effect, however, is relevant only at very low $z$, when $z$ is comparable to $\Delta z$.} To have a sense for these choices of $\theta_c$ and $\Delta z$, in Fig.~\ref{fig:cone_volume} we plot  $V_c^{1/3}$ as a function of $z$,  where $V_c$ is the comoving volume of a truncated cone with  $\theta_c=5^{\circ}$ and redshift between $z-(\Delta z/2)$ and $z+(\Delta z/2)$, for the two choices $\Delta z=0.01$ and $\Delta z=0.05$. The cosmic web has typical structures on scales up to ${\cal O}(100)$~Mpc, with the largest cosmic voids having a  size  
$\sim 50 h_0^{-1}$~Mpc, and filaments stretching for tens of Mpc~\cite{Pan:2011hx,Libeskind:2017tun}. Therefore,  
it is not surprising that the  choice $\Delta z=0.01$, for which, for instance,  $V_c(z=1)\simeq (190\, {\rm Mpc})^3$, is not sufficient to average over these structures. In contrast, for
$\Delta z=0.05$  we have $V_c(z=1)\simeq (326\, {\rm Mpc})^3$, which appears more adequate to smooth out structures that can extend up to the 100~Mpc scale.

Apart from obtaining an insight on the criterion for the optimal choice of $\Delta z$, what we learn from   Fig.~\ref{fig:Mollview_z02} is that, when using the weighting $w_{\alpha}=1$, the completeness of the GLADE catalog is quite poor  already at $z\simeq 
0.05$ (below $10\%$ for most directions, as shown by the $4\pi$ average, even having set $\bar{n}_{\rm gal}=0.1 \, {\rm Mpc}^{-3}$, which is on the lower side of the estimate for $\bar{n}_{\rm gal}$), and at $z=0.10$ the catalog, with this weighting, is so incomplete that it is basically useless for our purposes. Therefore, except for events at very close distances, such as GW170817, we cannot get much information by correlating the localization of GW events with GLADE galaxies weighted with $w_{\alpha}=1$. In the following, we will therefore  not use this weighting, when using GLADE.\footnote{The situation is different for catalogs, such as DES~Y1, that cover only a part of the sky, where they can be  very complete  up to some redshift, and in fact the $w_{\alpha}=1$ weighting has been used in
\cite{Soares-Santos:2019irc,Palmese:2020aof}.}

\begin{figure}[t]
\centering
\includegraphics[width=0.48\textwidth]{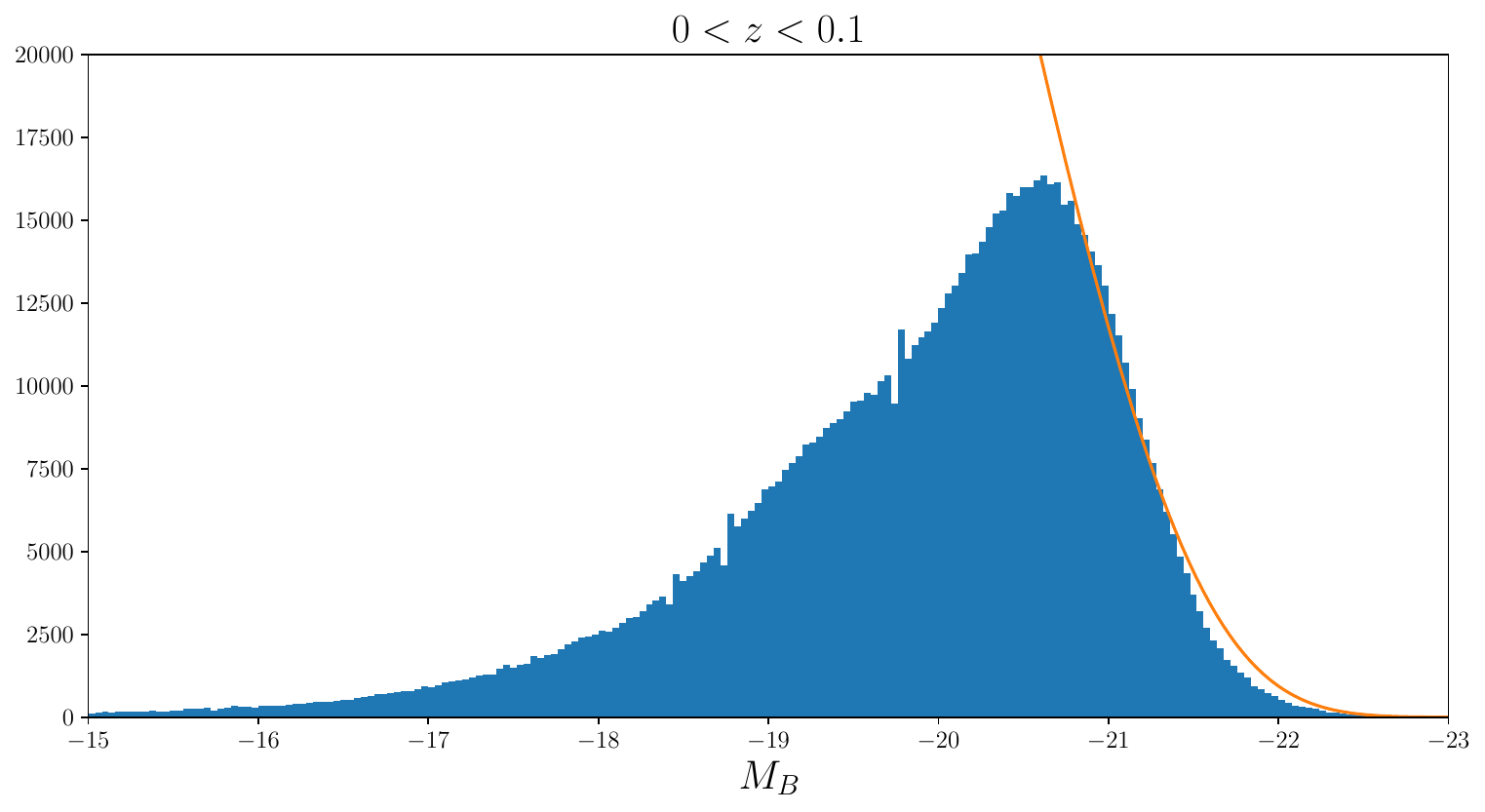}
\includegraphics[width=0.48\textwidth]{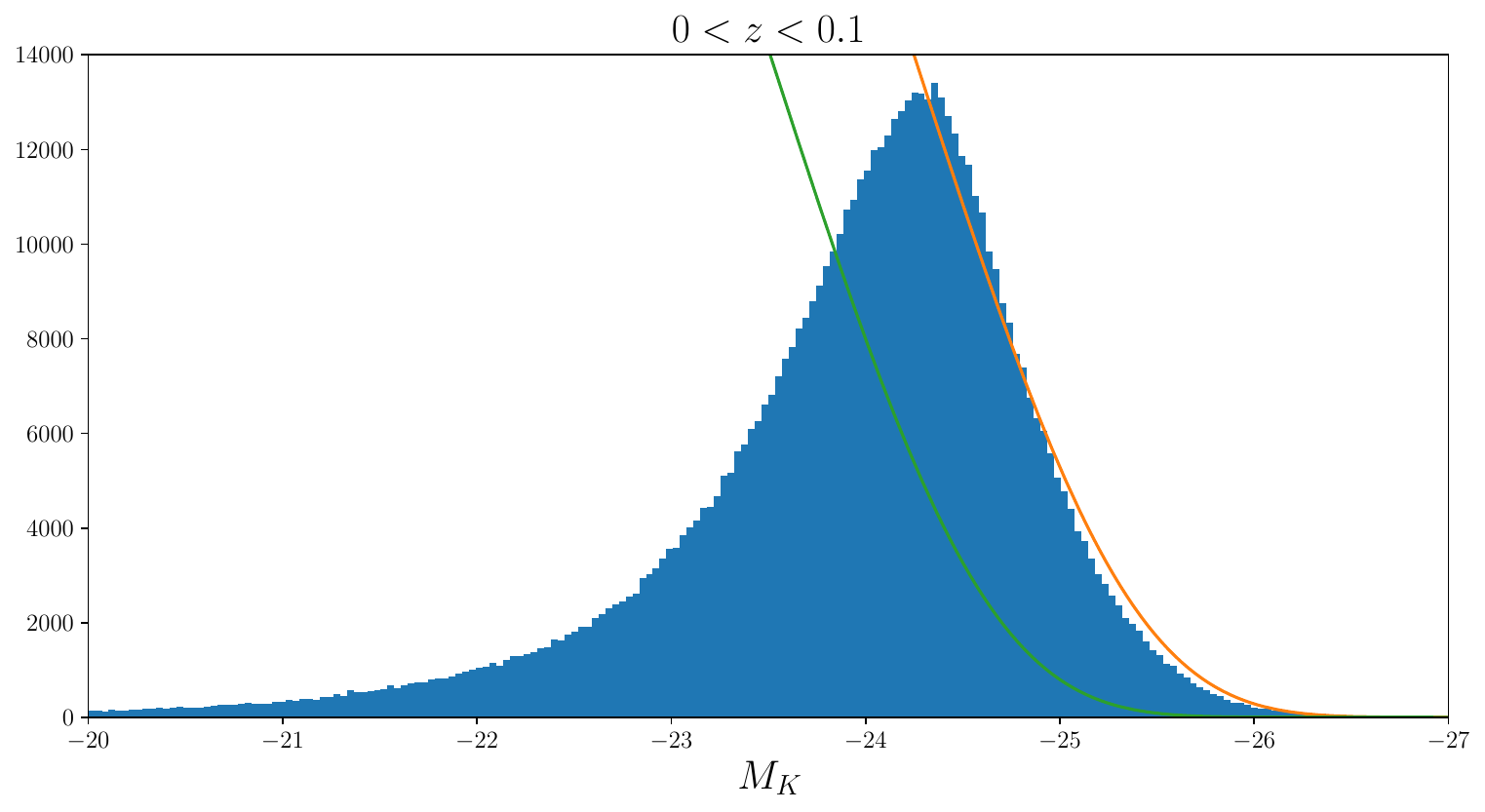}
\caption{Left: the distribution in B-magnitude of the galaxies in the GLADE catalog  compared to the Schechter function 
with parameters given in \eq{SchparB}
(yellow curve).  Right:  the same for K-band, with the Schechter function computed using  the two choices of parameters discussed in the text,
$M_K^*=-24.29$ (yellow curve) and $M_K^*=-23.55$ (green curve).}
\label{fig:Magnitude_vs_Sch}
\end{figure}

We then turn to luminosity weighting. First of all we observe that, for B band, the GLADE catalog provides both the measured apparent magnitudes of 1.6 millions galaxies, as well as their absolute magnitudes, that have been obtained, mostly from HyperLEDA,  correcting for Galactic extinction, internal extinction (i.e., extinction within the  galaxy in question)
and K-correction (which corrects for  the fact that the signal received in a given band was emitted in a different frequency range because of the frequency shift  of the signal due to  the cosmological or peculiar velocity of the source). In the left panel of Fig.~\ref{fig:Magnitude_vs_Sch} we show the distribution of GLADE galaxies with respect to the absolute B-band magnitude  $M_B$ (considering for definiteness the range $0<z<0.1$). The yellow curve is the  Schechter function (multiplied by the comoving volume out to redshift $z=0.1$)
written in terms of magnitudes,
\be\label{SchfunctM}
\phi(M_B)=(0.4\ln 10)\phi_B^*\,  \[10^{0.4(M_B^*-M_B)}\]^{(1+\alpha_B)}
\, \exp\[ -10^{0.4(M_B^*-M_B)} \]\, ,
\ee 
which is obtained from \eq{Schfun} using the standard relation between luminosity and absolute magnitude
(in this case written for  B-band quantities),
\be\label{L10M}
L_B=L_B^*\, 10^{0.4 (M_B^*-M_B)}\, ,
\ee
and requiring that $\phi(M)dM=\Phi(L)dL$, with $\Phi(L)dL$ given by 
\eq{Schfun}. For the   
Schechter parameters we used the values~\cite{Gehrels:2015uga,Arcavi:2017vbi,Dalya:2018cnd}
\be\label{SchparB}
M_B^*=-20.47+5\log_{10} h_{0.7}\, ,
\qquad
\phi_B^*=5.5\times 10^{-3} h_{0.7}^3\, {\rm Mpc}^{-3}\, ,\qquad 
\alpha_B=-1.07\, ,
\ee 
where $h_{0.7}$ is $H_0$ in units of  $70 \, {\rm km}\, {\rm s}^{-1}\, {\rm Mpc}^{-1}$ (in the plots, we set $h_{0.7}=1$). Note that,
plugging $L=L_{B,\odot}$ and $M=M_{B,\odot}$ in \eq{L10M}, 
and using the value  $M_{B,\odot}\simeq 5.498$ for the solar luminosity  in B band, this 
corresponds to
\be
L_B^*=2.45\, h_{0.7}^{-2}\, \times 10^{10} L_{B,\odot}\, .
\ee 
We see from the figure that the Schechter function  with these parameters
(and  $h_{0.7}=1$) fits reasonably  well the high-luminosity tail of the distribution while, as expected, the low-luminosity part of the distribution of the galaxies in the GLADE catalog is well below the Schechter function, as a consequence of the fact that most low-luminosity galaxies are missed. This result is consistent  with the  plots in Fig.~3 of \cite{Dalya:2018cnd}.

For K band, GLADE only provides the apparent magnitudes, from which we can obtain the absolute magnitude using
$M_K=m_K-5\log_{10}(d_L/{\rm Mpc}) -25$, which neglects extinction and K-corrections. 
However, in the case of B-band, we have checked that the distribution of absolute magnitudes shown in the left panel of Fig.~\ref{fig:Magnitude_vs_Sch}, which includes these corrections, is basically undistinguishable from the distribution that would be obtained by using the uncorrected relation $M_B=m_B-5\log_{10}(d_L/{\rm Mpc}) -25$. This indicates that, at the redshifts of interest, extinction and K-corrections are negligible, at least statistically, and we will then use the uncorrected absolute magnitudes for the K band.
The corresponding distribution is shown in the right panel 
of Fig.~\ref{fig:Magnitude_vs_Sch}  and compared with the result obtained from the Schechter function, for two different set of parameters used in the literature. The yellow curve is obtained using the values~\cite{Kochanek:2000im,2005IAUS..216..170H,Crook:2006sw,Lambert:2020dwg}
\be\label{MKstar}
M_K^*=-24.29+5\log_{10}h_{0.7}\, ,\qquad
\phi_K^*=3.70\times 10^{-3}  h_{0.7}^3\, {\rm Mpc}^{-3}\, ,\qquad
\alpha_K=-1.02\, ,
\ee 
while the green curve corresponds to the choice
$M_K^*=-23.55+5\log_{10}h_{0.7}$ (with the same values of  $\phi_K^*$ and $\alpha_K$),  proposed in \cite{Lu:2016vmu}.\footnote{Note that the value of $\phi_K^*$ given in \cite{Lu:2016vmu}, $\log\phi_K^*=1.08\times 10^{-2}$ is the same as that given in \eq{MKstar}, except that it  suffers of two clear typos: the log should not be there, and the units $h^3 {\rm Mpc}^{-3}$  have been forgotten. Writing 
$h=0.7 h_{0.7}$, this  leads to a value $\phi_K^*=1.08\times 10^{-2} (0.7)^3  h_{0.7}^3{\rm Mpc}^{-3}=
3.70\times 10^{-3}  h_{0.7}^3\, {\rm Mpc}^{-3}$.} 
We see that the yellow curve  fits reasonably well the high-luminosity tail, while the green curve misses completely, so we will use the parameters given in \eq{MKstar}.\footnote{Note that, in ref.~\cite{Fishbach:2018gjp}, has rather been used the value of $M_K^*$ from \cite{Lu:2016vmu}. This implies that the K-band completeness of GLADE computed  \cite{Fishbach:2018gjp} is actually  an overestimate, since it has been obtained comparing the catalog to a Schechter function that is  lower than the correct one.} Using
$M_{K,\odot}=3.27$ (in the Vegamax system), the value of $M_K^*$ in \eq{MKstar} corresponds to
\be
L_K^*=10.56\, h_{0.7}^{-2}\, \times 10^{10} L_{K,\odot}\, ,
\ee 
while the value in \cite{Lu:2016vmu} would give 
$L_K^*=5.35\, h_{0.7}^{-2}\, \times 10^{10} L_{K,\odot}$.

\begin{figure}[t]
\centering
\includegraphics[width=0.48\textwidth]{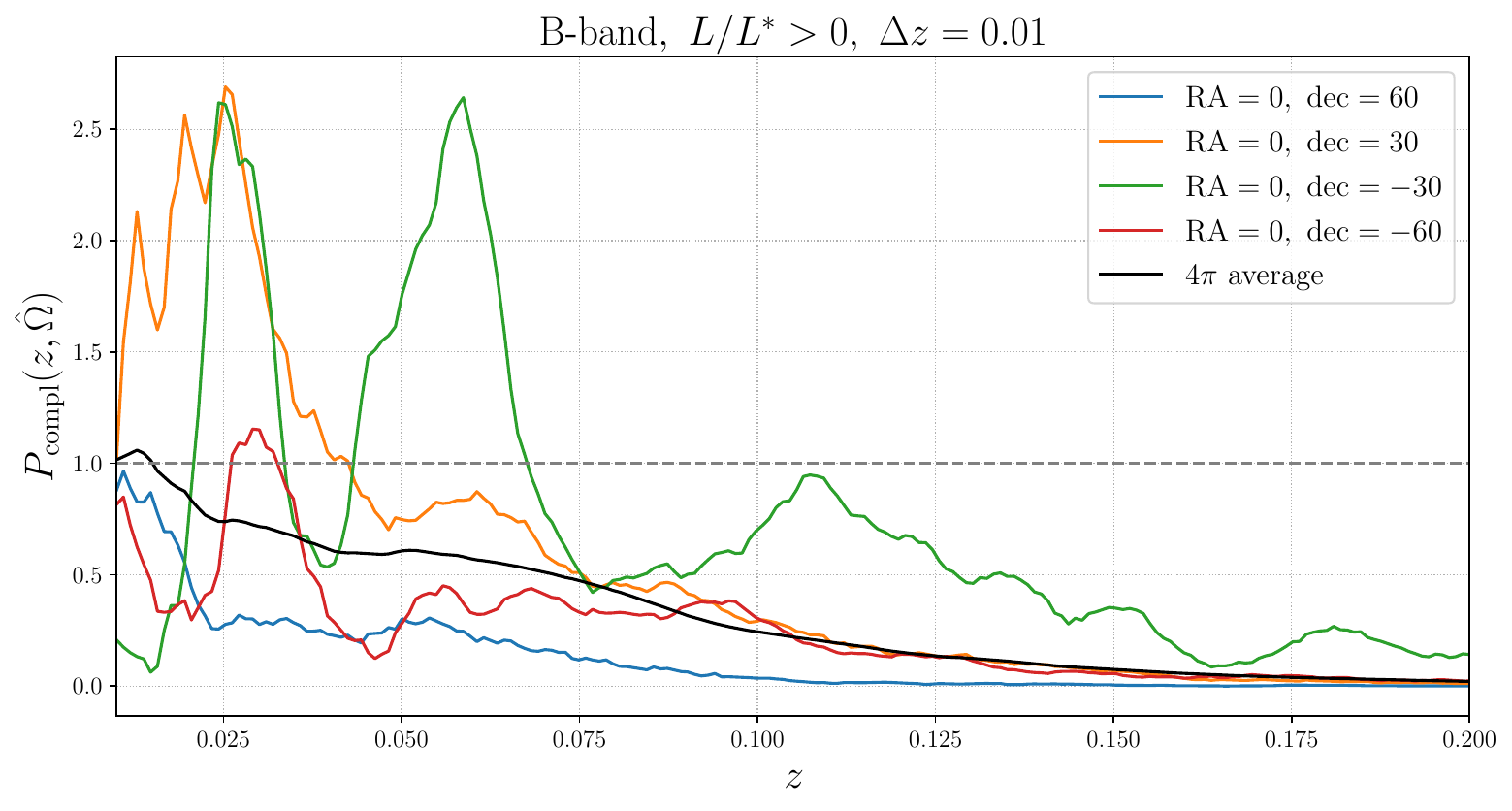}
\includegraphics[width=0.48\textwidth]{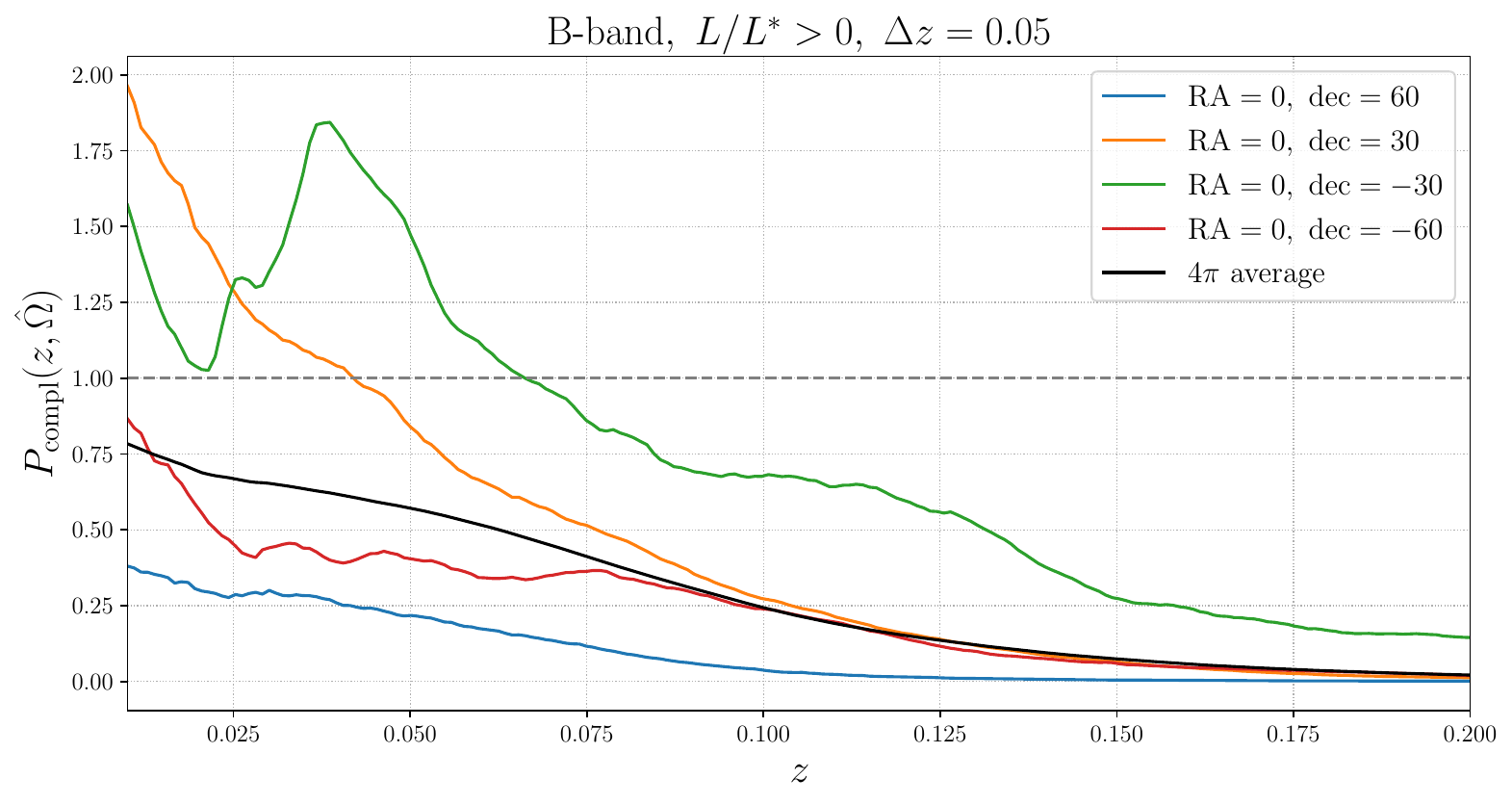}
\includegraphics[width=0.48\textwidth]{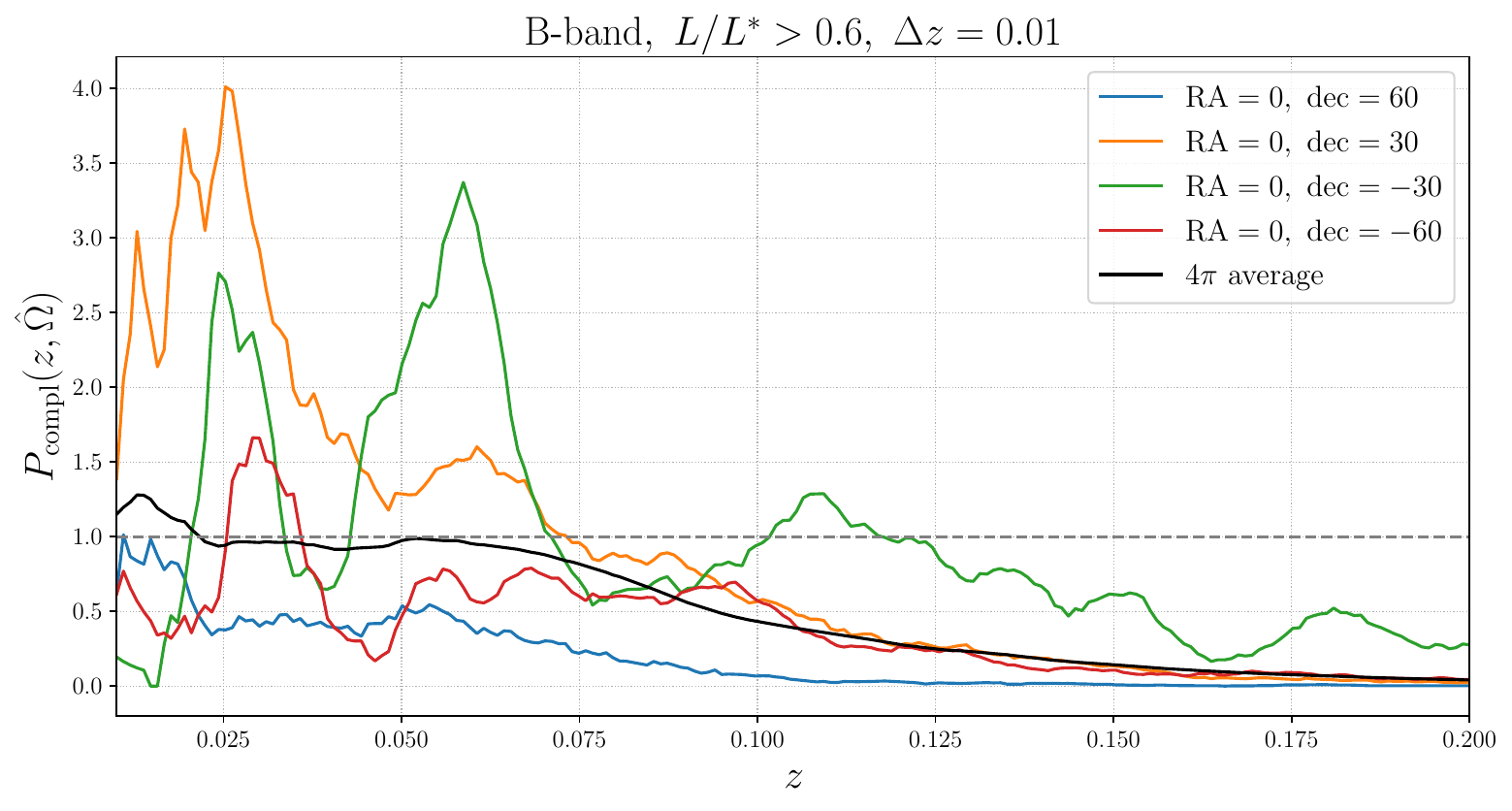}
\includegraphics[width=0.48\textwidth]{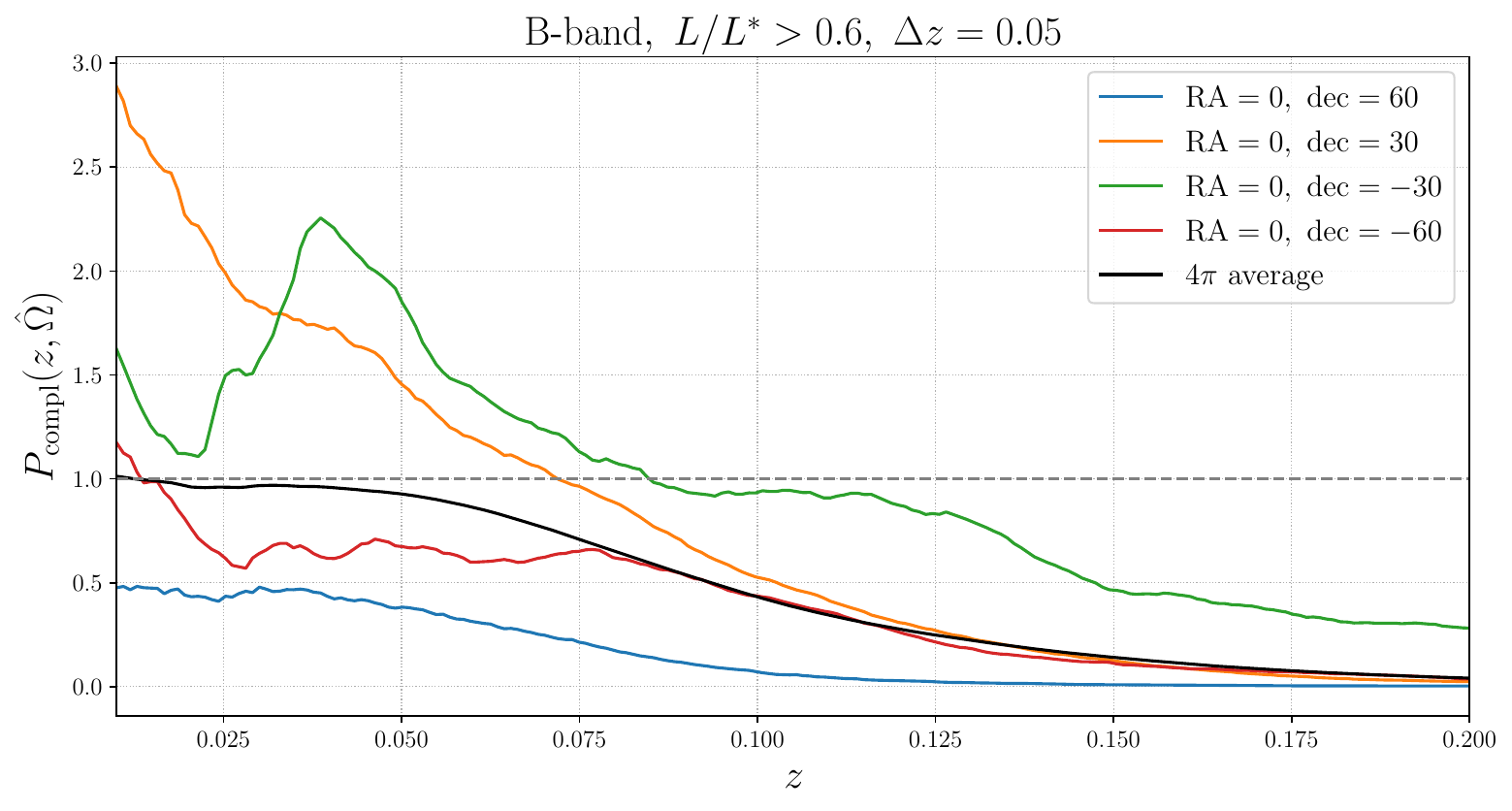}
\caption{The function $P_{\rm compl}(z,\hatO)$  with B-band luminosity weighting, for different choices of
$\Delta z$ and $L_{\rm cut}$ for  four sky directions, and for the $4\pi$ average $P_{\rm compl}(z)$. Upper panels: with no luminosity cut and
$\Delta z=0.01$ (left) and $\Delta z=0.05$ (right). Lower panels: the same, restricting to galaxies 
with $L>0.6 L^*_B$.}
\label{fig:Bcompleteness}
\end{figure}

\begin{figure}[t]
\centering
\includegraphics[width=0.48\textwidth]{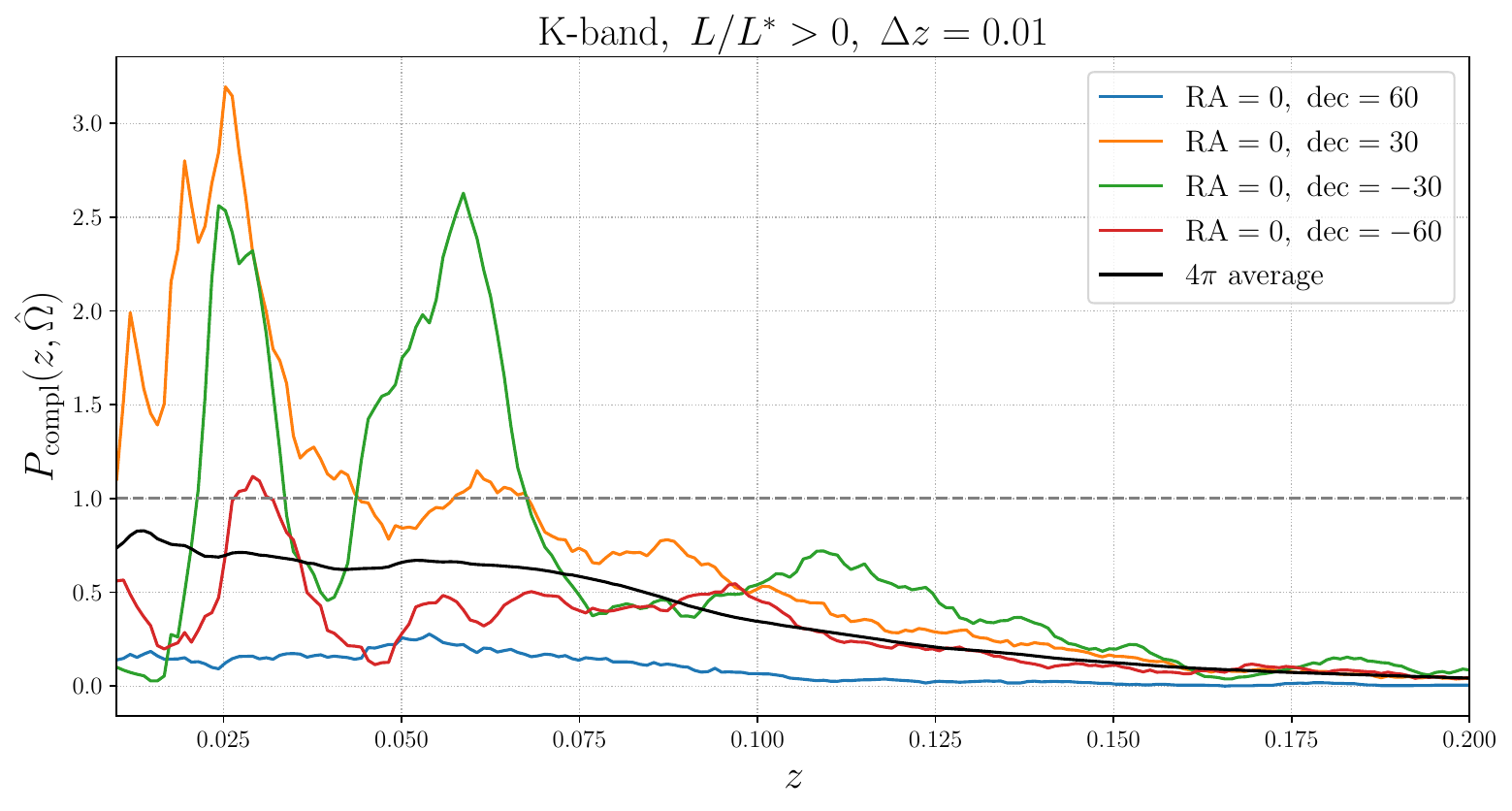}
\includegraphics[width=0.48\textwidth]{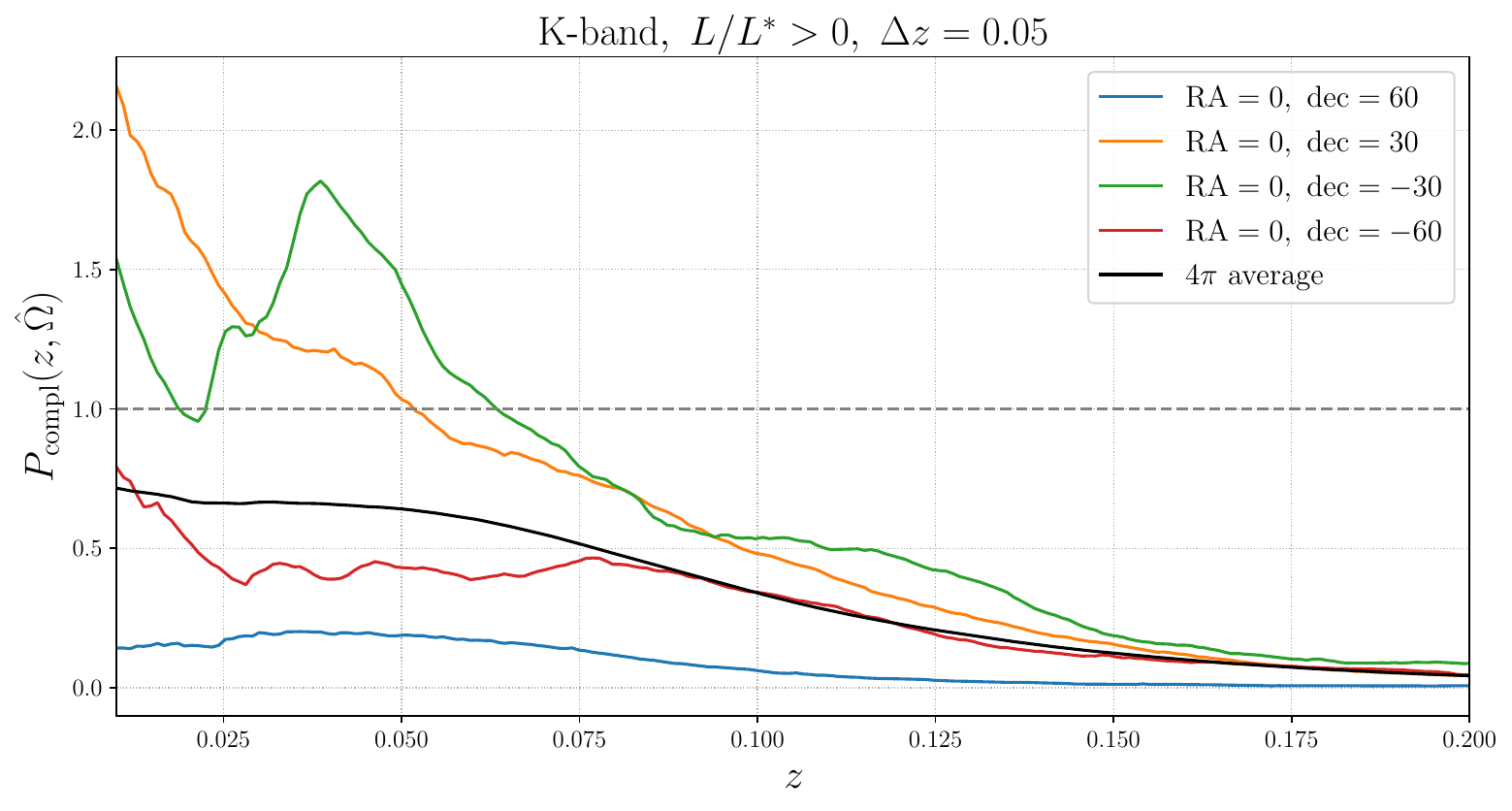}
\includegraphics[width=0.48\textwidth]{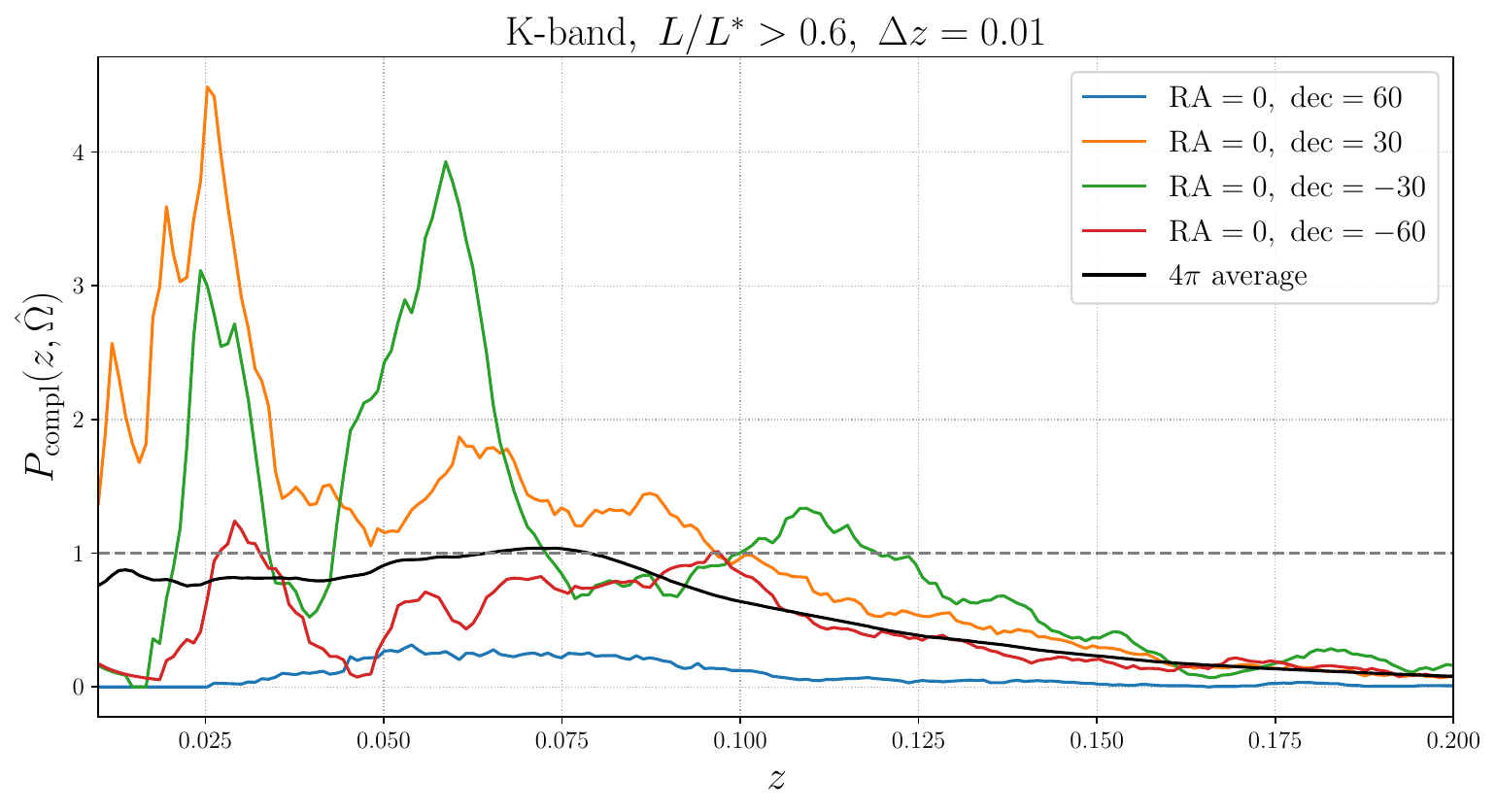}
\includegraphics[width=0.48\textwidth]{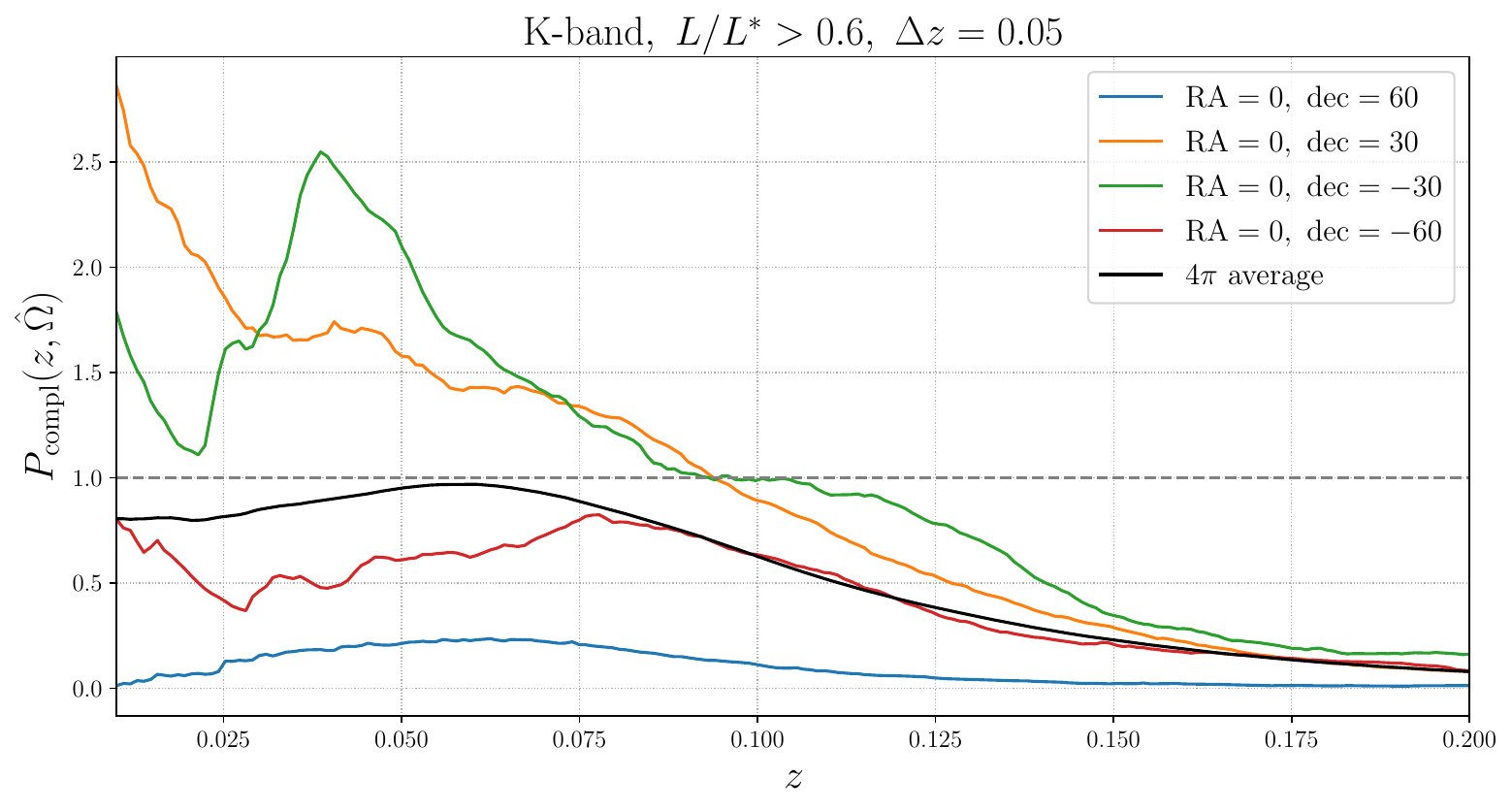}
\caption{As in Fig.~\ref{fig:Bcompleteness}, for K band.}
\label{fig:Kcompleteness}
\end{figure}

We can now study the completeness of GLADE with luminosity weighting, in the B and K bands. As we will see, compared  to the case $w_{\alpha}=1$,  the completeness improves drastically when using luminosity weighting, since the galaxies that are missed are typically those with low luminosity, that now weight less. We can further enhance this effect by setting a 
lower limit  $L_{\rm cut}$  on the luminosity of the sample considered. Here, however, a trade-off enters: restricting to high luminosity galaxies we get, of course, a more complete sample, because it is more difficult to miss a high-luminosity galaxy; however, the price that we pay is that, increasing $L_{\rm cut}$, we decrease the number of galaxies that we can use for performing a correlation with the GW signal, so a compromise between these two effects must be found.\footnote{Of course, if the completion procedure were `exact', this would not be an issue. In practice, any completion procedure will be unavoidably an approximation, and will in general introduce uncertainties and biases.}
To this purpose,
we compare the results for $P_{\rm compl}(z,\hatO;\Delta z,\Delta\Omega;L_{\rm cut})$
for different choices of $\Delta z$ and of $L_{\rm cut}$,  for B-band and for K-band.
The four panels of Fig.~\ref{fig:Bcompleteness} show the results  for B band when no luminosity cut is applied (upper panels) and with $L>0.6 L^*_B$ (lower panels). In each row we show the result  for the two choices $\Delta z=0.01$ (left) and $\Delta z=0.05$ (right). In each plot we show again the same four directions as in Fig.~\ref{fig:ngal} (with $\theta_c=5^{\circ}$), as well as the $4\pi$ average. Fig.~\ref{fig:Kcompleteness} shows the same quantities for the K band.

All plots on the left panels confirm that the choice $\Delta z=0.01$ is too small, since the actual over- and 
under-densities are clearly seen, and we do not want to artificially compensate them with the completion. 
We therefore restrict to the panels with $\Delta z=0.05$.
We first observe that, up to some critical $z$, the completeness  is larger than one.
This over-completeness at small distances is due to local over-densities. These results agree with similar plots shown in  refs.~\cite{Dalya:2018cnd,Fishbach:2018gjp}. Clearly, it would be meaningless to `compensate' these over-densities by assigning a negative luminosity density that brings back
$P_{\rm compl}(z,\hatO;\Delta z,\Delta\Omega)$ to the `expected' value of one. Then, as already mentioned in sect.~\ref{sect:hom_multi_compl}, once fixed
$\Delta z$ and $\Delta\Omega$, for each direction $\hatO$
our strategy will be to look for the highest value of $z$ for which $P_{\rm compl}(z,\hatO)=1$, that we denote by $z_*(\hatO)$. For $z<z_*(\hatO)$ we will use the galaxy catalog as it is, considering it complete
[which formally amounts to setting $P_{\rm compl}(z,\hatO)=1$ in eqs.~(\ref{p0isocompl}) or (\ref{p0multipcompl1})] while, for $z>z_*(\hatO)$, we will complete the catalog according to the 
homogeneous completion  or to the 
multiplicative completion discussed in  sect.~\ref{sect:hom_multi_compl}.\footnote{This is the same procedure that has been adopted in \cite{Fishbach:2018gjp}, that however considers only  the $4\pi$ average of the completeness, and homogeneous completion.}
From the plots in the right panels we see that, in all cases, 
for $\Delta z=0.05$ and $z>z_*(\hatO)$,  $P_{\rm compl}(z,\hatO)$ is basically a smoothly decreasing function of redshift, and therefore $\Delta z=0.05$ is an adequate choice.\footnote{Some minor wiggles remain in some directions, e.g. in the red curve corresponding to ${\rm RA}=0, {\rm dec}=-60^{\circ}$. However, the smoothing of these structures induced by the completion procedure, performed using 
$\Delta z=0.05$, is quite negligible compared to the actual amplitude of these fluctuations, which can be appreciated from the panels with $\Delta z=0.01$, and which would would be even larger if resolved with a smaller $\Delta z$.}

\begin{figure}[t]
\centering
\includegraphics[width=0.48\textwidth]{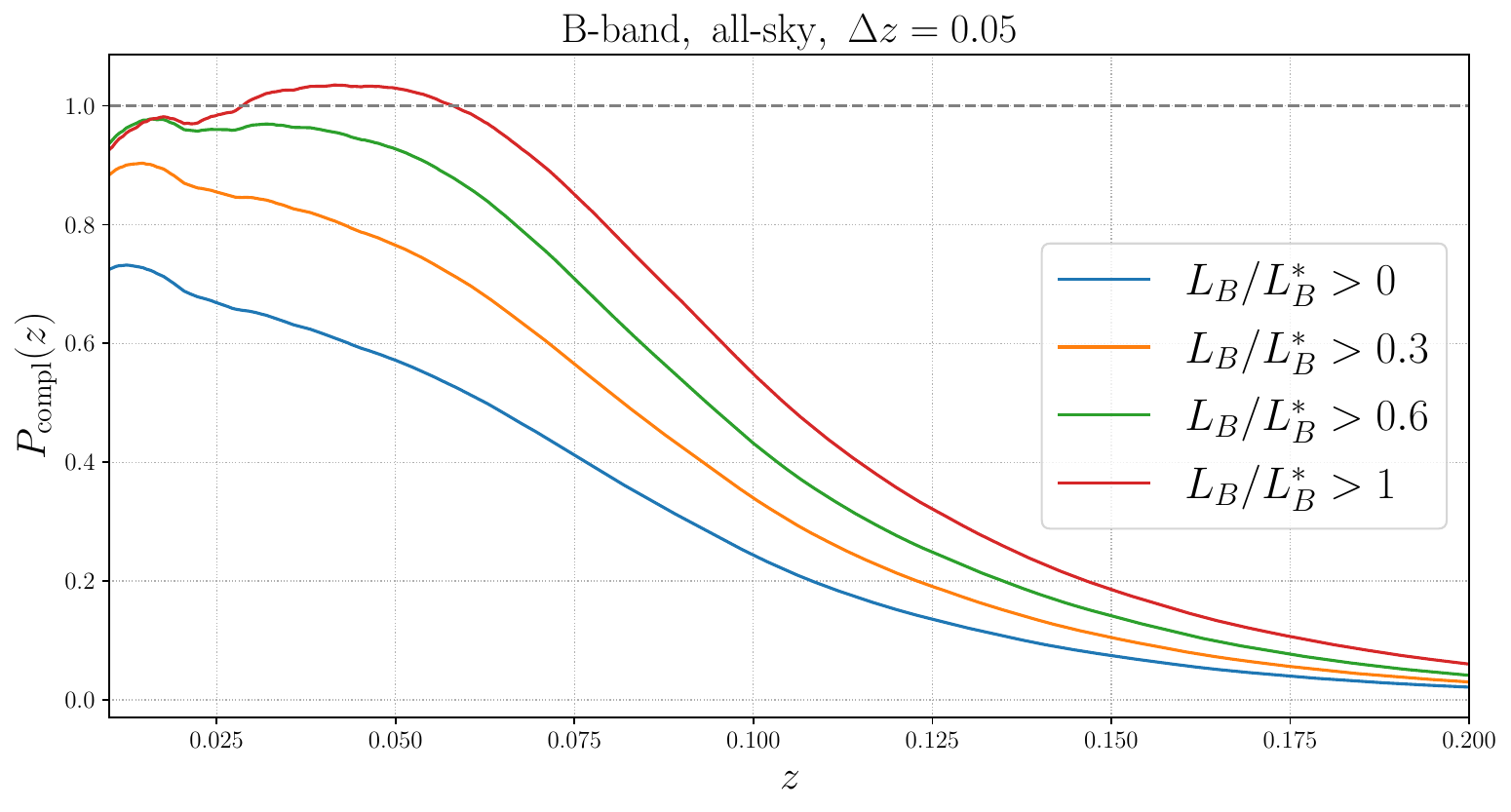}
\includegraphics[width=0.48\textwidth]{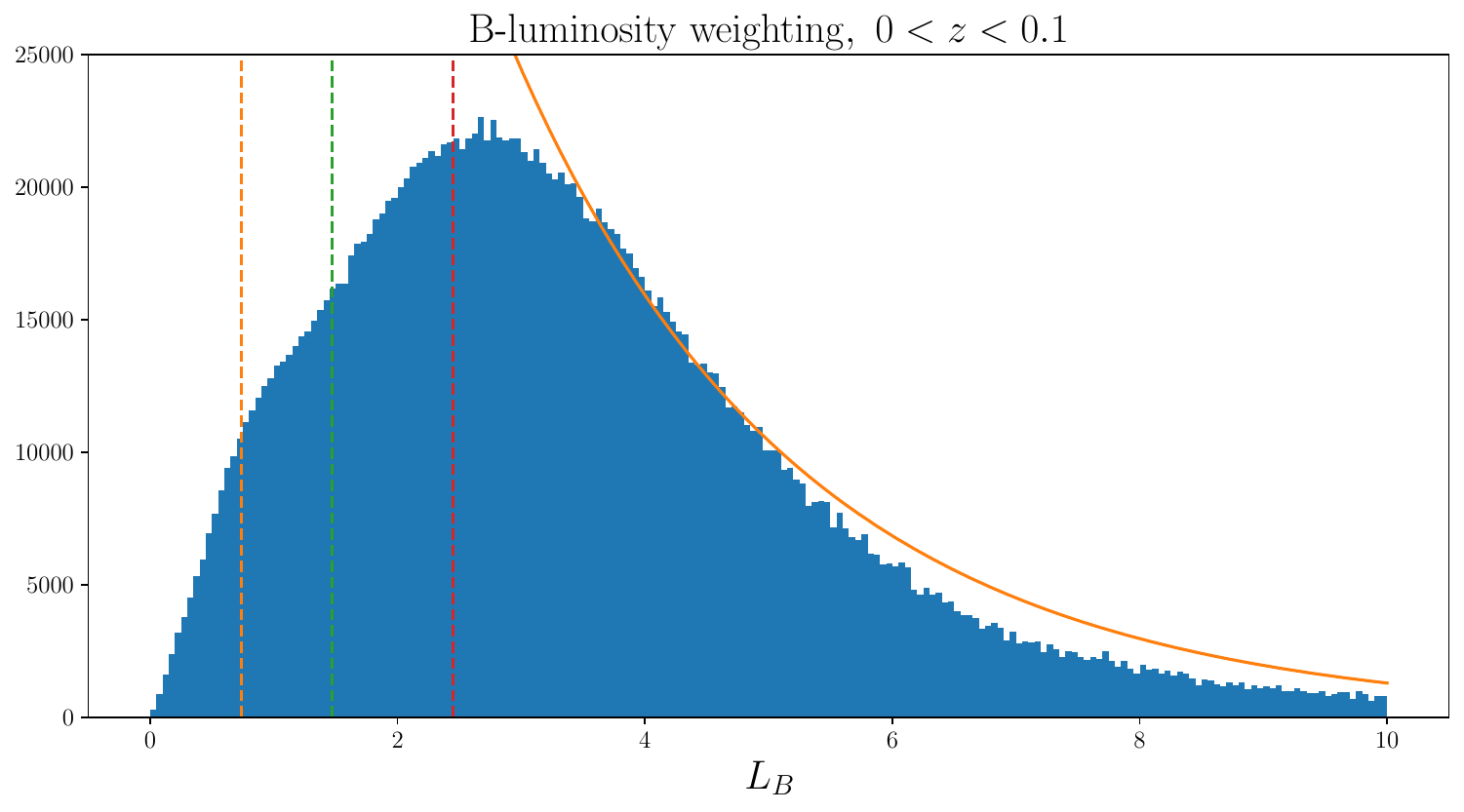}
\caption{Left: the function  $P_{\rm compl}(z)$, for $\Delta z=0.05$ and different luminosity cuts. Right: the number of galaxy per luminosity bin in the GLADE catalog, weighting each galaxy by its luminosity, in B-band. The yellow, green and red and dashed vertical lines mark the values $L_B=0.3L_B^*$,
$L_B=0.6L_B^*$ and $L_B=L_B^*$, respectively. The yellow continuous line is the fit to the Schechter function, as in Fig.~\ref{fig:Magnitude_vs_Sch}, now in terms of luminosity rather than absolute magnitude.}
\label{fig:PcomplB_different_cuts}
\end{figure}

\begin{figure}[t]
\centering
\includegraphics[width=0.48\textwidth]{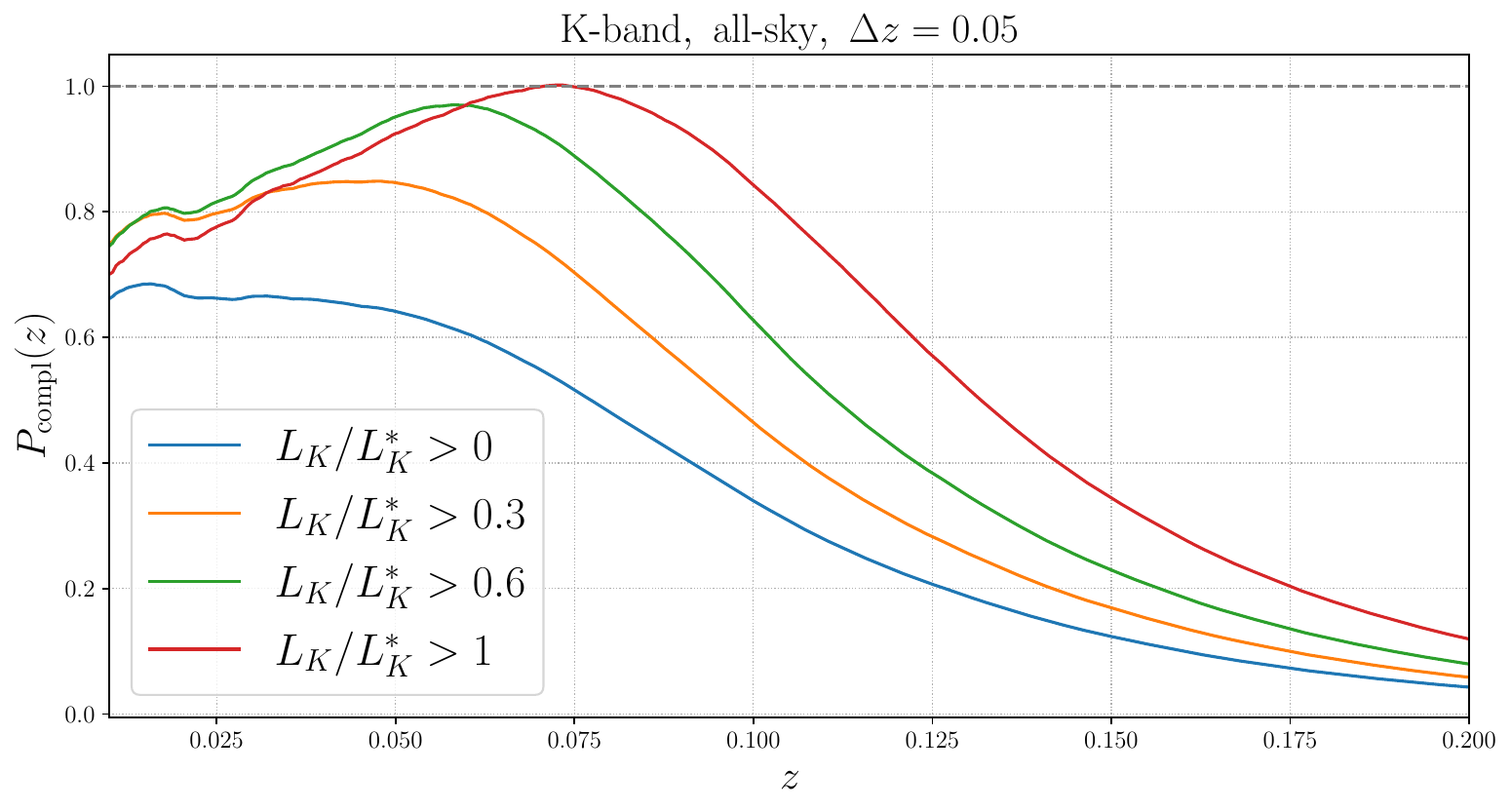}
\includegraphics[width=0.48\textwidth]{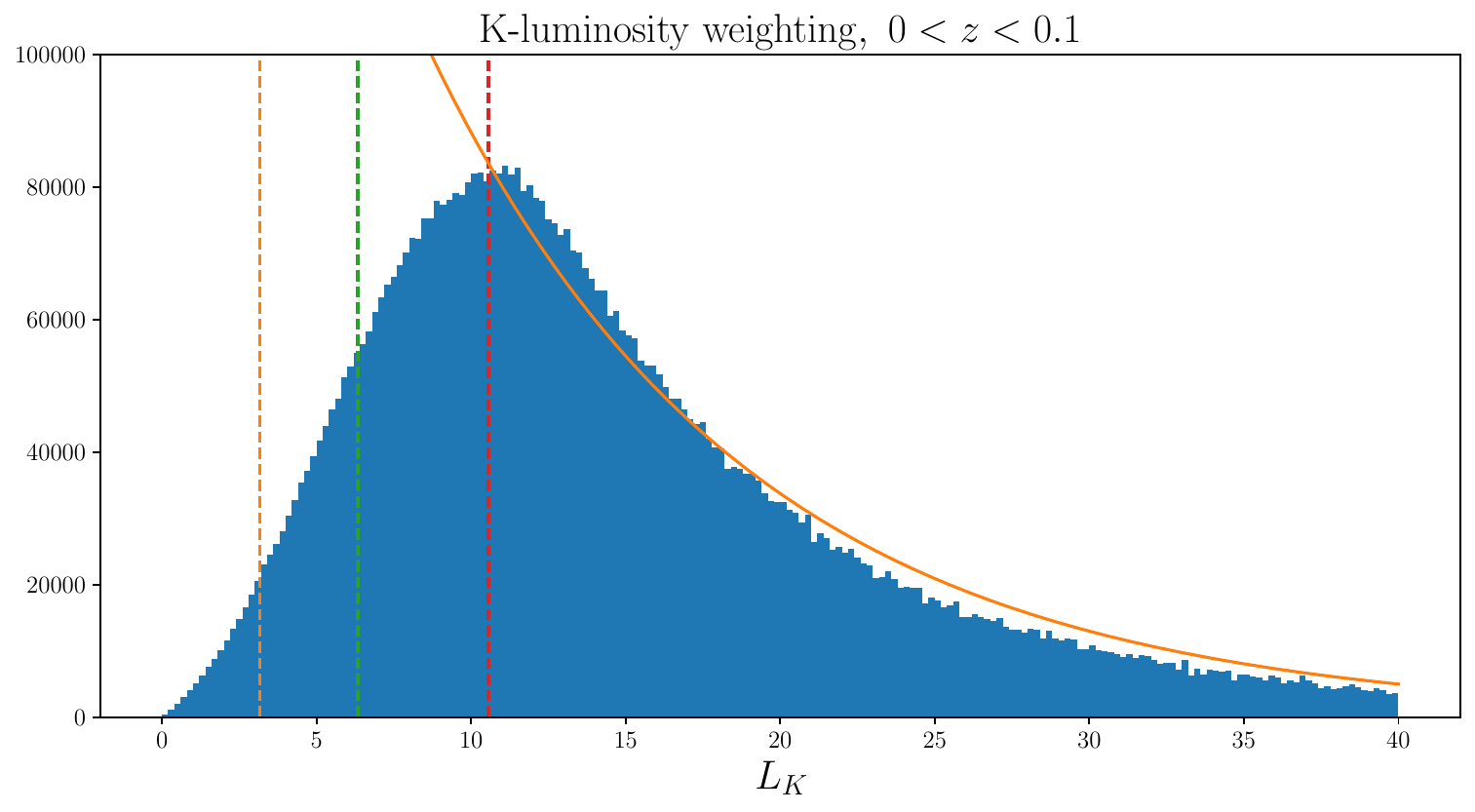}
\caption{As in Fig.~\ref{fig:PcomplB_different_cuts}, for K band.}
\label{fig:PcomplK_different_cuts}
\end{figure}

Comparison with Fig.~\ref{fig:ngal} shows that luminosity weighting (both in B band and in K band) improves  the completeness dramatically.  To understand the best choice for the luminosity cut,
in the left panel of Fig.~\ref{fig:PcomplB_different_cuts} we  show  the angular average
$P_{\rm compl}(z)$ for   $\Delta z=0.05$ and 
several different choices of $L_{\rm cut}$,  for B-band.
As expected, the completeness increases with the cut. On the other hand, the number of galaxies in the catalog, even if luminosity weighted, decreases as we increase  the cut. In the right panel Fig.~\ref{fig:PcomplB_different_cuts} we show the distribution of B-band luminosities in GLADE, weighting the galaxies by their luminosity (which is the relevant quantity when we use luminosity weighting). The yellow, green and red and dashed vertical lines mark the values $L_B=0.3L_B^*$,
$L_B=0.6L_B^*$ and $L_B=L_B^*$, respectively. We see that, raising the  cut 
from $L_{\rm cut}=0$ to $L_{\rm cut}=0.3L_B^*$, or from
$L_{\rm cut}=0.3L_B^*$ to $L_{\rm cut}=0.6L_B^*$ we lose a limited fraction of (luminosity-weighted) galaxies, while we have a good increase in completeness; in contrast, passing from 
$L_{\rm cut}=0.6L_B^*$ to  $L_{\rm cut}=L_B^*$, we begin to lose a more significant fraction of the integrated luminosity, because we approach  the peak of the distribution, while the increase in completeness, especially at large $z$, is not so significant.
On this ground, we choose $L_{\rm cut}=0.6L_B^*$. A rather similar situation takes place for K-band, as we see from Fig.~\ref{fig:PcomplK_different_cuts}, and will choose again $L_{\rm cut}=0.6L_K^*$.

To summarize, \label{page:summary} the results of this appendix indicate that: (1) the use of all galaxies without luminosity weighting, for the GLADE catalog, is not appropriate because of the very limited completeness. We will therefore not use the $w_{\alpha}=1$ weighting for GLADE. 
(2)  B-band and K-band luminosity weighting, with a lower cut on the luminosity, produces a good level of completeness up to redshifts $z\simeq 0.1$ and a still potentially useful level of completeness (i.e., at least larger than  $10\%$ in most directions) up to $z\simeq 0.2$. Setting the cut in B band at $L_{\rm cut}=0.6L_B^*$, and in K band 
at $L_{\rm cut}=0.6L_K^*$,  
appears to be a good compromise between having a  sample of galaxies as complete as possible and as large as possible.  (3). For  the region ${\cal S}$ used to define cone completeness, a good choice  is given by a section of a cone, with half-opening angle $\theta_c=5^{\circ}$ extending, in the transverse size, between redshifts $z\pm\Delta z/2$ with $\Delta z=0.05$.
We will therefore use these choices in  our  analysis.

A final technical point concerns the treatment of the error on the redshifts of the galaxies.
The GLADE catalog does not provide the error on the redshift of the individual galaxies. 
The simplest option is to assign to all galaxies a gaussian uncertainty
$\Delta v\simeq 200\, {\rm km/s}$, corresponding to $\Delta z= \Delta v/c\simeq 6.7\times 10^{-4}$, as in
\cite{Fishbach:2018gjp,Vasylyev:2020hgb}. This reflects a typical value of the peculiar velocities of the galaxies with respect to the Hubble flow. However,  in the GLADE catalog the redshifts have already been corrected for peculiar velocities, using an algorithm described in \cite{Dalya:2018cnd}. \red{Thus, what really matters here is the  residual   error in the measurement of the galaxy redshift, which depends on whether the redshift has been determined spectroscopically or photometrically. In the former case, for the GLADE catalog, $\Delta z$ is dominated by the uncertainty in the correction for peculiar velocities, and can be taken to be 
$\Delta z\simeq 1.5\times 10^{-4}$, while, for photometric redshifts, this is dominated by the error in the measurement of the redshift, which in this case is typically of order 
$\Delta z\simeq 1.5\times 10^{-2}$~\cite{Dalya:2018cnd}.} The GLADE catalog (v2.4) has a flag that indicates whether the redshift has been obtained photometrically or spectroscopically, so we can assign the corresponding average error to each galaxy. 
We then use, in \eq{pcatN},  a gaussian with the variance chosen according to whether the galaxy has a spectroscopic of photometric redshift, using $\Delta z\simeq 1.5\times 10^{-4}$ and redshifts 
$\Delta z\simeq 1.5\times 10^{-2}$, respectively. We finally convolve the gaussian with a prior given by $p(z)=dV_c/dz$, as discussed below \eq{pcatN}.

\section{Dependence of the prior on the region ${\cal R}$}\label{app:R}

As we saw in sect.~\ref{sect:cat}, upon completion the prior can be written in the form (\ref{p0pcatpmiss}).
At this level, it is necessary to specify a region ${\cal R}$ over which we compare the galaxies in the catalog to those that are missing. In this appendix we show that, in the end, the estimate of the parameters such as $H_0$ or $\Xi_0$ is independent of ${\cal R}$, as long as it is sufficiently larger than the localization region of the ensemble of all GW events (of the type that we are considering in the correlation, e.g. BBH, or BNS) detectable by the GW detector network.

This can be shown  directly starting from \eq{p0assum}. To keep the notation more compact, we write 
\be\label{wgal}
\sum_{\alpha=1}^{N_{\rm gal}({\cal R})} w_{\alpha }\equiv w_{\rm gal}({\cal R})\, ,\qquad
\sum_{\alpha=1}^{N_{\rm cat}({\cal R})} w_{\alpha }\equiv w_{\rm cat}({\cal R})\, ,
\ee
and we use the shorthand $\delta(z-z_{\alpha})\delta^{(2)}(\hatO-\hatO_{\alpha})
\equiv\delta(x-x_{\alpha})$. We consider for simplicity a single GW event (the argument is trivially generalized) and we
take ${\cal R}_1$ to be the localization region of the GW event, say at $99.9\%$~c.l., and 
${\cal R}_2$ a much larger region, that includes ${\cal R}_1$. Then
\be
p_0(x;{\cal R}_1)=\frac{1}{ w_{\rm gal}({\cal R}_1) } \[
 \sum_{\alpha_1=1}^{N_{\rm cat}({\cal R}_1) }    w_{\alpha_1}\delta(x-x_{\alpha_1})+
 \sum_{\alpha_2=1}^{N_{\rm miss}({\cal R}_1) } w_{\alpha_2}\delta(x-x_{\alpha_2})
 \]\, ,
\ee
\be\label{p0R2}
p_0(x;{\cal R}_2)=\frac{1}{ w_{\rm gal}({\cal R}_2) } \[
 \sum_{\alpha_1=1}^{N_{\rm cat}({\cal R}_2) }    w_{\alpha_1}\delta(x-x_{\alpha_1})+
 \sum_{\alpha_2=1}^{N_{\rm miss}({\cal R}_2) } w_{\alpha_2}\delta(x-x_{\alpha_2})
 \]\, .
\ee
We now observe that, when in \eq{pDizO} we multiply $p_0(z,\hatO;{\cal R}_2)$ by the GW likelihood 
$p({\cal D}^i_{\rm GW}|z,\hatO,\theta',H_0)$, all terms in the sums in \eq{p0R2} that fall outside the localization region ${\cal R}_1$ do not contribute, since there $p({\cal D}^i_{\rm GW}|z,\hatO,H_0)$ is
zero to all practical purposes. Therefore both sums in  \eq{p0R2} are effectively restricted to the region 
${\cal R}_1$ and,  in \eq{pDizO}, to use $p_0(x;{\cal R}_2)$ is the same as using a prior
$\tilde{p}_0(x;{\cal R}_2)$ given by 
\bees\label{p0R2bis}
\tilde{p}_0(x;{\cal R}_2)&=&\frac{1}{ w_{\rm gal}({\cal R}_2) } \[
 \sum_{\alpha_1=1}^{N_{\rm cat}({\cal R}_1) }    w_{\alpha_1}\delta(x-x_{\alpha_1})+
 \sum_{\alpha_2=1}^{N_{\rm miss}({\cal R}_1) } w_{\alpha_2}\delta(x-x_{\alpha_2})
 \]\nn\\
 &=&\frac{w_{\rm gal}({\cal R}_1)}{ w_{\rm gal}({\cal R}_2) }\, p_0(x;{\cal R}_1)\, .
\ees
The overall constant $w_{\rm gal}({\cal R}_1)/w_{\rm gal}({\cal R}_2)$ cancels between numerator and denominator in \eq{pDizO}, with the denominator given by \eq{betadL}. Therefore, in the end, 
while it is necessary to introduce a region ${\cal R}$ so that the expressions in \eqst{p0Ntot}{fR} are well defined, its exact definition is irrelevant, as long as it covers the localization region of the GW 
event.

\section{Semi-analytic evaluation of $\beta$ including the distribution of source parameters}\label{app:betafit}

\red{In this appendix we discuss an approach to the computation of $\beta(H_0)$ and $\beta(\Xi_0)$, that allows us to include the effect of the distribution of source parameters, with a relatively simple semi-analytic technique.  We stress that, for producing our final results, we have used  only our MC computation of $\beta$ (to be briefly discussed in app.~\ref{sect:ourMC}). However, the semi-analytic computation that we discuss here provides simple and ready to use formulas,  a physical understanding of some effects, and a check for the MC itself, so we find useful to present it here.
}

We will limit ourselves to the distribution of component masses and orbit inclination, although  the technique could in principle be generalized to other parameters. For the  orbit inclination we  assume a uniform prior for the source population. 
We keep  for the moment generic the distribution of  masses for the BH-BH  population,  and we denote it as $\tilde{p}_0(m_1,m_2)$. We will later specify it to the broken power-law distribution (\ref{brokenpowerlaw}) proposed in \cite{Abbott:2020niy}.

To compute the SNR, for this simple semi-analytic estimate we work  in the  approximation in which one includes only the inspiral part of waveform, and  to lowest-order in the post-Newtonian expansion. 
As discussed in \cite{Abadie:2010cf,Wysocki:2018mpo},  the  result of the actual detection pipelines is well approximated by the criterion that the SNR  in the second loudest detector is larger than 8. For O2, we checked this assumption by computing the selection effects with ${\rm SNR} > 8$  using inspiral-only waveforms, and comparing with publicly available LVC injections (we use the data in the file injections\_O1O2an\_spin.h5 at \url{https://dcc.ligo.org/LIGO-P2000434-v1/public}), and we found good agreement. 
\red{For O3, this  becomes a poorer approximation because of the heavier BBH masses of many events, which implies that these signals stay in the bandwidth for a small number of inspiral cycles.
For this reason, our final results are obtained with a MC, where  we have used exclusively full inspiral-merger-ringdown waveforms, both for O1-O2 and for O3a data.
The semi-analytic technique that we are discussing could, in principle, be generalized beyond the inspiral waveform, but, in any case, our aim  here is simply to improve on the analytic understanding provided by the lowest-order computation in presented in \cite{Chen:2017rfc,Fishbach:2018gjp}.}

In the approximation in which only the inspiral part of the waveform is included, the SNR  in a single detector is given by~\cite{Finn:1992xs}
(see, e.g., sects.~4.1.4 and 7.7.2 of \cite{Maggiore:1900zz} for review)
\be\label{eq:SNR}
({\rm SNR})^2 = \frac{5}{6} \frac{(G{\cal M}_c)^{5/3} |Q(\cos\iota,\hatO)|^2}{c^3 \pi^{4/3} d_L^2} 
\, I_{7/3}({\cal M}_{\rm tot})\, ,
\ee
where
\be\label{I73}
I_{7/3}({\cal M}_{\rm tot})\equiv \int_{f_{\mathrm{min}}}^{f_{\rm insp}(z)} df \,  \frac{f^{-7/3}}{S_{n}(f)}\, ,
\ee
and ${\cal M}_c(z)=(1+z)M_c$ and ${\cal M}_{\rm tot}(z)=(1+z)M_{\rm tot}$ are
the redshifted (`detector-frame') chirp mass  and total mass of the system, respectively
[with $M_c=(m_1m_2)^{3/5}/M_{\rm tot}^{1/5}$ and $M_{\rm tot}=m_1+m_2$]; the factor 
\be
|Q(\cos\iota,\hatO)|^2=F_+^2(\hatO)\, \(\frac{1+\cos^2\iota}{2}\)^2 +
F_{\times}^2(\hatO)\cos^2\iota\, ,
\ee
[where $F_{+,\times}(\hatO)$ are the pattern functions of the interferometer] 
encodes the detector response to the inclination angle $\iota$ of the orbit and to the source direction;
$S_n(f)$ is the noise spectral density of the detector;
$f_{\min}$ is the low frequency limit of the detector bandwidth, and  
$f_\mathrm{insp}(z)=f_\mathrm{insp}/(1+z)$ is the observed (redshifted) gravitational-wave frequency at the end of the inspiral phase. In the approximation in which we only consider the contribution to the SNR from the inspiral, the latter can be approximated as twice the orbital frequency at the innermost stable circular orbit, which gives  
$f_\mathrm{insp}\simeq 4.4\, {\rm kHz}\, (\msun/M_{\rm tot})$~\cite{Maggiore:1900zz}, so in the end
\be
f_\mathrm{insp}(z)\simeq 4.4\, {\rm kHz}\, \frac{\msun}{{\cal M}_{\rm tot}(z) }\, .
\ee
Including higher orders in the PN expansion, as well as the part of the waveform describing merger and ringdown, brings in the dependence on other parameters, such as the BH spins.
Note that, in the context of modified gravity,  in \eq{eq:SNR} $d_L$ must be replaced by the GW luminosity distance $\dgw$.  

To evaluate \eq{I73} we  use an expression for $S_h(f)$ typical of the detectors at the time of detection. This changes between runs, such as the O2 and O3 run, and therefore also the the beta functions depends on the run considered.\footnote{We illustrate our results using the O2 sensitivity, using the table given in 
\url{https://dcc.ligo.org/LIGO-G1801952/public}, which represents the best O2 sensitivity.
We have checked that our results for $\beta(H_0)$ are essentially unchanged if we rather use the table in 
\url{https://dcc.ligo.org/LIGO-G1700086/public}, which is representative of early O2.  When using the O3a data, for the sensitivity we use the table given in \url{https://dcc.ligo.org/LIGO-G1802165/public}.
\label{foot:O2strain} }
As detection criterion, we set a threshold value ${\rm SNR}_t=8$. In practice, the searches employed by LIGO/Virgo rather use a threshold on the false alarm rate of the detector network to identify the detections. However, at least at the level of O2 data,
in terms of SNR this search is well approximated by the condition that the threshold in the inspiral SNR of the second-most-sensitive detector is greater than 8~\cite{Abadie:2010cf,Wysocki:2018mpo}.

Adding to the list of $\theta'$ parameters  in \eq{betadL} the  two BH source-frame masses $m_1$ and $m_2$ (with $m_1\geq m_2$)  and the inclination angle $\iota$,  \eq{betadLR} is replaced by
\bees\label{betam1m2iota}
\beta(H_0)&=&\int_0^{z_{\cal R}}dz \int d\Omega \,  p_0(z,\hatO)\int dm_1dm_2\, \tilde{p}_0(m_1,m_2)\, \nn\\
&&\times\int_{-1}^{1}\frac{d\cos\iota}{2}\, P_{\rm det}\[d_L(z,H_0),\hatO,\cos\iota,{\cal M}_1(z),{\cal M}_2(z)\]\,  ,
\ees
where  $ \tilde{p}_0(m_1,m_2)$ is the normalized distribution of BH  masses, and ${\cal M}_{1,2}(z)=(1+z)m_{1,2}$  are the corresponding detector-frame
masses.   Notice that, according to \eq{eq:SNR} (as well as to its generalization beyond the lowest-order inspiral), the SNR, and therefore the detection probability, is a function of the detector-frame masses, rather than of the source-frame masses, while the distribution probability for the BH masses is at first more naturally expressed in terms of the intrinsic, source-frame, masses. The distribution in the inclination angle has been taken flat in $\cos\iota$, and the factor $1/2$ normalizes it to one.  For definiteness, we illustrate the procedure for $\beta(H_0)$ but all the results below are trivially adapted to the case of $\beta(\Xi_0)$.

We begin by evaluating this expression for the homogeneous completion, when 
$p_0(z,\hatO)$ is  given by \eq{p0isocompldimensionless}. The result can be written as
\be\label{betabetacatbetacompl}
\beta^{\rm hom}(H_0)=\beta_{\rm cat}(H_0)+\beta_{\rm compl}(H_0)\, ,
\ee
where  $\beta_{\rm cat}(H_0)$ is the term that depends on the catalog, while
the term $\beta_{\rm compl}(H_0)$ reflects the homogeneous component added by the completion procedure. The former is given by
\be\label{betacat}
\beta_{\rm cat}(H_0)=
\(\frac{\int_0^{z_{\cal R}}dz'\, j(z')P_{\rm compl}(z')}{\int_0^{z_{\cal R}}dz'\, j(z')} \)\, 
\frac{\sum_{\alpha=1}^{N_{\rm cat}({\cal R})} w_{\alpha} \beta_{\alpha}(H_0)}
{\sum_{\alpha=1}^{N_{\rm cat}({\cal R})} w_{\alpha}}\, ,
\ee
where
\be\label{betaalpha}
\beta_{\alpha}(H_0)=\int dm_1dm_2\, \tilde{p}_0(m_1,m_2)\, \int_{-1}^{1}\frac{d\cos\iota}{2}\, 
P_{\rm det}\[d_L(z_{\alpha},H_0),\hatO_{\alpha},\cos\iota,{\cal M}_1(z_{\alpha}),{\cal M}_2(z_{\alpha})\]\, .
\ee
The term $\beta_{\rm compl}(H_0)$ is given
\bees
\beta_{\rm compl}(H_0)&=& \frac{1}{4\pi \int_0^{z_{\cal R}}dz'\, j(z')}\, 
\int_0^{z_{\cal R}} dz\,j(z)\int d\Omega  \, \[ 1-P_{\rm compl}(z,\hat{\Omega})\] \label{betacompl}\\
&&\times
\int dm_1dm_2\, \tilde{p}_0(m_1,m_2)\int_{-1}^{1}\frac{d\cos\iota}{2}\, P_{\rm det}\[d_L(z,H_0),\hatO,\cos\iota,{\cal M}_1(z),{\cal M}_2(z)\]\,  .\nn
\ees
In the semi-analytic approach, a significant simplification can be obtained approximating $P_{\rm det}$ by its average over the solid angle. This can be justified observing that the orientation of a given  galaxy with respect to the interferometer arms changes with the Earth rotation. Adding  to the list of $\theta'$ parameters the time of the day at which  the GW  signal from a source in a given galaxy arrives, with a uniform prior over 24h, is equivalent to performing, for each galaxy, an angular average over a great circle in the sky. To a first approximation,   especially when we are interested in  an ensemble of sources, this can be replaced by an angular average over the full sphere. To compute the angular average  of $P_{\rm det}$
we take the angular average of \eq{eq:SNR}. 
For interferometers, the angular average over $\hatO$ gives $\langle F_{+}^2(\hatO)\rangle=\langle F_{\times}^2(\hatO)\rangle=1/5$, so 
$\langle |Q(\cos\iota,\hatO)|^2\rangle =(1/5)g(\cos\iota)$,
where
\be\label{gcosiota}
g(\cos\iota)= \(\frac{1+\cos^2\iota}{2}\)^2 +\cos^2\iota\, , 
\ee
and \eq{eq:SNR} becomes
\be\label{eq:SNRaveOa}
{\rm SNR}^2(d_L,\cos\iota,{\cal M}_1,{\cal M}_2) = g(\cos\iota)\, 
\frac{(G{\cal M}_c)^{5/3} }{6  \pi^{4/3} c^3 d_L^2} 
\, I_{7/3}({\cal M}_{\rm tot})\, .
\ee
We set a threshold value
${\rm SNR}_t$ in the signal-to-noise ratio as a criterion for detection, and we
observe that the average of $g(\iota)$  over $\cos\iota$  is equal to $4/5$. This motivates
the definition of
$d_H({\cal M}_1,{\cal M}_2)$, given by
(see e.g. eq.~(7.182) of \cite{Maggiore:1900zz})
\be \label{SNRt2}
{\rm SNR}_t^2=\frac{4}{5}\, 
\frac{(G{\cal M}_c)^{5/3} }{6  \pi^{4/3} c^3 d_H^2({\cal M}_1,{\cal M}_2)} 
\, I_{7/3}({\cal M}_{\rm tot})\, .
\ee
The physical meaning of $d_H({\cal M}_1,{\cal M}_2)$ is that it is the distance at which a compact binary with (detector-frame) masses ${\cal M}_1,{\cal M}_2$ can be detected, in a single interferometer, with a threshold 
${\rm SNR}_t$ in the signal-to-noise ratio, after averaging over direction and inclination (this quantity is usually called the `range'  to the compact binary in question). 
In terms of  $d_H({\cal M}_1,{\cal M}_2)$, \eq{eq:SNRaveOa} can be rewritten more compactly as
\be\label{eq:SNRaveOb}
{\rm SNR}(d_L,\cos\iota,{\cal M}_1,{\cal M}_2) = {\rm SNR}_t\, \(\frac{g(\cos\iota)}{4/5}\)^{1/2}\, 
\frac{d_H({\cal M}_1,{\cal M}_2)}{d_L} 
\, .
\ee
For a generic  ${\rm SNR}_t$, with our detection criterion we have
\be\label{eq:pdetaveOc}
P_{\rm det}(d_L,\cos\iota,{\cal M}_1,{\cal M}_2)=
\theta\[{\rm SNR}(d_L,\cos\iota,{\cal M}_1,{\cal M}_2)-{\rm SNR}_t\]\, ,
\ee
which, using \eq{eq:SNRaveOb}, can be rewritten as
\be\label{eq:pdetaveO}
P_{\rm det}(d_L,\cos\iota,{\cal M}_1,{\cal M}_2)=
\theta\[ g(\cos\iota)-\frac{4}{5}\, \frac{d^2_L}{d^2_H({\cal M}_1,{\cal M}_2)}\]\, .
\ee
In the following we will
set ${\rm SNR}_t=8$ and therefore, from \eq{SNRt2},
\be\label{dHM1M2}
d_H({\cal M}_1,{\cal M}_2)= \frac{1}{20}\, \(\frac{5}{6}\)^{1/2}\, \frac{1}{\pi^{2/3}c^{3/2}}\,
(G {\cal M}_c)^{5/6} I_{7/3}^{1/2}({\cal M}_{\rm tot})\, .
\ee
Having dropped the dependence of $P_{\rm det}$ on $\hatO$,
\eq{betaalpha}  simplifies to
\be\label{betaalpha1}
\beta_{\alpha}(H_0)=\int dm_1dm_2\, \tilde{p}_0(m_1,m_2)\, \int_{0}^{1}d\cos\iota\, 
P_{\rm det}\[d_L(z_{\alpha},H_0),\cos\iota,{\cal M}_1(z_{\alpha}),{\cal M}_2(z_{\alpha})\]\, .
\ee
(Note that, since $g(\cos\iota)$ is even under $\cos\iota\ra -\cos\iota$, the integral over $\cos\iota$ from $-1$ to $1$ becomes twice the integral from 0 to 1.)  Similarly, dropping the dependence of $P_{\rm det}$ on $\hatO$, the term $\beta_{\rm compl}(H_0)$ in \eq{betacompl} simplifies to
\bees
\beta_{\rm compl}(H_0)&=&
 \frac{1}{\int_0^{z_{\cal R}}dz'\, j(z')}\, 
\int_0^{z_{\cal R}} dz\,j(z)  \, \[ 1-P_{\rm compl}(z)\]\label{betacomplaveO}\\
&&\times
\int dm_1dm_2\, \tilde{p}_0(m_1,m_2)\int_0^{1}d\cos\iota\, P_{\rm det}\[d_L(z,H_0),\cos\iota,{\cal M}_1(z),{\cal M}_2(z)\]\,  .\nn
\ees
It is convenient to express also the mass distribution function in terms of the detector-frame masses, defining
$\tilde{p}({\cal M}_1,{\cal M}_2;z)$ from 
\be 
dm_1dm_2\, \tilde{p}_0(m_1,m_2)=d{\cal M}_1d{\cal M}_2\, \tilde{p}({\cal M}_1,{\cal M}_2;z)\, ,
\ee
Using \eqs{gcosiota}{eq:pdetaveO}, the integration over  $\cos\iota$ in \eqs{betaalpha1}{betacomplaveO}
can be done explicitly. Observing that $g(\cos\iota)$ is a monotonically increasing function of $\cos\iota$, such that $g(0)=1/4$ and $g(1)=2$, we see that the theta function in \eq{eq:pdetaveO} is always satisfied if
$(4/5)d_L^2/d_H^2<1/4$, in which case the integral over $\cos\iota$ is equal to one, and is never satisfied if
$(4/5)d_L^2/d_H^2>2$. In the intermediate regime there is one value $\cos\iota_*$ such that
$g(\cos\iota_*)=(4/5)d_L^2/d_H^2$, and the integral is equal to $1-\cos\iota_*$. In this way we get
\be
\int_0^{1}d\cos\iota\, P_{\rm det}\[d_L,\cos\iota,{\cal M}_1(z),{\cal M}_2(z)\]=
f\left(\frac{4}{5}\, \frac{d^2_L}{d^2_H({\cal M}_1,{\cal M}_2)}
\right)\, ,
\ee
where
\be\label{fofx}
f(x)\equiv\left\{\begin{array}{lcc}
1 & \quad{\rm for }& \quad x\leq 1/4\\
1-\sqrt{2\sqrt{x+2}-3} &\quad {\rm for} &\quad 1/4\leq x\leq 2\\
0 &\quad {\rm for} &\quad x\geq 2\,.\\
\end{array}\right.
\ee
It is convenient to define the function
\be\label{defofB}
B(d_L;z)\equiv \int d{\cal M}_1d{\cal M}_2\, \tilde{p}({\cal M}_1,{\cal M}_2;z)\,  f\left(\frac{4}{5}\, \frac{d^2_L}{d^2_H({\cal M}_1,{\cal M}_2)}
\right)
\, .
\ee 
Note that $B(d_L;z)$ depends both on $d_L$, which is itself a function of $z$ and $H_0$, and also explicitly on $z$ because of the dependence on $z$ in  $\tilde{p}({\cal M}_1,{\cal M}_2;z)$. The latter enters only through the $z$ dependence of ${\cal M}_{\rm min}(z)=(1+z)m_{\rm min}$ and
${\cal M}_{\rm max}(z)=(1+z)m_{\rm max}$. 

Then
putting everything together, for homogeneous completion
\bees
\beta(H_0)&=&\(\frac{\int_0^{z_{\cal R}}dz'\, j(z')P_{\rm compl}(z')}{\int_0^{z_{\cal R}}dz'\, j(z')} \)\, 
\frac{\sum_{\alpha=1}^{N_{\rm cat}({\cal R})} w_{\alpha} B[d_L(z_{\alpha}, H_0);z_{\alpha} ]}
{\sum_{\alpha=1}^{N_{\rm cat}({\cal R})} w_{\alpha}}\nn\\
&&+ 
 \frac{1}{\int_0^{z_{\cal R}}dz'\, j(z')}\, 
\int_0^{z_{\cal R}} dz\,j(z)  \, \[ 1-P_{\rm compl}(z)\]\, B[d_L(z, H_0);z]\, .
\label{betaH0fulldetectionmodel}
\ees
Comparing with \eq{betaH0simpledetectionmodel} we see that the function $B[d_L(z,H_0);z]$ plays the role of a smooth cutoff function, that replaces the  theta function $\theta[z_{\rm max}(H_0;d_{\rm max})-z]$, with $z_{\max}(H_0)$ determined by 
$d_L(z_{\rm max};H_0)= d_{\rm max}$,
that was coming from the simplified detection model, and that would become quite arbitrary for a population of sources, for which there is no single notion of maximum detection distance. Similarly, \eq{betaXi0simpledetectionmodel} is replaced by
\bees
\beta(\Xi_0)&=&\(\frac{\int_0^{z_{\cal R}}dz'\, j(z')P_{\rm compl}(z')}{\int_0^{z_{\cal R}}dz'\, j(z')} \)\, 
\frac{\sum_{\alpha=1}^{N_{\rm cat}({\cal R})} w_{\alpha} B[\dgw(z_{\alpha}, \Xi_0);z_{\alpha}]}
{\sum_{\alpha=1}^{N_{\rm cat}({\cal R})} w_{\alpha}}\nn\\
&&+ 
 \frac{1}{\int_0^{z_{\cal R}}dz'\, j(z')}\, 
\int_0^{z_{\cal R}} dz\,j(z)  \, \[ 1-P_{\rm compl}(z)\]\, B[\dgw(z,\Xi_0);z]\, ,
\label{betaXi0fulldetectionmodel}
\ees
with $\dgw(z,\Xi_0)$ replacing $d_L(z,H_0)\equiv \dem(z,H_0)$ in the argument of $B(d_L)$ (and, as usual, we have not written explicitly the dependence of $\dgw$ on $H_0$ when we study $\Xi_0$).

It is straightforward to repeat  this computation for multiplicative completion. The results are
\be\label{betaH0multiplicativeFull}
\beta^{\rm multi}(H_0)=
\(\frac{\int_0^{z_{\cal R}}dz'\, j(z')P_{\rm compl}(z')}{\int_0^{z_{\cal R}}dz'\, j(z')} \)\, 
\frac{\sum_{\alpha=1}^{N_{\rm cat}({\cal R})}w_{\alpha}B[d_L(z_{\alpha}, H_0);z_{\alpha} ]/P_{\alpha}}{\sum_{\alpha=1}^{N_{\rm cat}({\cal R})}w_{\alpha}}\, ,
\ee
and
\be\label{betaXi0multiplicativeFull}
\beta^{\rm multi}(\Xi_0)=
\(\frac{\int_0^{z_{\cal R}}dz'\, j(z')P_{\rm compl}(z')}{\int_0^{z_{\cal R}}dz'\, j(z')} \)\, 
\frac{\sum_{\alpha=1}^{N_{\rm cat}({\cal R})} 
w_{\alpha}B[\dgw(z_{\alpha}, \Xi_0);z_{\alpha}]/P_{\alpha}}{\sum_{\alpha=1}^{N_{\rm cat}({\cal R})}w_{\alpha}}\, ,
\ee
which generalize \eq{betamultiplicative2} and the corresponding result for $\beta(\Xi_0)$,  again with $B[d_L(z, H_0);z]$ or $B[\dgw(z, \Xi_0);z]$, respectively, replacing the theta functions. The same therefore happens for the mixture of homogeneous and multiplicative completion discussed in the text, and for the extention to $\lambda\neq 1$ discussed in \eqst{niH0lambda}{nilambdalast}.

Observe that the above results, that have been obtained using just the expression for the SNR with the lowest-order inspiral waveform, can be generalized to more accurate computations of the waveform. Indeed, as long as we work within the so-called restricted post-Newtonian (PN) approximation, where one considers corrections to the phase of the waveform of a coalescing binaries, but keeps only the lowest order in the amplitude, the dependence of the SNR on $\cos\iota$ remains the one that we have used, and therefore the integration over $\cos\iota$ goes through in the same way (in contrast, going beyond the restricted  PN approximation the dependence on $\cos\iota$ changes, see sect.~5.6.4 of \cite{Maggiore:1900zz}). Then, in the above results we can simply replace the horizon distance $d_H({\cal M}_1,{\cal M}_2)$, that in our lowest-order inspiral-only approximation is given by \eq{dHM1M2}, with the expression obtained from the waveform used. If the latter includes the spins of the compact bodies (or other variables such as eccentricity. etc.), these variables must then be added to the list of integration variables in \eq{defofB}.

We now evaluate these expression for the choice  (\ref{brokenpowerlaw})  of mass distribution.
The function $B(d_L;z)$ can be computed by direct numerical evaluation of the double integral in
\eq{defofB}.\footnote{The computation can be performed very efficiently using importance sampling with respect to the distribution $\tilde{p}({\cal M}_1,{\cal M}_2;z)$. The need of this numerical step, together with the need of  evaluating numerically  $\beta(H_0)$ in \eq{betaH0fulldetectionmodel} or \eq{betaH0multiplicativeFull}, once obtained numerically $B(d_L;z)$, is the reason why we refer to this method as `semi-analytic', rather than just `analytic'.} The left panel of Fig.~\ref{fig:B_vs_dL}, on page~\pageref{fig:B_vs_dL} of the main text, shows the result for $B(d_L)\equiv B[d_L;z(d_L,H_0)]$, using the O2 strain sensitivity to compute the integral 
$I_{7/3}({\cal M}_{\rm tot})$ that enters in 
\eq{dHM1M2},   for a population of BBHs, using the reference values (\ref{mminbmax}), (\ref{gamma1gamma2})
in the mass distribution (\ref{brokenpowerlaw}), and 
setting 
$H_0=70\,\, {\rm km}\, {\rm s}^{-1}\, {\rm Mpc}^{-1}$. First of all we see that, for $d_L$ smaller than a value
$d_0$,  $B(d_L)=1$. This is easily understood: with the mass distribution
 that we are using, the BBH which has the lowest SNR, for a given distance, is the lightest system in our distribution,  $m_1=m_2=2.2\msun$, when it has the most unfavorable orientation, $\cos\iota =0$, so that 
$g(\cos\iota)$ attains its its minimum value of $1/4$. When $d_L$  is so small that even this system goes above the detection threshold, all the BBHs in our distribution are above threshold, so the  function $f(x)$ in
\eq{fofx} is always equal to one, and then also 
$B(d_L)=1$, since the mass distribution is normalized to one. From \eq{eq:pdetaveO}, the value of $d_0$ is therefore given by
\be
d_0=\frac{\sqrt{5}}{4}\, d_H(2.2\msun,2.2\msun)\, ,
\ee
which, using the O2 strain sensitivity to compute $ d_H(m_1,m_2)$, gives
$d_0\simeq 71.69\, {\rm Mpc}$.
Above this value, we find that the numerical result is very well fitted by an exponential with a linear and a quadratic term in $(d_L-d_0)$. Therefore
\be\label{BdL1}
B(d_L)=1\, ,\qquad (d_L\leq d_0)\, ,
\ee
and
\be\label{BdL2}
B(d_L)
\simeq e^{ -a_0 \frac{d_L-d_0}{d_0}-a_1  \(\frac{d_L-d_0}{d_0}\)^2 }\, ,\qquad (d_L> d_0)\, .
\ee
Using  $\gamma_1 =1.05$ in the mass distribution (\ref{brokenpowerlaw}) for BBHs, together with the other reference values given in \eqs{mminbmax}{gamma1gamma2}, and the best O2 sensitivity curve (see footnote~\ref{foot:O2strain}) the fit gives $a_0\simeq 0.215$, $a_1\simeq 0.004$. 
Similar considerations hold for the  right panel of Fig.~\ref{fig:B_vs_dL}, that  shows the corresponding result for a BNS population with a distribution for the source-frame masses flat in the range $[1-3] \msun$. In this case,  the fit gives $a_0\simeq 0.106$, $a_1\simeq 0.144$, and the corresponding value of $d_0$ is $d_0=(\sqrt{5}/4) d_H(\msun,\msun)\simeq 37.2\, {\rm Mpc}$.
Fig.~\ref{fig:B_vs_dL} nicely shows how the naive theta function (\ref{pdetTheta}) is modified when we deal with a population of BBHs with an extended  mass distribution, rather than a monochromatic  population.\footnote{Eventually, $B(d_L)$ becomes identically zero when even the BBH of our distribution  
with the best combination of total mass and chirp mass, and with optimal orientation, goes below threshold. Notice that, eventually, increasing the total mass, the cutoff frequency decreases, so that the range is not a monotonic function of the masses.}

\begin{figure}[t]
\centering
\includegraphics[width=0.48\textwidth]{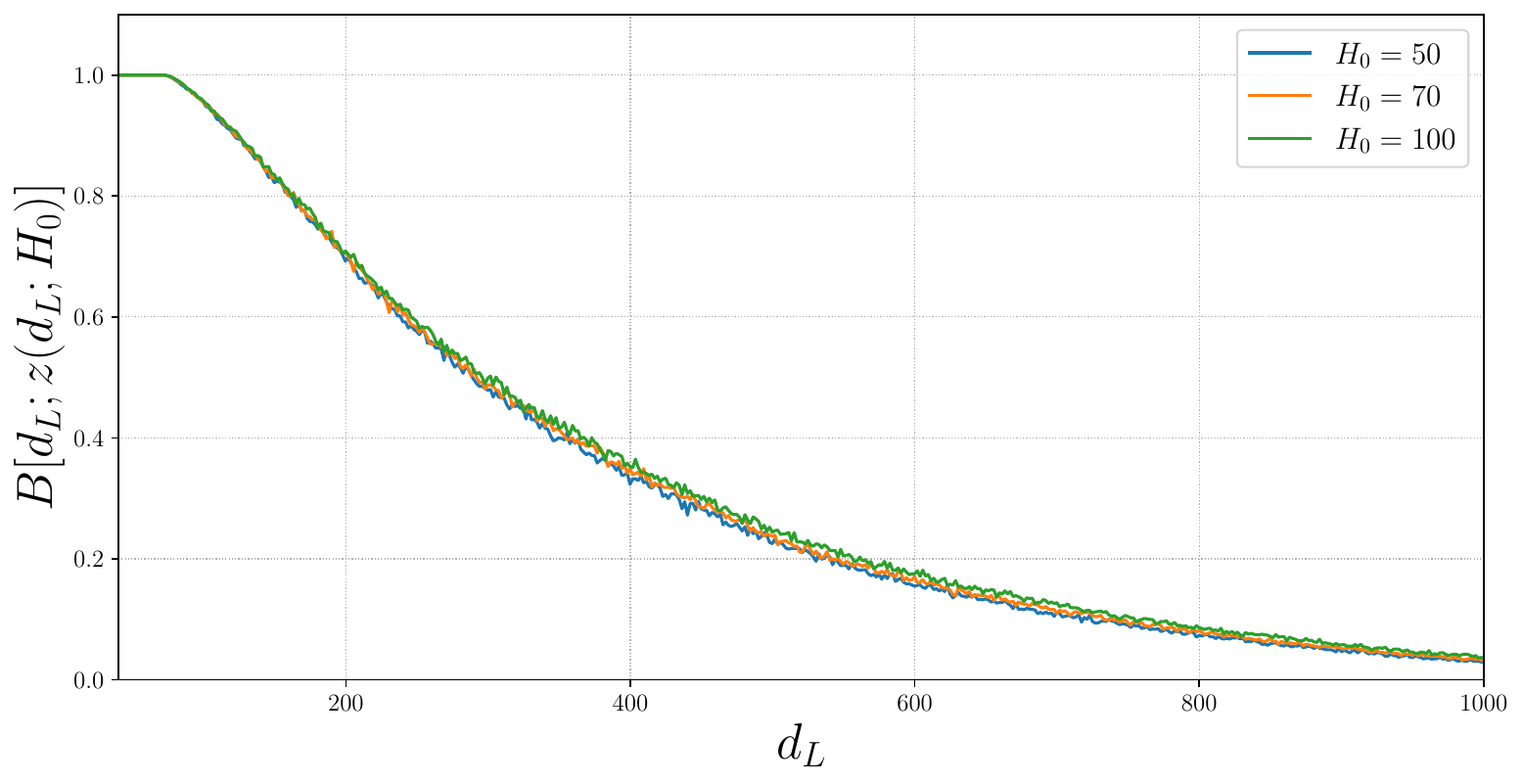}
\includegraphics[width=0.48\textwidth]{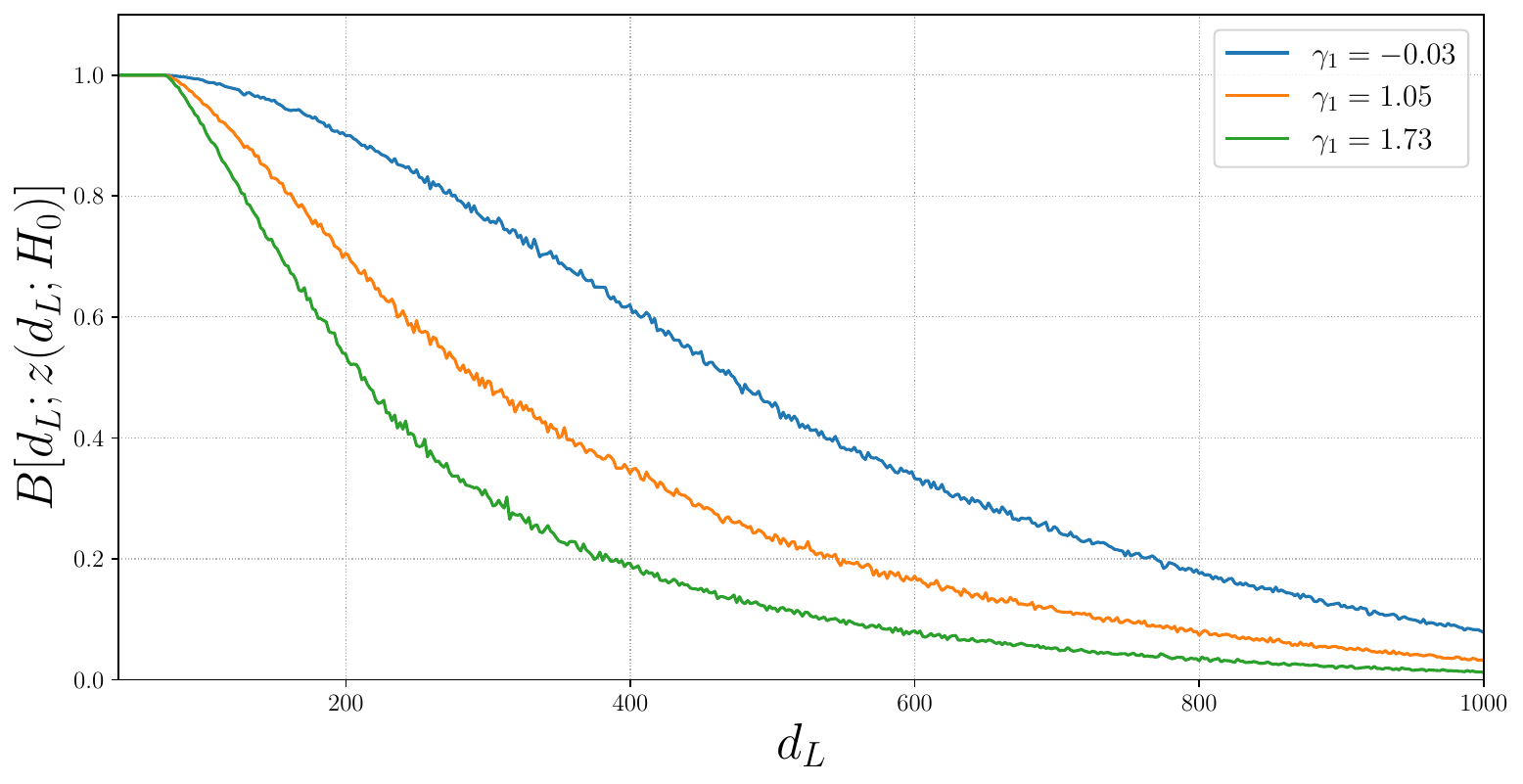}
\caption{Left panel: the function $B[d_L;z(d_L,H_0)]$, as a function of $d_L$, for three different values of $H_0$  (in units of ${\rm km}\, {\rm s}^{-1}\, {\rm Mpc}^{-1}$), and 
the value given in \eqs{mminbmax}{gamma1gamma2} for the other parameters. The three curves  are basically indistinguishable.
Right panel: the result for different values of $\gamma_1$, with the other parameters fixed.}
\label{fig:B_vs_dL_severalH0}
\end{figure}

Notice that,  in principle, $B$ is a function of two independent variables, that can be taken to be $d_L$ and $z$; alternatively, using the relation $d_L=d_L(H_0,z)$, it can be taken to be a function of $z$ and $H_0$, or of  $d_L$ and $H_0$.  One might then be worried that we should compute it numerically on a two-dimensional grid, for instance in  $d_L$ and $H_0$ (while 
Fig.~\ref{fig:B_vs_dL} was obtained computing it as a function of $d_L$, at fixed $H_0$). Fortunately, this is not the case and, when using  $d_L$ and $H_0$ as independent variables, $B(d_L,H_0)$ is basically independent of $H_0$. This is shown in the upper left panel of Fig.~\ref{fig:B_vs_dL_severalH0} where, even taking a very broad range of values of $H_0$, 
the three curves shown are basically indistinguishable.\footnote{This is due to the fact that, when 
deriving  $\tilde{p}({\cal M}_1,{\cal M}_2;z)$ from  $\tilde{p}_0(m_1,m_2)$, 
using $d_L$ and $z$ as independent variables, the dependence of $B(d_L;z)$ on $z$ only enters  through the fact that  $z$ gives a small correction to  the minimum and maximum values of the mass distribution, through the relations 
${\cal M}_{\rm min}(z)=(1+z) m_{\rm min}$ and ${\cal M}_{\rm max}(z)=(1+z) m_{\rm max}$,
and to the change of slope, that in terms of the variable  ${\cal M}_1$ takes place at 
${\cal M}_{\rm break}(z)=(1+z) m_{\rm break}$, and this gives only a  very minor effect when we integrate a function over this distribution.}
Thus $B(d_L,H_0)$ can be extremely well approximated by a function of $d_L$ only, with no explicit dependence on $H_0$, and given by \eqs{BdL1}{BdL2}. When we insert it into \eqst{betaH0fulldetectionmodel}{betaXi0multiplicativeFull} we need to write it as a function of $z$, to compare with the redshifts of the galaxies, and this is done writing $d_L=d_L(z,H_0)$, so 
in those equations $B[d_L(z, H_0);z]$ can be simply written as $B[d_L(z, H_0)]$, with no further explicit dependence on $z$, and with $B(d_L)$ approximated by \eqs{BdL1}{BdL2}. Then
\eqst{betaH0fulldetectionmodel}{betaXi0multiplicativeFull} simplify to 
\eqs{betaH0homfinal_text}{betaH0multfinal_text} of the main text.

Once computed $B(d_L)$, the corresponding numerical results for  $\beta(H_0)$ or $\beta(\Xi_0)$ can be obtained immediately from a numerical evaluation of \eq{betaH0fulldetectionmodel} or \eq{betaH0multiplicativeFull}.  These results  then allow us to compute a first approximation to $\beta(H_0)$ or $\beta(\Xi_0)$, including the mass distribution, and can be used to estimate the dependence of the result on the choice of parameters, such as $\gamma_1$,
without the need of performing each time an MC. Actually,  we found that the limit $P_{\rm compl}(z)=0$ of \eq{betaH0homfinal_text}, given by the simple expression
\be\label{betafitPcomplzero}
\left.\beta^{\rm hom}(H_0)\right|_{P_{\rm compl}=0}=\frac{\int_0^{z_{\cal R}} dz\,j(z)  \,\, B[d_L(z, H_0)]}{\int_0^{z_{\cal R}}dz'\, j(z')}\, ,
\ee
is already accurate to better than $5\%$ compared to a MC computation that also employs only the inspiral waveform (and without selection effect on the completeness of the events) over the most relevant range of values of $H_0$, and is much faster to compute.
However, 
when selection effects are important, an accurate evaluation of $\beta(H_0)$ requires a MC computation. Furthermore, the use of full inspiral-merger-waveform is also essential, particularly at the level of O3 data.
Therefore, while the above expressions give some useful insight, for instance showing how the theta function is replaced by a smoothed cutoff function $B(d_L)$ and how the latter is related to the mass distribution, eventually accurate numerical work still requires an MC evaluation, and all the resukts that we present are obtained with such a MC.

\section{Monte Carlo evaluation of $\beta(H_0)$ and $\beta(\Xi_0)$}\label{sect:ourMC}

In this appendix we give a short description of our MC evaluation of $\beta(H_0)$ and $\beta(\Xi_0)$, which is part of
our code, publicly available at  \url{https://github.com/CosmoStatGW/DarkSirensStat}.

For each value of $H_0$ (or $\Xi_0$) we generate a random sample of merger events by drawing the two source frame masses, inclination angle and detection time (which is used to evaluate the orientation in space of the arms of the interferometers), as well as the position in redshift space. For the latter, we split the computation with our completed galaxy catalog prior into two separate MC integrations as follows. For the part of the prior that depends directly on the galaxy positions in the catalog,
 we sample galaxies from the luminosity weighted catalog by drawing from the discrete distribution $w^{\text{gal}}_\alpha/\sum_\alpha w^{\text{gal}}_\alpha $ where
$w^{\text{gal}}_\alpha= w_\alpha (1+z_\alpha)^{\lambda-1} $, and each $z_\alpha$ in turn has been sampled from the galaxy posterior of the galaxy denoted by $\alpha$. For the homogenous part that is added to complete the catalog, we importance sample the redshift from $dV_c/dz (1+z)^{\lambda-1} / \mathcal{N}(z_\text{max})$ up to the maximal detector visibility $z_\text{max}$ for the current cosmological parameters at optimal detector-frame masses, where 
 $\mathcal{N}(z_\text{max}) = \int_0^{z_\text{max}} (dV_c/dz) (1+z)^{\lambda-1} dz$. 
We then discard all generated GW events that do not satisfy the completeness threshold cut. For the remaining events we evaluate the corresponding completeness factors that appear in our prior, and store them as weights $W_i$, where the index $i$ labels the GW event. We then compute, for each event, the theoretical SNR with the PyCBC software\footnote{\url{https://pycbc.org/}}
and the IMRPhenomXAS waveform model~\cite{Pratten:2020fqn}, and multiply the weights $W_i$  by the probability of detection, assuming a gaussian distribution for the detection SNR, and a detection threshold of 8 in the less loud detector among LIGO Hanford and LIGO Livingston (with their position on Earth at observation time taken into account).  
The $W_i$ for the galaxy term and for the completion term are then multiplied by the factors
$\sum_\alpha w^{\text{gal}}_\alpha$
and $\mathcal{N}(z_\text{max})$, respectively, that compensate for the same factors  implicitly present at the denominator, due to the fact that the MC samples from normalized distributions.\footnote{The term $\mathcal{N}(z_\text{max})$ is particularly interesting since it depends on $H_0$ (or, respectively, $\Xi_0$).  The analytical extraction of this term   greatly improves the convergence of the MC estimator and, apart from a constant normalization, this term is the same as  the analytical formula for $\beta$ in  \eq{Chenbeta},  generalized to the case of $\lambda \neq 1$. The actual MC integration of the homogenous part is therefore just computing corrections to the expression for $\beta$ given by  \eq{Chenbeta}. 
} 
We also need to include factors reflecting the two sample sizes used for the galaxy term and for the completion term, respectively, which we adapt to an estimate of their expected relative sizes to reduce variance of the estimator.  
Finally, the value of $\beta$ is proportional to the sum of the resulting weights $W_i$. This results in an unbiased estimator of $\beta$ at a single value of $H_0$ (or $\Xi_0$). We repeat this for $\mathcal{O}(1000)$ values of $H_0$ (or $\Xi_0$), resulting in total to the use of  $\mathcal{O}(10^8)$ sampled events, and we finally filter the resulting function to create a smoother result.

\section{Dependence of the results on different choices}\label{sect:dependence_results}

In this appendix we discuss how our results change when making different choices with respect to our baseline model that, as discussed in sect.~\ref{sect:results}, is defined by the fact of using 
mask completion to quantify the quasi-local completeness level of the catalog, multiplicative completion to fill in  the missing galaxies, K-band luminosity weighting, and a completeness threshold 
$P_{\rm th}=0.7$ for the selection of the GW events.

\subsection{Dependence on the completeness threshold}\label{sect:dependencePth}

We begin by studying the effect of the completeness threshold $P_{\rm th}$. This choice is determined by a balance between two different considerations.  Lowering $P_{\rm th}$ we include more events, and we therefore have a larger statistical sample. On the other hand, when we correlate a GW event with a galaxy catalog, if the GW event falls in a region where the galaxy catalog is very incomplete, the result is eventually dominated by the completion procedure. As we have discussed, see in particular the discussion on page~\pageref{foot:compens} and footnote~\ref{foot:compens}, 
as long as we know the `exact' functions $\beta(H_0)$ appropriate to the completion procedure used, events that falls into very incomplete region should give, on average (i.e, after averaging over many such low-completeness events), a flat contribution to the posterior. However, this comes from a  delicate compensation of potentially large effects between the numerator and the denominator of the posterior (\ref{pDinolambda}), and would require a very tight control over the function $\beta(H_0)$ [or $\beta(\Xi_0)$], including the precise BBH mass function, details of the detector sensitivities at time of detection, etc. Any error in the computation of $\beta(H_0)$ or $\beta(\Xi_0)$ would then systematically bias the contribution from events coming from incomplete regions.

To minimize such systematic biases, a conservative choice is to increase the value of the 
threshold $P_{\rm th}$, which makes the results less sensitive to the completion procedure, at the price of reducing the statistical sample. The determination of the optimal range of values of $P_{\rm th}$ cannot be made a priori, and requires some experimentation. Here we can be guided by the fact that, even if, in sect.~\ref{sect:H0results}, for $H_0$ we use a very broad prior, $H_0\in [30, 140] \,\, {\rm km}\, {\rm s}^{-1}\, {\rm Mpc}^{-1}$, we actually know from cosmological observations that the prior range could be significantly restricted. For instance, any peak at $H_0\sim 40 \,\, {\rm km}\, {\rm s}^{-1}\, {\rm Mpc}^{-1}$  should clearly be spurious. If such a peak happens to disappear increasing $P_{\rm th}$, it is natural to suspect that this spurious peak is due to a partially incorrect completion procedure. We will use this information to obtain some understanding on the optimal choice of $P_{\rm th}$, and we will  then use the same  value of $P_{\rm th}$ also in the study of $\Xi_0$, for which the prior information is less stringent.

\begin{figure}[t]
\centering
\includegraphics[width=0.48\textwidth]{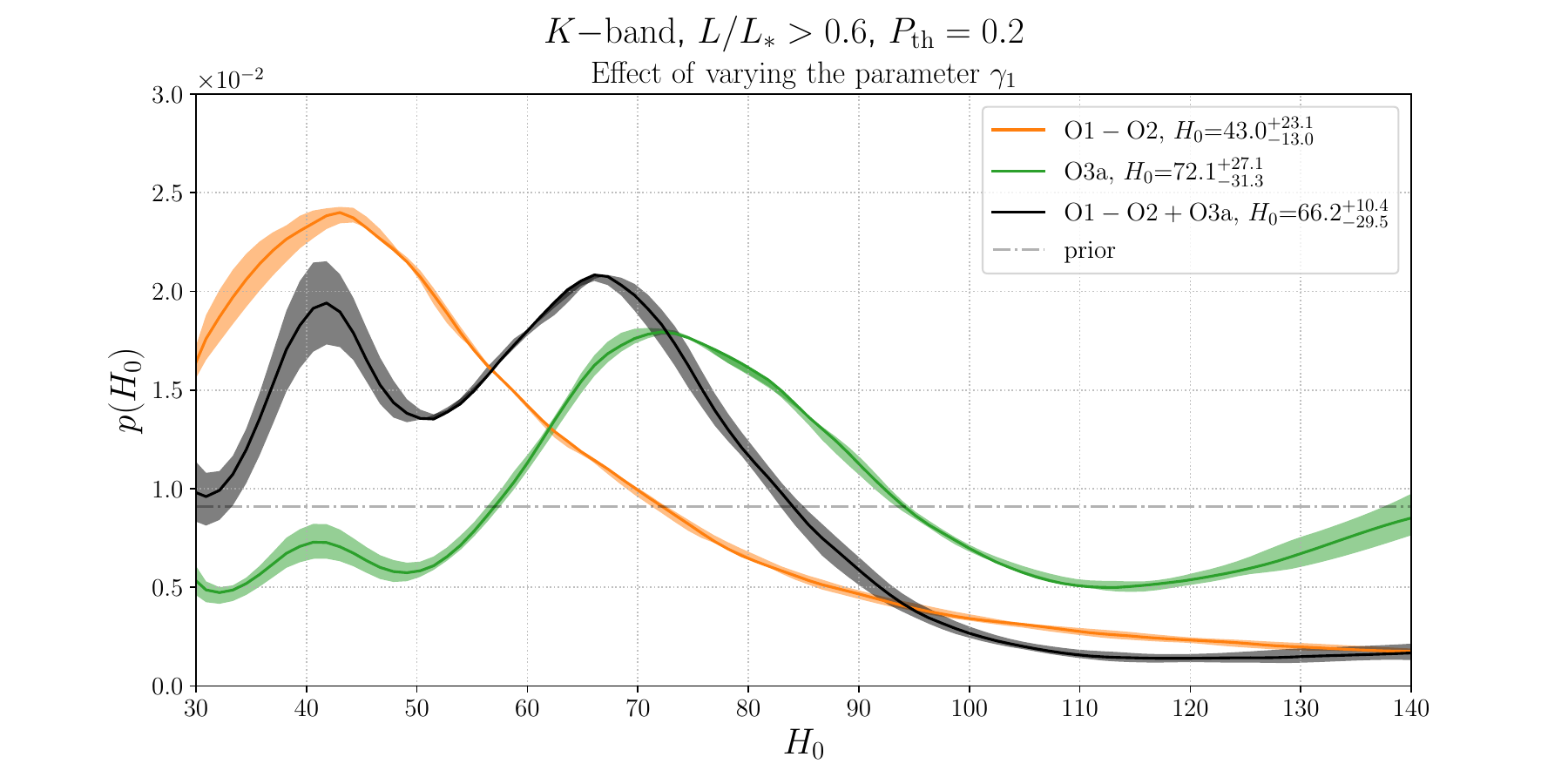}
\includegraphics[width=0.48\textwidth]{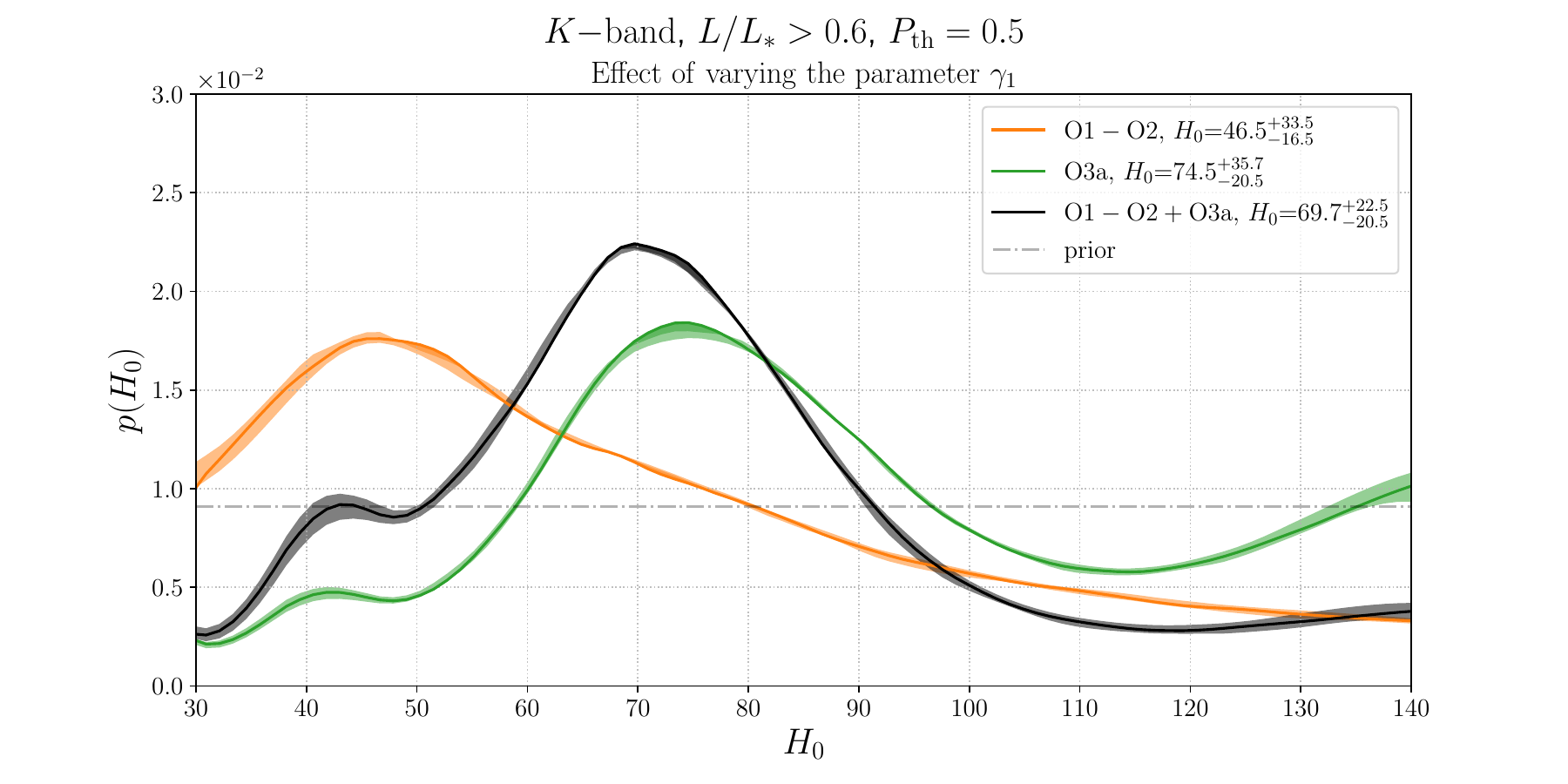}
\caption{The same as Fig.~\ref{fig:varyingGamma_H0} of the main text, using however $P_{\rm th}=0.2$ (left) and $P_{\rm th}=0.5$ (right). We use mask completion, multiplicative completion, and K-band luminosity weighting.}
\label{fig:H0_K_pth0205}
\end{figure}

Fig.~\ref{fig:H0_K_pth0205} shows the posterior of $H_0$ for $P_{\rm th}=0.2$ (left) and $P_{\rm th}=0.5$ (right), while the result for $P_{\rm th}=0.7$ was shown in Fig.~\ref{fig:varyingGamma_H0} on page \pageref{fig:varyingGamma_H0} of the main text. Comparing these three plots, we note first of all that the dependence on $P_{\rm th}$ is very significant. We  see that, in the total O1+O2+O3a posterior, for $P_{\rm th}=0.2$ there are two comparable peaks, one  at   $H_0\sim 40 \,\, {\rm km}\, {\rm s}^{-1}\, {\rm Mpc}^{-1}$,  and one at  $H_0\sim 70 \,\, {\rm km}\, {\rm s}^{-1}\, {\rm Mpc}^{-1}$, where we know that the correct result should be. Both the O1+O2 and the O3a posterior separately show a peak at   $H_0\sim 40 \,\, {\rm km}\, {\rm s}^{-1}\, {\rm Mpc}^{-1}$, suggesting a systematic effect rather than a statistical fluctuations.
For $P_{\rm th}=0.5$ the peak at low $H_0$  has just become a minor bump and, as we see from Fig.~\ref{fig:varyingGamma_H0}, this secondary peak basically disappears when we further raise the completeness threshold to $P_{\rm th}=0.7$.  
This indicates that the peak at $H_0\sim 40 \,\, {\rm km}\, {\rm s}^{-1}\, {\rm Mpc}^{-1}$  is an artifact of the completion procedure. The value $P_{\rm th}=0.7$ appears sufficient to basically get rid of this artifact, while leaving still enough statistics to have an informative, although broad, posterior. For this reason, we have adopted $P_{\rm th}=0.7$ as our reference choice, in the main text. In the following, we will then restrict to $P_{\rm th}=0.7$, both for $H_0$ and $\Xi_0$. With a larger statistical sample, as one could obtain in the near future from further LVC data, one could afford to set an even larger threshold, which likely will be advisable for obtaining accurate unbiased results.

\subsection{Comparison of B band and K band luminosity weighting}

\begin{figure}[t]
\centering
\includegraphics[width=0.48\textwidth]{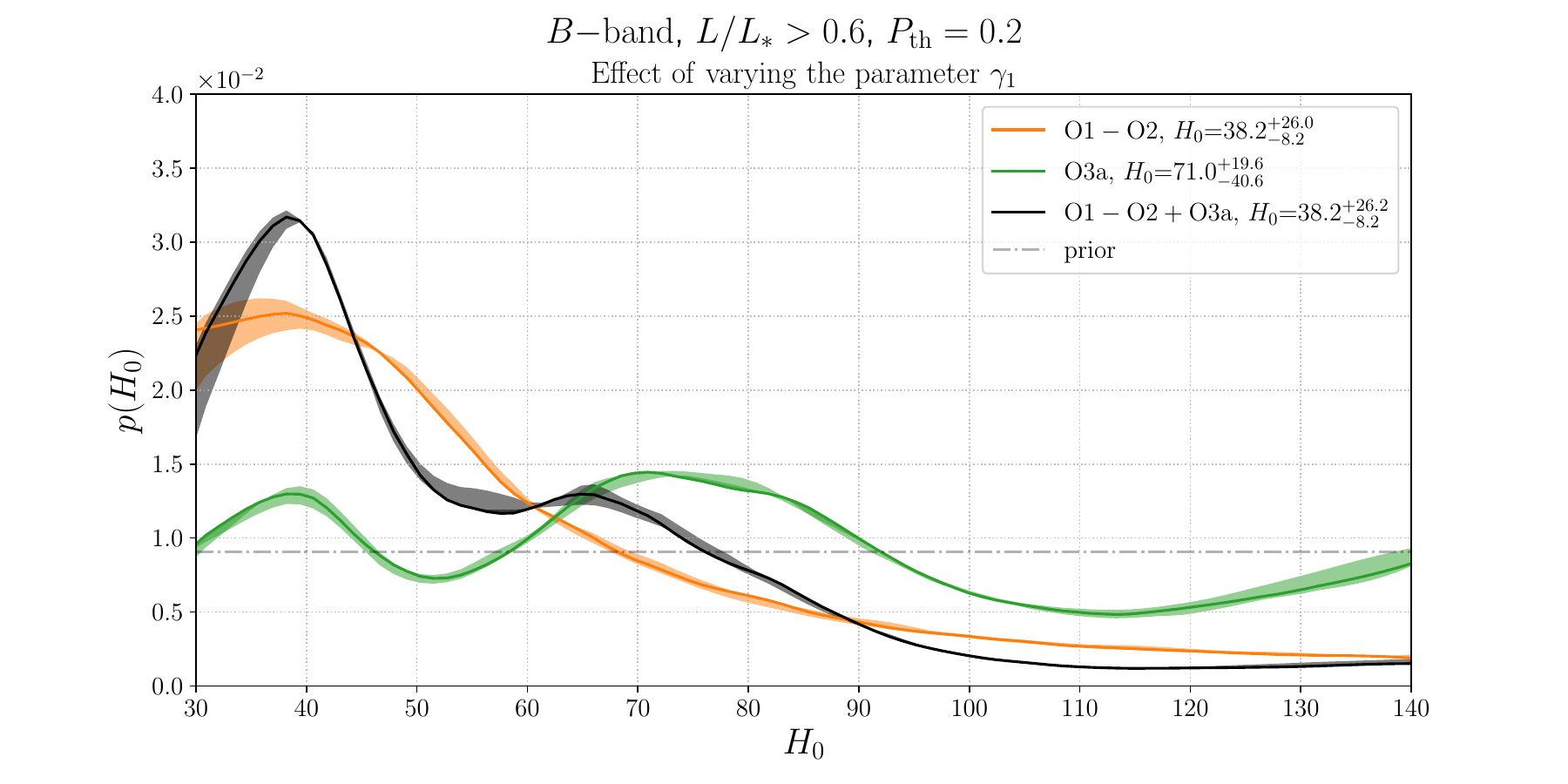}
\includegraphics[width=0.48\textwidth]{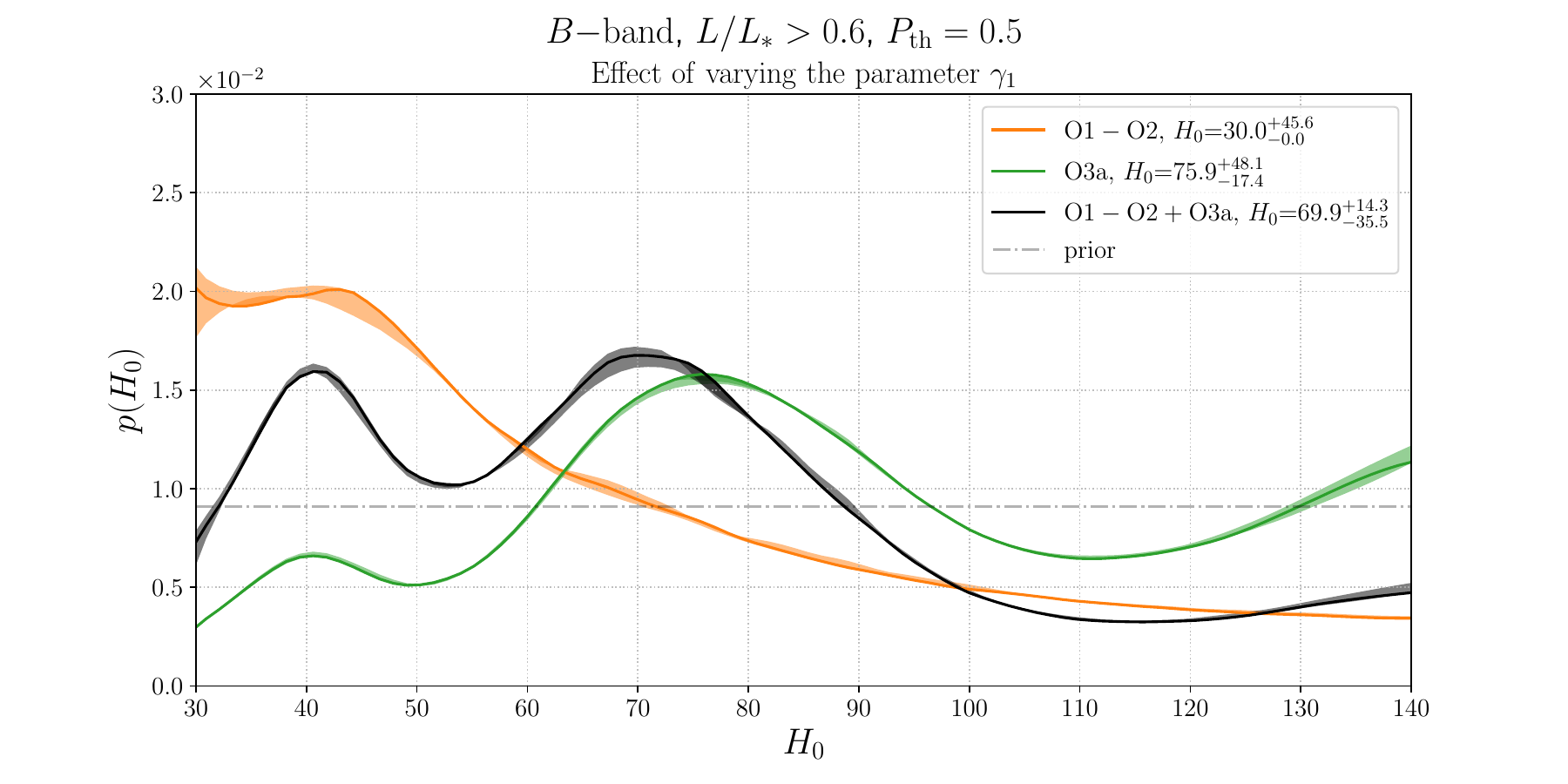}
\includegraphics[width=0.48\textwidth]{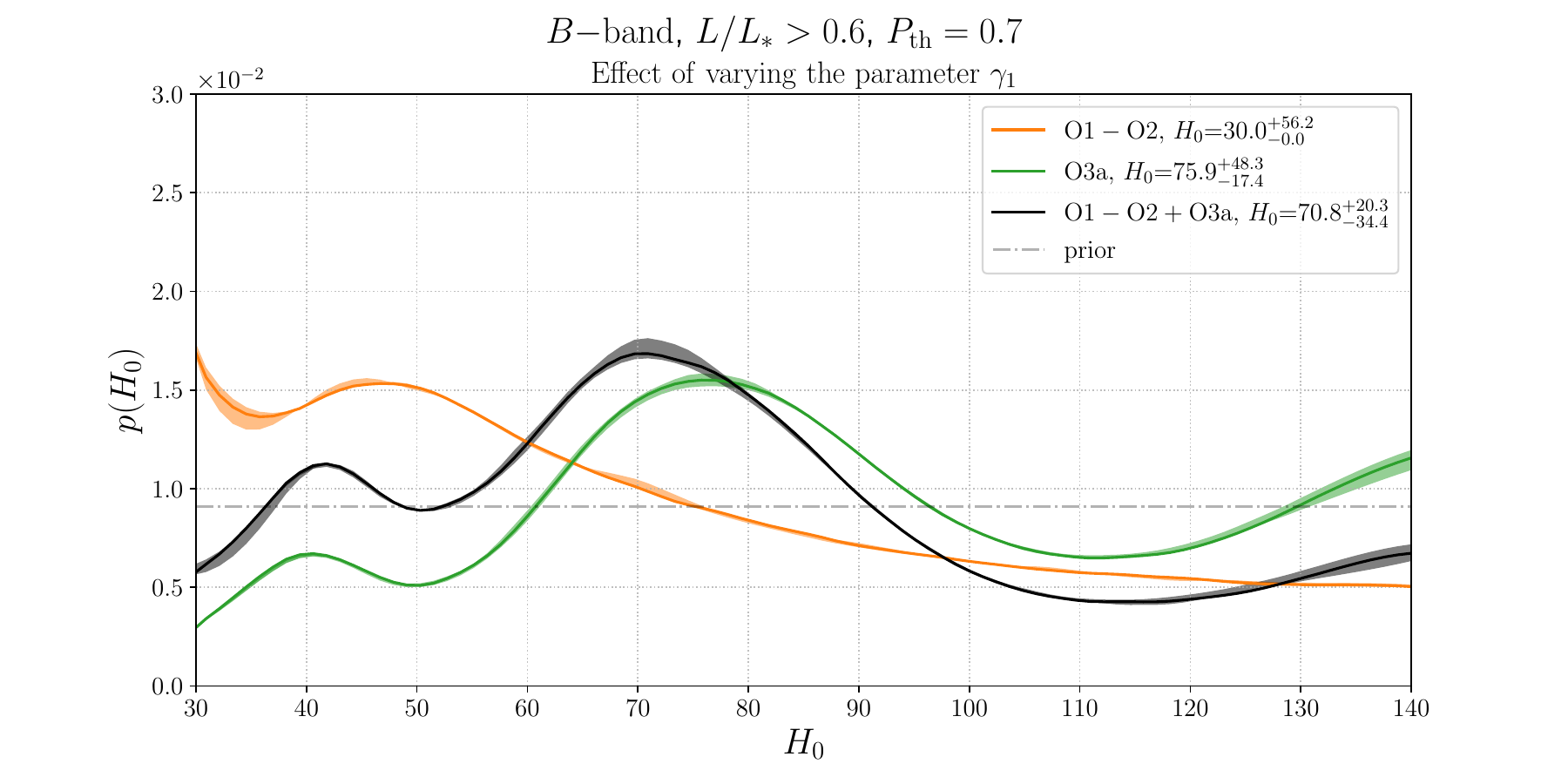}
\caption{The posterior on $H_0$ in B~band for $P_{\rm th}=0.2$ (upper left),  $P_{\rm th}=0.5$ (upper right) and  $P_{\rm th}=0.7$ (lower panel), using mask completeness and  multiplicative completion.}
\label{fig:H0_B_pth020507}
\end{figure}

Fig.~\ref{fig:H0_B_pth020507} shows the analogous plot for B band,  using $P_{\rm th}=0.2$ (upper left),  $P_{\rm th}=0.5$ (upper right) and  $P_{\rm th}=0.7$ (lower panel).
We see that the situation for the dependence on  $P_{\rm th}$ is similar to that in K band, with a spurious peak at $H_0\sim 40 \,\, {\rm km}\, {\rm s}^{-1}\, {\rm Mpc}^{-1}$ that significantly decreases  as we increase $P_{\rm th}$. However, in B~band, for the same value of $P_{\rm th}$, the spurious peak is  stronger than in K~band. For 
$P_{\rm th}=0.2$ it is the dominant peak, and at
$P_{\rm th}=0.5$ the peak at
low $H_0$ is still basically as large as that at $H_0\sim 70 \,\, {\rm km}\, {\rm s}^{-1}\, {\rm Mpc}^{-1}$ (while, in K band, it was just a small bump), and it is still well visible for $P_{\rm th}=0.7$, when in K~band it had basically disappeared. Thus,  to mitigate this bias, in B~band one should set  a higher value of $P_{\rm th}$, compared to K~band (for instance, the plot with $P_{\rm th}=0.2$ in K~band looks quite similar to that with
$P_{\rm th}=0.5$ in B~band, and that  with $P_{\rm th}=0.5$ in K~band looks quite similar to that with
$P_{\rm th}=0.7$ in B~band). For this reason, we use K~band in our baseline model. Still, once a relatively high threshold such as $P_{\rm th}=0.7$ has been chosen, the results for the maximum a posteriori value and the corresponding highest density interval in B~band become already  comparable to those in K~band. In particular, for $P_{\rm th}=0.7$, in B band, we get 
\be
H_0=70.8^{+20.3}_{-34.4} \,\qquad\qquad (\mbox{\rm B band})\, .
\ee
which is quite consistent with the K~band result (\ref{H0KbandPth07}), although with a somewhat larger error.

\subsection{Comparison of different completeness measures and completion procedures}

We finally compare the results using our two ways of quantifying the quasi-completeness of a catalog, namely mask completeness and cone completeness, and our two ways of adding the missing galaxy, homogeneous and multiplicative completion (as explained in the text, what we call simply multiplicative completion in the result section,  is actually  the interpolation between  multiplicative and homogeneous completion).

\begin{figure}[t]
\centering
\includegraphics[width=0.48\textwidth]{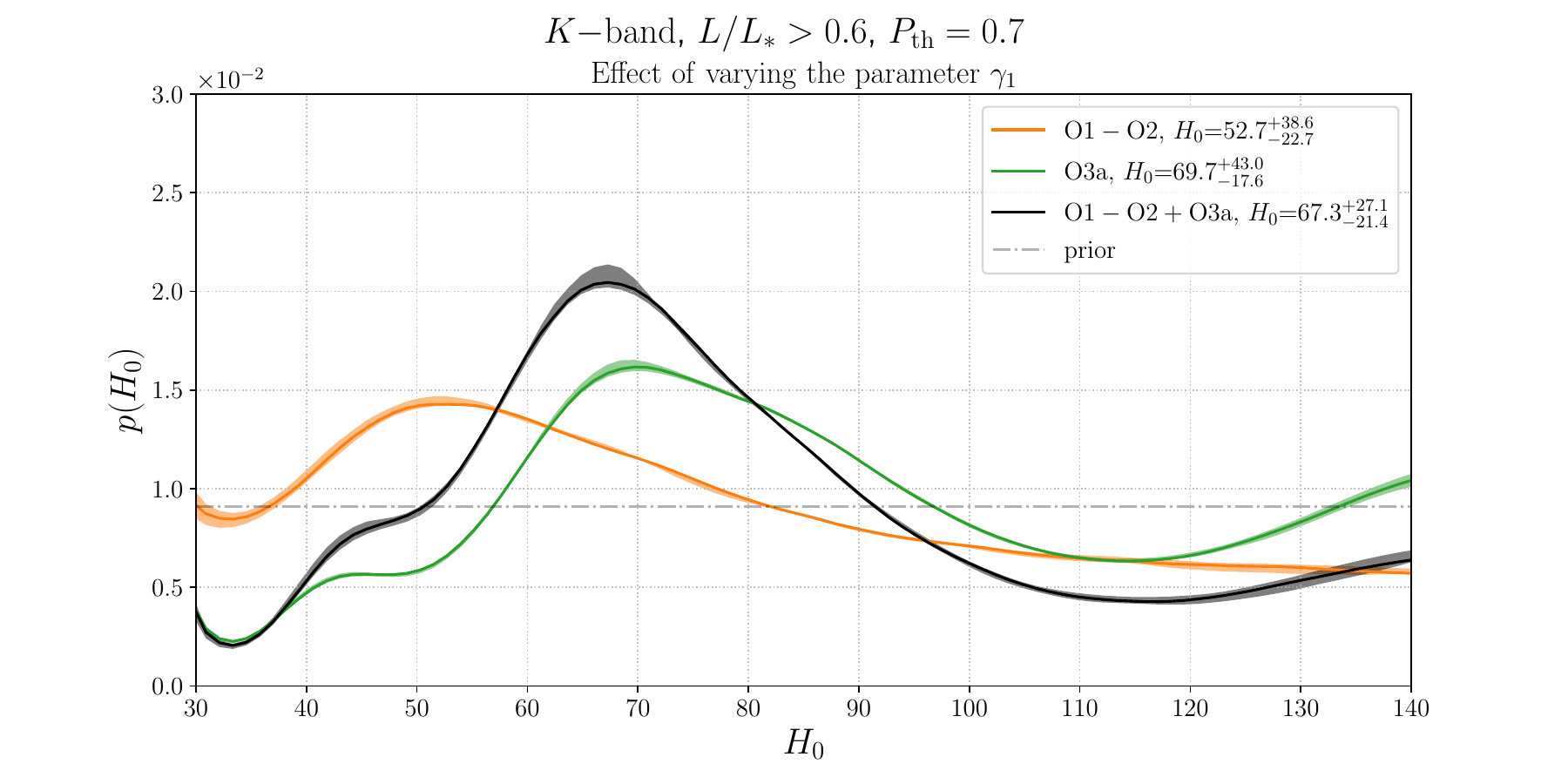}
\includegraphics[width=0.48\textwidth]{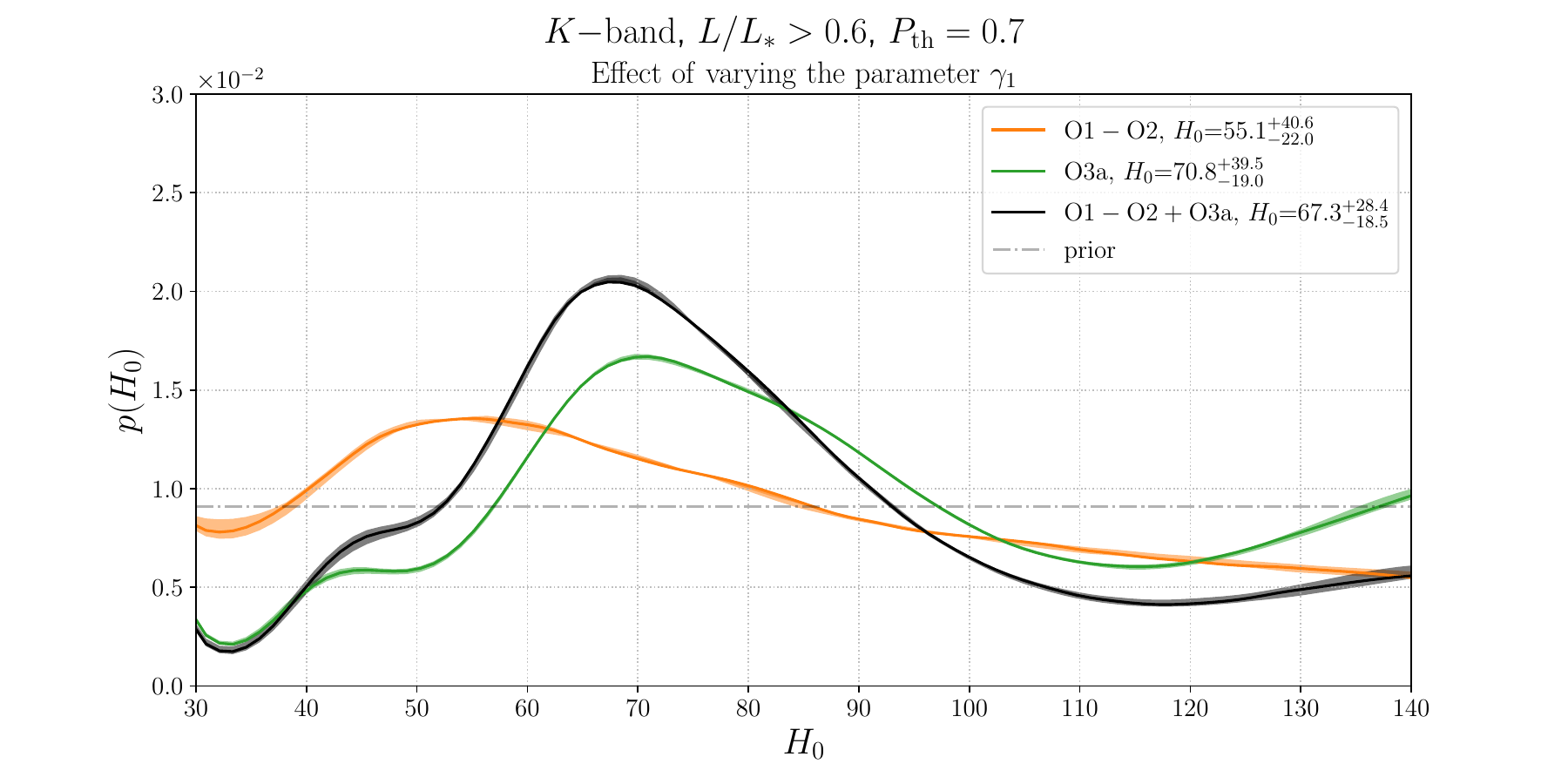}
\caption{The posterior for $H_0$ using cone completeness and homogeneous completion (left) and   cone completeness and multiplicative completion (right),  while still keeping our default choices,  K~band and $P_{\rm th}=0.7$.}
\label{fig:H0_cone}
\end{figure}

The results using  cone completeness are shown in Fig.~\ref{fig:H0_cone}, for  homogeneous completion (left) and   multiplicative completion (right), where the shaded bands represent the effect of varying $\gamma_1$. The corresponding values for the maximum a posteriori value and the  highest density interval, using the combined O1+O2+O3a data, are
\bees
H_0&=&67.3^{+27.1}_{-21.4} 
\,\qquad\qquad (\mbox{\rm cone completeness, homogeneous completion})\, ,\\
H_0&=&67.3^{+28.4}_{-18.5} 
\,\qquad\qquad (\mbox{\rm cone completeness, multiplicative completion})\, ,
\ees
These should  be compared with our baseline result using mask completion and multiplicative completion, given by Fig.~\ref{fig:varyingGamma_H0} and \eq{H0KbandPth07}. We see that the overall shapes of the posteriors in Fig.~\ref{fig:H0_cone} are visually almost identical to those in Fig.~\ref{fig:varyingGamma_H0} and the maximum a posteriori value of $H_0$ is exactly the same, within the digits given. The highest density intervals are also very similar, although slightly broader, compared to that given in \eq{H0KbandPth07}. 
The same happens using our default choice of mask completion as  a measure of completeness, but filling the missing galaxies according to homogeneous completion, that gives again a very similar posterior, and 
\be
H_0= 67.3^{+26.7}_{-22.3} 
\,\qquad\qquad (\mbox{\rm mask completeness, homogeneous completion})\, ,
\ee
This shows that the different prescriptions that we have explored, both for quantifying the semi-local completeness  of the catalog, and for distributing the missing galaxies, give very consistent results, at least for the relatively large completeness threshold $P_{\rm th}=0.7$ that we must anyhow chose, according to the discussion in sect.~\ref{sect:dependencePth}.

\begin{figure}[t]
\centering
\includegraphics[width=0.48\textwidth]{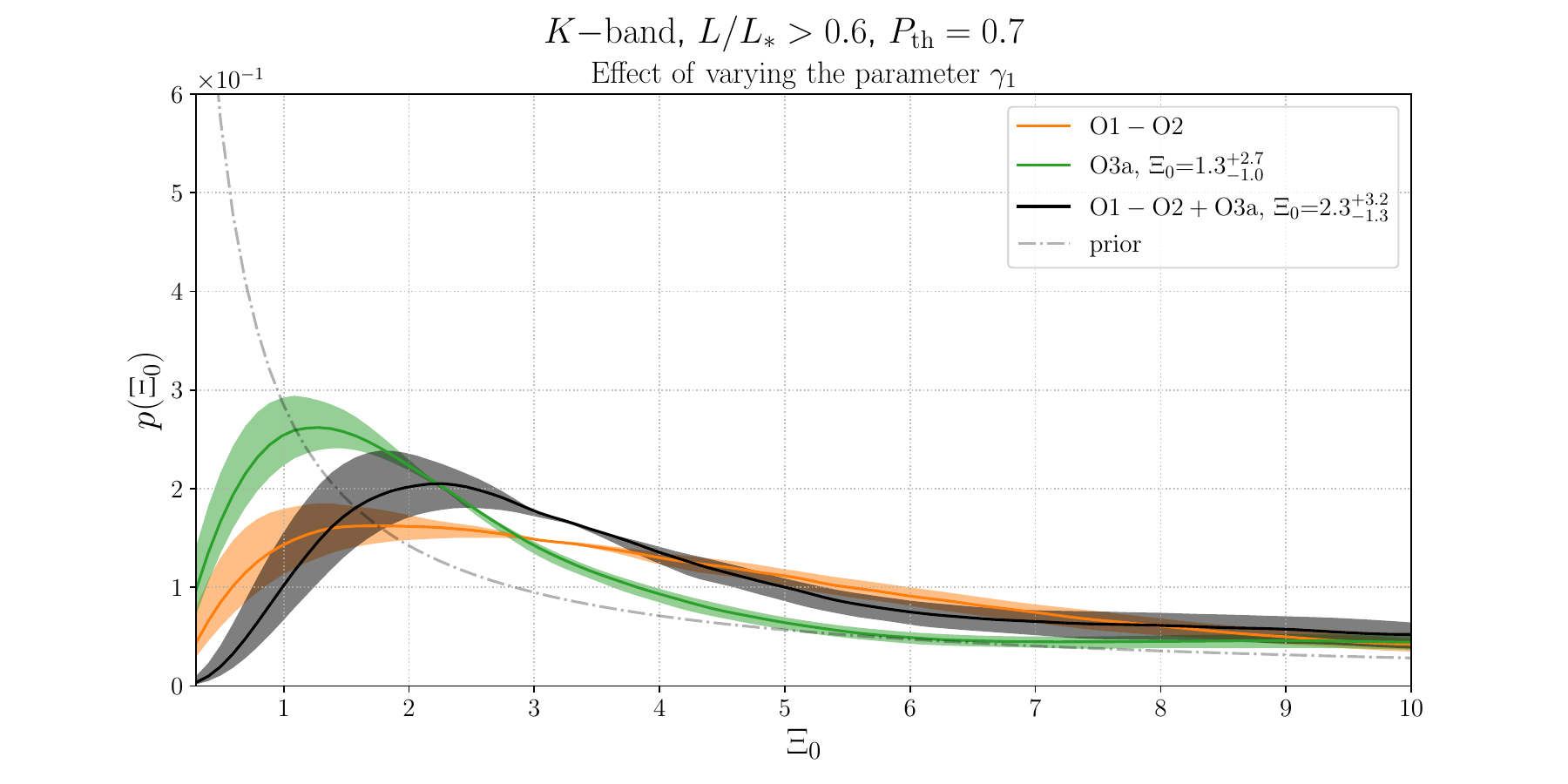}
\includegraphics[width=0.48\textwidth]{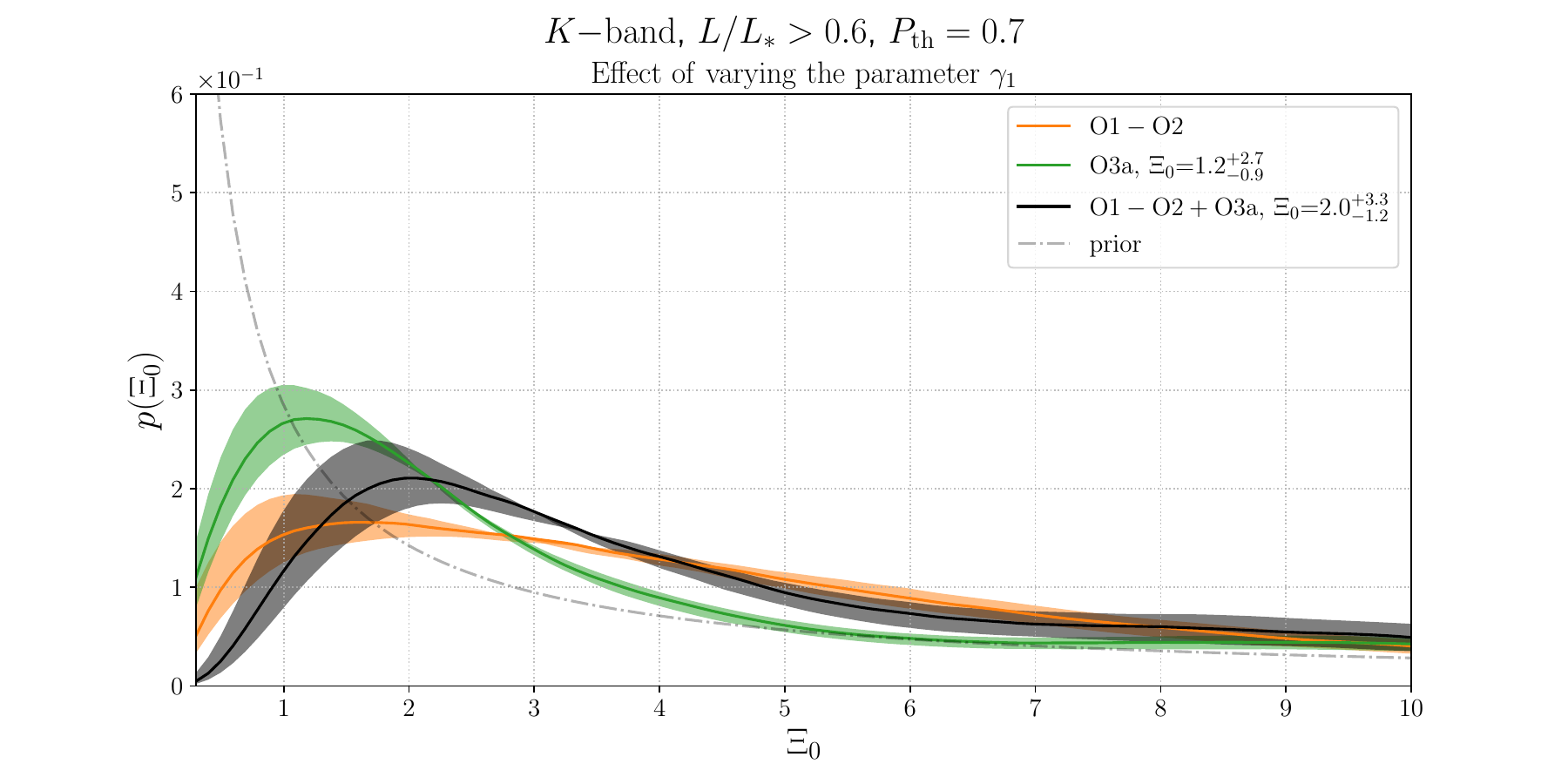}
\caption{The posterior for $\Xi_0$ using cone completeness and homogeneous completion (left) and   cone completeness and multiplicative completion (right),  while still keeping our default choices,  K~band and $P_{\rm th}=0.7$.}
\label{fig:Xi0_cone}
\end{figure}

The same situation takes place for $\Xi_0$.  In Fig.~\ref{fig:Xi0_cone} we show the posteriors for $\Xi_0$ using cone completeness and homogeneous completion (left) and   cone completeness and multiplicative completion (right), which are visually identical to that in Fig.~\ref{fig:varyingGamma_Xi0}. The corresponding values for the maximum a posteriori value and the  highest density interval, using the combined O1+O2+O3a data, are
\bees
\Xi_0&=&2.3^{+3.2}_{-1.3} 
\,\qquad\qquad (\mbox{\rm cone completeness, homogeneous completion})\, ,\\
\Xi_0&=&2.0^{+3.3}_{-1.2} 
\,\qquad\qquad (\mbox{\rm cone completeness, multiplicative completion})\, ,
\ees
again very consistent among them and with the result given in \eq{Xi0limitdarkK02}. Similarly, for
mask completeness and homogeneous completion, we get 
\be
\Xi_0=2.2^{+3.3}_{-1.2} \, .
\ee

\bibliographystyle{utphys}
\bibliography{myrefs}

\providecommand{\href}[2]{#2}\begingroup\raggedright\begin{thebibliography}{100}

\bibitem{Abbott:2016blz}
B.~P. Abbott {\em et~al.}, ``{Observation of Gravitational Waves from a Binary
  Black Hole Merger},''
  \href{http://dx.doi.org/10.1103/PhysRevLett.116.061102}{{\em Phys. Rev.
  Lett.} {\bfseries 116} (2016) 061102},
\href{http://arxiv.org/abs/1602.03837}{{\ttfamily arXiv:1602.03837 [gr-qc]}}.

\bibitem{TheLIGOScientific:2017qsa}
B.~Abbott {\em et~al.}, ``{GW170817: Observation of Gravitational Waves from a
  Binary Neutron Star Inspiral},''
  \href{http://dx.doi.org/10.1103/PhysRevLett.119.161101}{{\em Phys. Rev.
  Lett.} {\bfseries 119} (2017) 161101},
\href{http://arxiv.org/abs/1710.05832}{{\ttfamily arXiv:1710.05832 [gr-qc]}}.

\bibitem{Goldstein:2017mmi}
A.~Goldstein {\em et~al.}, ``{An Ordinary Short Gamma-Ray Burst with
  Extraordinary Implications: Fermi-GBM Detection of GRB 170817A},''
  \href{http://dx.doi.org/10.3847/2041-8213/aa8f41}{{\em Astrophys. J.}
  {\bfseries 848} (2017) L14},
\href{http://arxiv.org/abs/1710.05446}{{\ttfamily arXiv:1710.05446
  [astro-ph.HE]}}.

\bibitem{Savchenko:2017ffs}
V.~Savchenko {\em et~al.}, ``{INTEGRAL Detection of the First Prompt Gamma-Ray
  Signal Coincident with the Gravitational-wave Event GW170817},''
  \href{http://dx.doi.org/10.3847/2041-8213/aa8f94}{{\em Astrophys. J.}
  {\bfseries 848} (2017) L15},
\href{http://arxiv.org/abs/1710.05449}{{\ttfamily arXiv:1710.05449
  [astro-ph.HE]}}.

\bibitem{Monitor:2017mdv}
B.~P. Abbott {\em et~al.}, ``{Gravitational Waves and Gamma-rays from a Binary
  Neutron Star Merger: GW170817 and GRB 170817A},''
  \href{http://dx.doi.org/10.3847/2041-8213/aa920c}{{\em Astrophys. J.}
  {\bfseries 848} (2017) L13},
\href{http://arxiv.org/abs/1710.05834}{{\ttfamily arXiv:1710.05834
  [astro-ph.HE]}}.

\bibitem{Abbott:2020niy}
{\bfseries LIGO Scientific, Virgo} Collaboration, R.~Abbott {\em et~al.},
  ``{GWTC-2: Compact Binary Coalescences Observed by LIGO and Virgo During the
  First Half of the Third Observing Run},''
  \href{http://arxiv.org/abs/2010.14527}{{\ttfamily arXiv:2010.14527 [gr-qc]}}.

\bibitem{Schutz:1986gp}
B.~F. Schutz, ``{Determining the Hubble Constant from Gravitational Wave
  Observations},''
\href{http://dx.doi.org/10.1038/323310a0}{{\em Nature} {\bfseries 323} (1986)
  310--311}.

\bibitem{Holz:2005df}
D.~E. Holz and S.~A. Hughes, ``{Using gravitational-wave standard sirens},''
  \href{http://dx.doi.org/10.1086/431341}{{\em Astrophys. J.} {\bfseries 629}
  (2005) 15--22},
\href{http://arxiv.org/abs/astro-ph/0504616}{{\ttfamily arXiv:astro-ph/0504616
  [astro-ph]}}.

\bibitem{Dalal:2006qt}
N.~Dalal, D.~E. Holz, S.~A. Hughes, and B.~Jain, ``{Short GRB and binary black
  hole standard sirens as a probe of dark energy},''
  \href{http://dx.doi.org/10.1103/PhysRevD.74.063006}{{\em Phys. Rev.}
  {\bfseries D74} (2006) 063006},
\href{http://arxiv.org/abs/astro-ph/0601275}{{\ttfamily arXiv:astro-ph/0601275
  [astro-ph]}}.

\bibitem{MacLeod:2007jd}
C.~L. MacLeod and C.~J. Hogan, ``{Precision of Hubble constant derived using
  black hole binary absolute distances and statistical redshift information},''
  \href{http://dx.doi.org/10.1103/PhysRevD.77.043512}{{\em Phys. Rev.}
  {\bfseries D77} (2008) 043512},
\href{http://arxiv.org/abs/0712.0618}{{\ttfamily arXiv:0712.0618 [astro-ph]}}.

\bibitem{Nissanke:2009kt}
S.~Nissanke, D.~E. Holz, S.~Hughes, N.~Dalal, and J.~L. Sievers, ``{Exploring
  short gamma-ray bursts as gravitational-wave standard sirens},''
  \href{http://dx.doi.org/10.1088/0004-637X/725/1/496}{{\em Astrophys. J.}
  {\bfseries 725} (2010) 496--514},
\href{http://arxiv.org/abs/0904.1017}{{\ttfamily arXiv:0904.1017
  [astro-ph.CO]}}.

\bibitem{Cutler:2009qv}
C.~Cutler and D.~E. Holz, ``{Ultra-high precision cosmology from gravitational
  waves},'' \href{http://dx.doi.org/10.1103/PhysRevD.80.104009}{{\em Phys.
  Rev.} {\bfseries D80} (2009) 104009},
\href{http://arxiv.org/abs/0906.3752}{{\ttfamily arXiv:0906.3752
  [astro-ph.CO]}}.

\bibitem{DelPozzo:2011yh}
W.~Del~Pozzo, ``{Inference of the cosmological parameters from gravitational
  waves: application to second generation interferometers},''
  \href{http://dx.doi.org/10.1103/PhysRevD.86.043011}{{\em Phys. Rev.}
  {\bfseries D86} (2012) 043011},
\href{http://arxiv.org/abs/1108.1317}{{\ttfamily arXiv:1108.1317
  [astro-ph.CO]}}.

\bibitem{Taylor:2012db}
S.~R. Taylor and J.~R. Gair, ``{Cosmology with the lights off: standard sirens
  in the Einstein Telescope era},''
  \href{http://dx.doi.org/10.1103/PhysRevD.86.023502}{{\em Phys. Rev.}
  {\bfseries D86} (2012) 023502},
\href{http://arxiv.org/abs/1204.6739}{{\ttfamily arXiv:1204.6739
  [astro-ph.CO]}}.

\bibitem{Chen:2017rfc}
H.-Y. Chen, M.~Fishbach, and D.~E. Holz, ``{A two per cent Hubble constant
  measurement from standard sirens within five years},''
  \href{http://dx.doi.org/10.1038/s41586-018-0606-0}{{\em Nature} {\bfseries
  562} (2018) 545}, \href{http://arxiv.org/abs/1712.06531}{{\ttfamily
  arXiv:1712.06531 [astro-ph.CO]}}.

\bibitem{Feeney:2018mkj}
S.~M. Feeney, H.~V. Peiris, A.~R. Williamson, S.~M. Nissanke, D.~J. Mortlock,
  J.~Alsing, and D.~Scolnic, ``{Prospects for resolving the Hubble constant
  tension with standard sirens},''
\href{http://arxiv.org/abs/1802.03404}{{\ttfamily arXiv:1802.03404
  [astro-ph.CO]}}.

\bibitem{Gray:2019ksv}
R.~Gray {\em et~al.}, ``{Cosmological Inference using Gravitational Wave
  Standard Sirens: A Mock Data Challenge},''
  \href{http://dx.doi.org/10.1103/PhysRevD.101.122001}{{\em Phys. Rev. D}
  {\bfseries 101} (2020) 122001},
  \href{http://arxiv.org/abs/1908.06050}{{\ttfamily arXiv:1908.06050 [gr-qc]}}.

\bibitem{Abbott:2017xzu}
B.~P. Abbott {\em et~al.}, ``{A gravitational-wave standard siren measurement
  of the Hubble constant},'' \href{http://dx.doi.org/10.1038/nature24471}{{\em
  Nature} {\bfseries 551} no.~7678, (2017) 85--88},
\href{http://arxiv.org/abs/1710.05835}{{\ttfamily arXiv:1710.05835
  [astro-ph.CO]}}.

\bibitem{Riess:2019cxk}
A.~G. Riess, S.~Casertano, W.~Yuan, L.~M. Macri, and D.~Scolnic, ``{Large
  Magellanic Cloud Cepheid Standards Provide a 1\% Foundation for the
  Determination of the Hubble Constant and Stronger Evidence for Physics Beyond
  LambdaCDM},'' \href{http://dx.doi.org/10.3847/1538-4357/ab1422}{{\em
  Astrophys. J.} {\bfseries 876} (2019) 85},
\href{http://arxiv.org/abs/1903.07603}{{\ttfamily arXiv:1903.07603
  [astro-ph.CO]}}.

\bibitem{Wong:2019kwg}
K.~C. Wong {\em et~al.}, ``{H0LiCOW \textendash{} XIII. A 2.4 per cent
  measurement of H0 from lensed quasars: 5.3\ensuremath{\sigma} tension between
  early- and late-Universe probes},''
  \href{http://dx.doi.org/10.1093/mnras/stz3094}{{\em Mon. Not. Roy. Astron.
  Soc.} {\bfseries 498} no.~1, (2020) 1420--1439},
  \href{http://arxiv.org/abs/1907.04869}{{\ttfamily arXiv:1907.04869
  [astro-ph.CO]}}.

\bibitem{Aghanim:2018eyx}
{\bfseries Planck} Collaboration, N.~Aghanim {\em et~al.}, ``{Planck 2018
  results. VI. Cosmological parameters},''
  \href{http://dx.doi.org/10.1051/0004-6361/201833910}{{\em Astron. Astrophys.}
  {\bfseries 641} (2020) A6}, \href{http://arxiv.org/abs/1807.06209}{{\ttfamily
  arXiv:1807.06209 [astro-ph.CO]}}.

\bibitem{Abbott:2018xao}
{\bfseries DES} Collaboration, T.~M.~C. Abbott {\em et~al.}, ``{Dark Energy
  Survey Year 1 Results: Constraints on Extended Cosmological Models from
  Galaxy Clustering and Weak Lensing},''
  \href{http://dx.doi.org/10.1103/PhysRevD.99.123505}{{\em Phys. Rev.}
  {\bfseries D99} (2019) 123505},
\href{http://arxiv.org/abs/1810.02499}{{\ttfamily arXiv:1810.02499
  [astro-ph.CO]}}.

\bibitem{Nair:2018ign}
R.~Nair, S.~Bose, and T.~D. Saini, ``{Measuring the Hubble constant:
  Gravitational wave observations meet galaxy clustering},''
  \href{http://dx.doi.org/10.1103/PhysRevD.98.023502}{{\em Phys. Rev. D}
  {\bfseries 98} (2018) 023502},
  \href{http://arxiv.org/abs/1804.06085}{{\ttfamily arXiv:1804.06085
  [astro-ph.CO]}}.

\bibitem{Mukherjee:2019wcg}
S.~Mukherjee, B.~D. Wandelt, and J.~Silk, ``{Probing the theory of gravity with
  gravitational lensing of gravitational waves and galaxy surveys},''
  \href{http://dx.doi.org/10.1093/mnras/staa827}{{\em Mon. Not. Roy. Astron.
  Soc.} {\bfseries 494} no.~2, (2020) 1956--1970},
  \href{http://arxiv.org/abs/1908.08951}{{\ttfamily arXiv:1908.08951
  [astro-ph.CO]}}.

\bibitem{Yu:2020vyy}
J.~Yu, Y.~Wang, W.~Zhao, and Y.~Lu, ``{Hunting for the host galaxy groups of
  binary black holes and the application in constraining Hubble constant},''
  \href{http://dx.doi.org/10.1093/mnras/staa2465}{{\em Mon. Not. Roy. Astron.
  Soc.} {\bfseries 498} no.~2, (2020) 1786--1800},
  \href{http://arxiv.org/abs/2003.06586}{{\ttfamily arXiv:2003.06586
  [astro-ph.CO]}}.

\bibitem{Vijaykumar:2020pzn}
A.~Vijaykumar, M.~Saketh, S.~Kumar, P.~Ajith, and T.~R. Choudhury, ``{Probing
  the large scale structure using gravitational-wave observations of binary
  black holes},'' \href{http://arxiv.org/abs/2005.01111}{{\ttfamily
  arXiv:2005.01111 [astro-ph.CO]}}.

\bibitem{Mukherjee:2020hyn}
S.~Mukherjee, B.~D. Wandelt, S.~M. Nissanke, and A.~Silvestri, ``{Accurate and
  precision Cosmology with redshift unknown gravitational wave sources},''
  \href{http://arxiv.org/abs/2007.02943}{{\ttfamily arXiv:2007.02943
  [astro-ph.CO]}}.

\bibitem{Bera:2020jhx}
S.~Bera, D.~Rana, S.~More, and S.~Bose, ``{Incompleteness Matters Not:
  Inference of $H_0$ from Binary Black Hole\textendash{}Galaxy
  Cross-correlations},'' \href{http://dx.doi.org/10.3847/1538-4357/abb4e0}{{\em
  Astrophys. J.} {\bfseries 902} (2020) 79},
  \href{http://arxiv.org/abs/2007.04271}{{\ttfamily arXiv:2007.04271
  [astro-ph.CO]}}.

\bibitem{Mukherjee:2020mha}
S.~Mukherjee, B.~D. Wandelt, and J.~Silk, ``{Testing the general theory of
  relativity using gravitational wave propagation from dark standard sirens},''
  \href{http://arxiv.org/abs/2012.15316}{{\ttfamily arXiv:2012.15316
  [astro-ph.CO]}}.

\bibitem{Borhanian:2020vyr}
S.~Borhanian, A.~Dhani, A.~Gupta, K.~Arun, and B.~Sathyaprakash, ``{Dark Sirens
  to Resolve the Hubble-Lema\^itre Tension},''
  \href{http://dx.doi.org/10.3847/2041-8213/abcaf5}{{\em Astrophys. J. Lett.}
  {\bfseries 905} (2020) L28},
  \href{http://arxiv.org/abs/2007.02883}{{\ttfamily arXiv:2007.02883
  [astro-ph.CO]}}.

\bibitem{Fishbach:2018gjp}
M.~Fishbach {\em et~al.}, ``{A Standard Siren Measurement of the Hubble
  Constant from GW170817 without the Electromagnetic Counterpart},''
  \href{http://dx.doi.org/10.3847/2041-8213/aaf96e}{{\em Astrophys. J. Lett.}
  {\bfseries 871} (2019) L13},
  \href{http://arxiv.org/abs/1807.05667}{{\ttfamily arXiv:1807.05667
  [astro-ph.CO]}}.

\bibitem{Soares-Santos:2019irc}
{\bfseries DES, LIGO Scientific, Virgo} Collaboration, M.~Soares-Santos {\em
  et~al.}, ``{First Measurement of the Hubble Constant from a Dark Standard
  Siren using the Dark Energy Survey Galaxies and the LIGO/Virgo
  Binary--Black-hole Merger GW170814},''
  \href{http://dx.doi.org/10.3847/2041-8213/ab14f1}{{\em Astrophys. J. Lett.}
  {\bfseries 876} (2019) L7}, \href{http://arxiv.org/abs/1901.01540}{{\ttfamily
  arXiv:1901.01540 [astro-ph.CO]}}.

\bibitem{Palmese:2020aof}
A.~Palmese {\em et~al.}, ``{A statistical standard siren measurement of the
  Hubble constant from the LIGO/Virgo gravitational wave compact object merger
  GW190814 and Dark Energy Survey galaxies},''
  \href{http://arxiv.org/abs/2006.14961}{{\ttfamily arXiv:2006.14961
  [astro-ph.CO]}}.

\bibitem{Abbott:2019yzh}
{\bfseries LIGO Scientific, Virgo} Collaboration, B.~Abbott {\em et~al.}, ``{A
  gravitational-wave measurement of the Hubble constant following the second
  observing run of Advanced LIGO and Virgo},''
  \href{http://arxiv.org/abs/1908.06060}{{\ttfamily arXiv:1908.06060
  [astro-ph.CO]}}.

\bibitem{Belgacem:2017ihm}
E.~Belgacem, Y.~Dirian, S.~Foffa, and M.~Maggiore, ``{The gravitational-wave
  luminosity distance in modified gravity theories},'' {\em Phys. Rev.}
  {\bfseries D97} (2018) 104066,
\href{http://arxiv.org/abs/1712.08108}{{\ttfamily arXiv:1712.08108
  [astro-ph.CO]}}.

\bibitem{Belgacem:2018lbp}
E.~Belgacem, Y.~Dirian, S.~Foffa, and M.~Maggiore, ``{Modified
  gravitational-wave propagation and standard sirens},''
  \href{http://dx.doi.org/10.1103/PhysRevD.98.023510}{{\em Phys. Rev.}
  {\bfseries D98} (2018) 023510},
\href{http://arxiv.org/abs/1805.08731}{{\ttfamily arXiv:1805.08731 [gr-qc]}}.

\bibitem{deRham:2018red}
C.~de~Rham and S.~Melville, ``{Gravitational Rainbows: LIGO and Dark Energy at
  its Cutoff},'' \href{http://dx.doi.org/10.1103/PhysRevLett.121.221101}{{\em
  Phys. Rev. Lett.} {\bfseries 121} (2018) 221101},
\href{http://arxiv.org/abs/1806.09417}{{\ttfamily arXiv:1806.09417 [hep-th]}}.

\bibitem{Creminelli:2017sry}
P.~Creminelli and F.~Vernizzi, ``{Dark Energy after GW170817 and GRB170817A},''
  \href{http://dx.doi.org/10.1103/PhysRevLett.119.251302}{{\em Phys. Rev.
  Lett.} {\bfseries 119} (2017) 251302},
\href{http://arxiv.org/abs/1710.05877}{{\ttfamily arXiv:1710.05877
  [astro-ph.CO]}}.

\bibitem{Sakstein:2017xjx}
J.~Sakstein and B.~Jain, ``{Implications of the Neutron Star Merger GW170817
  for Cosmological Scalar-Tensor Theories},''
  \href{http://dx.doi.org/10.1103/PhysRevLett.119.251303}{{\em Phys. Rev.
  Lett.} {\bfseries 119} (2017) 251303},
\href{http://arxiv.org/abs/1710.05893}{{\ttfamily arXiv:1710.05893
  [astro-ph.CO]}}.

\bibitem{Ezquiaga:2017ekz}
J.~M. Ezquiaga and M.~Zumalac\'arregui, ``{Dark Energy After GW170817: Dead
  Ends and the Road Ahead},''
  \href{http://dx.doi.org/10.1103/PhysRevLett.119.251304}{{\em Phys. Rev.
  Lett.} {\bfseries 119} (2017) 251304},
\href{http://arxiv.org/abs/1710.05901}{{\ttfamily arXiv:1710.05901
  [astro-ph.CO]}}.

\bibitem{Boran:2017rdn}
S.~Boran, S.~Desai, E.~O. Kahya, and R.~P. Woodard, ``{GW170817 Falsifies Dark
  Matter Emulators},'' \href{http://dx.doi.org/10.1103/PhysRevD.97.041501}{{\em
  Phys. Rev. D} {\bfseries 97} (2018) 041501},
  \href{http://arxiv.org/abs/1710.06168}{{\ttfamily arXiv:1710.06168
  [astro-ph.HE]}}.

\bibitem{Baker:2017hug}
T.~Baker, E.~Bellini, P.~G. Ferreira, M.~Lagos, J.~Noller, and I.~Sawicki,
  ``{Strong constraints on cosmological gravity from GW170817 and GRB
  170817A},'' \href{http://dx.doi.org/10.1103/PhysRevLett.119.251301}{{\em
  Phys. Rev. Lett.} {\bfseries 119} (2017) 251301},
\href{http://arxiv.org/abs/1710.06394}{{\ttfamily arXiv:1710.06394
  [astro-ph.CO]}}.

\bibitem{Saltas:2014dha}
I.~D. Saltas, I.~Sawicki, L.~Amendola, and M.~Kunz, ``{Anisotropic Stress as a
  Signature of Nonstandard Propagation of Gravitational Waves},''
  \href{http://dx.doi.org/10.1103/PhysRevLett.113.191101}{{\em Phys. Rev.
  Lett.} {\bfseries 113} (2014) 191101},
\href{http://arxiv.org/abs/1406.7139}{{\ttfamily arXiv:1406.7139
  [astro-ph.CO]}}.

\bibitem{Lombriser:2015sxa}
L.~Lombriser and A.~Taylor, ``{Breaking a Dark Degeneracy with Gravitational
  Waves},'' \href{http://dx.doi.org/10.1088/1475-7516/2016/03/031}{{\em JCAP}
  {\bfseries 1603} (2016) 031},
\href{http://arxiv.org/abs/1509.08458}{{\ttfamily arXiv:1509.08458
  [astro-ph.CO]}}.

\bibitem{Nishizawa:2017nef}
A.~Nishizawa, ``{Generalized framework for testing gravity with
  gravitational-wave propagation. I. Formulation},''
  \href{http://dx.doi.org/10.1103/PhysRevD.97.104037}{{\em Phys. Rev.}
  {\bfseries D97} (2018) 104037},
\href{http://arxiv.org/abs/1710.04825}{{\ttfamily arXiv:1710.04825 [gr-qc]}}.

\bibitem{Arai:2017hxj}
S.~Arai and A.~Nishizawa, ``{Generalized framework for testing gravity with
  gravitational-wave propagation. II. Constraints on Horndeski theory},''
  \href{http://dx.doi.org/10.1103/PhysRevD.97.104038}{{\em Phys. Rev.}
  {\bfseries D97} (2018) 104038},
\href{http://arxiv.org/abs/1711.03776}{{\ttfamily arXiv:1711.03776 [gr-qc]}}.

\bibitem{Amendola:2017ovw}
L.~Amendola, I.~Sawicki, M.~Kunz, and I.~D. Saltas, ``{Direct detection of
  gravitational waves can measure the time variation of the Planck mass},''
  \href{http://dx.doi.org/10.1088/1475-7516/2018/08/030}{{\em JCAP} {\bfseries
  1808} (2018) 030},
\href{http://arxiv.org/abs/1712.08623}{{\ttfamily arXiv:1712.08623
  [astro-ph.CO]}}.

\bibitem{Belgacem:2019pkk}
{\bfseries LISA Cosmology Working Group} Collaboration, E.~Belgacem {\em
  et~al.}, ``{Testing modified gravity at cosmological distances with LISA
  standard sirens},''
  \href{http://dx.doi.org/10.1088/1475-7516/2019/07/024}{{\em JCAP} {\bfseries
  1907} (2019) 024},
\href{http://arxiv.org/abs/1906.01593}{{\ttfamily arXiv:1906.01593
  [astro-ph.CO]}}.

\bibitem{Maggiore:1900zz}
M.~Maggiore, {\em {Gravitational Waves. Vol. 1: Theory and Experiments}}.
\newblock Oxford Master Series in Physics. Oxford University Press, 2007.
\newblock
\url{http://www.oup.com/uk/catalogue/?ci=9780198570745}.
\newblock

\bibitem{Chevallier:2000qy}
M.~Chevallier and D.~Polarski, ``{Accelerating universes with scaling dark
  matter},'' \href{http://dx.doi.org/10.1142/S0218271801000822}{{\em
  Int.J.Mod.Phys.} {\bfseries D10} (2001) 213--224},
\href{http://arxiv.org/abs/gr-qc/0009008}{{\ttfamily arXiv:gr-qc/0009008
  [gr-qc]}}.

\bibitem{Linder:2002et}
E.~V. Linder, ``{Exploring the expansion history of the universe},''
  \href{http://dx.doi.org/10.1103/PhysRevLett.90.091301}{{\em Phys.Rev.Lett.}
  {\bfseries 90} (2003) 091301},
\href{http://arxiv.org/abs/astro-ph/0208512}{{\ttfamily arXiv:astro-ph/0208512
  [astro-ph]}}.

\bibitem{Belgacem:2019tbw}
E.~Belgacem, Y.~Dirian, S.~Foffa, E.~J. Howell, M.~Maggiore, and T.~Regimbau,
  ``{Cosmology and dark energy from joint gravitational wave-GRB
  observations},'' \href{http://dx.doi.org/10.1088/1475-7516/2019/08/015}{{\em
  JCAP} {\bfseries 1908} (2019) 015},
\href{http://arxiv.org/abs/1907.01487}{{\ttfamily arXiv:1907.01487
  [astro-ph.CO]}}.

\bibitem{Farr:2019twy}
W.~M. Farr, M.~Fishbach, J.~Ye, and D.~Holz, ``{A Future Percent-Level
  Measurement of the Hubble Expansion at Redshift 0.8 With Advanced LIGO},''
  \href{http://dx.doi.org/10.3847/2041-8213/ab4284}{{\em Astrophys. J. Lett.}
  {\bfseries 883} no.~2, (2019) L42},
  \href{http://arxiv.org/abs/1908.09084}{{\ttfamily arXiv:1908.09084
  [astro-ph.CO]}}.

\bibitem{Punturo:2010zz}
M.~Punturo {\em et~al.}, ``{The Einstein Telescope: A third-generation
  gravitational wave observatory},''
\href{http://dx.doi.org/10.1088/0264-9381/27/19/194002}{{\em Class. Quant.
  Grav.} {\bfseries 27} (2010) 194002}.

\bibitem{Maggiore:2019uih}
M.~Maggiore {\em et~al.}, ``{Science Case for the Einstein Telescope},''
  \href{http://dx.doi.org/10.1088/1475-7516/2020/03/050}{{\em JCAP} {\bfseries
  03} (2020) 050}, \href{http://arxiv.org/abs/1912.02622}{{\ttfamily
  arXiv:1912.02622 [astro-ph.CO]}}.

\bibitem{Reitze:2019iox}
D.~Reitze {\em et~al.}, ``{Cosmic Explorer: The U.S. Contribution to
  Gravitational-Wave Astronomy beyond LIGO},'' {\em Bull. Am. Astron. Soc.}
  {\bfseries 51} (2019) 035,
\href{http://arxiv.org/abs/1907.04833}{{\ttfamily arXiv:1907.04833
  [astro-ph.IM]}}.

\bibitem{Sathyaprakash:2009xt}
B.~S. Sathyaprakash, B.~F. Schutz, and C.~Van Den~Broeck, ``{Cosmography with
  the Einstein Telescope},''
  \href{http://dx.doi.org/10.1088/0264-9381/27/21/215006}{{\em Class. Quant.
  Grav.} {\bfseries 27} (2010) 215006},
\href{http://arxiv.org/abs/0906.4151}{{\ttfamily arXiv:0906.4151
  [astro-ph.CO]}}.

\bibitem{Zhao:2010sz}
W.~Zhao, C.~Van Den~Broeck, D.~Baskaran, and T.~G.~F. Li, ``{Determination of
  Dark Energy by the Einstein Telescope: Comparing with CMB, BAO and SNIa
  Observations},'' \href{http://dx.doi.org/10.1103/PhysRevD.83.023005}{{\em
  Phys. Rev.} {\bfseries D83} (2011) 023005},
\href{http://arxiv.org/abs/1009.0206}{{\ttfamily arXiv:1009.0206
  [astro-ph.CO]}}.

\bibitem{Audley:2017drz}
P.~Amaro-Seoane {\em et~al.}, ``{Laser Interferometer Space Antenna},''
\href{http://arxiv.org/abs/1702.00786}{{\ttfamily arXiv:1702.00786
  [astro-ph.IM]}}.

\bibitem{Belgacem:2019lwx}
E.~Belgacem, Y.~Dirian, A.~Finke, S.~Foffa, and M.~Maggiore, ``{Nonlocal
  gravity and gravitational-wave observations},''
  \href{http://dx.doi.org/10.1088/1475-7516/2019/11/022}{{\em JCAP} {\bfseries
  1911} (2019) 022},
\href{http://arxiv.org/abs/1907.02047}{{\ttfamily arXiv:1907.02047
  [astro-ph.CO]}}.

\bibitem{Belgacem:2020pdz}
E.~Belgacem, Y.~Dirian, A.~Finke, S.~Foffa, and M.~Maggiore, ``{Gravity in the
  infrared and effective nonlocal models},''
  \href{http://dx.doi.org/10.1088/1475-7516/2020/04/010}{{\em JCAP} {\bfseries
  04} (2020) 010}, \href{http://arxiv.org/abs/2001.07619}{{\ttfamily
  arXiv:2001.07619 [astro-ph.CO]}}.

\bibitem{Cantiello:2018ffy}
M.~Cantiello {\em et~al.}, ``{A Precise Distance to the Host Galaxy of the
  Binary Neutron Star Merger GW170817 Using Surface Brightness Fluctuations},''
  \href{http://dx.doi.org/10.3847/2041-8213/aaad64}{{\em Astrophys. J. Lett.}
  {\bfseries 854} no.~2, (2018) L31},
  \href{http://arxiv.org/abs/1801.06080}{{\ttfamily arXiv:1801.06080
  [astro-ph.GA]}}.

\bibitem{Lagos:2019kds}
M.~Lagos, M.~Fishbach, P.~Landry, and D.~E. Holz, ``{Standard sirens with a
  running Planck mass},''
  \href{http://dx.doi.org/10.1103/PhysRevD.99.083504}{{\em Phys. Rev.}
  {\bfseries D99} (2019) 083504},
\href{http://arxiv.org/abs/1901.03321}{{\ttfamily arXiv:1901.03321
  [astro-ph.CO]}}.

\bibitem{Bellini:2014fua}
E.~Bellini and I.~Sawicki, ``{Maximal freedom at minimum cost: linear
  large-scale structure in general modifications of gravity},''
  \href{http://dx.doi.org/10.1088/1475-7516/2014/07/050}{{\em JCAP} {\bfseries
  07} (2014) 050}, \href{http://arxiv.org/abs/1404.3713}{{\ttfamily
  arXiv:1404.3713 [astro-ph.CO]}}.

\bibitem{Pardo:2018ipy}
K.~Pardo, M.~Fishbach, D.~E. Holz, and D.~N. Spergel, ``{Limits on the number
  of spacetime dimensions from GW170817},''
  \href{http://dx.doi.org/10.1088/1475-7516/2018/07/048}{{\em JCAP} {\bfseries
  1807} (2018) 048},
\href{http://arxiv.org/abs/1801.08160}{{\ttfamily arXiv:1801.08160 [gr-qc]}}.

\bibitem{Bertacca:2017vod}
D.~Bertacca, A.~Raccanelli, N.~Bartolo, and S.~Matarrese, ``{Cosmological
  perturbation effects on gravitational-wave luminosity distance estimates},''
  \href{http://dx.doi.org/10.1016/j.dark.2018.03.001}{{\em Phys. Dark Univ.}
  {\bfseries 20} (2018) 32--40},
  \href{http://arxiv.org/abs/1702.01750}{{\ttfamily arXiv:1702.01750 [gr-qc]}}.

\bibitem{Kalomenopoulos:2020klp}
M.~Kalomenopoulos, S.~Khochfar, J.~Gair, and S.~Arai, ``{Mapping the
  inhomogeneous Universe with Standard Sirens: Degeneracy between inhomogeneity
  and modified gravity theories},''
  \href{http://arxiv.org/abs/2007.15020}{{\ttfamily arXiv:2007.15020
  [astro-ph.CO]}}.

\bibitem{Abbott:2020khf}
{\bfseries LIGO Scientific, Virgo} Collaboration, R.~Abbott {\em et~al.},
  ``{GW190814: Gravitational Waves from the Coalescence of a 23 Solar Mass
  Black Hole with a 2.6 Solar Mass Compact Object},''
  \href{http://dx.doi.org/10.3847/2041-8213/ab960f}{{\em Astrophys. J.}
  {\bfseries 896} (2020) L44},
  \href{http://arxiv.org/abs/2006.12611}{{\ttfamily arXiv:2006.12611
  [astro-ph.HE]}}.

\bibitem{Morgan:2020wvf}
{\bfseries DES} Collaboration, R.~Morgan {\em et~al.}, ``{Constraints on the
  Physical Properties of GW190814 through Simulations Based on DECam Follow-up
  Observations by the Dark Energy Survey},''
  \href{http://dx.doi.org/10.3847/1538-4357/abafaa}{{\em Astrophys. J.}
  {\bfseries 901} no.~1, (2020) 83},
  \href{http://arxiv.org/abs/2006.07385}{{\ttfamily arXiv:2006.07385
  [astro-ph.HE]}}.

\bibitem{Graham:2020gwr}
M.~Graham {\em et~al.}, ``{Candidate Electromagnetic Counterpart to the Binary
  Black Hole Merger Gravitational Wave Event S190521g},''
  \href{http://dx.doi.org/10.1103/PhysRevLett.124.251102}{{\em Phys. Rev.
  Lett.} {\bfseries 124} no.~25, (2020) 251102},
  \href{http://arxiv.org/abs/2006.14122}{{\ttfamily arXiv:2006.14122
  [astro-ph.HE]}}.

\bibitem{Ashton:2020kyr}
G.~Ashton, K.~Ackley, I.~M. Hernandez, and B.~Piotrzkowski, ``{Current
  observations are insufficient to confidently associate the binary black hole
  merger GW190521 with AGN J124942.3+344929},''
  \href{http://arxiv.org/abs/2009.12346}{{\ttfamily arXiv:2009.12346
  [astro-ph.HE]}}.

\bibitem{Mastrogiovanni:2020mvm}
S.~Mastrogiovanni, L.~Haegel, C.~Karathanasis, I.~Magana-Hernandez, and
  D.~Steer, ``{Gravitational wave friction in light of GW170817 and
  GW190521},'' \href{http://arxiv.org/abs/2010.04047}{{\ttfamily
  arXiv:2010.04047 [gr-qc]}}.

\bibitem{Maggiore:2013mea}
M.~Maggiore, ``{Phantom dark energy from nonlocal infrared modifications of
  general relativity},'' {\em Phys.Rev.} {\bfseries D89} (2014) 043008,
\href{http://arxiv.org/abs/1307.3898}{{\ttfamily arXiv:1307.3898 [hep-th]}}.

\bibitem{Maggiore:2016gpx}
M.~Maggiore, ``{Nonlocal Infrared Modifications of Gravity. A Review},''
  \href{http://dx.doi.org/10.1007/978-3-319-51700-1_16}{{\em Fundam. Theor.
  Phys.} {\bfseries 187} (2017) 221--281},
\href{http://arxiv.org/abs/1606.08784}{{\ttfamily arXiv:1606.08784 [hep-th]}}.

\bibitem{Foffa:2013vma}
S.~Foffa, M.~Maggiore, and E.~Mitsou, ``{Cosmological dynamics and dark energy
  from non-local infrared modifications of gravity},'' {\em Int.J.Mod.Phys.}
  {\bfseries A29} (2014) 1450116,
\href{http://arxiv.org/abs/1311.3435}{{\ttfamily arXiv:1311.3435 [hep-th]}}.

\bibitem{Kehagias:2014sda}
A.~Kehagias and M.~Maggiore, ``{Spherically symmetric static solutions in a
  non-local infrared modification of General Relativity},''
  \href{http://dx.doi.org/10.1007/JHEP08(2014)029}{{\em JHEP} {\bfseries 1408}
  (2014) 029},
\href{http://arxiv.org/abs/1401.8289}{{\ttfamily arXiv:1401.8289 [hep-th]}}.

\bibitem{Maggiore:2014sia}
M.~Maggiore and M.~Mancarella, ``{Non-local gravity and dark energy},''
  \href{http://dx.doi.org/10.1103/PhysRevD.90.023005}{{\em Phys.Rev.}
  {\bfseries D90} (2014) 023005},
\href{http://arxiv.org/abs/1402.0448}{{\ttfamily arXiv:1402.0448 [hep-th]}}.

\bibitem{Nesseris:2014mea}
S.~Nesseris and S.~Tsujikawa, ``{Cosmological perturbations and observational
  constraints on nonlocal massive gravity},''
  \href{http://dx.doi.org/10.1103/PhysRevD.90.024070}{{\em Phys.Rev.}
  {\bfseries D90} (2014) 024070},
\href{http://arxiv.org/abs/1402.4613}{{\ttfamily arXiv:1402.4613
  [astro-ph.CO]}}.

\bibitem{Dirian:2014ara}
Y.~Dirian, S.~Foffa, N.~Khosravi, M.~Kunz, and M.~Maggiore, ``{Cosmological
  perturbations and structure formation in nonlocal infrared modifications of
  general relativity},''
  \href{http://dx.doi.org/10.1088/1475-7516/2014/06/033}{{\em JCAP} {\bfseries
  1406} (2014) 033},
\href{http://arxiv.org/abs/1403.6068}{{\ttfamily arXiv:1403.6068
  [astro-ph.CO]}}.

\bibitem{Barreira:2014kra}
A.~Barreira, B.~Li, W.~A. Hellwing, C.~M. Baugh, and S.~Pascoli, ``{Nonlinear
  structure formation in Nonlocal Gravity},''
  \href{http://dx.doi.org/10.1088/1475-7516/2014/09/031}{{\em JCAP} {\bfseries
  1409} (2014) 031},
\href{http://arxiv.org/abs/1408.1084}{{\ttfamily arXiv:1408.1084
  [astro-ph.CO]}}.

\bibitem{Dirian:2014bma}
Y.~Dirian, S.~Foffa, M.~Kunz, M.~Maggiore, and V.~Pettorino, ``{Non-local
  gravity and comparison with observational datasets},''
  \href{http://dx.doi.org/10.1088/1475-7516/2015/04/044}{{\em JCAP} {\bfseries
  1504} (2015) 044},
\href{http://arxiv.org/abs/1411.7692}{{\ttfamily arXiv:1411.7692
  [astro-ph.CO]}}.

\bibitem{Dirian:2016puz}
Y.~Dirian, S.~Foffa, M.~Kunz, M.~Maggiore, and V.~Pettorino, ``{Non-local
  gravity and comparison with observational datasets. II. Updated results and
  Bayesian model comparison with $\Lambda$CDM},''
  \href{http://dx.doi.org/10.1088/1475-7516/2016/05/068}{{\em JCAP} {\bfseries
  1605} (2016) 068},
\href{http://arxiv.org/abs/1602.03558}{{\ttfamily arXiv:1602.03558
  [astro-ph.CO]}}.

\bibitem{Dirian:2017pwp}
Y.~Dirian, ``{Changing the Bayesian prior: Absolute neutrino mass constraints
  in nonlocal gravity},''
  \href{http://dx.doi.org/10.1103/PhysRevD.96.083513}{{\em Phys. Rev.}
  {\bfseries D96} (2017) 083513},
\href{http://arxiv.org/abs/1704.04075}{{\ttfamily arXiv:1704.04075
  [astro-ph.CO]}}.

\bibitem{Belgacem:2017cqo}
E.~Belgacem, Y.~Dirian, S.~Foffa, and M.~Maggiore, ``{Nonlocal gravity.
  Conceptual aspects and cosmological predictions},''
  \href{http://dx.doi.org/10.1088/1475-7516/2018/03/002}{{\em JCAP} {\bfseries
  1803} (2018) 002},
\href{http://arxiv.org/abs/1712.07066}{{\ttfamily arXiv:1712.07066 [hep-th]}}.

\bibitem{Belgacem:2018wtb}
E.~Belgacem, A.~Finke, A.~Frassino, and M.~Maggiore, ``{Testing nonlocal
  gravity with Lunar Laser Ranging},''
  \href{http://dx.doi.org/10.1088/1475-7516/2019/02/035}{{\em JCAP} {\bfseries
  1902} (2019) 035},
\href{http://arxiv.org/abs/1812.11181}{{\ttfamily arXiv:1812.11181 [gr-qc]}}.

\bibitem{Ishak:2019aay}
M.~Ishak {\em et~al.}, ``{Modified Gravity and Dark Energy models Beyond
  $w(z)$CDM Testable by LSST},''
\href{http://arxiv.org/abs/1905.09687}{{\ttfamily arXiv:1905.09687
  [astro-ph.CO]}}.

\bibitem{Mandel:2018mve}
I.~Mandel, W.~M. Farr, and J.~R. Gair, ``{Extracting distribution parameters
  from multiple uncertain observations with selection biases},''
  \href{http://dx.doi.org/10.1093/mnras/stz896}{{\em Mon. Not. Roy. Astron.
  Soc.} {\bfseries 486} (2019) 1086--1093},
  \href{http://arxiv.org/abs/1809.02063}{{\ttfamily arXiv:1809.02063
  [physics.data-an]}}.

\bibitem{Vasylyev:2020hgb}
S.~Vasylyev and A.~Filippenko, ``{A Measurement of the Hubble Constant using
  Gravitational Waves from the Neutron-Star Black-Hole Candidate GW190814},''
  \href{http://arxiv.org/abs/2007.11148}{{\ttfamily arXiv:2007.11148
  [astro-ph.CO]}}.

\bibitem{Loredo:2004nn}
T.~J. Loredo, ``{Accounting for source uncertainties in analyses of
  astronomical survey data},'' \href{http://dx.doi.org/10.1063/1.1835214}{{\em
  AIP Conf. Proc.} {\bfseries 735} (2004) 195},
  \href{http://arxiv.org/abs/astro-ph/0409387}{{\ttfamily
  arXiv:astro-ph/0409387}}.

\bibitem{Adams:2012qw}
M.~R. Adams, N.~J. Cornish, and T.~B. Littenberg, ``{Astrophysical Model
  Selection in Gravitational Wave Astronomy},''
  \href{http://dx.doi.org/10.1103/PhysRevD.86.124032}{{\em Phys. Rev. D}
  {\bfseries 86} (2012) 124032},
  \href{http://arxiv.org/abs/1209.6286}{{\ttfamily arXiv:1209.6286 [gr-qc]}}.

\bibitem{Thrane:2018qnx}
E.~Thrane and C.~Talbot, ``{An introduction to Bayesian inference in
  gravitational-wave astronomy: parameter estimation, model selection, and
  hierarchical models},'' \href{http://dx.doi.org/10.1017/pasa.2019.2}{{\em
  Publ. Astron. Soc. Austral.} {\bfseries 36} (2019) e010},
  \href{http://arxiv.org/abs/1809.02293}{{\ttfamily arXiv:1809.02293
  [astro-ph.IM]}}.

\bibitem{Vitale:2020aaz}
S.~Vitale, D.~Gerosa, W.~M. Farr, and S.~R. Taylor, ``{Inferring the properties
  of a population of compact binaries in presence of selection effects},''
  \href{http://arxiv.org/abs/2007.05579}{{\ttfamily arXiv:2007.05579
  [astro-ph.IM]}}.

\bibitem{Taylor:2011fs}
S.~R. Taylor, J.~R. Gair, and I.~Mandel, ``{Hubble without the Hubble:
  Cosmology using advanced gravitational-wave detectors alone},''
  \href{http://dx.doi.org/10.1103/PhysRevD.85.023535}{{\em Phys. Rev.}
  {\bfseries D85} (2012) 023535},
\href{http://arxiv.org/abs/1108.5161}{{\ttfamily arXiv:1108.5161 [gr-qc]}}.

\bibitem{Mastrogiovanni:2021wsd}
S.~Mastrogiovanni, K.~Leyde, C.~Karathanasis, E.~Chassande-Mottin, D.~A. Steer,
  J.~Gair, A.~Ghosh, R.~Gray, S.~Mukherjee, and S.~Rinaldi, ``{Cosmology in the
  dark: On the importance of source population models for gravitational-wave
  cosmology},'' \href{http://arxiv.org/abs/2103.14663}{{\ttfamily
  arXiv:2103.14663 [gr-qc]}}.

\bibitem{Singer:2016eax}
L.~P. Singer {\em et~al.}, ``{Going the Distance: Mapping Host Galaxies of LIGO
  and Virgo Sources in Three Dimensions Using Local Cosmography and Targeted
  Follow-up},'' \href{http://dx.doi.org/10.3847/2041-8205/829/1/L15}{{\em
  Astrophys. J. Lett.} {\bfseries 829} (2016) L15},
  \href{http://arxiv.org/abs/1603.07333}{{\ttfamily arXiv:1603.07333
  [astro-ph.HE]}}.

\bibitem{Singer:2016erz}
L.~P. Singer {\em et~al.}, ``{Supplement: Going the Distance: Mapping Host
  Galaxies of LIGO and Virgo Sources in Three Dimensions Using Local
  Cosmography and Targeted Follow-up},''
  \href{http://dx.doi.org/10.3847/0067-0049/226/1/10}{{\em Astrophys. J.
  Suppl.} {\bfseries 226} (2016) 10},
  \href{http://arxiv.org/abs/1605.04242}{{\ttfamily arXiv:1605.04242
  [astro-ph.IM]}}.

\bibitem{Gorski:2004by}
K.~Gorski, E.~Hivon, A.~Banday, B.~Wandelt, F.~Hansen, M.~Reinecke, and
  M.~Bartelman, ``{HEALPix - A Framework for high resolution discretization,
  and fast analysis of data distributed on the sphere},''
  \href{http://dx.doi.org/10.1086/427976}{{\em Astrophys. J.} {\bfseries 622}
  (2005) 759}, \href{http://arxiv.org/abs/astro-ph/0409513}{{\ttfamily
  arXiv:astro-ph/0409513}}.

\bibitem{Fishbach:2021yvy}
M.~Fishbach, Z.~Doctor, T.~Callister, B.~Edelman, J.~Ye, R.~Essick, W.~M. Farr,
  B.~Farr, and D.~E. Holz, ``{When are LIGO/Virgo's Big Black-Hole Mergers?},''
  \href{http://arxiv.org/abs/2101.07699}{{\ttfamily arXiv:2101.07699
  [astro-ph.HE]}}.

\bibitem{Dalya:2018cnd}
G.~D\'alya, G.~Galg\'oczi, L.~Dobos, Z.~Frei, I.~S. Heng, R.~Macas,
  C.~Messenger, P.~Raffai, and R.~S. de~Souza, ``{GLADE: A galaxy catalogue for
  multimessenger searches in the advanced gravitational-wave detector era},''
  \href{http://dx.doi.org/10.1093/mnras/sty1703}{{\em Mon. Not. Roy. Astron.
  Soc.} {\bfseries 479} (2018) 2374--2381},
  \href{http://arxiv.org/abs/1804.05709}{{\ttfamily arXiv:1804.05709
  [astro-ph.HE]}}.

\bibitem{Conselice2016}
C.~J. {Conselice}, A.~{Wilkinson}, K.~{Duncan}, and A.~{Mortlock}, ``{The
  Evolution of Galaxy Number Density at $z < 8$ and its Implications},''
  \href{http://dx.doi.org/10.3847/0004-637X/830/2/83}{{\em \apj} {\bfseries
  830} (2016) 83}, \href{http://arxiv.org/abs/1607.03909}{{\ttfamily
  arXiv:1607.03909 [astro-ph.GA]}}.

\bibitem{Fishbach:2018edt}
M.~Fishbach, D.~E. Holz, and W.~M. Farr, ``{Does the Black Hole Merger Rate
  Evolve with Redshift?},''
  \href{http://dx.doi.org/10.3847/2041-8213/aad800}{{\em Astrophys. J. Lett.}
  {\bfseries 863} (2018) L41},
  \href{http://arxiv.org/abs/1805.10270}{{\ttfamily arXiv:1805.10270
  [astro-ph.HE]}}.

\bibitem{Callister:2020arv}
T.~Callister, M.~Fishbach, D.~Holz, and W.~Farr, ``{Shouts and Murmurs:
  Combining Individual Gravitational-Wave Sources with the Stochastic
  Background to Measure the History of Binary Black Hole Mergers},''
  \href{http://dx.doi.org/10.3847/2041-8213/ab9743}{{\em Astrophys. J. Lett.}
  {\bfseries 896} no.~2, (2020) L32},
  \href{http://arxiv.org/abs/2003.12152}{{\ttfamily arXiv:2003.12152
  [astro-ph.HE]}}.

\bibitem{Madau:2014bja}
P.~Madau and M.~Dickinson, ``{Cosmic Star Formation History},''
  \href{http://dx.doi.org/10.1146/annurev-astro-081811-125615}{{\em Ann. Rev.
  Astron. Astrophys.} {\bfseries 52} (2014) 415--486},
\href{http://arxiv.org/abs/1403.0007}{{\ttfamily arXiv:1403.0007
  [astro-ph.CO]}}.

\bibitem{Madau:2016jbv}
P.~Madau and T.~Fragos, ``{Radiation Backgrounds at Cosmic Dawn: X-Rays from
  Compact Binaries},'' \href{http://dx.doi.org/10.3847/1538-4357/aa6af9}{{\em
  Astrophys. J.} {\bfseries 840} (2017) 39},
  \href{http://arxiv.org/abs/1606.07887}{{\ttfamily arXiv:1606.07887
  [astro-ph.GA]}}.

\bibitem{Abbott:2020gyp}
{\bfseries LIGO Scientific, Virgo} Collaboration, R.~Abbott {\em et~al.},
  ``{Population Properties of Compact Objects from the Second LIGO-Virgo
  Gravitational-Wave Transient Catalog},''
  \href{http://arxiv.org/abs/2010.14533}{{\ttfamily arXiv:2010.14533
  [astro-ph.HE]}}.

\bibitem{LIGOScientific:2018mvr}
{\bfseries LIGO Scientific, Virgo} Collaboration, B.~Abbott {\em et~al.},
  ``{GWTC-1: A Gravitational-Wave Transient Catalog of Compact Binary Mergers
  Observed by LIGO and Virgo during the First and Second Observing Runs},''
  \href{http://dx.doi.org/10.1103/PhysRevX.9.031040}{{\em Phys. Rev. X}
  {\bfseries 9} (2019) 031040},
  \href{http://arxiv.org/abs/1811.12907}{{\ttfamily arXiv:1811.12907
  [astro-ph.HE]}}.

\bibitem{Pratten:2020fqn}
G.~Pratten, S.~Husa, C.~Garcia-Quiros, M.~Colleoni, A.~Ramos-Buades,
  H.~Estelles, and R.~Jaume, ``{Setting the cornerstone for a family of models
  for gravitational waves from compact binaries: The dominant harmonic for
  nonprecessing quasicircular black holes},''
  \href{http://dx.doi.org/10.1103/PhysRevD.102.064001}{{\em Phys. Rev. D}
  {\bfseries 102} (2020) 064001},
  \href{http://arxiv.org/abs/2001.11412}{{\ttfamily arXiv:2001.11412 [gr-qc]}}.

\bibitem{Chen:2014yla}
H.-Y. Chen and D.~E. Holz, ``{The Loudest Gravitational Wave Events},''
  \href{http://arxiv.org/abs/1409.0522}{{\ttfamily arXiv:1409.0522 [gr-qc]}}.

\bibitem{Calabrese:2016bnu}
E.~Calabrese, N.~Battaglia, and D.~N. Spergel, ``{Testing Gravity with
  Gravitational Wave Source Counts},''
  \href{http://dx.doi.org/10.1088/0264-9381/33/16/165004}{{\em Class. Quant.
  Grav.} {\bfseries 33} (2016) 165004},
  \href{http://arxiv.org/abs/1602.03883}{{\ttfamily arXiv:1602.03883 [gr-qc]}}.

\bibitem{Garcia-Bellido:2016zmj}
J.~Garc\'ia-Bellido, S.~Nesseris, and M.~Trashorras, ``{Gravitational wave
  source counts at high redshift and in models with extra dimensions},''
  \href{http://dx.doi.org/10.1088/1475-7516/2016/07/021}{{\em JCAP} {\bfseries
  07} (2016) 021}, \href{http://arxiv.org/abs/1603.05616}{{\ttfamily
  arXiv:1603.05616 [astro-ph.CO]}}.

\bibitem{MariaEzquiaga:2021lli}
J.~Mar\'\i{}a~Ezquiaga, ``{Hearing gravity from the cosmos: GWTC-2 probes
  general relativity at cosmological scales},''
  \href{http://arxiv.org/abs/2104.05139}{{\ttfamily arXiv:2104.05139
  [astro-ph.CO]}}.

\bibitem{doi:10.1287/deca.2016.0338}
T.~W. Keelin, ``The metalog distributions,''
  \href{http://dx.doi.org/10.1287/deca.2016.0338}{{\em Decision Analysis}
  {\bfseries 13} no.~4, (2016) 243--277},
  \href{http://arxiv.org/abs/https://doi.org/10.1287/deca.2016.0338}{{\ttfamily
  https://doi.org/10.1287/deca.2016.0338}}.

\bibitem{Crook:2006sw}
A.~C. Crook, J.~P. Huchra, N.~Martimbeau, K.~L. Masters, T.~Jarrett, and L.~M.
  Macri, ``{Groups of Galaxies in the Two Micron All-Sky Redshift Survey},''
  \href{http://dx.doi.org/10.1086/510201}{{\em Astrophys. J.} {\bfseries 655}
  (2007) 790}, \href{http://arxiv.org/abs/astro-ph/0610732}{{\ttfamily
  arXiv:astro-ph/0610732}}.

\bibitem{Mukherjee:2019qmm}
S.~Mukherjee, G.~Lava~ux, F.~R. Bouchet, J.~Jasche, B.~D. Wandelt, S.~M.
  Nissanke, F.~Leclercq, and K.~Hotokezaka, ``{Velocity correction for Hubble
  constant measurements from standard sirens},''
  \href{http://arxiv.org/abs/1909.08627}{{\ttfamily arXiv:1909.08627
  [astro-ph.CO]}}.

\bibitem{Chen:2020gek}
H.-Y. Chen, C.-J. Haster, S.~Vitale, W.~M. Farr, and M.~Isi, ``{A Standard
  Siren Cosmological Measurement from the Potential GW190521 Electromagnetic
  Counterpart ZTF19abanrhr},''
  \href{http://arxiv.org/abs/2009.14057}{{\ttfamily arXiv:2009.14057
  [astro-ph.CO]}}.

\bibitem{Mooley:2018qfh}
K.~Mooley, A.~Deller, O.~Gottlieb, E.~Nakar, G.~Hallinan, S.~Bourke, D.~Frail,
  A.~Horesh, A.~Corsi, and K.~Hotokezaka, ``{Superluminal motion of a
  relativistic jet in the neutron-star merger GW170817},''
  \href{http://dx.doi.org/10.1038/s41586-018-0486-3}{{\em Nature} {\bfseries
  561} no.~7723, (2018) 355--359},
  \href{http://arxiv.org/abs/1806.09693}{{\ttfamily arXiv:1806.09693
  [astro-ph.HE]}}.

\bibitem{Mastrogiovanni:2020ppa}
S.~Mastrogiovanni, R.~Duque, E.~Chassande-Mottin, F.~Daigne, and
  R.~Mochkovitch, ``{What role will binary neutron star merger afterglows play
  in multimessenger cosmology?},''
  \href{http://arxiv.org/abs/2012.12836}{{\ttfamily arXiv:2012.12836
  [astro-ph.HE]}}.

\bibitem{Khetan:2020hmh}
N.~Khetan {\em et~al.}, ``{A new measurement of the Hubble constant using Type
  Ia supernovae calibrated with surface brightness fluctuations},''
  \href{http://arxiv.org/abs/2008.07754}{{\ttfamily arXiv:2008.07754
  [astro-ph.CO]}}.

\bibitem{Abbott:2020tfl}
{\bfseries LIGO Scientific, Virgo} Collaboration, R.~Abbott {\em et~al.},
  ``{GW190521: A Binary Black Hole Merger with a Total Mass of $150
  M_{\odot}$},'' \href{http://dx.doi.org/10.1103/PhysRevLett.125.101102}{{\em
  Phys. Rev. Lett.} {\bfseries 125} no.~10, (2020) 101102},
  \href{http://arxiv.org/abs/2009.01075}{{\ttfamily arXiv:2009.01075 [gr-qc]}}.

\bibitem{Abbott:2020mjq}
{\bfseries LIGO Scientific, Virgo} Collaboration, R.~Abbott {\em et~al.},
  ``{Properties and Astrophysical Implications of the 150 M$_\odot$ Binary
  Black Hole Merger GW190521},''
  \href{http://dx.doi.org/10.3847/2041-8213/aba493}{{\em Astrophys. J. Lett.}
  {\bfseries 900} no.~1, (2020) L13},
  \href{http://arxiv.org/abs/2009.01190}{{\ttfamily arXiv:2009.01190
  [astro-ph.HE]}}.

\bibitem{McKernan:2019hqs}
B.~McKernan, K.~E.~S. Ford, I.~Bartos, M.~J. Graham, W.~Lyra, S.~Marka,
  Z.~Marka, N.~P. Ross, D.~Stern, and Y.~Yang, ``{Ram-pressure stripping of a
  kicked Hill sphere: Prompt electromagnetic emission from the merger of
  stellar mass black holes in an AGN accretion disk},''
  \href{http://dx.doi.org/10.3847/2041-8213/ab4886}{{\em Astrophys. J. Lett.}
  {\bfseries 884} no.~2, (2019) L50},
  \href{http://arxiv.org/abs/1907.03746}{{\ttfamily arXiv:1907.03746
  [astro-ph.HE]}}.

\bibitem{Fishbach:2020qag}
M.~Fishbach and D.~E. Holz, ``{Minding the gap: GW190521 as a straddling
  binary},'' \href{http://dx.doi.org/10.3847/2041-8213/abc827}{{\em Astrophys.
  J. Lett.} {\bfseries 904} no.~2, (2020) L26},
  \href{http://arxiv.org/abs/2009.05472}{{\ttfamily arXiv:2009.05472
  [astro-ph.HE]}}.

\bibitem{Gayathri:2020coq}
V.~Gayathri, J.~Healy, J.~Lange, B.~O'Brien, M.~Szczepanczyk, I.~Bartos,
  M.~Campanelli, S.~Klimenko, C.~Lousto, and R.~O'Shaughnessy, ``{GW190521 as a
  Highly Eccentric Black Hole Merger},''
  \href{http://arxiv.org/abs/2009.05461}{{\ttfamily arXiv:2009.05461
  [astro-ph.HE]}}.

\bibitem{Hinshaw:2008kr}
{\bfseries WMAP} Collaboration, G.~Hinshaw {\em et~al.}, ``{Five-Year Wilkinson
  Microwave Anisotropy Probe (WMAP) Observations: Data Processing, Sky Maps,
  and Basic Results},''
  \href{http://dx.doi.org/10.1088/0067-0049/180/2/225}{{\em Astrophys. J.
  Suppl.} {\bfseries 180} (2009) 225--245},
  \href{http://arxiv.org/abs/0803.0732}{{\ttfamily arXiv:0803.0732
  [astro-ph]}}.

\bibitem{Pan:2011hx}
D.~C. Pan, M.~S. Vogeley, F.~Hoyle, Y.-Y. Choi, and C.~Park, ``{Cosmic Voids in
  Sloan Digital Sky Survey Data Release 7},''
  \href{http://dx.doi.org/10.1111/j.1365-2966.2011.20197.x}{{\em Mon. Not. Roy.
  Astron. Soc.} {\bfseries 421} (2012) 926--934},
  \href{http://arxiv.org/abs/1103.4156}{{\ttfamily arXiv:1103.4156
  [astro-ph.CO]}}.

\bibitem{Libeskind:2017tun}
N.~I. Libeskind {\em et~al.}, ``{Tracing the cosmic web},''
  \href{http://dx.doi.org/10.1093/mnras/stx1976}{{\em Mon. Not. Roy. Astron.
  Soc.} {\bfseries 473} (2018) 1195--1217},
  \href{http://arxiv.org/abs/1705.03021}{{\ttfamily arXiv:1705.03021
  [astro-ph.CO]}}.

\bibitem{Gehrels:2015uga}
N.~Gehrels, J.~K. Cannizzo, J.~Kanner, M.~M. Kasliwal, S.~Nissanke, and L.~P.
  Singer, ``{Galaxy Strategy for LIGO-Virgo Gravitational Wave Counterpart
  Searches},'' \href{http://dx.doi.org/10.3847/0004-637X/820/2/136}{{\em
  Astrophys. J.} {\bfseries 820} no.~2, (2016) 136},
  \href{http://arxiv.org/abs/1508.03608}{{\ttfamily arXiv:1508.03608
  [astro-ph.HE]}}.

\bibitem{Arcavi:2017vbi}
I.~Arcavi {\em et~al.}, ``{Optical Follow-up of Gravitational-wave Events with
  Las Cumbres Observatory},''
  \href{http://dx.doi.org/10.3847/2041-8213/aa910f}{{\em Astrophys. J. Lett.}
  {\bfseries 848} (2017) L33},
  \href{http://arxiv.org/abs/1710.05842}{{\ttfamily arXiv:1710.05842
  [astro-ph.HE]}}.

\bibitem{Kochanek:2000im}
C.~Kochanek, M.~Pahre, E.~Falco, J.~Huchra, J.~Mader, T.~Jarrett, T.~Chester,
  R.~Cutri, and S.~Schneider, ``{The k-band galaxy luminosity function},''
  \href{http://dx.doi.org/10.1086/322488}{{\em Astrophys. J.} {\bfseries 560}
  (2001) 566--579}, \href{http://arxiv.org/abs/astro-ph/0011456}{{\ttfamily
  arXiv:astro-ph/0011456}}.

\bibitem{2005IAUS..216..170H}
J.~{Huchra}, N.~{Martimbeau}, T.~{Jarrett}, R.~{Cutri}, M.~{Skrutskie},
  S.~{Schneider}, R.~{Steining}, L.~{Macri}, J.~{Mader}, and T.~{George},
  \href{http://dx.doi.org/10.1017/S0074180900196603}{``{2MASS and the Nearby
  Universe},''} in {\em Maps of the Cosmos}, M.~{Colless}, L.~{Staveley-Smith},
  and R.~A. {Stathakis}, eds., vol.~216 of {\em IAU Symposium}, pp.~170--179.
\newblock Jan., 2005.

\bibitem{Lambert:2020dwg}
T.~Lambert, R.~Kraan-Korteweg, T.~Jarrett, and L.~Macri, ``{The 2MASS redshift
  survey galaxy group catalogue derived from a graph-theory based
  friends-of-friends algorithm},''
  \href{http://dx.doi.org/10.1093/mnras/staa1946}{{\em Mon. Not. Roy. Astron.
  Soc.} {\bfseries 497} no.~3, (2020) 2954--2973},
  \href{http://arxiv.org/abs/2007.00581}{{\ttfamily arXiv:2007.00581
  [astro-ph.GA]}}.

\bibitem{Lu:2016vmu}
Y.~Lu, X.~Yang, F.~Shi, H.~Mo, D.~Tweed, H.~Wang, Y.~Zhang, S.~Li, and S.~Lim,
  ``{Galaxy groups in the 2MASS Redshift Survey},''
  \href{http://dx.doi.org/10.3847/0004-637X/832/1/39}{{\em Astrophys. J.}
  {\bfseries 832} (2016) 39}, \href{http://arxiv.org/abs/1607.03982}{{\ttfamily
  arXiv:1607.03982 [astro-ph.GA]}}.

\bibitem{Abadie:2010cf}
{\bfseries LIGO Scientific, VIRGO} Collaboration, J.~Abadie {\em et~al.},
  ``{Predictions for the Rates of Compact Binary Coalescences Observable by
  Ground-based Gravitational-wave Detectors},''
  \href{http://dx.doi.org/10.1088/0264-9381/27/17/173001}{{\em Class. Quant.
  Grav.} {\bfseries 27} (2010) 173001},
  \href{http://arxiv.org/abs/1003.2480}{{\ttfamily arXiv:1003.2480
  [astro-ph.HE]}}.

\bibitem{Wysocki:2018mpo}
D.~Wysocki, J.~Lange, and R.~O'Shaughnessy, ``{Reconstructing phenomenological
  distributions of compact binaries via gravitational wave observations},''
  \href{http://dx.doi.org/10.1103/PhysRevD.100.043012}{{\em Phys. Rev. D}
  {\bfseries 100} (2019) 043012},
  \href{http://arxiv.org/abs/1805.06442}{{\ttfamily arXiv:1805.06442 [gr-qc]}}.

\bibitem{Finn:1992xs}
L.~S. Finn and D.~F. Chernoff, ``{Observing binary inspiral in gravitational
  radiation: One interferometer},''
  \href{http://dx.doi.org/10.1103/PhysRevD.47.2198}{{\em Phys. Rev. D}
  {\bfseries 47} (1993) 2198--2219},
  \href{http://arxiv.org/abs/gr-qc/9301003}{{\ttfamily arXiv:gr-qc/9301003}}.

\end{thebibliography}\endgroup

\end{document}